\documentclass{ws-rv961x669}
\usepackage[square]{ws-rv-van} 
\usepackage{graphicx}
\usepackage{slashed}
\usepackage{xcolor}
\usepackage{verbatim}
\usepackage{upgreek}
\usepackage[utf8]{inputenc}
\usepackage{hyperref}
\usepackage{mathrsfs,mathtools, wasysym,amsmath, amsfonts,amssymb, enumitem, amsbsy, multirow}

\newcommand{\zzz}{z}

\def\cropnote{\relax}

\begin{document}
%\markboth{Authors' Names}{Instructions for typing manuscripts (paper's title)}

%\catchline{}{}{}{}{}
%
%%%%%%%%%%%%%%%%%%%%%%%%%%%%%%%%%%%%%%%%%%%%%%%%%%%%%%%%%%%%%%%%%%%%

\chapter*{Probing Leptogenesis \label{Ch5}}

\author[]{E.~J.~Chun$^*$, G.~Cveti\v{c}$^\dagger$, P.~S.~B. Dev$^\ddagger$, M.~Drewes$^{\S, \parallel}$, C.~S.~Fong$^\P$, B.~Garbrecht$^\parallel$ T.~Hambye$^{\ast\ast}$, J.~Harz$^{\dagger\dagger}$\footnote{Corresponding Author.}, P.~Hern\'{a}ndez$^{\ddagger\ddagger}$, C.~S.~Kim$^{\S\S}$, E.~Molinaro$^{\P\P}$, E.~Nardi$^{\parallel\parallel}$, J.~Racker$^{\ast\ast\ast}$, N.~Rius$^{\dagger\dagger\dagger}$, J.~Zamora-Saa$^{\ddagger\ddagger\ddagger}$}

\address{$^*$Korea Institute for Advanced Study, Seoul 02455, Korea
\\[3pt]
$^\dagger$Department of Physics, Universidad T\'ecnica Federico Santa Mar\'ia, Valpara\'iso, Chile
\\[3pt]
$^\ddagger$Department of Physics and McDonnell Center for the Space Sciences,
Washington University, St. Louis, MO 63130, USA
\\[3pt]
$^{\S}$ Centre for Cosmology, Particle Physics and Phenomenology, Universit\'{e} catholique de Louvain, Louvain-la-Neuve B-1348, Belgium \\[3pt]
$^\P$Instituto de F\'{i}sica, Universidade de S\~{a}o Paulo, C. P. 66.318, 05315-970 S\~{a}o Paulo, Brazil
\\[3pt]
$^\parallel$ Physik Department T70, Technische Universit\"at M\"unchen, James Franck Straße 1, 85748 Garching, Germany
\\[3pt]
$^{\ast\ast}$Service  de  Physique  Th\'{e}orique  -  Universit\'{e}  Libre  de  Bruxelles, Boulevard  du  Triomphe,  CP225,  1050  Brussels,  Belgium
\\[3pt]
$^{\dagger\dagger}$Institut  Lagrange  de  Paris  (ILP), Sorbonne  Universit\`{e}s,  98 bis  Boulevard  Arago,  75014  Paris,  France\\
$^{\dagger\dagger}$LPTHE, UPMC  Univ  Paris  06,  Sorbonne  Universit\`{e}s, UMR  7589,  75005  Paris,  France\\
$^{\dagger\dagger}$LPTHE, CNRS, UMR  7589, 75005  Paris,  France
\\[3pt]
$^{\ddagger\ddagger}$Instituto  de  F\'{i}sica  Corpuscular,  Universidad  de  Valencia  and  CSIC,  Edificio  Institutos  Investigaci\'{o}n, Catedr\'{a}tico Jos\'{e} Beltr\'{a}n 2, 46980,  Spain\\
$^{\ddagger\ddagger}$Theory Division, CERN 1211 Geneve 23, Switzerland
\\[3pt]
$^{\S\S}$Department of Physics and IPAP, Yonsei University, Seoul 120-749, Korea
\\[3pt]
$^{\P\P}$CP$^3$-Origins, University of Southern Denmark, Campusvej 55,\\ DK-5230 Odense M, Denmark
\\[3pt]
$^{\parallel\parallel}$INFN, Laboratori Nazionali di Frascati, C.P. 13, I-00044 Frascati, Italy
\\[3pt]
$^{\ast\ast\ast}$Instituto  de  Astronom\'{i}a  Te\'{o}rica y Experimental (IATE), Universidad Nacional de C\'{o}rdoba (UNC) - Consejo Nacional de Investigaciones Cient\'{i}ficas y T\'{e}cnicas (CONICET), C\'{o}rdoba, Argentina
\\[3pt]
$^{\dagger\dagger\dagger}$Instituto  de  F\'{i}sica  Corpuscular  (IFIC), CSIC-Universitat  de  Val\`{e}ncia, Apartado  de  Correos  22085, 46071  Valencia, Spain
\\[3pt]
$^{\ddagger\ddagger\ddagger}$Dzhelepov Laboratory of Nuclear Problems, Joint Institute for Nuclear Research, \\ Dubna 141980, Russia.%\\author@domain\_name
}

\vspace{5cm}

\begin{abstract}
\textbf{Abstract}: The focus of this chapter lies on the possible experimental tests of leptogenesis scenarios. We consider both leptogenesis generated from oscillations, as well as leptogenesis from out-of-equilibrium decays. As the Akhmedov-Rubakov-Smirnov (ARS) mechanism allows for heavy neutrinos in the GeV range, this opens up a plethora of possible experimental tests, e.g. at neutrino oscillation experiments, neutrinoless double beta decay, and direct searches for neutral heavy leptons at future facilities. In contrast, testing leptogenesis from out-of-equilibrium decays is a quite difficult task. We comment on the necessary conditions for having successful leptogenesis at the TeV-scale. We further discuss possible realizations and their model specific testability in extended seesaw models, models with extended gauge sectors, and supersymmetric leptogenesis. Not being able to test high-scale leptogenesis directly, we present a way to falsify such scenarios by focusing on their washout processes. This is discussed specifically for the left-right symmetric model and the observation of a heavy  $W_R$, as well as model independently when measuring $\Delta L = 2$ washout processes at the LHC or neutrinoless double beta decay.
\end{abstract}

\body

\newpage

\tableofcontents

\newpage
\section{Introduction}	
\label{ch5:sec0:intro}
The baryon asymmetry of our Universe is precisely determined by the baryon-to-photon ratio\cite{Ade:2015xua}
\begin{align}
\label{eq0:etaBobs}
  \eta_B^\text{obs} = \frac{n_B - n_{\bar{B}}}{n_\gamma} = \left(6.09 \pm 0.06\right) \times 10^{-10}.
\end{align}
or in terms of the yield
\begin{align}
\label{eq0:etaBobsY}
Y_{\Delta B}  = \frac{n_B - n_{\bar{B}}}{s} = (8.65\pm 0.09)\times 10^{-11}\,.
\end{align}
As the Standard Model (SM) is not able to explain the asymmetry by itself, this observation is a clear indication for the existence of new physics \cite{Gavela:1993ts,Huet:1994jb,Gavela:1994dt,Gurtler:1997hr,Bochkarev:1987wf,Kajantie:1995kf,Laine:1998jb,Csikor:1998eu,Aoki:1999fi}.  While different ideas exist to create such an asymmetry (e.g. electroweak baryogenesis \cite{
Trodden:1998ym, Morrissey:2012db, Dolgov:1991fr,Turok:1992jp,Cohen:1993nk,Rubakov:1996vz}, Affleck-Dine mechanism \cite{Affleck:1984fy,Affleck:1984xz}, etc.), the most popular scenario is baryogenesis via leptogenesis \cite{Fukugita:1986hr,Davidson:2008bu}, as it allows one to address additionally open questions of neutrino physics like the underlying mechanism of neutrino mass generation and mixing.

While the SM charged fermions can only acquire Dirac masses, as they carry charges of the unbroken electromagnetic/color groups, the SM neutrinos $\nu_{L}$, being electric and color charge neutral, can lead another type of masses,\footnote{Throughout this chapter we use doublets $\ell=(\nu_L, e_L)^T$ with $\nu_L, e_L, N$ denoting four-component Dirac spinors. $\nu_L, e_L$ are assumed to be left-handed, i.e. $P_L e_L=e_L, P_L \nu_L=\nu_L$, while $N$ is right-handed, $P_R N = N$. We define $\phi^c \equiv \epsilon
\phi^*$ with $(\epsilon_{12} = 1)$.}
%%%
\begin{equation}
M_{\nu} \bar \nu_{L} \nu^c_{L}, \label{eq0:Majorana} 
\end{equation} 
%%% 
with the flavor index being suppressed. This was pointed out for the first time by Ettore Majorana. The Majorana mass term can be described by the dimension-5 Weinberg operator\cite{Weinberg:1979sa,Weinberg:1980bf}
%%%
\begin{equation}
% {\cal O}_{\alpha\beta}^{d=5} 
{\cal O}^{d=5} 
= \frac{y}{\Lambda_{\rm seesaw}}\,
(\bar \ell \phi^c)( \ell^c \phi^c)\,, \label{eq0:Weinberg} 
\end{equation} 
%%%
where $y$ is some dimensionless coupling and $\Lambda_{\rm seesaw}$ is the seesaw scale at which new degrees of freedom set in. There are three possible ultraviolet completions of the Weinberg operator at tree-level, known as type I, type II and type III seesaw. The respective new degrees of freedom are singlet fermions, $SU(2)_L$ triplet scalars and $SU(2)_L$ triplet fermions.

The most minimal realization is hereby the type I seesaw of which the Lagrangian reads
\begin{equation}
\mathcal{L} =  \bar{N}_k\,  i \slashed\partial \, N_k -  \left( \frac{1}{2} M_k \, \bar{N}^c_k N_k + \lambda_{\alpha k}\, \bar{\ell}_\alpha \phi^c N_{k} + \mathrm{h.c.}\right) \,,
\label{eq0:lagrangiantypeI_seesaw}
\end{equation}
where at least two heavy singlet fermions $N_k$, also known as right-handed (RH) neutrinos, are introduced. With this simple extension two problems can be solved at the same time: {\it light neutrino masses}
\begin{equation}
M_\nu \approx - \frac{v^2}{2} \lambda M_N^{-1} \lambda^T\,,
\label{eq0:lightneutrinomasses}
\end{equation}
with $v=246~\mathrm{GeV}$, as well as {\it leptogenesis}. The standard approach for leptogenesis is via 
\begin{enumerate}
%%%
\item \emph{leptogenesis from out-of-equilibrium decays.} In order to feature small enough neutrino masses and at the same time a large enough baryon asymmetry, the right handed (RH) neutrino mass is constrained. This was firstly discussed in \cite{Davidson:2002qv} in the context of type I seesaw and is known as \emph{Davidson-Ibarra bound}. For thermal production, the heavy neutrino mass is bounded from below as
\begin{equation}
M_N \gtrsim 10^8 \left( \frac{\eta_B}{5\times10^{-11}} \right) \left(\frac{0.06 \mathrm{eV}}{m_3}\right) \mathrm{GeV}\,,
\label{eq0:davidsonibarra}
\end{equation}
with $m_{3}$ being the heaviest of the SM neutrinos.
While this bound holds for vanilla seesaw models, it does not apply to the inverse seesaw model or radiative neutrino mass generation. However, the tight connection of the CP-violating decays with the washout processes, forces the heavy neutrino mass to similar high values in order to achieve successful baryogenesis. Due to this, the vanilla seesaw scenarios as mechanism behind leptogenesis are difficult to probe.  These problems (and possible solutions) will be discussed in more detail in \sref{ch5:sec1:concepts}. The main features of the thermal leptogenesis scenario including its possible connection to CP violation in the lepton sector are reviewed in the accompanying chapters of this review \cite{leptogenesis:A01,leptogenesis:A04,leptogenesis:A06}. 

\parindent=5mm In the following sections, we will discuss models which lead to successful leptogenesis and can be tested for at different experiments. In \sref{ch5:sec3:extended_seesaw}, seesaw models with additional new matter fields will be addressed. Seesaw models with an extended gauge sector will be considered in \sref{ch5:sec4:extendedgauge}. Testability of leptogenesis in the supersymmetric context will be discussed in \sref{ch5:sec5:soft-leptogenesis}. If the energy scale of the new physics is too high to be tested, possibilities to falsify high-scale models exist. A model independent approach will be outlined in \sref{ch5:sec6:highscale}.

\indent Besides leptogenesis from out-of-equilibrium decays, another approach exists, namely
%%%
\item \emph{leptogenesis from oscillations}. This scenario is also called Akhmedov-Rubakov-Smirnov (ARS) mechanism, being firstly discussed in Ref.~\cite{Akhmedov:1998qx}. In this approach, Yukawa couplings have to be small and feature a certain hierarchy such that at least one of the sterile neutrinos does not reach thermal equilibrium until the critical temperature of sphaleron decoupling ($t_c$), whereas at least another one thermally equilibrates before $t_c$. While the total lepton number is approximately conserved, flavor oscillations create an asymmetry in between the different flavor sectors. The singlets that are in thermal equilibrium are able to transfer their asymmetry to the active neutrino sector before sphaleron decoupling and thus create a baryon asymmetry. The ones having not reached thermal equilibrium before $t_c$ will have, however, no effect on the final baryon asymmetry as they translate their asymmetry only after sphaleron decoupling. For further details on this setup, we refer to \cite{leptogenesis:A02}. With respect to testability, this scenario is highly interesting. As the Yukawa couplings are generally small, heavy neutrinos with much smaller masses, i.e. around the GeV scale, are allowed
\begin{equation}
M_\nu \approx - \frac{v^2}{2} \lambda M_N^{-1} \lambda^T \simeq 0.3 \left(\frac{\mathrm{GeV}}{M_N}\right) \left(\frac{\lambda^2}{10^{-14}}\right) \mathrm{eV}\,.
\label{eq0:lightneutrinomassesARS}
\end{equation}
This opens up a plethora of possible experimental tests, which will be the topic of the following \sref{ch5:sec2:ARS}.
\end{enumerate}
\section{Testability of GeV-scale leptogenesis}
\label{ch5:sec2:ARS}
In chapter \cite{leptogenesis:A02} we have reviewed the Akhmedov-Rubakov-Smirnov mechanism of leptogenesis 
in the context of the type I seesaw model at low scales, i.e. for Majorana neutrino masses
between $0.1-100$ GeV. We have seen that the type I seesaw model in \eref{eq0:lagrangiantypeI_seesaw} can explain neutrino masses provided that a number of Majorana 
singlets $n \geq 2$ is added to the SM.  In spite of being a relatively mild extension of the SM, 
 these models add an enormous complexity to the lepton flavor sector, which increases rapidly with $n$. 
The type I seesaw model with $n=2$ involves four mixing angles, four mass eigenstates and three CP phases. The model with $n=3$ instead involves six angles as well as six masses and six CP phases. However, at present only two mass differences
and three angles have been determined (see \cite{leptogenesis:A06} for a summary of the current status of data on  neutrino masses and lepton mixing). 

The flavor complexity opens up in full blossom when considering observables such as the baryon asymmetry. 
For the $n=2$ model, for example, it has been shown  that this observable is sensitive to  all flavor parameters 
of the lepton sector \cite{Shaposhnikov:2008pf,Canetti:2010aw,Canetti:2012kh,Hernandez:2016kel}.  The question is therefore, whether there is any hope that all these 
flavor parameters can be determined experimentally, at least in principle, and thus the baryon asymmetry predicted. 

A number of experiments will explore the lepton flavor sector with more accuracy in the future. Particularly important
to test the ARS leptogenesis scenario will be the following experiments:

\begin{itemize}

 \item {\it Neutrino oscillation experiments}

These experiments will search for leptonic CP violation in light neutrino mixing.
 The oscillation probabilities are sensitive to the light neutrino mixing angles, mass differences and the
CP phase, $\delta$, of the PMNS matrix, see Eq.~2 in \cite{leptogenesis:A06}. Although existing accelerator neutrino experiments such as
T2K \cite{Abe:2013hdq} and NO$\nu$A \cite{Adamson:2016tbq}, as well as atmospheric neutrino experiments, are sensitive to 
$\delta$,  this sensitivity is rather limited and will be improved significantly in the future with the 
experiments DUNE \cite{Acciarri:2015uup} and HyperKamiokande \cite{Abe:2011ts}. For a maximal CP phase, they could measure the phase $\delta$ with 
$\approx$ 0.17 rad accuracy. \\

 \item {\it Neutrinoless double beta ($0\nu \beta \beta$) decay searches }

This measurement is very sensitive to low-scale seesaw scenarios \cite{Blennow:2010th,Ibarra:2011xn,Mitra:2011qr,LopezPavon:2012zg,Lopez-Pavon:2015cga}, as the heavy neutrinos 
can give a significant contribution to the amplitude  for masses below 1~GeV. Besides, 
the light neutrino contribution to this amplitude is very sensitive to the CP phases of the PMNS matrix, in particular the Majorana phases
that decouple from neutrino oscillations, but  nevertheless are relevant for the baryon asymmetry. For a review of current and future searches of $0\nu \beta \beta$ decay we refer to \cite{leptogenesis:A06}.\\

 \item {\it Future direct searches for neutral heavy leptons}

Experiments such as SHiP \cite{Gorbunov:2007ak,Anelli:2015pba} or high intensity $e^+ e^-$ colliders \cite{Blondel:2014bra,Antusch:2015mia,Antusch:2017pkq} will improve significantly the sensitivity to neutral heavy leptons  in the 
range of masses relevant for ARS leptogenesis. If neutral heavy leptons are discovered and their masses and mixings to the different flavors
measured, these observables provide very useful information on CP phases and other flavor parameters. 
\end{itemize}

In the next section we will show that in the context of the minimal model  with $n=2$, these three inputs might suffice 
to actually predict the baryon asymmetry if the Majorana masses lie in  the GeV range.

Other phenomenological implications of seesaw models include lepton flavor violating processes, such as $\mu \rightarrow e \gamma$, $\mu \rightarrow 3e$ or $\mu - e$ conversion in nuclei \cite{Minkowski:1977sc,Alonso:2012ji,Dinh:2012bp}, lepton  electric dipole moments \cite{Ng:1995cs,Archambault:2004td,Abada:2016awd},  electroweak corrections \cite{Langacker:1988up,Antusch:2006vwa,Akhmedov:2013hec,Basso:2013jka,Fernandez-Martinez:2015hxa,Fernandez-Martinez:2016lgt}, or CP-violating meson decays as will be covered in \sref{ch5:sec2:CPMesonDecays}.  Any such measurement could of course add very valuable information on the flavor parameters, but it is unlikely that the parameter space relevant for 
successful baryogenesis will 
give measurable contributions to these observables in the near future, since typically this would require larger mixings.

\subsection{Testing the minimal seesaw model at the GeV-scale}
\label{ch5:sec2:ARS-seesaw}
In the following, we constrain ourselves to the $n=2$ type I seesaw model, the simplest extension of the SM that can explain neutrino masses. 
A convenient parameterization of the Yukawa matrix in the Lagrangian \eref{eq0:lagrangiantypeI_seesaw} is given by the Casas-Ibarra parameterization (adapted to the case of $n=2$) \cite{Casas:2001sr}:
\begin{eqnarray}\label{CasasIbarra}
\lambda = - i U^*_\nu \sqrt{M^{\mathrm{diag}}_\nu}~ P_{NO} ~R^T(z) \sqrt{M_N}{\frac{\sqrt{2}}{ v}},
\end{eqnarray}
where $U_\nu$ is the PMNS matrix of Eq.~2 in \cite{leptogenesis:A06},  $M^{\mathrm{diag}}_\nu$ is the diagonal matrix of the light neutrino masses  (note that the lightest neutrino is massless because only two Majorana singlets are included), $M_N=M^{\mathrm{diag}}_N={\rm diag}(M_1,M_2)$, where $M_1,M_2$ are the heavy neutrino masses, $P_{\rm NO}$ is a  $3\times 2$ matrix that depends on the neutrino ordering (NH, IH) 
\begin{eqnarray}
 P_{NH} =\left(\begin{array}{cc} 0& 0 \\
 1 & 0 \\
 0 & 1 \end{array}\right),~ P_{IH} =\left(\begin{array}{cc} 1 & 0\\ 0& 1\\ 0 & 0\end{array}\right),
 \end{eqnarray}
and finally $R(z)$ is a generic two dimensional orthogonal complex matrix that depends on one complex angle $z \equiv \theta + i \gamma$. 
The mixings of the heavy states are given by
\begin{eqnarray}
U_{\alpha k} = - i \left(U^*_\nu \sqrt{M^{\mathrm{diag}}_\nu}~ P_{NO} ~R^T(z) {\frac{1}{ \sqrt{M^{\mathrm{diag}}_N}}}\right)_{\alpha k}.
\end{eqnarray} 
This parameterization 
assumes that non-unitarity effects are sufficiently small, which is a good approximation in the mass range relevant for leptogenesis, given the existing constraints reviewed in the following \sref{ch5:sec2:nhlcons}.

It is important to stress that the model with $n=2$ is also a good representation of the model with $n\geq 3$ in which one of the states is very weakly coupled and effectively decouples from the physics relevant in baryogenesis. This happens e.g. in the $\nu$MSM\cite{Asaka:2005pn}, where the almost-decoupled state could play the role of warm dark matter. The realization of leptogenesis in this minimal model, which is, in practice, equivalent to the $\nu$MSM, has been studied by many authors. As the mechanism has been described in detail in \cite{leptogenesis:A02}, we refer to that chapter for a full list of references.

\begin{figure}[h]
 \begin{center}
\includegraphics[scale=0.4]{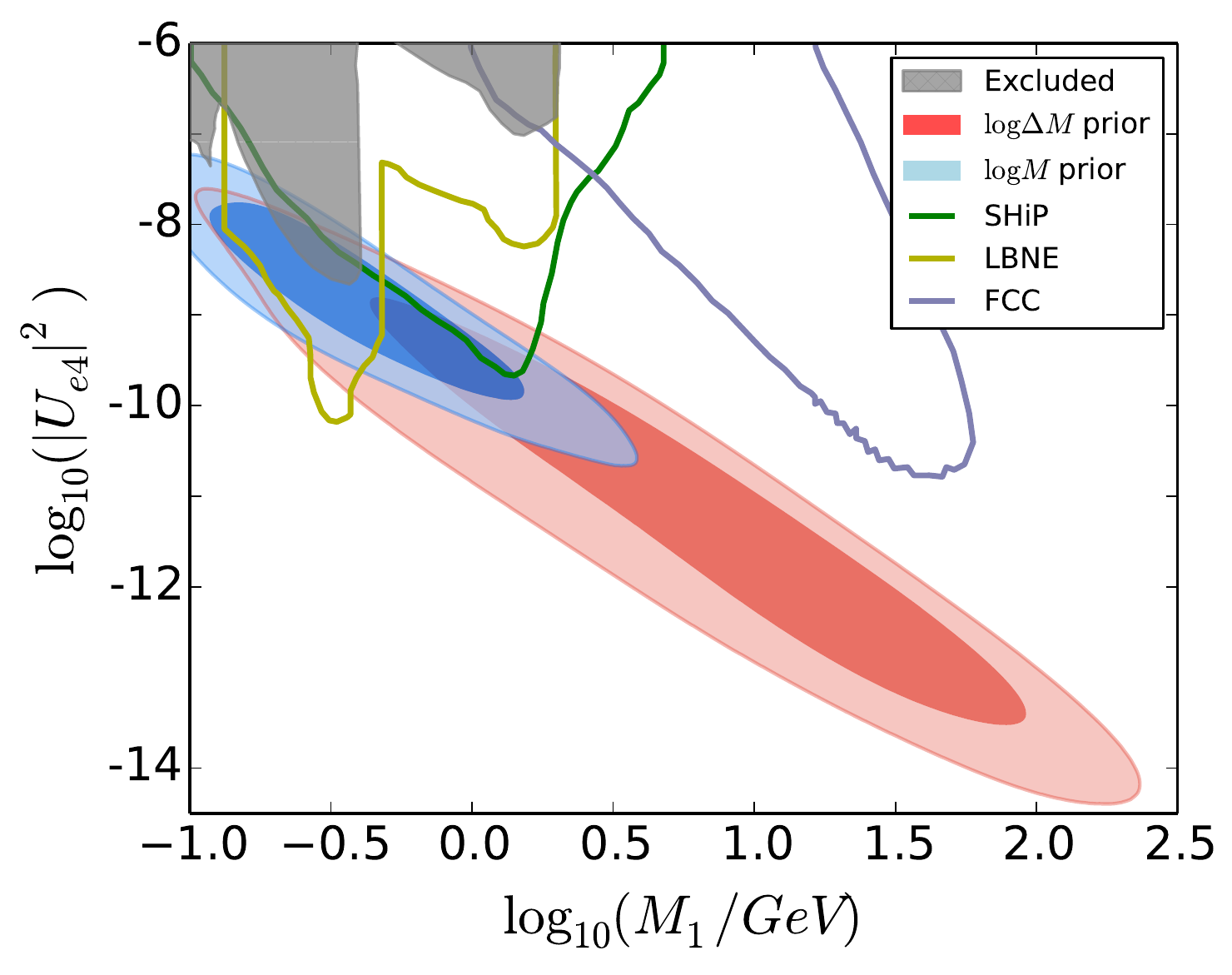} \includegraphics[scale=0.4]{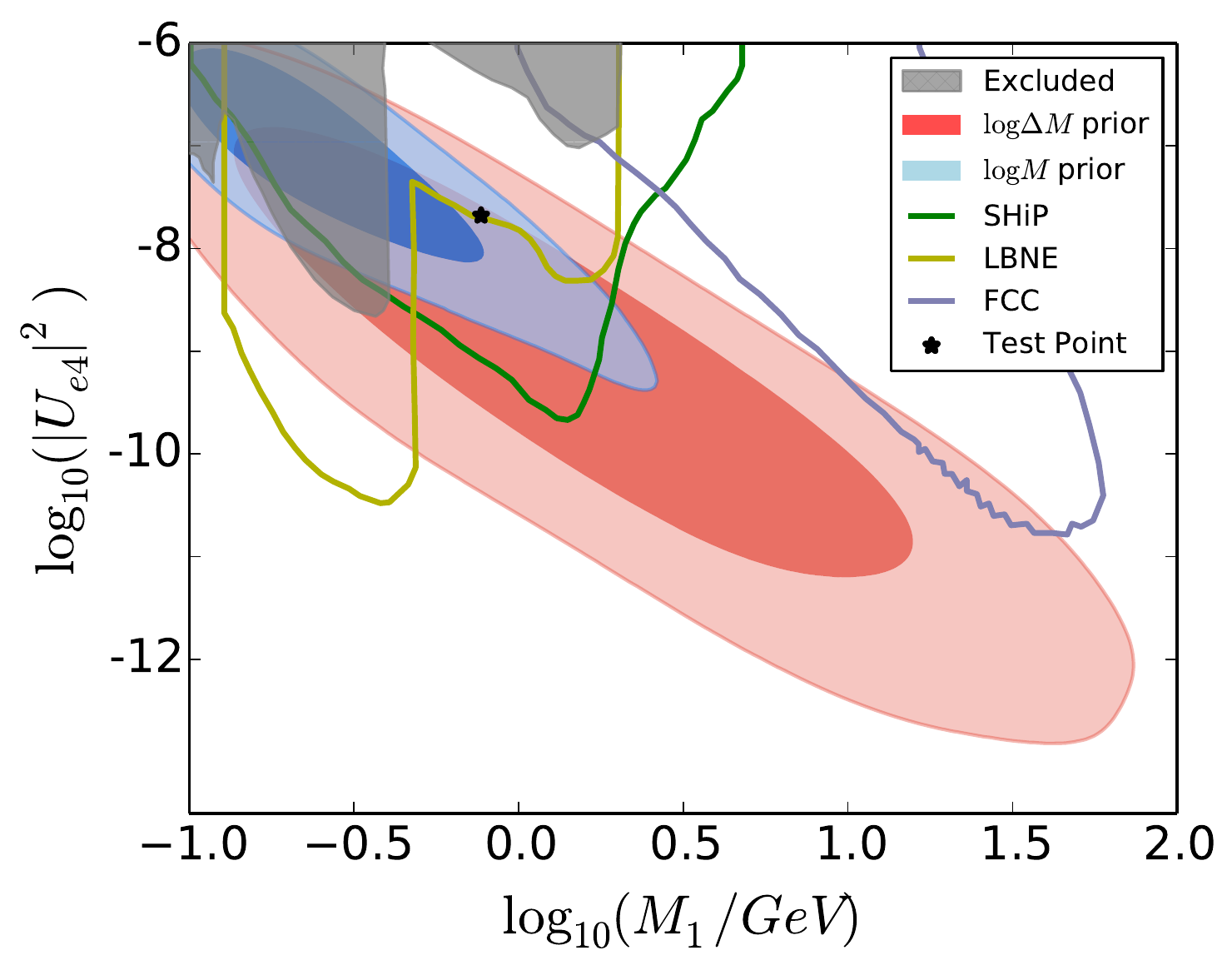} \includegraphics[scale=0.4]{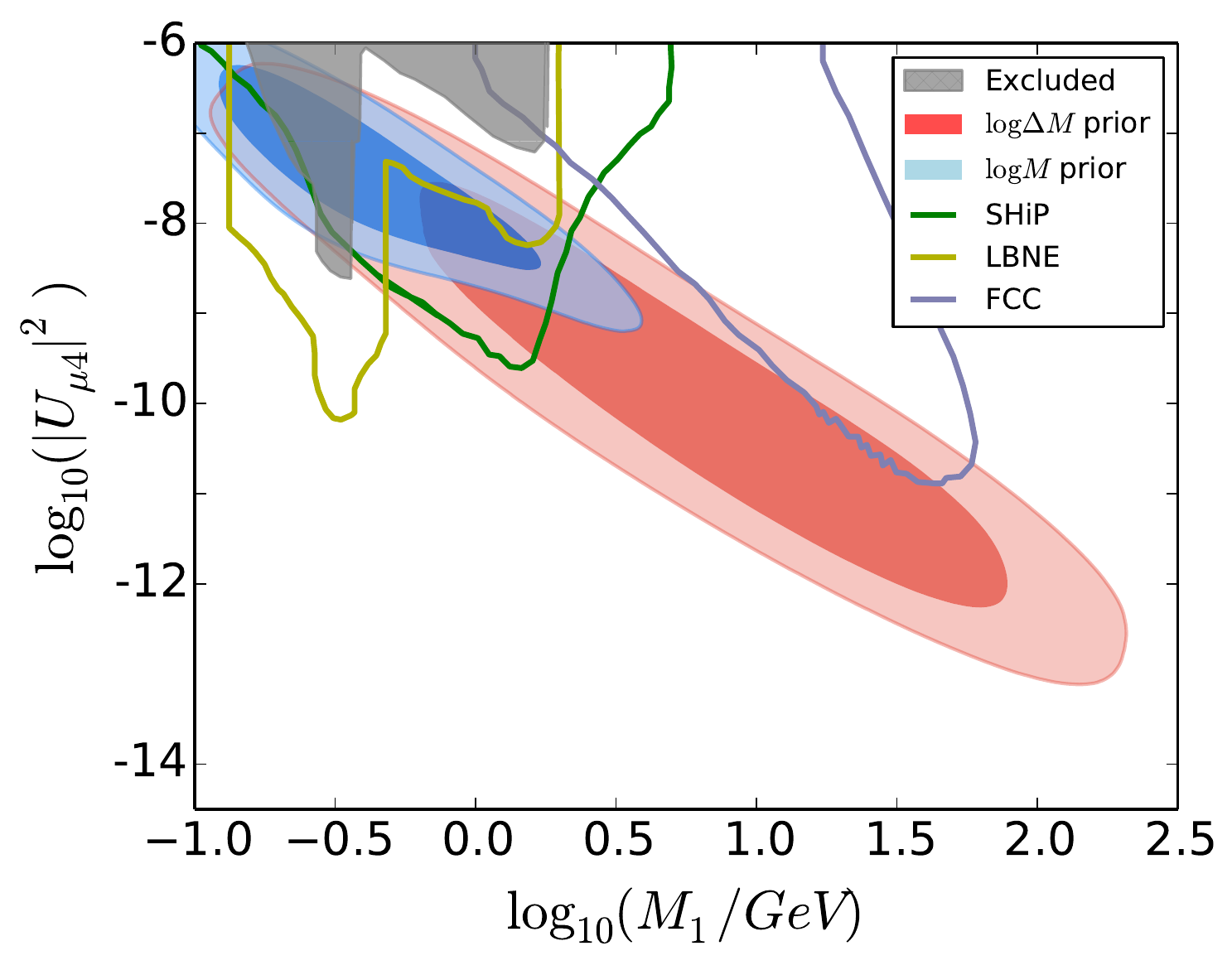} 
\includegraphics[scale=0.4]{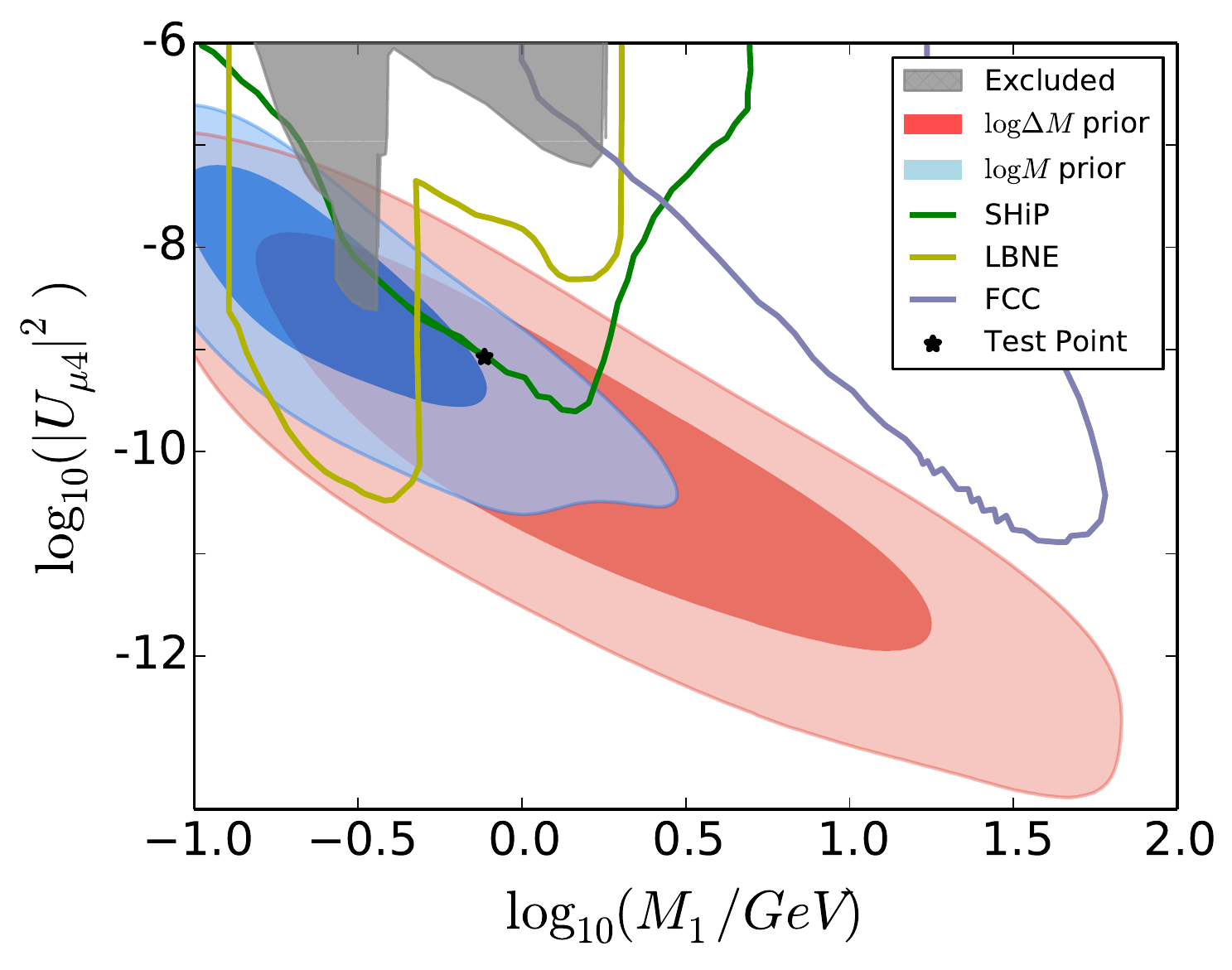} \includegraphics[scale=0.4]{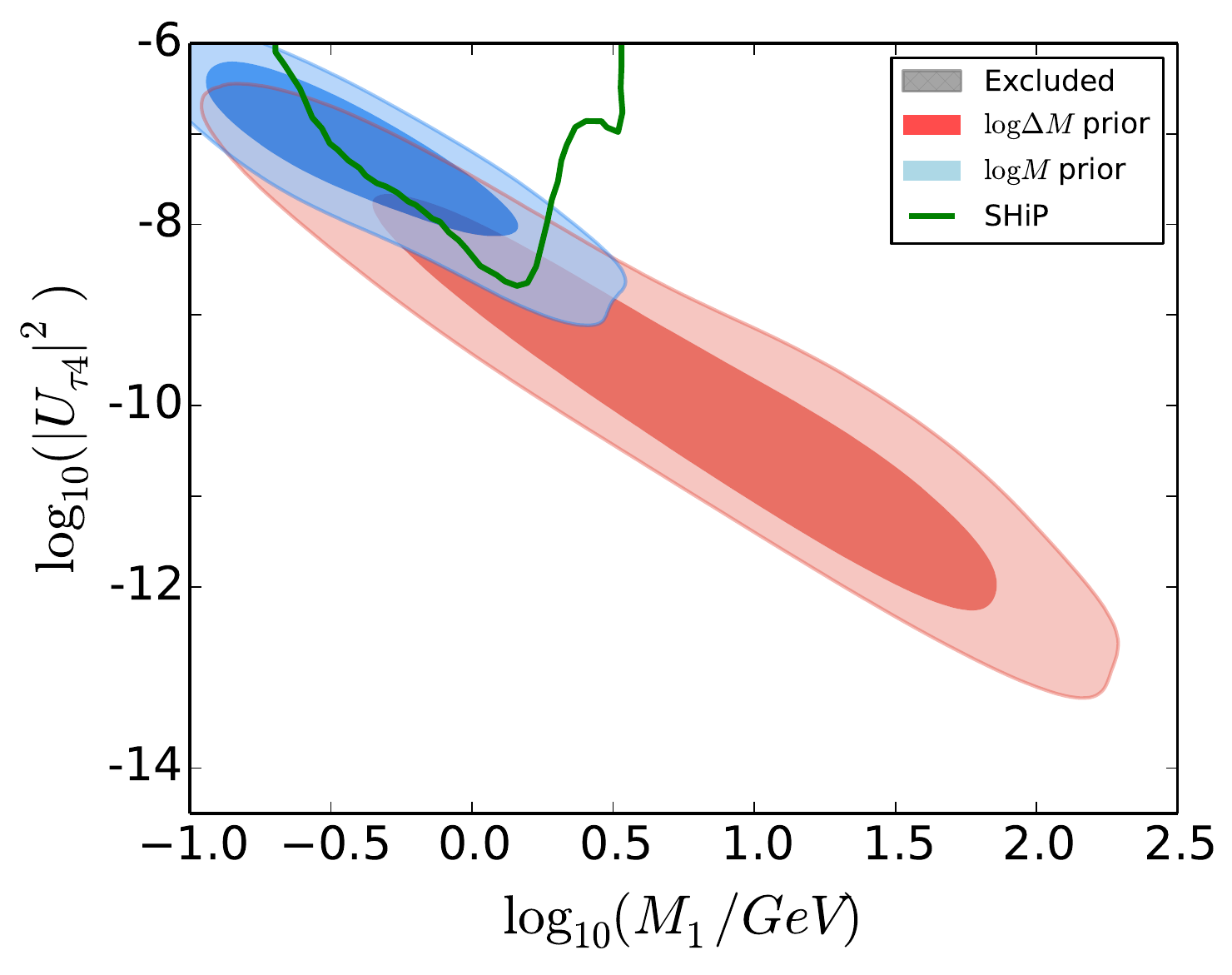} \includegraphics[scale=0.4]{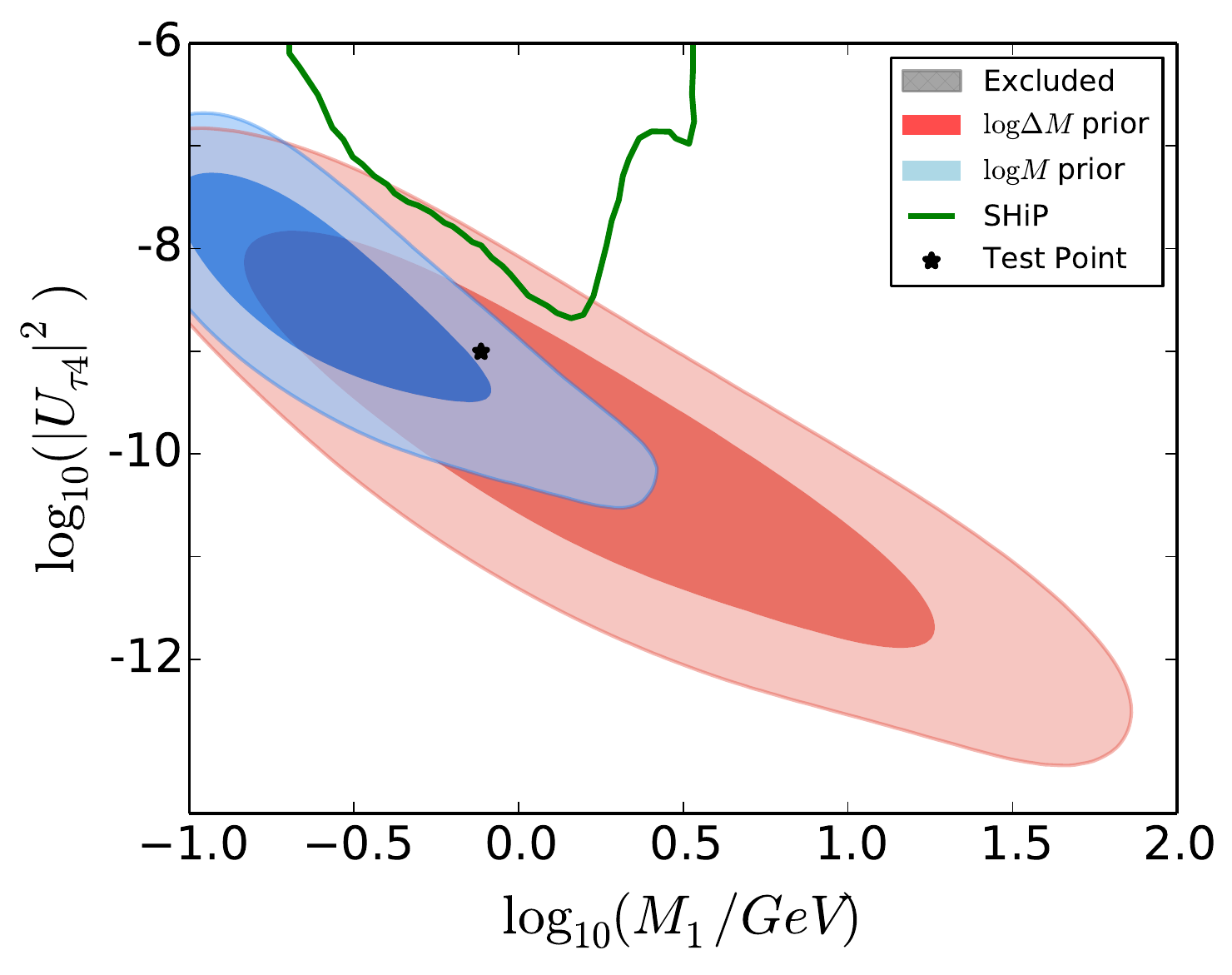} 
\caption{\label{fig2:bayes} Comparison of the posterior probability contours at 68$\%$ and 90$\%$ from successful matter-antimatter asymmetry on the planes mixings with $e,\mu,\tau$ versus masses, with the present (shaded region) and future constraints from DUNE, FCC and SHiP for NH (left) and IH (right). Figure taken from Ref.~\cite{Hernandez:2016kel}.}
\end{center}
\end{figure}

\begin{figure}[h]
 \begin{center}
\includegraphics[scale=0.6]{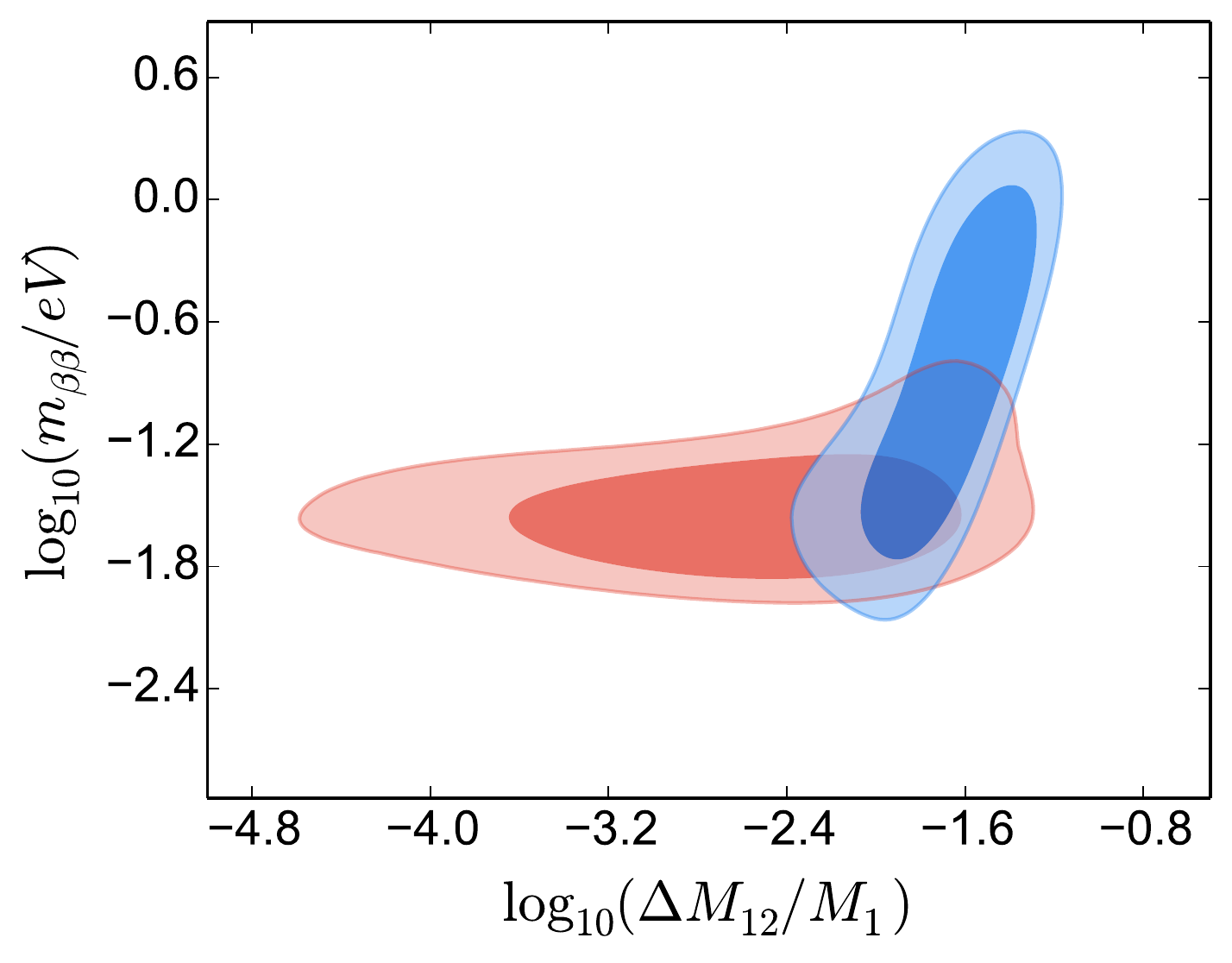} 
\caption{\label{fig2:zoom}  68$\%$ and 90$\%$  posterior probabilities from successful matter-antimatter asymmetry for the amplitude of $0 \nu \beta \beta$ decay versus the mass degeneracy. Figure taken from Ref.~\cite{Hernandez:2016kel}. }
\end{center}
\end{figure}

In a recent work, the Bayesian posterior probabilities from a successful 
prediction of the baryon asymmetry have been evaluated \cite{Hernandez:2016kel}. The results for the heavy state mixing $|U_{\alpha 4}|^2$ with $\alpha =\tau, \mu, e$ versus the mass $M_1$ is reproduced in \fref{fig2:bayes} for normal (left) and inverted (right) hierarchy. The result for the $0 \nu \beta \beta$ decay amplitude $m_{\beta\beta}$ versus the mass degeneracy $\Delta M_{12}/M_1$ is shown in \fref{fig2:zoom}.  
 The blue/red regions correspond to considering flat priors in $\log_{10} M_1$ and $\log_{10} M_2$/$\log_{10} |M_2-M_1|$, respectively. Thus, the red regions allow for larger tuning in the degeneracy of the masses. Interestingly, the less tuned blue regions are within the reach of SHiP, and seem to be correlated with a sizeable non-standard contribution to the amplitude of $0 \nu \beta \beta$ decay, see \fref{fig2:zoom}. This has been observed as well in other studies \cite{Drewes:2016lqo,Asaka:2016zib}.

 On general grounds, we expect that, in the weak washout regime, where a perturbative expansion in Yukawas is a good approximation, the baryon asymmetry is proportional to the most general weak-basis invariant:
\begin{eqnarray}
Y_{\Delta B} \propto \Delta_{CP}= \sum_{\alpha, k} |\lambda_{\alpha k} |^2 \,  \Delta_\alpha,
 \label{eq2:DeltaCP1}
\end{eqnarray}
with
\begin{eqnarray}
\label{eq2:DeltaCP2}
\Delta_\alpha =
\sum_{i,j}  {\rm Im} [ \lambda_{\alpha i}  \lambda^*_{\alpha j} (\lambda^\dagger \lambda)_{ij} ] f(M_i,M_j),
\end{eqnarray}
where $f$ is some arbitrary function.
The ${\mathcal O}(\lambda^6)$ dependence holds only in the weak washout regime, while at equilibration time it reaches ${\mathcal O}(\lambda^4)$, as discussed in \cite{leptogenesis:A02}. 

 The dependence of $Y_{\Delta B}$  on the flavor parameters is well represented by this combination. In terms of the Casas-Ibarra parameters,  the exact formulae are not very illuminating, but simpler expressions can be obtained by expanding in the following small parameters:
 \begin{eqnarray}
\mathcal{O}\left(\epsilon \right): r\equiv\sqrt{\frac{\Delta m^2_{sol}}{\Delta m^2_{atm}}}\sim  \theta_{13}  \sim e^{-{\frac{\gamma}{ 2}}},
 \end{eqnarray}
 where $\gamma$ is assumed to be positive.\footnote{
Note that $\gamma$ can also be negative, but there is an approximate symmetry $\gamma\rightarrow -\gamma$, that would lead to very similar results by expanding in $e^{-{\frac{|\gamma|}{ 2}}}$ in this case. } Note that, although $\gamma$ is unknown, the region of sensitivity of SHiP and high intensity colliders requires that 
$|\gamma| \geq 2$.

At leading order in $\epsilon$ expansion, the CP asymmetry in this regime can be approximated by 
\begin{eqnarray}
\begin{split}
\left|{\Delta_{CP}\over g(M_1,M_2)}\right|_{IH}=&e^{4\gamma}\,\frac{(\Delta m^2_{atm})^{3/2}}{4v^6} M_1M_2(M_1+M_2)\\
&\times \Big[( \sin 2\theta \cos2\theta_{12}-\cos\phi_1 \cos 2\theta\sin2\theta_{12} )\\
 &\times \left(\sin^22\theta_{23}+(4 + \cos 4 \theta_{23})\sin\phi_1\sin2\theta_{12}\right)+ {\mathcal O}(\epsilon)\Big],\\
\left|\Delta_{CP}\over g(M_1,M_2)\right|_{NH} =&e^{4\gamma}\,\frac{\, (\Delta m^2_{atm})^{3/2}}{4v^6} \, M_1 M_2(M_1+M_2)\Big[{\sqrt{r}\over 2} \sin4\theta_{23} c_{12}\cos(\phi_1-2\theta)
\\
+&
 \,r \Big(\sin^22\theta_{23}\left[c_{12}^2\sin2(\phi_1-\theta)+(2+\cos2\theta_{12})\sin2\theta\right]-2 \Big)
\\
+&
\sqrt{r}\,\theta_{13}\,s_{12}(1 + \cos^22\theta_{23})\cos(\delta +\phi_1- 2 \theta)
 +{\mathcal O}(\epsilon^2)\Big],
\\
\end{split}
\end{eqnarray}
where $g$ is some function of the masses that can be found in \cite{Hernandez:2016kel}. In the case of NH we have included NLO corrections, because the LO contributions cancel for maximal atmospheric mixing. 
We see that for both neutrino hierarchies, the baryon asymmetry depends on all flavor parameters of the model: the light neutrino masses and mixings, the heavy neutrino masses,  as well as the real and imaginary part of the Casas-Ibarra complex angle $\theta$, $\gamma$. A very important question is how many of these parameters could be measured in the future experiments listed above, at least in principle. 

Future searches for heavy neutral leptons in the GeV range are obviously most promising. Ideally, such experiments could discover  the two heavy states of the minimal model, measure their splitting, and their mixings to individual SM flavors. Defining 
\begin{eqnarray}
A\equiv \frac{e^{2\gamma}\sqrt{\Delta m^2_{atm}}}{4},
\end{eqnarray}
the perturbative expansion in $\epsilon$ gives for their mixing the following approximate expressions in IH:
\begin{eqnarray}
|U_{e4}|^2 M_1 \simeq |U_{e5}|^2 M_2 &\simeq& A \Big[ (1 + \sin\phi_1\sin2\theta_{12})(1-\theta_{13}^2)  \nonumber\\
&&+ {1\over 2} r^2 s_{12} (c_{12} \sin\phi_1+s_{12})+{\mathcal O}(\epsilon^3)\Big] ,\nonumber\\
|U_{\mu4}|^2 M_1 \simeq |U_{\mu5}|^2 M_2 &\simeq&A \Big[ \left
(1 - \sin\phi_1\sin2\theta_{12} \left(1+ {1\over 4}r^2 \right)+ {1\over 2} r^2 c^2_{12} \right) c_{23}^2 \nonumber\\
&&+ \theta_{13} (\cos \phi_1 \sin \delta - \sin\phi_1 \cos 2 \theta_{12} \cos\delta) \sin 2 \theta_{23}\nonumber\\
&&+ \theta_{13}^2 (1+\sin\phi_1 \sin 2 \theta_{12}) s_{23}^2  +{\mathcal O}(\epsilon^3)\Big],\nonumber\\
|U_{\tau4}|^2 M_1 \simeq |U_{\tau5}|^2 M_2 &\simeq&A \Big[ \left
(1 - \sin\phi_1\sin2\theta_{12} \left(1+ {1\over 4} r^2\right)+ {1\over 2} r^2 c^2_{12} \right) s_{23}^2  \nonumber\\
&&- \theta_{13} (\cos \phi_1 \sin \delta - \sin\phi_1 \cos 2 \theta_{12} \cos\delta) \sin 2 \theta_{23}\nonumber\\
&&+ \theta_{13}^2 (1+\sin\phi_1 \sin 2 \theta_{12}) c_{23}^2  +{\mathcal O}(\epsilon^3)\Big],\nonumber\\
\label{eq2:uihanal}
\end{eqnarray}
while the result for NH is as follows:
\begin{eqnarray}
|U_{e4}|^2 M_1 \simeq |U_{e5}|^2 M_2 &\simeq& A\Big[  r s_{12}^2  -2  \sqrt{r} \theta_{13} \sin(\delta+\phi_1) s
_{12}+ \theta_{13}^2 +{\mathcal O}(\epsilon^{5/2})\Big] ,\nonumber\\
|U_{\mu4}|^2 M_1 \simeq |U_{\mu5}|^2 M_2 &\simeq&A\Big[ s_{23}^2  - \sqrt{r}~  c_{12} \sin\phi_1 \sin 2\theta_{23}  + r c_{12}^2 c_{23}^2 \nonumber\\
&&+ 2  \sqrt{r} ~\theta_{13} \sin(\phi_1+\delta) s_{12} s_{23}^2 - \theta_{13}^2 s_{23}^2 +{\mathcal O}(\epsilon^{5/2})\Big],\nonumber\\
|U_{\tau4}|^2 M_1 \simeq |U_{\tau5}|^2 M_2 &\simeq& A \Big[ c_{23}^2  + \sqrt{r}  c_{12} \sin\phi_1 \sin 2\theta_{23}  + r c_{12}^2 s_{23}^2\nonumber\\
&&+ 2  \sqrt{r} ~\theta_{13}\sin(\phi_1+\delta) s_{12} c_{23}^2 -\theta_{13}^2 c_{23}^2 +{\mathcal O}(\epsilon^{5/2})\Big].\nonumber\\
\label{eq2:unhanal}
\end{eqnarray}
We see that all the mixings are inversely proportional to the mass and exponentially dependent on the parameter $\gamma$. However the 
ratio of the mixings to the electron and muon flavor depends only on the light neutrino masses and the PMNS mixing angles and phases.  
In particular it depends on the two unknown CP phases: $\delta$ which can be determined in future
neutrino oscillation experiments and the Majorana phase $\phi_1$, that modifies the amplitude for $0\nu \beta \beta$ decay.
 A recent study \cite{Caputo:2016ojx} has evaluated the sensitivity of this observable to leptonic CP violation. 

Note that there is no dependence at this order on
 the angle $\theta$ and therefore  the measurement of the mixings and masses of the heavy states does not provide sufficient information to predict the baryon asymmetry. There is 
always a choice of $\theta$ for which the asymmetry vanishes in this approximation. 

There is, however, what could be called a GeV-miracle: the heavy states in this mass range can give non-standard contributions to $0\nu \beta \beta$ decay, as we have seen in the posterior probabilities in \fref{fig2:zoom}.  Interestingly, the angle $\theta$ controls the interference between the light and heavy contributions to this observable. In the $\epsilon$ expansion, the amplitude for $0\nu \beta \beta$ decay is given by (see \cite{leptogenesis:A06} for the definition of $|m_{\beta\beta}|$)
\begin{eqnarray}
{|m_{\beta\beta}|_{IH}\over \sqrt{\Delta m^2_{atm}}} &\simeq&  \left|c_{13}^2\left(c_{12}^2 + e^{2i \phi_1 }s_{12}^2 \left(1+ {r^2\over 2}\right)
\right)\right.\nonumber
\\
&-& \left.\,e^{2i\theta }e^{2\gamma}(c_{12}-i e^{i\phi_1}s_{12})^2 (1-2 e^{i \delta} s_{23} \theta_{13})h(M_1,M_2)
\right|,\\
{|m_{\beta\beta}|_{NH}\over \sqrt{\Delta m^2_{atm}}}&\simeq& \left|  e^{2 i \phi_1}
c_{13}^2 s_{12}^2 r +e^{-2 i \delta}
s_{13}^2 \right. \nonumber
\\
&-&\left.\,e^{2i \theta}e^{2\gamma}s_{12}\,(r s_{12} e^{2 i \phi_1}-2 i \sqrt{r} \theta_{13}  e^{-i \delta}) 
h(M_1,M_2)\right|,
\end{eqnarray}
where
\begin{eqnarray}
h(M_1, M_2)\equiv f(A) \frac{\left(0.9\,\text{GeV}\right)^2}{ 4 M_1^2}
\left(1-\left(\frac{M_1}{M_1+\Delta M_{12}}\right)^2\right).
\end{eqnarray}
The two lines in each amplitude correspond respectively to the light and heavy contributions and $f(A)$ depends on the nucleus under consideration. For $^{48}$Ca, 
$^{76}$Ge, $^{82}$Se, $^{130}$Te and $^{136}$Xe, $f(A)\approx$ 0.035, 0.028, 0.028, 0.033 and 0.032, 
respectively \cite{Blennow:2010th,Ibarra:2010xw}. Since $f(A)$ is very small, we have neglected ${\mathcal O}(\epsilon^2)$ effects in the heavy contribution. 

\begin{figure}[t]
 \begin{center}
 \includegraphics[scale=0.6]{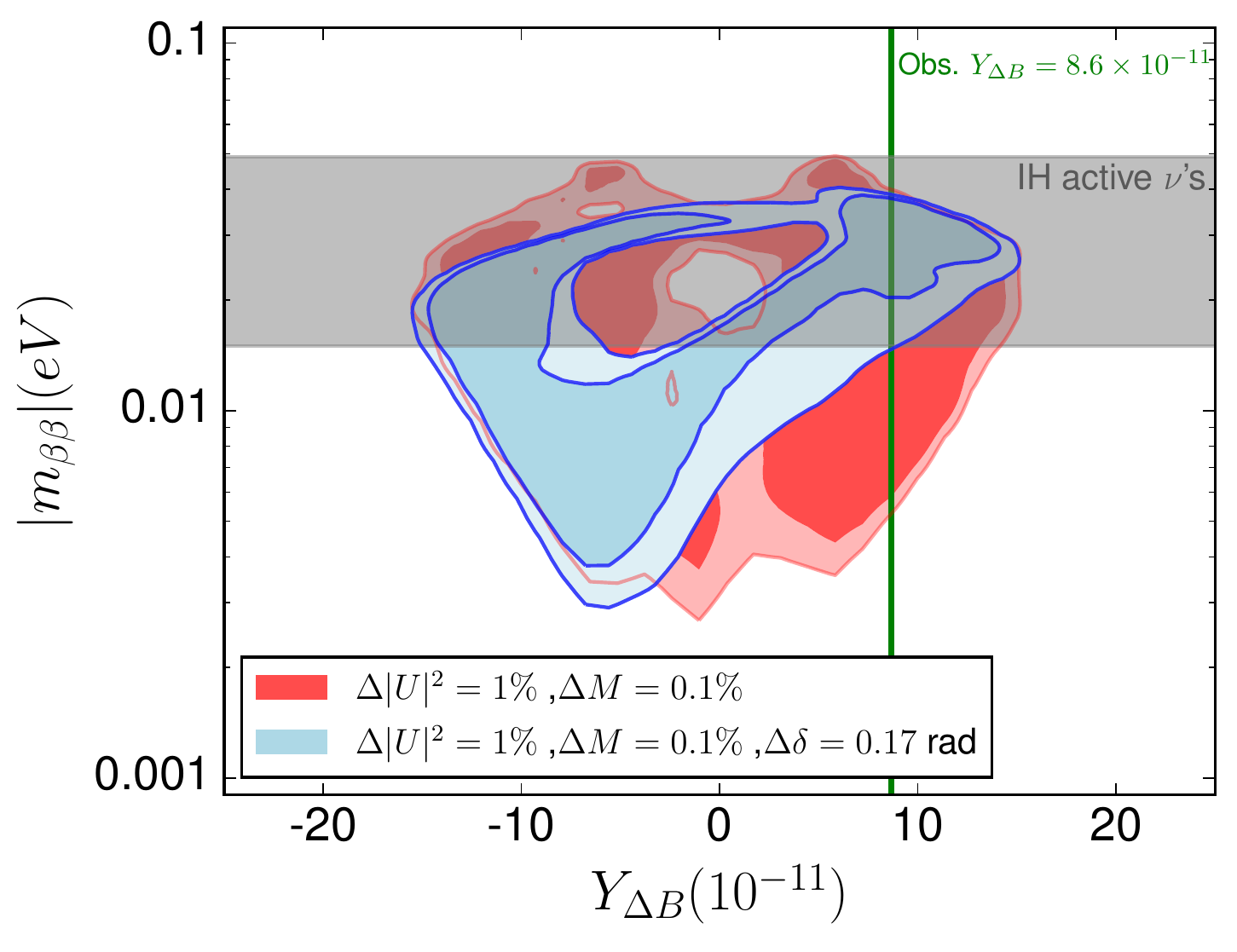} 
\caption{\label{fig2:YBvsmbb} Posterior probabilities in the $|m_{\beta\beta}|$ vs $Y_{\Delta B}$ plane from a putative measurement at SHiP, assuming
$0.1\%, 1\%$ uncertainties in the masses and mixings (red) or the latter with an additional measurement of $\delta$ with a 0.17 rad uncertainty  in DUNE and HyperK (blue). 
The  grey band indicates the standard $3\nu$ expectation. Figure taken from Ref.~\cite{Hernandez:2016kel}.}
\end{center}
\end{figure}

As a proof of principle, we have considered the model indicated by the star in \fref{fig2:bayes},  corresponding to an inverted neutrino hierarchy. Assuming a putative measurement of the two heavy states at SHiP, with $M_1, M_2$ determined with a $0.1\%$ accuracy, and their mixings to electrons and muons with a $1\%$ accuracy, the Bayesian posterior probabilities in the plane 
$(Y_{\Delta B}, |m_{\beta\beta}|)$ are shown in \fref{fig2:YBvsmbb}. The red contour corresponds to a measurement of SHiP only, while the blue includes
the assumption that the phase $\delta$ is determined in future neutrino oscillation experiments with a 0.17 rad accuracy \cite{Acciarri:2015uup}. The latter measurement is very important,
because it resolves the $(\delta, \phi_1)$ correlation in the mixings. The shape of the blue contour can be understood as resulting from this correlation \cite{Hernandez:2016kel}. Depending on the measurement of $0\nu \beta \beta$ decay, the baryon asymmetry might be predicted up to a sign.  

This exercise clearly demonstrates the synergy of the different experimental tests of this minimal model: long-baseline neutrino oscillation experiments, $0\nu \beta \beta$ decay and direct searches of neutral heavy leptons, are optimal for probing the GeV range, where a prediction of the matter-antimatter asymmetry in the Universe might be achievable. 

\subsection{Testing the symmetry protected scenario}\label{SymmetryProtectedScenario}
In much of the parameter space discussed in the previous sections the $|U_{\alpha i}|^2$ are too small to give sizeable branching ratios in existing experiments. 
This is mostly owed to the relation Eq.~(\ref{eq0:lightneutrinomasses}) and the fact that the light neutrino masses are very small.
If we neglect the matrix structure of Eq.~(\ref{eq0:lightneutrinomasses}), i.e., pretend to live in a world with only one generation of SM particles and one flavour of heavy neutrinos, then the magnitude of the Majorana mass $M_N$ uniquely fixes the size of the Yukawa coupling as 
$\lambda^2=2 M_N  m_\nu/v^2$. When the matrices $M_N$ and $\lambda$ have no special structure, then one would expect a similar scaling for their eigenvalues. In this case we can estimate the magnitude of the active-sterile mixing 
by the ``naive seesaw scaling"
\begin{equation}\label{NaiveSeesaw}
U^2 \sim \sqrt{\Delta m_{\rm atm}^2 + m_{\rm lightest}^2}/M_N,
\end{equation}
where $\Delta m_{\rm atm}$ and  $m_{\rm lightest}$ are the atmospheric mass splitting and the mass of the lightest neutrino, respectively. This corresponds to simply scaling the familiar relation between light and heavy neutrino masses in high-scale seesaw models down to a lower scale.
The smallness of the light neutrino masses in this scenario does not come from the familiar seesaw suppression $v/M_N\ll1$,\footnote{One may consider it questionable that the name ``seesaw'' is appropriate for scenarios with $M_N/v\gtrsim 1$. However, it is common to refer to the Lagrangian (\ref{eq0:lagrangiantypeI_seesaw}) with the parameter choice $M_i<$ TeV as \emph{low-scale seesaw}.} 
but is due to very small Yukawa couplings $\lambda^2 \sim 2  m_\nu M_N /v^2 \sim 10^{-8} M_N/v$.

However, many models that incorporate a low-scale seesaw exhibit additional symmetries, e.g. an approximate $B-L$ symmetry that makes the limit $M_N\rightarrow 0$ technically natural. 
The probably most prominent class of  models that realize this symmetry are the \emph{inverse seesaw models} \cite{Mohapatra:1986bd,Mohapatra:1986aw,GonzalezGarcia:1988rw}, and the \emph{symmetry protected scenario} that we will consider here and is also often called the \emph{inverse seesaw limit} of the standard type I seesaw Lagrangian.
In the Casas-Ibarra parameterization with $n=2$, an  approximate $B-L$ symmetry emerges in the limit where the parameters 
\begin{eqnarray}\label{B-L_parameters}
\upepsilon \equiv  e^{-2 \gamma } \ &{\rm and}& \ \upmu=\frac{M_2-M_1}{M_2+M_1} 
\end{eqnarray}
are small. 
Here $M_1$ and $M_2$ are the heavy neutrino masses, and $\gamma$ is the imaginary part of the complex angle $z=\theta + i \gamma$ in Eq.~(\ref{CasasIbarra}).
It is therefore instructive to use the variables
\begin{equation}\label{MassSplitting}
\bar{M}=\frac{M_2+M_1}{2} \ {\rm and} \ \Delta M = \frac{M_2-M_1}{2} = \upmu\bar{M}.
\end{equation}
$\upepsilon$ basically controls the overall interaction strength of the heavy neutrinos,
%$U^2\equiv \sum_{\alpha,i}|U_{\alpha i}^2$
\begin{eqnarray}\label{U2Definition}
U^2 \equiv \sum_{\alpha,i} |U_{\alpha i}|^2 = \upmu\frac{\bar{M}}{M_1 M_2}(m_i-m_j)\cos(2\theta) + \frac{\bar{M}}{M_1 M_2}(m_i-m_j)\frac{1}{2}(\upepsilon + \frac{1}{\upepsilon}),
\end{eqnarray}
while $\upmu$ determines the splitting between their masses. Here $m_i$ and $m_j$ are the two non-zero light neutrino masses, and we assumed $n=2$.
In this scenario, Yukawa couplings $|\lambda_{\alpha i}|$ that are larger than the electron Yukawa coupling can be made consistent with small neutrino masses $m_i$ for $M_N$ below the TeV scale. The smallness of the light neutrino $m_i$ is primarily a result of the smallness of $\upepsilon$ and $\upmu$.
Specific models that predict $\upmu, \upepsilon \ll 1$ typically establish relations between these parameters, i.e., specify a trajectory in the $\upepsilon$-$\upmu$ plane along which the limit $\upmu, \upepsilon \rightarrow 0$ should be taken. While there is nothing that prevents us from setting $\upmu=0$, $\upepsilon$ must remain finite in order to ensure that the Yukawa couplings $\lambda$ remain finite. This is nothing but a manifestation of the fact that the $B-L$ symmetry cannot be exact within the seesaw model with $M_i\gg m_i$ unless the light neutrinos were massless ($m_i=0$).
In the present discussion we prefer to remain agnostic with respect to the model building and treat $\upepsilon$ and $\upmu$ as independent parameters.
In the following we mostly focus on the minimal seesaw model with $n=2$ for illustrative purposes, and because this is by far the most studied scenario.

\subsubsection{Leptogenesis with large mixing angles} 
\begin{figure}
	\centering
	\includegraphics[width=0.44\textwidth]{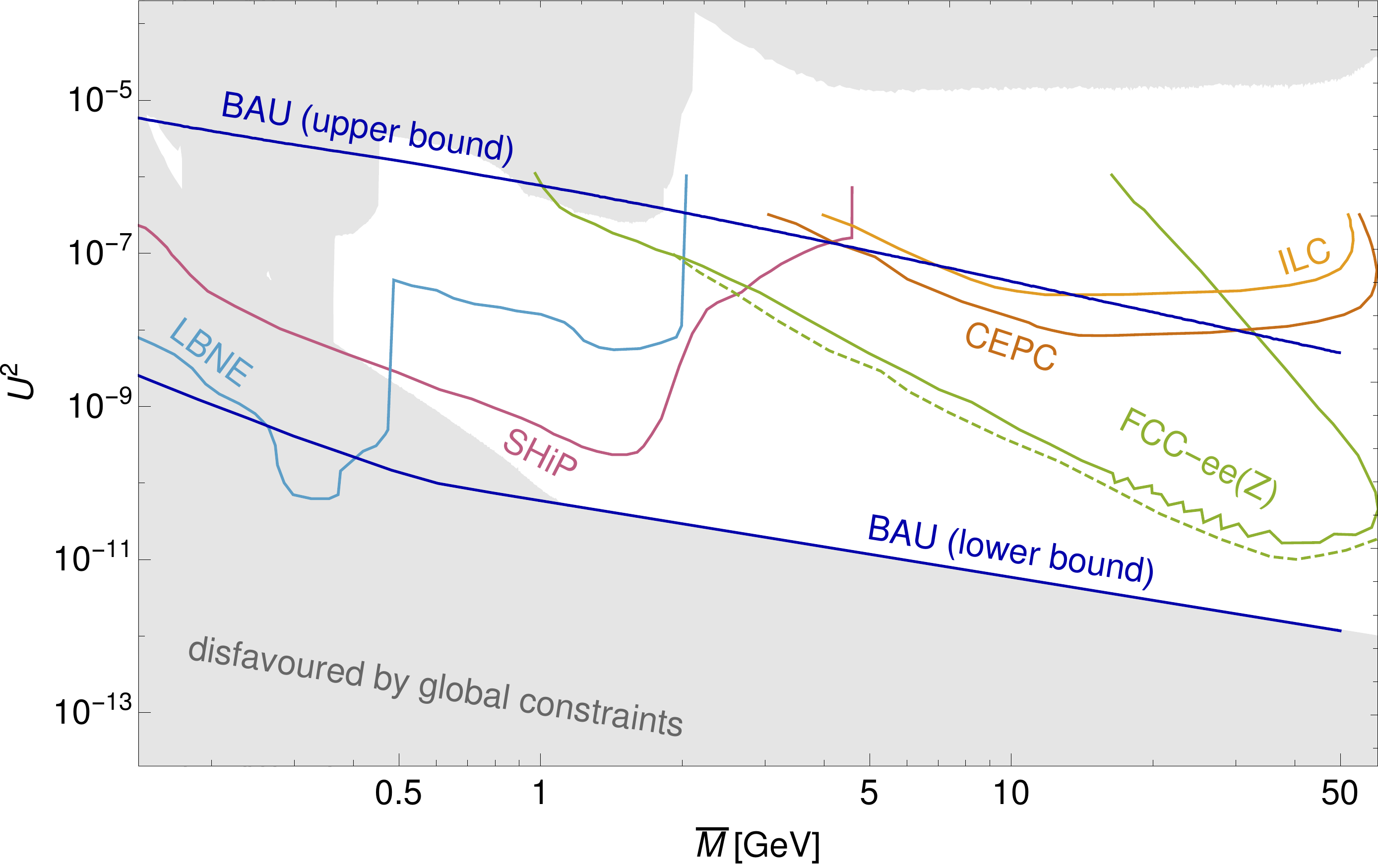}
	\includegraphics[width=0.44\textwidth]{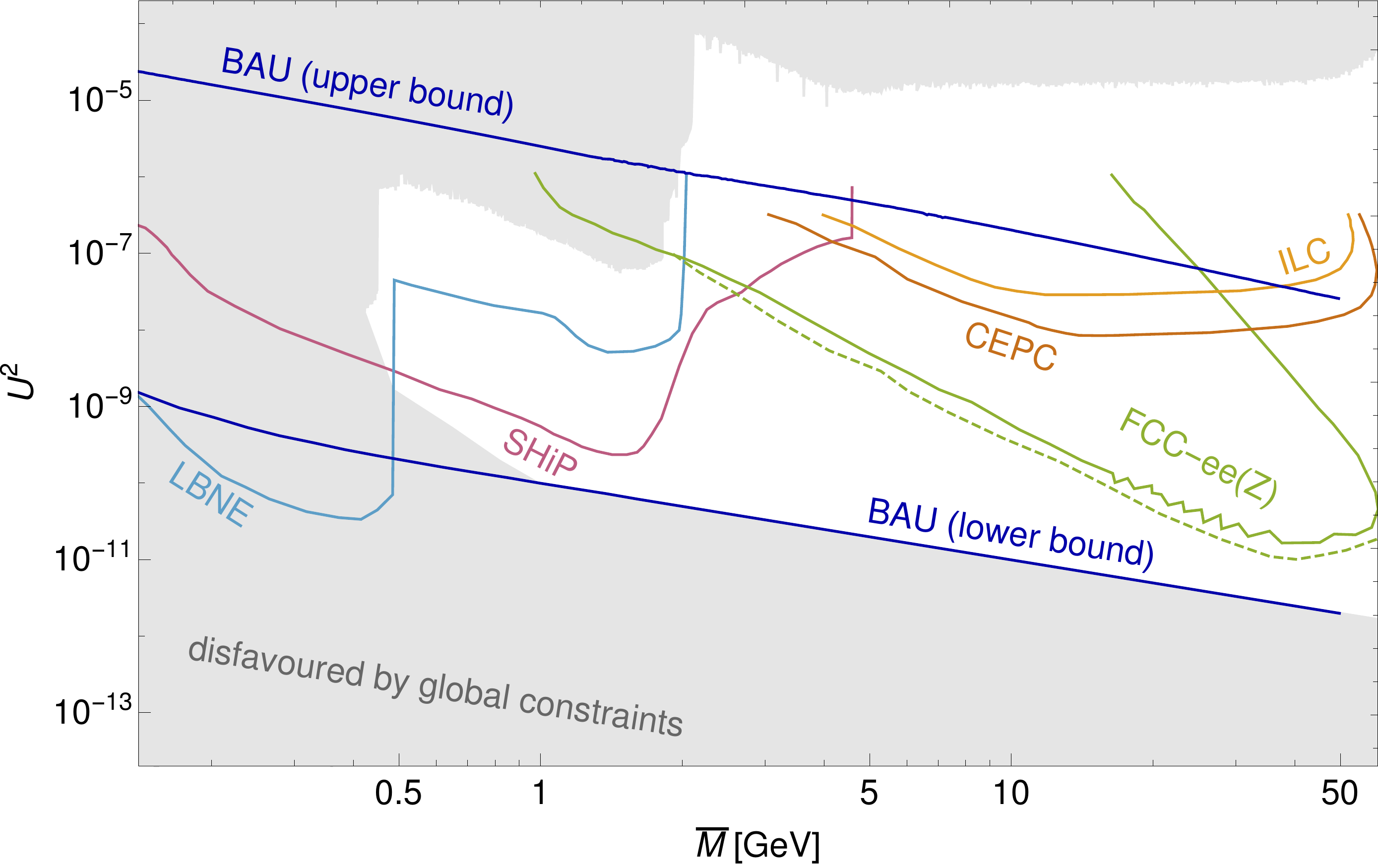}\\
	\includegraphics[width=0.44\textwidth]{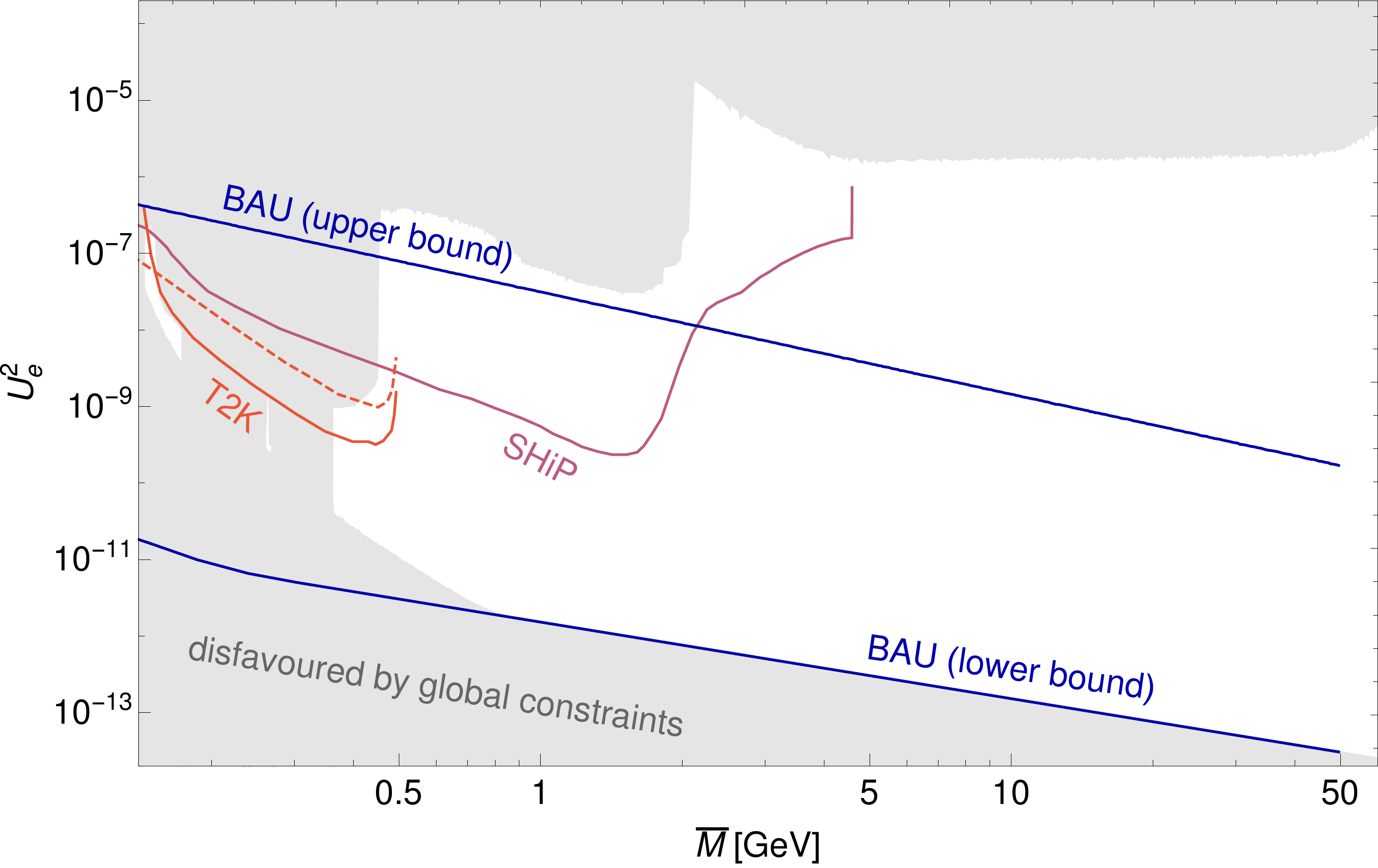}
	\includegraphics[width=0.44\textwidth]{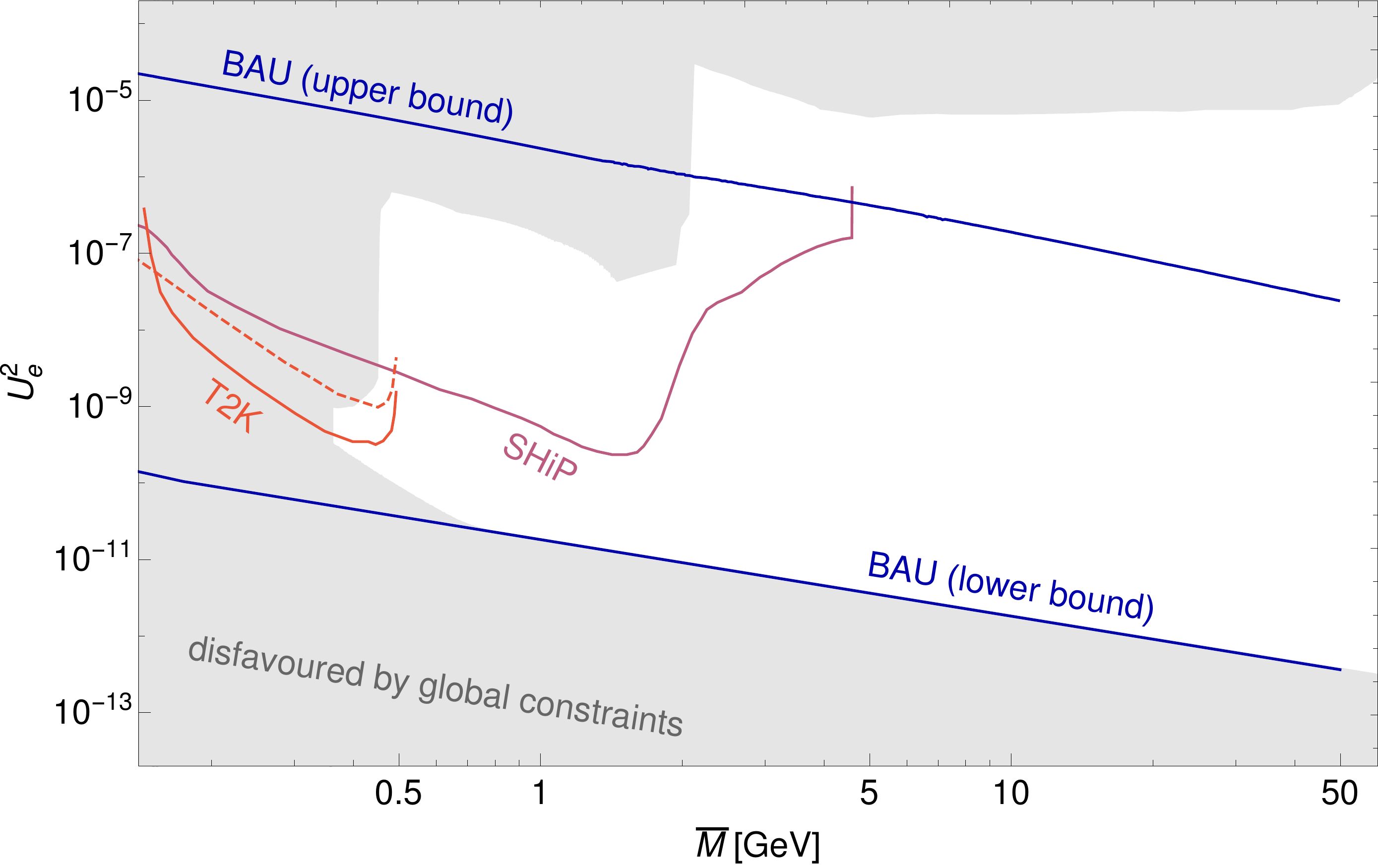}\\
	\includegraphics[width=0.44\textwidth]{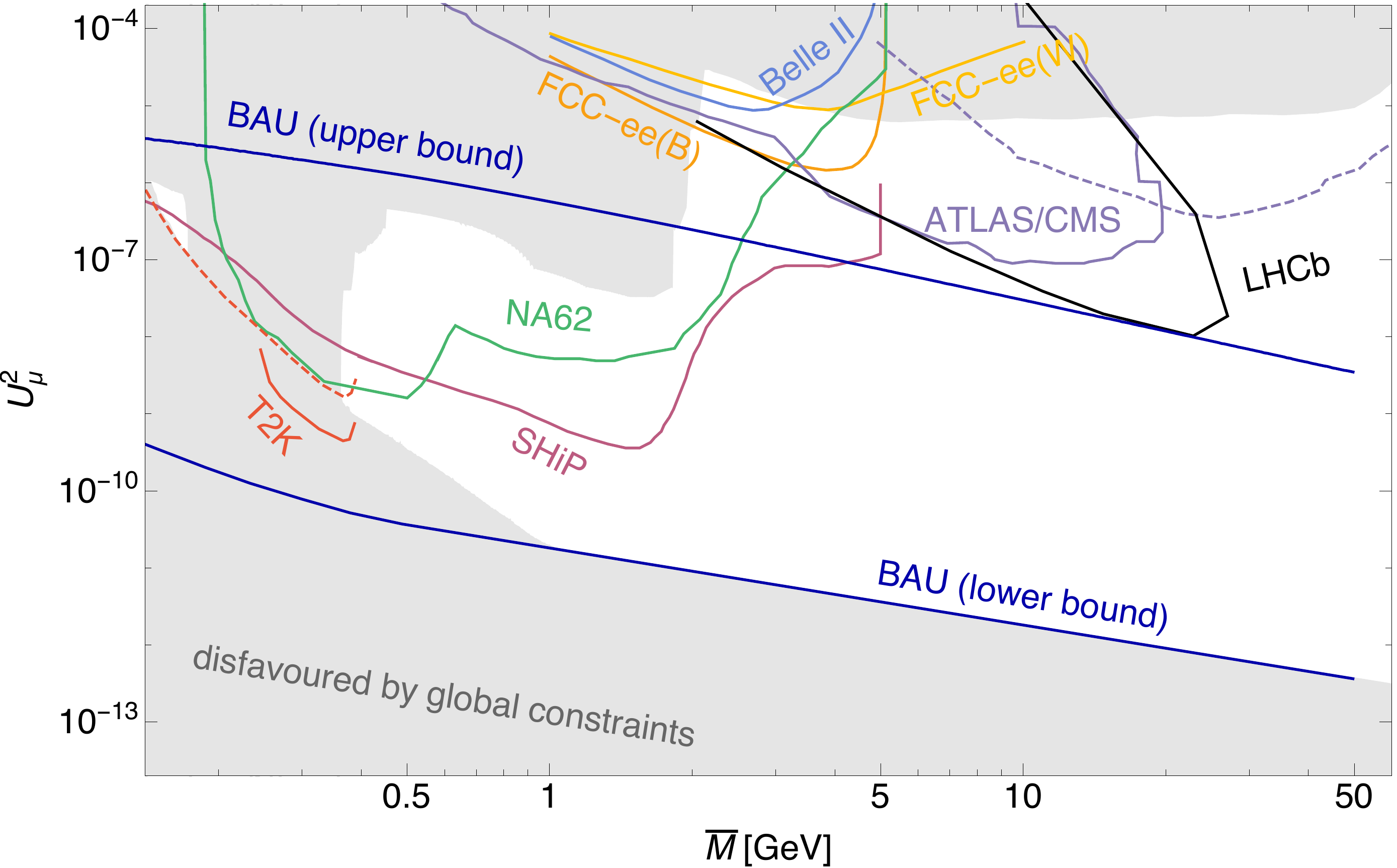}
	\includegraphics[width=0.44\textwidth]{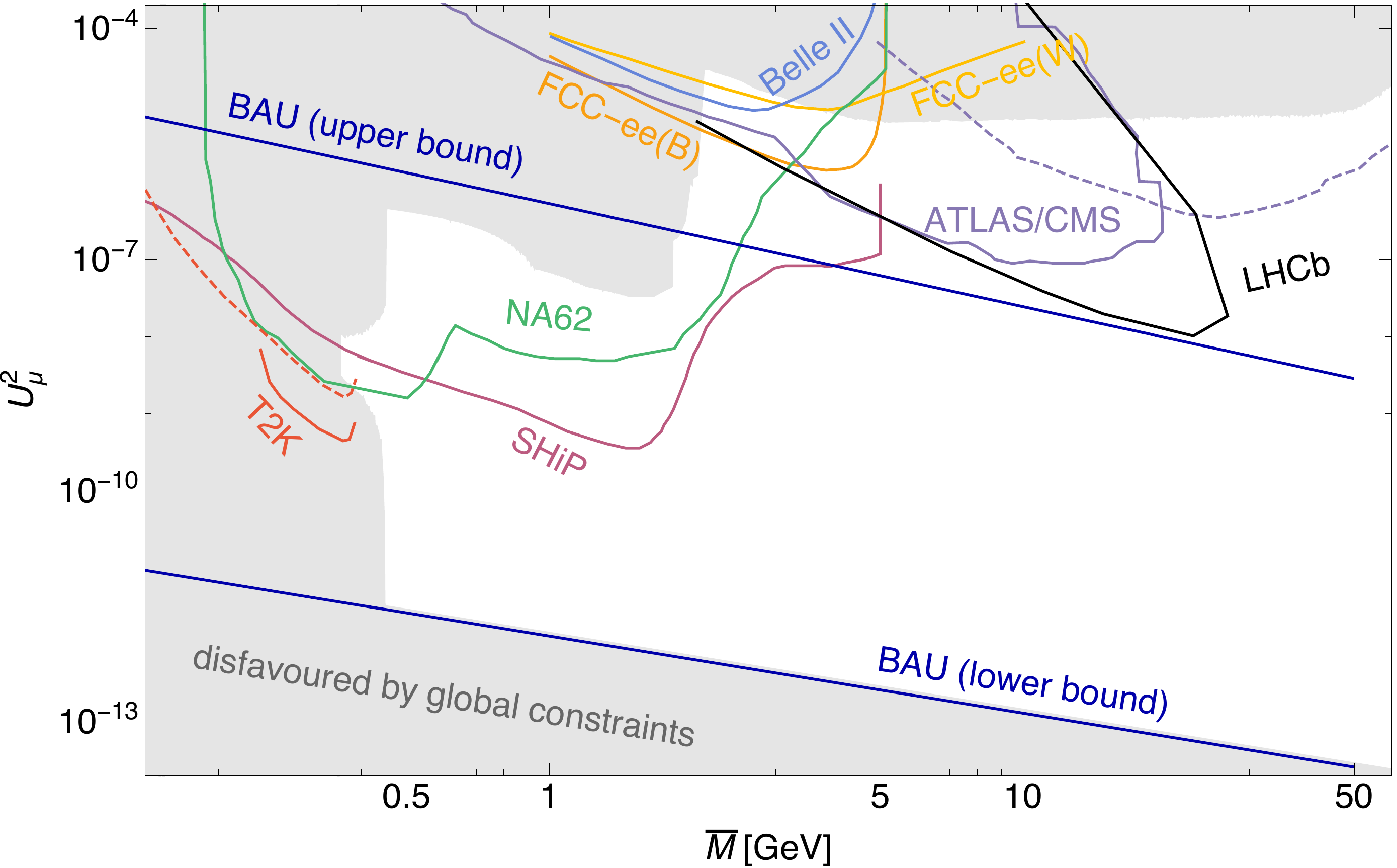}\\
	\includegraphics[width=0.44\textwidth]{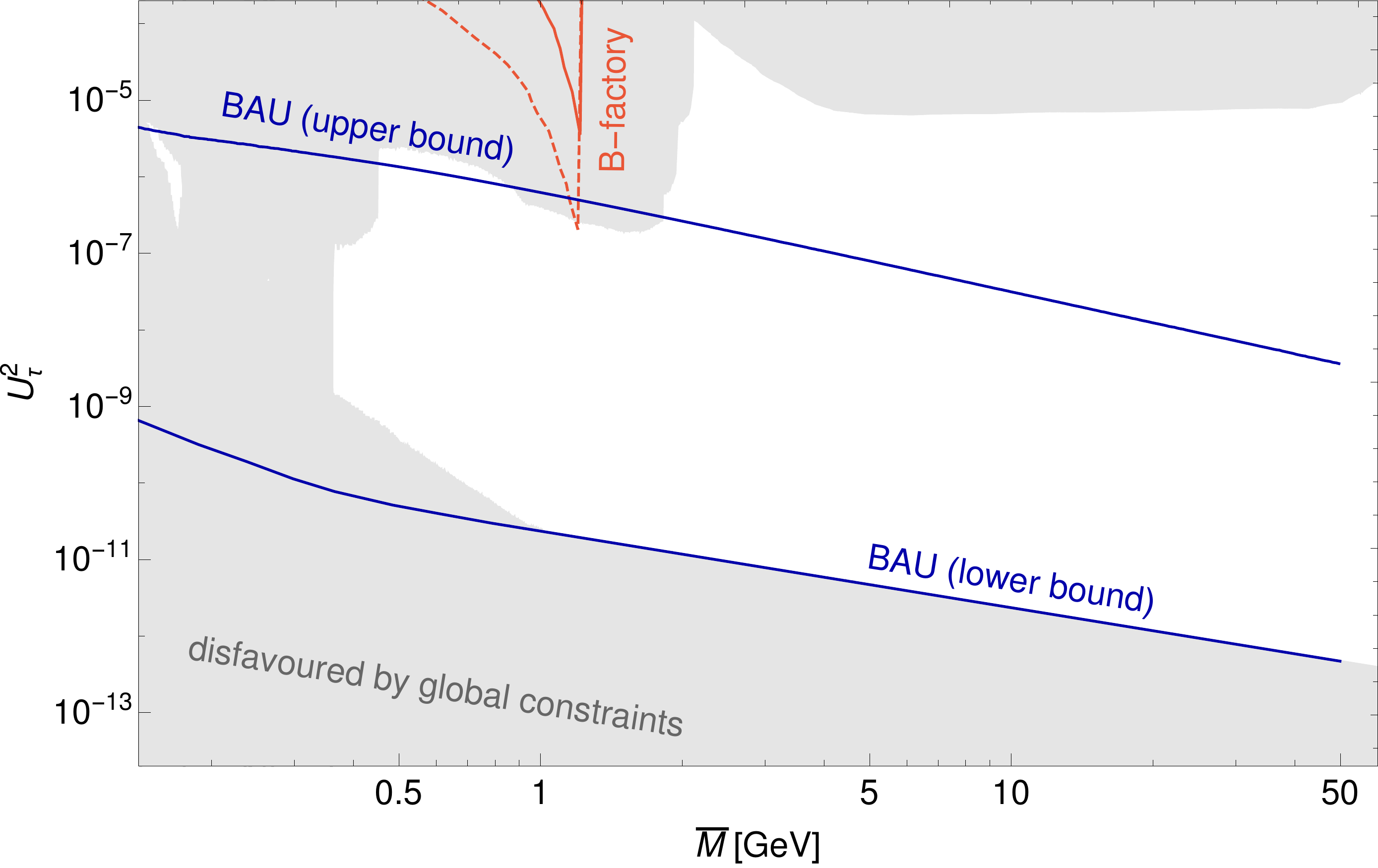}
	\includegraphics[width=0.44\textwidth]{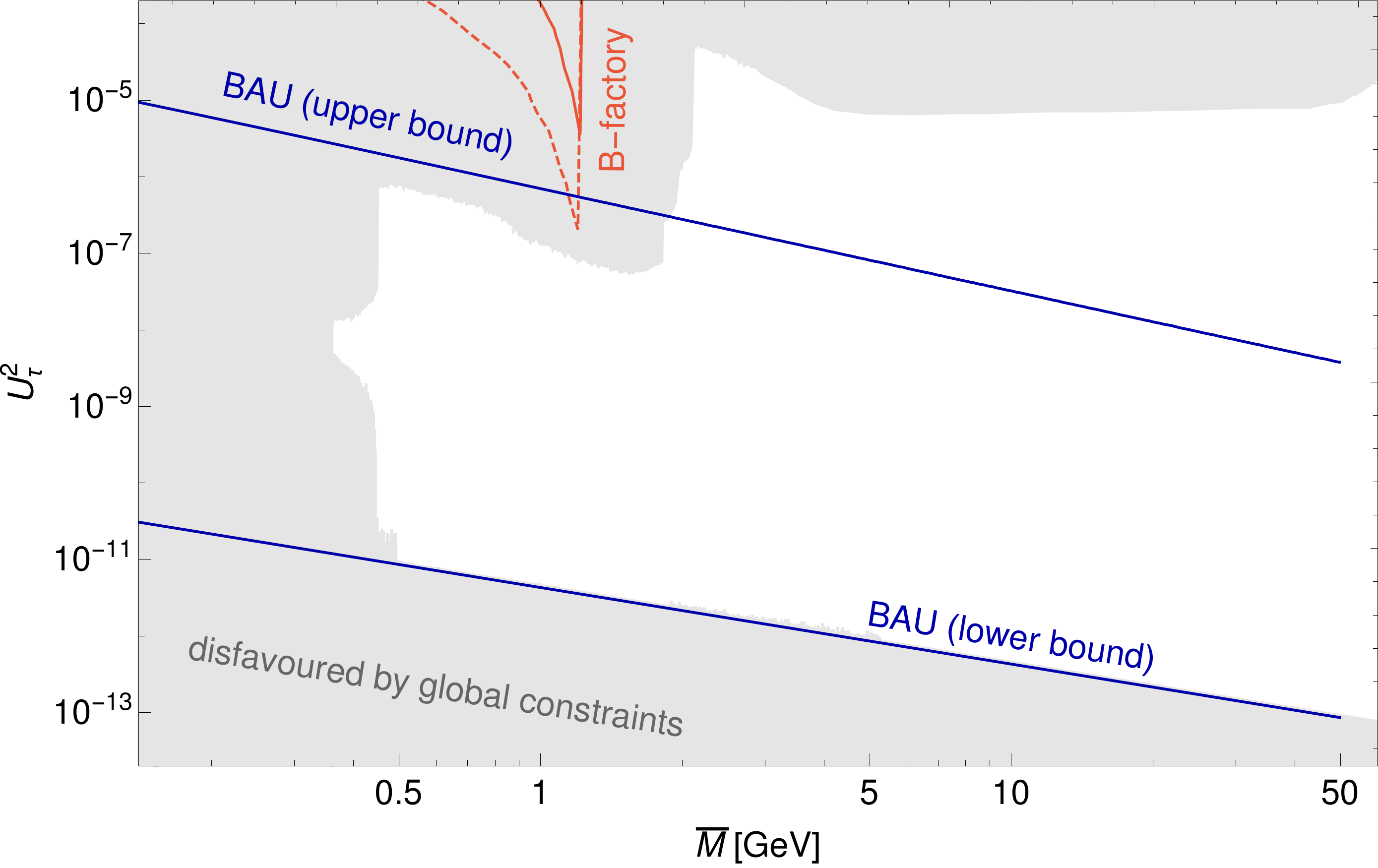}
\caption{
The viable leptogenesis parameter space for $n=2$ (between the blue lines) \cite{Drewes:2016jae}
compared to the constraints from past experiments (gray area) \cite{Drewes:2016jae} and the expected reach of 
NA62 \cite{Talk:Spadaro}, 
SHiP \cite{Anelli:2015pba,Graverini:2214085},
a similar detector at LBNE/DUNE \cite{Adams:2013qkq} or T2K \cite{Asaka:2012bb}
displaced vertices at FCC-ee \cite{Blondel:2014bra} (solid) \cite{Antusch:2016vyf} (dashed),
B decays at FCC-ee \cite{Asaka:2016rwd},
BELLE II \cite{Asaka:2016rwd,Kobach:2014hea},
LHCb \cite{Antusch:2017hhu},
ATLAS or CMS \cite{Izaguirre:2015pga} (cf. also \cite{Helo:2013esa,Gago:2015vma})
ILC \cite{Antusch:2016vyf} and 
CEPC \cite{Antusch:2016vyf}.
Here $U^2=\sum_\alpha U_\alpha^2$.
The left column corresponds to normal light neutrino mass ordering, the right column to inverted ordering. Figure taken from \cite{Drewes:2016jae}.
\label{MvsU2}
}
\end{figure}

The limit $\upmu, \upepsilon \ll 1$ is interesting for two different reasons. First, it allows to realize large active-sterile mixing angles while respecting constraints from light neutrino oscillation data. Second, it allows for at least three mechanisms of low-scale leptogenesis in the minimal model, see (\ref{eq0:lagrangiantypeI_seesaw}).
\begin{itemize}
\item[(a)] For $M_i$ above the electroweak scale, the baryon asymmetry can be generated in the decay of $N_i$ via \emph{resonant leptogenesis} \cite{Pilaftsis:2003gt}. The lower bound on the mass comes from the requirement that the heavy neutrinos freeze out and decay before sphalerons freeze out at $T_{\rm EWsp}\sim 130$ GeV \cite{DOnofrio:2014rug}. 
\item[(b)] For smaller masses $M_i$, the baryon asymmetry can be produced in the decay of Higgs bosons into heavy neutrinos and SM leptons. 
This decay is kinematically possible in the symmetric phase of the SM due to the ``thermal masses" that (quasi)particles obtain from forward scatterings in the primordial plasma. Both, the Higgs bosons and the SM leptons receive a thermal mass of equal size due to their gauge interactions. The Higgs boson receives an additional contribution to its thermal mass from the couplings to fermions (primarily the top quark) and its self-interaction, which makes the $1\rightarrow 2$ decay into a SM lepton $\ell_a$ and a heavy neutrino $N_i$ kinematically possible.
This kinematic possibility was already realized in Ref.~\cite{Giudice:2003jh} and briefly discussed in Ref.~\cite{Garbrecht:2010sz} (cf. also Ref.~\cite{Drewes:2015eoa} in the context of sterile neutrino Dark Matter production). In Refs.~\cite{Hambye:2016sby,Hambye:2017elz} it has been pointed out that it opens a new channel for low-scale leptogenesis. In addition to the decays and inverse decays considered in Ref.~\cite{Hambye:2016sby}, scatterings may play an important role for total lepton number violation.
This is known to be the case for purely flavor violating processes \cite{Anisimov:2010gy,Besak:2012qm}, see Refs.~\cite{Laine:2013lka,Garbrecht:2013urw,Ghisoiu:2014ena}.
\item[(c)] For $M_i$ below the electroweak scale, the baryon asymmetry may also be generated via CP-violating oscillations of the heavy neutrinos during their production (instead of their decay) \cite{Akhmedov:1998qx,Asaka:2005pn}. This mechanism of \emph{leptogenesis from neutrino oscillations}
is reviewed in detail in \cite{leptogenesis:A02}.
\end{itemize}
For $n=2$, all three mechanisms require $\upmu\ll1$ to achieve a resonant enhancement of the produced baryon asymmetry,\footnote{For $n>2$, a flavor asymmetric washout can allow for successful leptogenesis from neutrino oscillations (mechanism $(c)$) even without a degeneracy \cite{Drewes:2012ma}.}
while $\epsilon$ does not necessarily have to be small.
The mechanisms $(b)$ and $(c)$ in addition require that the number densities of the heavy neutrinos are negligibly small at the beginning of the radiation dominated epoch. They are then thermally produced from the primordial plasma. The first of them reaches thermal equilibrium at $T=T_{\rm eq}\simeq {\rm max}[\lambda^\dagger\lambda]  \gamma_{\rm av} M_{\mathrm{Pl}}\sqrt{45/(4\pi^3 g_*)}$. Here $\gamma_{\rm av}$ is a numerical factor $\sim 10^{-2}$ that appears in the matrix of thermal heavy neutrino damping rates 
\begin{equation}
\Gamma_N = \lambda^\dagger\lambda \gamma_{\rm av} T
\end{equation}
in the symmetric phase of the SM at $T\gg M_i$ \cite{Laine:2013lka,Garbrecht:2013urw,Ghisoiu:2014ena}.
In the following we focus on the symmetry protected scenario with $n=2$ and on the mass range below the electroweak scale.
In this region planned or proposed experiments have a realistic chance to enter the parameter region in which leptogenesis is possible, cf. Fig.~\ref{MvsU2}. 
In this regime, the mechanisms $(b)$ and $(c)$ generally coexist. 

To estimate the magnitude of the different contributions, we first consider the ``naive seesaw regime'' ($\upepsilon \sim 1$). 
In the context of low-scale leptogenesis, this regime is also referred to as the \emph{oscillatory regime} because the heavy neutrinos undergo a large number of oscillations between the temperature $T_{\rm osc}\simeq \big(|M_2^2-M_1^2|  M_{\mathrm{Pl}}\sqrt{45/(4\pi^3 g_*)}\big)^{1/3}$ when the heavy neutrinos perform their first oscillation
and the freeze-out of weak sphalerons at the temperature $T_{\rm EWsp}$. There is a clear separation of scales $T_{\rm osc} \gg T_{\rm eq}$, which is the basis for the derivation of the expressions presented in Sec.~\ref{ch5:sec2:ARS-seesaw}.
The source term in the kinetic equations can be separated into a lepton number violating part and a lepton flavour violating part.\footnote{Here \emph{lepton number violation} refers to the violation of a generalized lepton number to which the two heavy Majorana neutrinos' helicity states contribute with opposite sign. That is, for $T\gg \bar{M}$ the two helicity states behave like ``particles" and ``antiparticles". This generalized lepton number, which is approximately conserved during leptogenesis, is in general not identical to the generalized lepton charge that is conserved in the limit $\upmu,\upepsilon\rightarrow 0$. A detailed discussion of the different lepton numbers can e.g. be found in Ref.~\cite{Antusch:2017pkq}.
} 
Both of them are of order $\mathcal{O}[\lambda^4]$ in the Yukawa couplings.
It has often been argued that the total lepton number violating part of the source term (which drives the Higgs decay mechanism $(b)$) should be sub-dominant  compared to the purely flavor violating source (which drives the leptogenesis mechanism $(c)$ via neutrino oscillations) because the only lepton number violating term in the Lagrangian is the Majorana mass $M_N$, which leads to a suppression $\sim M_N^2/T^2$ of lepton number violating decays in the regime $T\gg M_N$. However, because the total lepton number is approximately conserved in the heavy neutrino oscillations at $T_{\rm osc}\gg M_N$, %the source term itself (which is $\propto \lambda^4$) does 
they cannot directly generate a net baryon asymmetry. This can only be achieved with the help of a flavor asymmetric washout. Since the washout is also mediated by the heavy neutrino Yukawa interactions, the final baryon asymmetry in the regime $\upepsilon \sim 1$ is $\mathcal{O}[\lambda^6]$ (cf. also Ref.~\cite{Shuve:2014zua} for a pedagogical discussion). This can also be seen in Eq.~(\ref{eq2:DeltaCP1}) and Eq.~(\ref{eq2:DeltaCP2}). 
In contrast to that, the Higgs decays directly violate lepton number and generate a baryon asymmetry $\mathcal{O}[\lambda^4 M_N^2/T^2]$. Since the seesaw relation (\ref{eq0:lightneutrinomasses}) in the regime $\upepsilon\sim 1$  predicts a ``naive seesaw scaling'' $\lambda^2 \propto  m_\nu M_N/v^2$, the Higgs decays can dominate if the asymmetry is primarily generated near the electroweak scale ($T\sim v$).\footnote{A more detailed comparison of the contributions from both mechanisms in oscillatory regime is given in Ref.~\cite{Hambye:2017elz}. In addition to the different dependence on the Yukawa couplings, also the heavy neutrino mass spectrum affects the asymmetry in a different way.}  
Delaying the oscillations until $T\sim T_{\rm EWsp}$, however, requires much smaller mass splittings than the values that are required for the mechanism $(c)$ ( $\upmu\lesssim 10^{-1}-10^{-3}$ with $n=2$ \cite{Canetti:2010aw,Canetti:2012kh,Hernandez:2016kel,Antusch:2017pkq} and $\mu\sim 1$ with $n=3$ \cite{Drewes:2012ma,Hernandez:2015wna}).
Such small splittings seem highly tuned if they cannot be explained by a symmetry. Such a protecting symmetry exists if we require $\upepsilon \ll 1$.

The behavior in the symmetry protected regime $\upepsilon,\upmu\ll1$ is qualitatively rather different from the oscillatory regime discussed above and in Sec.~\ref{ch5:sec2:ARS-seesaw}. The difference can best be understood in terms of the temperatures $T_{\rm osc}$ and $T_{\rm eq}$. The oscillatory regime is characterized by a clear hierarchy $T_{\rm osc}\gg T_{\rm eq}$, which implies that the oscillations that generate the initial flavored asymmetries occur long before the heavy neutrinos come into equilibrium and the washout is efficient. For $\upepsilon,\upmu\ll1$, however, the matrix $\lambda^\dagger\lambda$ (and hence matrix $\Gamma_N$ of damping rates) has two vastly different eigenvalues of size $\sim \upepsilon$ and $1/\upepsilon$. The large eigenvalue $\sim 1/\upepsilon$ implies that one heavy neutrino flavor eigenstate of $\Gamma_N$ reaches equilibrium before the oscillations commence, i.e., $T_{\rm eq}\gg T_{\rm osc}$. The other flavor-eigenstate evolves slowly and exhibits overdamped behavior. 
The \emph{overdamped regime} has first been discussed in the $\nu$MSM \cite{Shaposhnikov:2008pf}. 
The parametric dependence of the baryon asymmetry on the model parameters is in general non-polynomial.
A detailed discussion %including analytic approximations 
can be found in Ref.~\cite{Drewes:2016gmt}  and in  \cite{leptogenesis:A02}. 

The $1/\upepsilon$ enhancement of one eigenvalue of $\lambda^\dagger\lambda$ 
also implies that successful leptogenesis is possible for much larger mixing angles $U_{\alpha i}$ than in the oscillatory regime. 
This can be seen by comparing Figure~\ref{fig2:bayes} and Figure~\ref{MvsU2}.
Note that, in contrast to Fig.~\ref{fig2:bayes}, the upper limits on the mixings
in Fig.~\ref{MvsU2}  are the maximum values found compatible with leptogenesis. Instead, Fig.~\ref{fig2:bayes} shows the most 
probable regions (at 90$\%$CL) in a Bayesian analysis that  assumes a flat measure
 in the chosen parameter space.  Even though this measure does not disfavor the 
symmetry protected scenario, it does not favor it either. 
In contrast to the contours in Fig.~\ref{fig2:bayes}, the lines in Fig.~\ref{MvsU2} do not correspond to Bayesian likelihoods, but simply show the boundaries of the viable parameter region if one remains entirely agnostic about the values of the heavy neutrino parameters (in particular $\upepsilon$ and $\upmu$) and the unknown phases $\delta$ and $\alpha$ in the light neutrino mixing matrix $U_\nu$. 
Here $\delta$ is the Dirac phase and $\alpha$ the Majorana phase, see \cite{leptogenesis:A02}.
These results were obtained from the flavored source alone; total lepton number violating processes should be included in further studies \cite{Ghiglieri:2017gjz,Eijima:2017anv,Antusch:2017pkq}. 

\subsubsection{Testability} From a viewpoint of testability\footnote{By ``testability" we here mean the potential to actually discover the heavy neutrinos in near future experiments, which requires that their masses are below the TeV-scale. The potential to indirectly rule out leptogenesis with larger $M_i$ has been studied in Refs.~\cite{Deppisch:2013jxa,Deppisch:2015yqa}, cf. \sref{ch5:sec6:highscale}.} 
the symmetry protected scenario is very attractive because the $1/\upepsilon$ enhancement of the Yukawa coupling allows for larger mixing angles and hence larger branching ratios in experiments.
The quantities that determine the branching ratios in experiments are the mixing angles $|U_{\alpha i}|^2$.
The overall interaction strength of the heavy neutrinos can be characterized by the quantity $U^2={\rm tr}(U^\dagger U)$,
which can be expressed in terms of the model parameters by Eq.~(\ref{U2Definition})
without any approximations, which clearly shows the enhancement for small $\upepsilon$.

To relate the dynamics during leptogenesis to the behavior in the laboratory, we have to relate the heavy neutrino interaction eigenstates (i.e., the flavor eigenvectors of the matrix $\Gamma_N=\lambda^\dagger \lambda \gamma_{\rm av}T$) to the physical mass states.
In the very early Universe, the effective heavy neutrino masses are dominated by the thermal masses $\propto \lambda^\dagger \lambda T$. 
Hence, $\Gamma_N$ and the effective mass matrix are both proportional to the matrix $\lambda^\dagger \lambda$ and can be diagonalized simultaneously, so that the $N_i$ mass and interaction eigenstates are identical. 
As the temperature drops, the contribution from $M_N$ to their effective mass becomes increasingly important, and the gradual decrease of the thermal masses leads to a rotation of the effective mass matrix. Moreover, below the electroweak scale, the heavy neutrinos receive a mass contribution of $\mathcal{O}[U^2 M_N]$ from the Higgs mechanism, which roughly is of the same order as the light neutrino mass matrix. The Higgs contribution is tiny compared to the overall scale of the eigenvalues of $M_N$ and therefore has no effect on particle kinematics. However, it may not be negligible compared to the mass splitting between them in the symmetry protected limit of small $\upmu$, and it can therefore affect flavor oscillations of the heavy neutrinos in the laboratory and low energy CP violation \cite{Drewes:2016jae}.\footnote{This contribution also plays a crucial role in the $\nu$MSM \cite{Shaposhnikov:2008pf}: The interplay between the different contributions to the effective mass splitting allows for resonant production of lepton asymmetries below the electroweak scale \cite{Canetti:2012kh}, which is crucial to produce sufficiently cold sterile neutrino Dark Matter at temperatures $T\sim 100$ MeV \cite{Laine:2008pg}. } 
The complete physical mass matrix in the laboratory is given by
\begin{equation}
(M_N^{\rm phys})_{jk}=(M_N)_{jk} + \frac{1}{2}\left(U_{j \alpha}^{\dagger} U_{\alpha i} (M_N)_{ik} + (M_N)_{ij} U_{\alpha i} U^{\dagger}_{k\alpha}\right).
\end{equation}
In the symmetry protected scenario,  $\Gamma_N$ and the thermal mass on the one hand and the heavy neutrinos' vacuum Majorana mass matrix $M_N$ on the other 
are maximally misaligned, which can easily be verified by taking the limit $\upepsilon, \upmu\rightarrow 0$ in the Casas-Ibarra parameterization (\ref{CasasIbarra}), which corresponds to $M_1=M_2$ and $\gamma\rightarrow \infty$.
Since $M_N^{\rm phys}$ is dominated by $M_N$, the mass basis is strongly misaligned with the interaction basis. The two interaction eigenstates couple with very different strengths $\sim\upepsilon$ and $\sim1/\upepsilon$. The two mass eigenstates, on the other hand, couple with the same strengths, i.e., $\lambda_{\alpha 1}=\lambda_{\alpha 2}$ and $ U_{\alpha 1}= U_{\alpha 2}$. 
Moreover, for $\upmu\ll 1$ it is difficult to resolve their mass splitting experimentally, and experiments are only sensitive to 
%If the mass splitting of the two heavy neutrinos is too small to be resolved, then they will appear as a single Dirac particle with mixings 
\begin{equation}
U_\alpha^2=\sum_i |U_{\alpha i}|^2
\end{equation}
in the laboratory.
In the  $\upepsilon, \upmu\rightarrow 0$ limit, $N_1$ and $N_2$ effectively can be treated as one Dirac spinor $\Psi_N=(N_1+iN_2)/\sqrt{2}$ with couplings $U_\alpha^2$. Hence, the symmetry protected scenario predicts pseudo-Dirac heavy neutrinos in the laboratory.  
Finite values of $\upepsilon$ and $\upmu$ lead to deviations from the Dirac-like behaviour that can be treated as perturbations if they are sufficiently small. Some proposals have been made to look for these deviations even in the regime where the mass splitting cannot be resolved kinematically \cite{Dib:2016wge,Dib:2015oka,Anamiati:2016uxp}.

While the ratio $U_{\alpha 1}^2/U_{\alpha 2}^2=1$ is fixed in the $B-L$ conserving limit $\upepsilon,\upmu\rightarrow 0$, the ratios $|U_{\alpha i}|^2/|U_{\beta i}|^2$ of the heavy neutrino couplings to different flavors remain free. They are determined by the CP-violating phases in the light neutrino mass matrix $U_\nu$, cf. Eqns.~(\ref{eq2:uihanal}) and (\ref{eq2:unhanal}).\footnote{The relations between the individual $|U_{\alpha i}|^2$ and the model parameters for arbitrary values of $\upmu$ are e.g. given in the appendix of Ref.~\cite{Drewes:2016jae} (at leading order in $\upepsilon$).}
This point has previously been observed in Refs.~\cite{Gorbunov:2007ak,Shaposhnikov:2008pf,Ruchayskiy:2011aa,Asaka:2011pb}, and its importance for the full testability of the low-scale seesaw was pointed out in Refs.~\cite{Hernandez:2016kel,Drewes:2016jae}.
If $\Delta M$ is large enough to be resolved experimentally, then the mixings $U_{\alpha 1}^2$ and $U_{\alpha 2}^2$ can be measured independently. 
A measurement of all $|U_{\alpha i}|^2$ would fix the $\zzz$ and the phases in $U_\nu$ up to one discrete transformation, with $M_1$ and $M_2$ being extracted from the kinematics. The remaining discrete parameter degeneracy can be broken by an independent measurement of the Dirac phase $\delta$ in $U_\nu$ in neutrino oscillation experiments. Hence, the Lagrangian can at least in principle be fully reconstructed from experimental data, which allows to calculate the baryon asymmetry and compare it to the observed value. The parameter space can even be over-constrained if measurements of indirect observables that are sensitive to the heavy neutrino properties  are added, such as $0\nu \beta \beta$ decay, lepton universality violation or searches for CP-violation in the heavy neutrino decay (cf. Sec.~\ref{ch5:sec2:nhlcons}). This at least in principle makes the low-scale seesaw a fully testable model of neutrino masses and baryogenesis \cite{Drewes:2016jae}. 

\begin{figure}
	\includegraphics[width=0.5\textwidth]{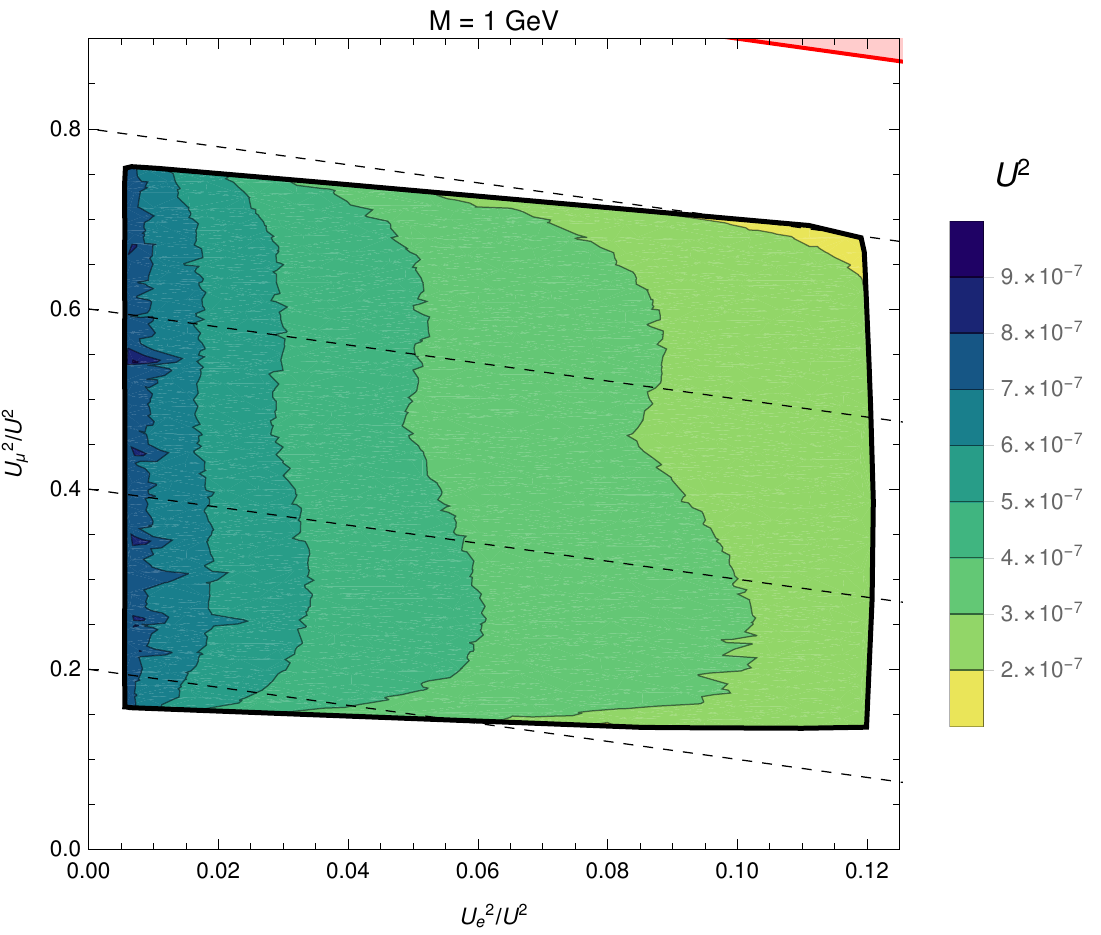}\includegraphics[width=0.52\textwidth]{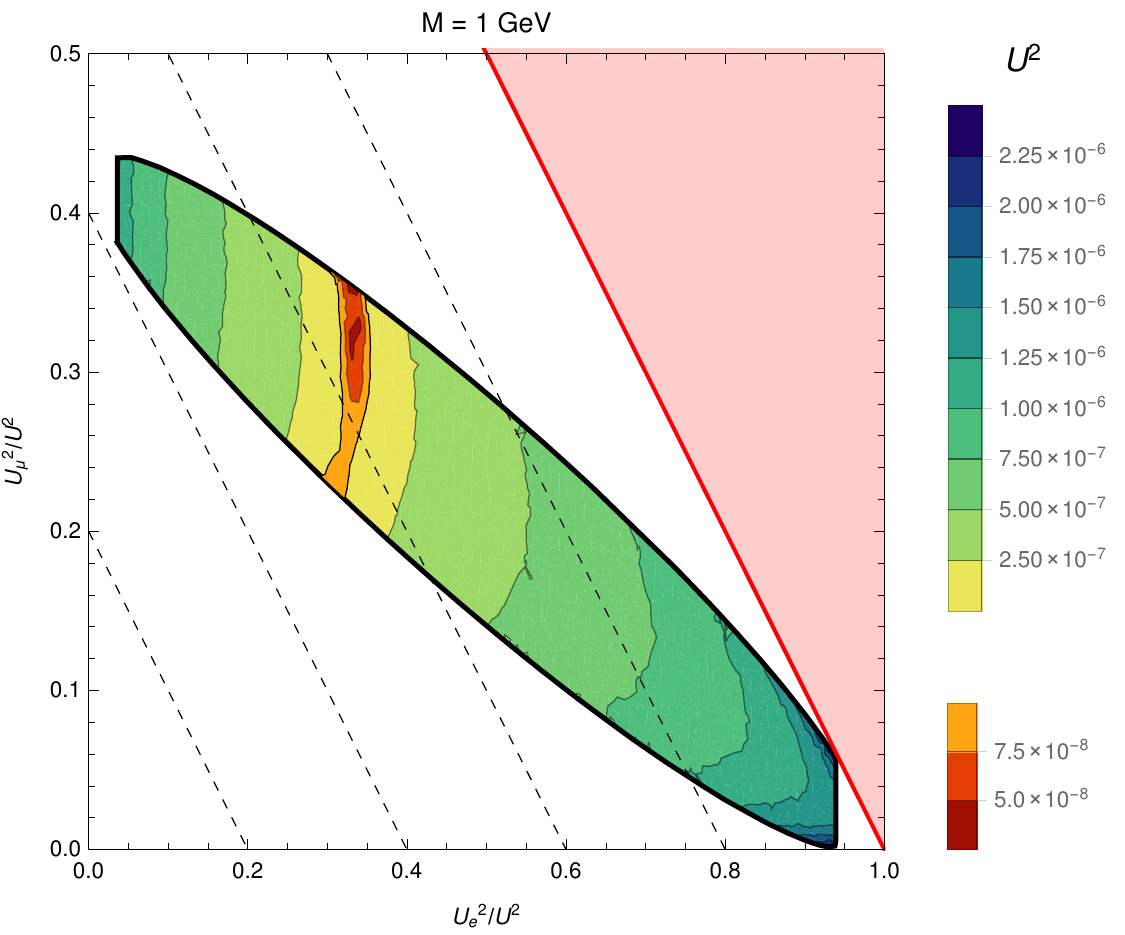}
	\caption{Values of $U_\alpha^2/U^2$ inside the black line are consistent with neutrino oscillation data for normal hierarchy (left) and inverted hierarchy (right) of light neutrino masses. The dashed lines correspond to constant $U_\tau^2$.
The light region marked in red is unphysical because it would imply $U_\tau^2<0$.  
The colored regions indicate the maximally allowed value of $U^2$ for given $U_\alpha^2/U^2$ if one requires that the the observed $\eta_B$ can be generated by leptogenesis with $\bar{M}=1$ GeV. Figure taken from \cite{Drewes:2016jae}.
	\label{Fig:Lepto_NO}}
\end{figure}

However, in practice it may be difficult to determine all parameters with sufficient accuracy to reliably predict the baryon asymmetry. In particular, it will be very difficult to resolve the heavy neutrino mass splitting $\Delta M$, which strongly affects the baryon asymmetry, at SHiP or  future lepton collider if it is smaller than about 10 MeV.\footnote{
Information about $\Delta M$ may still be obtained indirectly by 
an observation of $0\nu \beta \beta$ decay \cite{Drewes:2016lqo,Hernandez:2016kel,Asaka:2016zib,Drewes:2016jae}
comparing the rates of lepton number violating and conserving decays as well as the momentum distribution of the decay products \cite{Gluza:2015goa,Dev:2015pga,Anamiati:2016uxp,Dib:2016wge,Antusch:2017ebe,Das:2017hmg}
or by looking for the heavy neutrino oscillations in the detector \cite{Boyanovsky:2014una,Antusch:2017ebe}.
}
For the tiny mass splittings in the overdamped regime this is certainly not possible. This also means that the mixings $U_{\alpha 1}^2$ and $U_{\alpha 2}^2$ cannot be measured independently, and experiments are only sensitive to $U_\alpha^2=\sum_i |U_{\alpha i}|^2$. Compared to the complete set of $|U_{\alpha i}|^2$, the set $(U_e^2,U_\mu^2,U_\tau^2)$ is invariant under one more transformation of the model parameters that has no simple analytic form.
If in addition $\upepsilon\ll1$, then the ratios $U_\alpha^2/U^2$ are in good approximation independent of $\zzz$ and $\bar{M}$. 
As a result, one cannot put constraints on the two parameters $\Delta M$ and $\theta$, both of which are crucial for leptogenesis, if $\upmu$ is very small. 
In spite of this, consistency checks for both, the hypothesis that the $N_i$ generate the light neutrino masses via the seesaw mechanism and the BAU via leptogenesis, are still possible \cite{Hernandez:2016kel,Drewes:2016jae}.
In the approximation  $\upmu=0$, a measurement of all $U_\alpha^2$ allows to uniquely fix $\upepsilon$, cf. Eq.~(\ref{U2Definition}). The relative sizes of the $U_\alpha^2$ are then entirely determined by the phases $\alpha$ and $\delta$ alone. %This is illustrated in Fig.~\ref{fig:regions_NO.pdf}.
Since not every set $(U_e^2,U_\mu^2,U_\tau^2)$ can be realized by varying $(\alpha,\delta,\gamma)$, see Fig.~\ref{Fig:Lepto_NO}, 
the consistency of the $U_\alpha^2$ measurements with each other and with a possible determination of $\delta$ in neutrino oscillation experiments allow to rule out or support the minimal model with $n=2$. Moreover, for large $U^2$, leptogenesis requires $U_\alpha^2/U^2\ll1$ for at least one flavor in order to prevent a complete washout of the asymmetries before sphaleron freeze-out.  
This means that the requirement for successful leptogenesis allows to make additional predictions for the allowed range of $U_\alpha^2/U^2$, cf. Fig.~\ref{Fig:Lepto_NO}, which can be used as a test for leptogenesis.

Finally, it has been pointed out in Refs.~\cite{Hernandez:2016kel,Caputo:2017pit} that one may turn the tables and use the $U_\alpha^2$ measured in heavy neutrino decays at SHiP or a future collider to constrain the phases in $U_\nu$, in particular the Majorana phase $\alpha$, which is hard (if not impossible) to detect in light neutrino experiments.
\subsubsection{Parameter space in the minimal model} Based on these considerations, we can dissect the parameter space in the following way.
\begin{itemize}
\item In the ``naive seesaw" regime $\upepsilon\sim 1$, $\upmu\sim 1$, leptogenesis is not possible in the minimal model with $n=2$ because it requires a mass degeneracy.\footnote{Models with $n>2$ allow for successful leptogenesis without mass degeneracy, but the phenomenology of these models (which contain more free parameters) has not been studied in detail. First estimates indicate that leptogenesis can be achieved with larger $|U_{\alpha i}|^2$ than for $n=2$ \cite{Canetti:2014dka,Hernandez:2015wna}.} 
Moreover, the tiny coupling strengths $|U_{\alpha i}|^2$ of the heavy neutrinos in the regime $\upepsilon\sim 1$ leads to tiny branching ratios in experiments.
\item In the  regime $\upepsilon\sim 1$, $\upmu\gtrsim 10^{-3}$
the minimal low-scale seesaw is in principle a fully testable model of neutrino masses and baryogenesis. However, the overall coupling strength $U^2$ of the heavy neutrinos in this regime is too tiny to detect them in any near future experiment.
\item In the mildly symmetry protected regime $\upepsilon\ll1$, $\upmu\gtrsim 10^{-3}$, the chances for an experimental discovery of the heavy neutrinos are 
greatly improved because their overall coupling strength is enhances as $1/\upepsilon$.
However, this also implies an enhanced washout, so that leptogenesis with the largest $U^2$ relies on a hierarchy $U_\alpha^2/U^2\ll 1$ for at least one flavor.
It is in principle still possible to extract all model parameters from experiments, but the accuracy of near future experiments is probably not sufficient to uniquely determine the baryon asymmetry of the universe from such measurements. However, if any heavy neutral leptons are discovered at colliders, the relationships between the $|U_{\alpha i}|^2$ and other observables still provide a powerful tool to test the hypothesis that these particles are the origin of neutrino mass and the origin of matter. In particular, these relations allow to indirectly measure the Majorana phase $\alpha$ in $U_\nu$.
\item Deep in the symmetry protected regime $\upepsilon, \upmu\ll1$, the chances for an experimental discovery of the heavy neutrinos are the best because the $1/\upepsilon$ enhancement of the couplings $|U_{\alpha i}|^2$, and because the smaller  $\upmu$ allow for leptogenesis with larger $|U_{\alpha i}|^2$.\footnote{In the overdamped regime this is not primarily a consequence of the usual ``resonant enhancement", which is indeed regulated by the width $\Gamma_N\gg \Delta M$ in this regime. Instead, it is due to the fact that $T_{\rm osc}\simeq T_{\rm eq}$ for small $\upmu$, which means that the washout has not enough time to erase the asymmetries before sphaleron freeze-out even if the Yukawa couplings $\lambda$  are comparably large. }  
However, it is very difficult (practically impossible) to gain any information about the parameters $\Delta M$ and $\theta$, which are crucial for leptogenesis, and experiments can only determine $U_\alpha^2$ (rather than the individual $|U_{\alpha i}|^2$).
In spite of this, the comparison of independent measurements of all three $U_\alpha^2$
with indirect probes still provides a powerful test for the hypothesis that the heavy neutrinos generate the light neutrino masses.
It also provides a way to determine the Majorana phase $\alpha$ in the light neutrino mixing matrix $U_\nu$. 
For comparably large $U^2$, the requirement to reproduce the correct baryon asymmetry imposes additional constraints on the ratios $U_\alpha^2/U^2$, which can be used as a test for leptogenesis.
An independent measurement of the Dirac phase in the light neutrino mixing matrix would strongly modify the allowed leptogenesis parameter region in this regime because it would put strong constraints on the ratios $U_\alpha^2/U^2$. 
\end{itemize}

\subsection{Constraints on sterile neutrinos}\label{ch5:sec2:nhlcons}
If the heavy neutrinos $N_i$ generate the light neutrino masses via the seesaw mechanism, then Eq.~(\ref{eq0:lightneutrinomasses}) implies that they necessarily must mix with the SM neutrino flavor eigenstates $\nu_{L\alpha}$, 
\begin{equation}
\nu_{L \alpha} = (U_\nu)_{\alpha i}\nu_i +  U_{\alpha k}N^c_k,
\end{equation}
and therefore feel the weak interaction with a strength that is suppressed by the mixing angle 
\begin{equation}
U_{\alpha i}=v(\lambda M_N^{-1})_{ai}.
\end{equation}
Here $\nu_i$ are the light neutrino mass eigenstates and $U_\nu$ is the unitary matrix that diagonalizes the matrix $ M_\nu$ in (\ref{eq0:lightneutrinomasses}). It is related to the full light neutrino mixing matrix $V_\nu$ as 
\begin{equation}\label{nonunitary}
(V_\nu)_{\alpha k} = \left(\delta_{\alpha \beta} - \frac{1}{2}U_{\alpha i} U_{i \beta}^\dagger\right)(U_\nu)_{\beta k}\,.
\end{equation} 
On this basis, their properties can be constrained by the negative result of various searches for deviations from the SM \cite{Shrock:1980ct,Shrock:1981wq}. 
Which experiments and observations are sensitive to the $N_i$ properties strongly depends on the magnitude(s) of their masses $M_i$.
Though neutrino oscillation data currently provides the strongest indirect probe of heavy neutrino properties, their masses cannot be constrained from the relation (\ref{eq0:lightneutrinomasses}) alone for two reasons:
\begin{enumerate}
\item If one assumes that $M_N$ and $\lambda$ have no special structure, then the relation (\ref{eq0:lightneutrinomasses}) predicts $M_N \sim \frac{v^2}{ M_\nu} \frac{\lambda^2}{2}$, cf. also the discussion following Eq.~(\ref{NaiveSeesaw}). That is, any value of $M_i$ can be made consistent with the constraints on the light neutrino mass matrix $ M_\nu$ by adjusting $\lambda$. 

\item There is even more freedom if we make use of the full matrix structure of equation (\ref{eq0:lightneutrinomasses}). What is experimentally constrained to be small are the eigenvalues $m_i^2$ of $ M_\nu  M_\nu^\dagger$, not the individual entries of the matrix $ M_\nu$. For example, in the symmetry protected scenario described in Sec.~\ref{SymmetryProtectedScenario}, rather large individual entries can be made consistent with small $m_i$ in a natural way if the parameters $\upmu$ and $\upepsilon$ defined in Eq.~(\ref{B-L_parameters}) are chosen small.
\end{enumerate} 
Hence, the range of masses allowed by neutrino oscillation data in principle is very large and reaches from the eV-scale \cite{deGouvea:2005er} up to values above the suspected scale of grand unification.\footnote{A theoretical upper bound can be imposed if one requires that their Yukawa interactions can be described by perturbative quantum field theory; then their masses $M_i$ should be at least 1-2 orders of magnitude below the Planck mass \cite{Asaka:2015eda}.} 
We in the following restrict the discussion to the parameter region in which the heavy neutrinos that generate the baryon asymmetry can be found in existing or proposed experiments. This effectively restricts us to masses $M_i$ below the electroweak scale.\footnote{Heavier $N_i$ can of course in principle be produced at the LHC or future colliders, but the production rates are much smaller than those in weak gauge boson decays, and it is currently not clear whether searches can enter the range of $|U_{\alpha i}|^2$ for which leptogenesis is possible.} 
A more general discussion of right handed neutrinos with various different masses and their role in particle physics and cosmology can e.g. be found in the review \cite{Drewes:2013gca}.

All following considerations are based on the fact that the heavy neutrinos couple to the weak currents via a term
\begin{eqnarray}\label{WeakWW}
&&-\frac{g}{\sqrt{2}}\overline{N^c}_i U^\dagger_{i \alpha}\gamma^\mu e_{L \alpha} W^+_\mu
-\frac{g}{\sqrt{2}}\overline{e_{L \alpha}}\gamma^\mu U_{\alpha i} N_i^c W^-_\mu\nonumber\\  
&&- \frac{g}{2\cos\theta_W}\overline{N_i^c} U_{i \alpha}^\dagger\gamma^\mu \nu_{L \alpha} Z_\mu
- \frac{g}{2\cos\theta_W}\overline{\nu_{L \alpha}}\gamma^\mu U_{\alpha i} N_i^c Z_\mu.
\end{eqnarray}
In addition, they may have other (new physics) interactions that we do not discuss here because they are model dependent. A popular choice are e.g. SU(2) gauge interactions in left-right symmetric theories, which are e.g. discussed in Ref.~\cite{Deppisch:2015qwa} and references therein.
It is instructive to classify the experimental signatures in two qualitatively different categories. In \emph{direct searches} the $N_i$ appear as real particles and can therefore be discovered. In contrast to that, we refer to experiments that are indirectly affected by the heavy neutrinos as \emph{indirect searches}; this usually primarily happens via the modification of the light neutrino interactions due to the non-unitarity in (\ref{nonunitary}).
 
\subsubsection{Direct Searches}
Depending on their mass, heavy neutrinos are predominantly produced in meson decays ($M_i<5$ GeV), real weak gauge boson decays ($5 \ {\rm GeV} < M_i < 80 \ {\rm GeV}$ or exchange of virtual weak gauge bosons ($80 \ {\rm GeV} < M_i$).\footnote{The production involving t-channel exchange of photons \cite{Dev:2013wba,Alva:2014gxa},
Higgs decays \cite{BhupalDev:2012zg,Cely:2012bz,Das:2017zjc,Das:2017rsu} or vector boson fusion \cite{Andres:2017daw} are not relevant for the searches discussed in the following.}
Being neutral, the $N_i$ themselves do not leave any trace in the detector and can only be observed by studying the kinematics of the charged particles that are produced along with them and when they decay into charged particles. 
The best search strategy strongly depends on their mass because their lifetime scales as $\propto |U_{\alpha i}|^2 M_i^5$ \cite{Gorbunov:2007ak}. 

Heavy neutrinos that are light enough to be produced in meson decays can be so long lived that the number of $N_i$-decays within conventional detectors at colliders is very small.  They can, however,  be observed in fixed target experiments if one places a detector behind the target. Past \emph{beam dump experiments} 
include PS191 \cite{Bernardi:1987ek}, NuTeV \cite{Vaitaitis:1999wq}, CHARM \cite{Bergsma:1985is}, CHARM and CHARM II \cite{Vilain:1994vg}, NA3 \cite{Badier:1985wg},  E949 \cite{Artamonov:2014urb}, IHEP-JINR \cite{Baranov:1992vq}, BEBC \cite{CooperSarkar:1985nh}, FMMF \cite{Gallas:1994xp} and NOMAD \cite{Astier:2001ck}. 
The NA62 experiment has performed peak searches \cite{Lazzeroni:2017fza} and will perform a dedicated search for heavy neutrinos in dump mode (``SHADOWS run''). In the next decade, the proposed SHiP \cite{Bonivento:2013jag,Alekhin:2015byh,Anelli:2015pba} experiment could improve the existing constraints by several orders of magnitude \cite{Graverini:2214085}. 
For very small masses even fixed target experiments are limited by the length of their detector. However, the heavy neutrinos may still reveal their existence because their on-shell production causes a peak in the spectrum or charged leptons emitted in meson decays. \emph{Peak searches} have been performed using kaons and pions as initial mesons \cite{PIENU:2011aa,Britton:1992pg,Britton:1992xv,Bryman:1996xd,Abela:1981nf,Daum:1987bg,Yamazaki:1984sj,Hayano:1982wu}.
For larger masses of a few GeV, the experiments BELLE \cite{Liventsev:2013zz} LHCb \cite{Aaij:2014aba}\footnote{Note that the LHCb results presented in Ref.~\cite{Aaij:2014aba} have recently been corrected in Ref.~\cite{Shuve:2016muy}.} have looked for lepton number violation in meson decays, which can be mediated by heavy neutrinos. LHCb has also looked for displaced vertices \cite{Aaij:2016xmb}.
The interesting possibility to look for CP violation in meson decays mediated by heavy neutrinos is discussed in Sec.~\ref{ch5:sec2:CPMesonDecays}.

Heavy neutrinos with larger masses have been searched for by the experiments DELPHI \cite{Abreu:1996pa}, L3 \cite{Adriani:1992pq,Achard:2001qv}, ATLAS \cite{Aad:2015xaa} and CMS \cite{Khachatryan:2015gha,Oh:2016ead,CMS:2017uoz}. The LHC searches have so far been focused on \emph{lepton number violating signals}, such as same sign di-leptons in the final state. These are probably the most promising signals if the heavy neutrinos decay promptly and have been studied by a large number of authors (see e.g.  Refs.~\cite{Deppisch:2015qwa,Drewes:2015iva} for a partial list of references).
However, if the $N_i$ are lighter than the W boson, they may be long lived enough to be observed by the displacement between their production and decay. How light exactly they have to be depends on the experiment. \emph{Displaced vertex searches} are more powerful at masses below roughly 20-30 GeV at ATLAS and CMS \cite{Helo:2013esa,Izaguirre:2015pga,Gago:2015vma} or LHCb \cite{Antusch:2017hhu}. 
Based on the current estimates, none of these experiments can reach deeply in the leptogenesis region for $n=2$  (though at least LHCb may touch it in the high luminosity run of the LHC). 
An upgrade of the MATHUSLA-type \cite{Chou:2016lxi} could enable the LHC to probe at least part of the leptogenesis region.
For $n>3$ the leptogenesis region is larger and may be accessed by current LHC experiments or BELLE II \cite{Canetti:2014dka}.\footnote{The results presented in Ref.~\cite{Canetti:2014dka} should be treated with some care because the authors did not take the fast equilibration of some heavy neutrino interaction eigenstates in the large $|U_{\alpha i}|^2$ regime into account consistently.
However, the main observation made in this paper is that the constraints from neutrino oscillation data allow for larger hierarchies amongst the $N_i$ couplings to individual SM flavors (i.e., smaller $U_\alpha ^2/U^2$) for $n>2$, as compared to the minimal model with $n=2$. This allows to protect part of the asymmetries from washout even if some heavy neutrino interaction eigenstates have reached equilibrium. Hence, the conclusion that leptogenesis can be possible with (much) larger $|U_{\alpha i}|^2$ for $n>2$ remains true irrespectively of the details of the $N_i$ equilibration.}
The perspectives to search for heavy neutrinos are certainly much better at a future high energy collider. They have been studied by a number of authors, see e.g. \cite{Deppisch:2015qwa} for a partial list of references. A recent summary of signatures at different types of colliders can be found in Ref.~\cite{Antusch:2016ejd}. 
In Ref.~\cite{Antusch:2017pkq} the potential of future colliders to probe low-scale leptogenesis has been studied specifically.

\begin{figure}
\begin{center}
\includegraphics[width=0.6\textwidth]{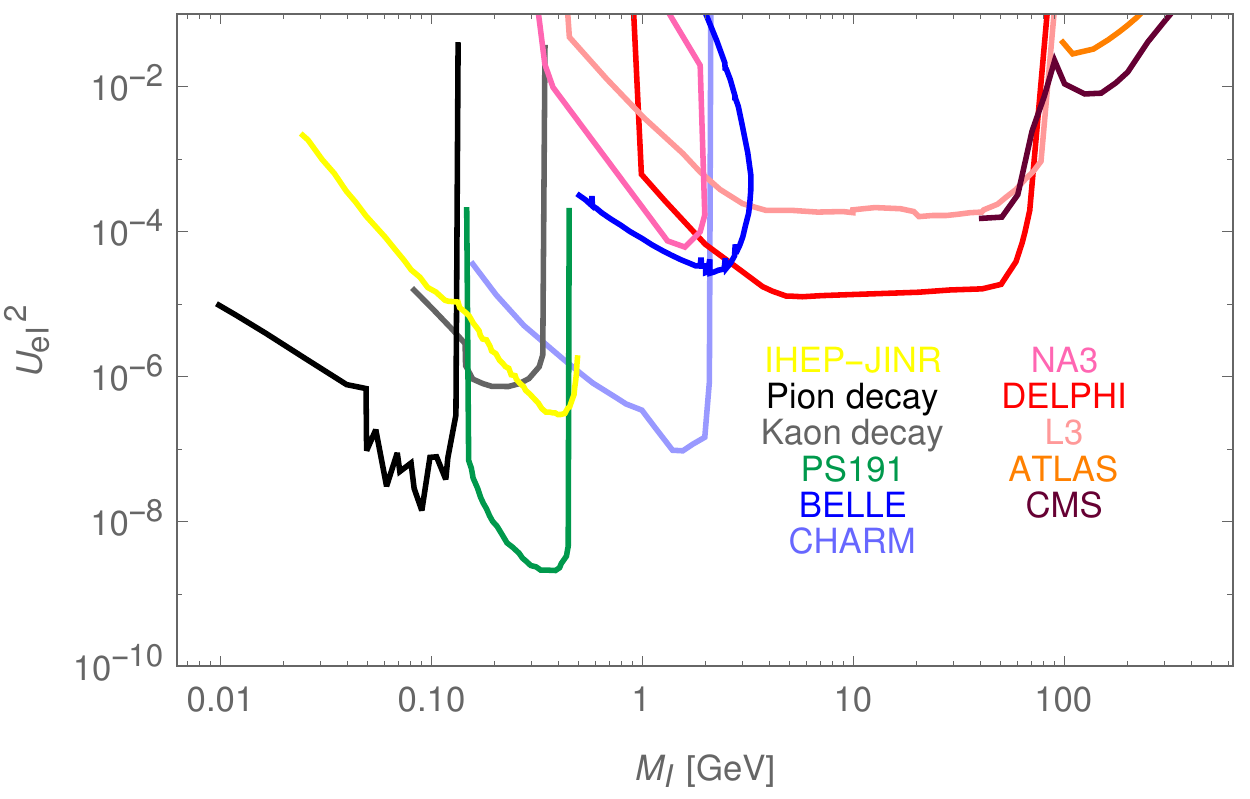}\\
\includegraphics[width=0.6\textwidth]{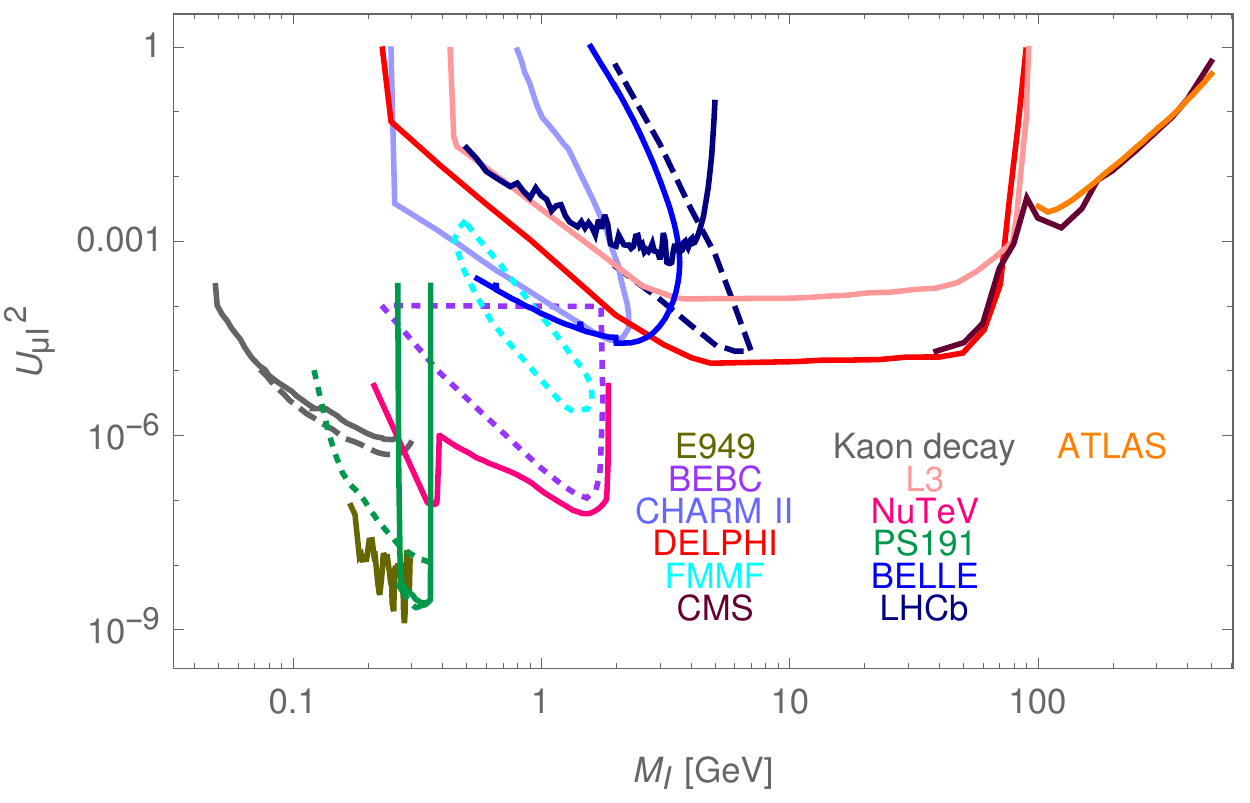}\\
\includegraphics[width=0.6\textwidth]{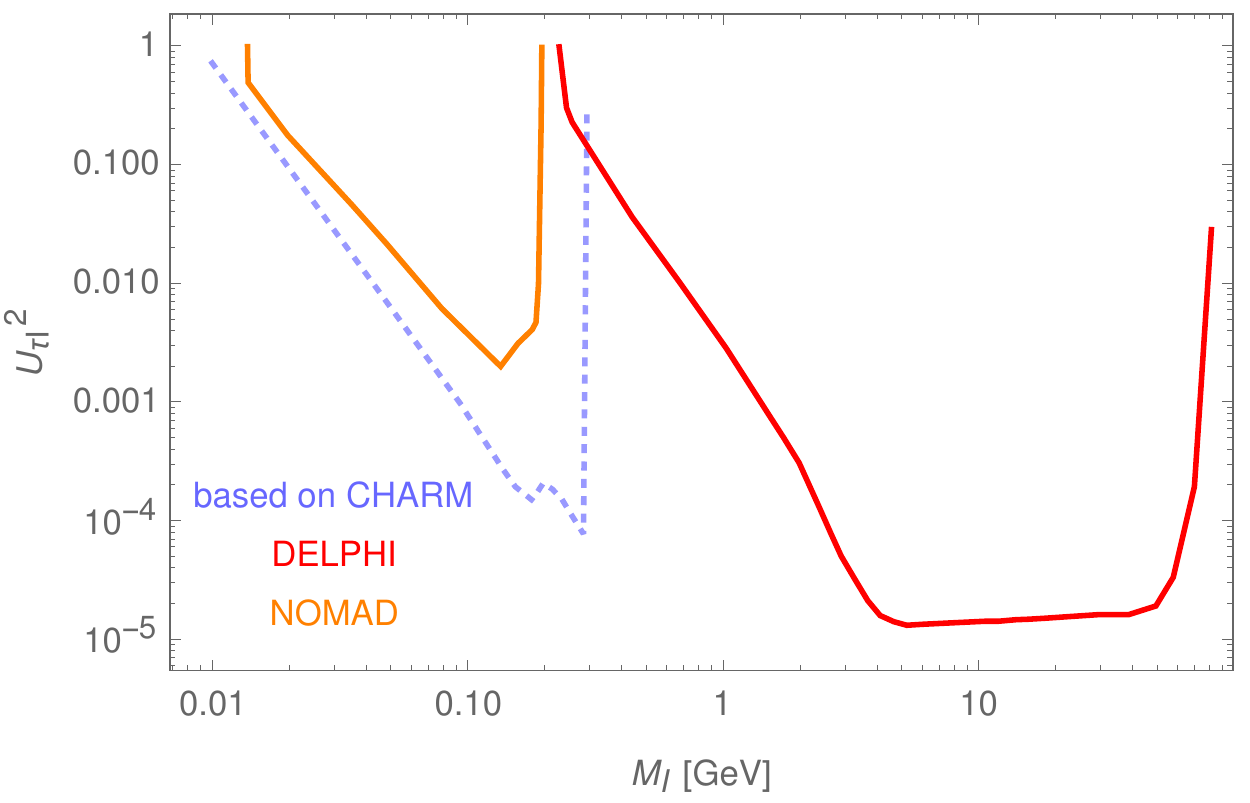}
\caption{\label{DirectSearches}
\emph{Upper panel}:
Constraints on $U_{e i}^2$ (versus $M_i$) from the experiments 
ATLAS \cite{Aad:2015xaa},
CMS \cite{Khachatryan:2016olu},
L3 \cite{Adriani:1992pq},
DELPHI \cite{Abreu:1996pa},
PIENU \cite{PIENU:2011aa},
BELLE \cite{Liventsev:2013zz} (as given in the erratum),
%TINA 
TRIUMF \cite{Britton:1992xv},
PS191 \cite{Bernardi:1987ek},
CHARM \cite{Bergsma:1985is},
NA3  \cite{Badier:1985wg}, %as estimated in Pascoli
and kaon decays \cite{Yamazaki:1984sj}. %andother list.
For peak searches below the kaon mass we show the summarized bound given in Ref.~\cite{Atre:2009rg}, 
for the PS191 experiment we show the re-interpretation given in \cite{Ruchayskiy:2011aa}.
\emph{Middle panel:}
Constraints on $U_{\mu i}^2$ from  
DELPHI \cite{Abreu:1996pa}, 
L3 \cite{Adriani:1992pq}, 
ATLAS \cite{Aad:2015xaa},
CMS \cite{Khachatryan:2015gha}, 
BELLE \cite{Liventsev:2013zz} (as given in the erratum), 
BEBC \cite{CooperSarkar:1985nh}, %as in Pascoli
FMMF \cite{Gallas:1994xp}, %as in Pascoli
E949 \cite{Artamonov:2014urb},
PIENU \cite{PIENU:2011aa}, 
TRIUMF %TINA 
\cite{Britton:1992xv},
PS191 \cite{Bernardi:1987ek},  
CHARMII \cite{Vilain:1994vg},
NuTeV \cite{Vaitaitis:1999wq}, 
NA3 \cite{Badier:1985wg}
and kaon decays in \cite{Yamazaki:1984sj,Hayano:1982wu}.
For LHCb we show the re-interpretation of the search for lepton number violating decays \cite{Aaij:2014aba} given in Ref.~\cite{Shuve:2016muy} (solid line) and the interpretation of the displaced vertex search \cite{Aaij:2016xmb} as given in Ref.~\cite{Antusch:2017hhu} (dashed line).
For the bounds from kaon decays we use the interpretation given in \cite{Kusenko:2004qc,Atre:2009rg}.
For NA3, BEBC and FMMF we use the estimates from \cite{Atre:2009rg}.
For the PS191 experiment we compare the re-interpretation in Ref.~\cite{Ruchayskiy:2011aa} (solid line) to that shown in Ref.~\cite{Artamonov:2014urb} for two different channels (dashed and dotted line). 
\emph{Lower Panel:}
The bounds on $U_{\tau i}^2$ are based on the interpretation of CHARM data given in \cite{Orloff:2002de}, NOMAD \cite{Astier:2001ck}, L3 \cite{Adriani:1992pq} and DELPHI \cite{Abreu:1996pa}.
Note that many of the experiments also constrain ratios of the $|U_{\alpha i}|^2$ or sums thereof in addition to what is shown here. Figure taken from \cite{Drewes:2015iva}.
}
\end{center}
\end{figure}

The results from different past searches are shown in Fig.~\ref{DirectSearches}.
In Fig.~\ref{MvsU2} we compare the sensitivity of different planned or proposed direct searches to the viable leptogenesis parameter region for $n=2$.

\subsubsection{Indirect Searches}

The properties of heavy neutrinos are not only constrained by the negative result of direct searches for them, but also by searches for deviations from the SM predictions in precision observables and rare processes. These are primarily sensitive to the properties of the $N_i$ because they modify the weak interactions of the SM leptons in Eq.~(\ref{WeakWW}). Various processes have been studied by a large number of different authors. A (possibly incomplete) list can e.g. be found in Refs.~\cite{Atre:2009rg,Kusenko:2009up,Antusch:2014woa,Gorbunov:2014ypa,Drewes:2015iva}, along with a (certainly incomplete) list of references.
The most important ones in the parameter regime considered here are 
\begin{itemize}
\item \emph{Neutrino oscillation data} - The strongest indirect constraints on $N_i$ properties come from neutrino oscillation data. The requirement to generate the observed light neutrino masses imposes a mass dependent lower bound on $U^2$, which provides a ``bottom line" for experimental searches.
The physical reason is that the right handed neutrinos cannot give mass to the SM neutrinos if they do not interact with them.
For $n=2$ this requirement also imposes a lower bound on the individual $U_\alpha^2$ \cite{Asaka:2011pb,Ruchayskiy:2011aa}, which can be seen in Fig.~\ref{MvsU2}. The lower  bound is considerably weaker for $n>2$ \cite{Gorbunov:2013dta,Drewes:2015iva}. 
In addition, neutrino oscillation data also constrains the relative size $U_\alpha^2/U^2$ of the heavy neutrino couplings to individual SM flavours, cf. Fig.~\ref{Fig:Lepto_NO}.
The use of the Casas-Ibarra parameterization (\ref{CasasIbarra}) automatically ensures consistency with neutrino oscillation data constraints at tree level. For mixing angles much larger than the ''naive seesaw estimate" (\ref{NaiveSeesaw}), loop corrections may be not negligible. They can be included by using the radiatively corrected Casas-Ibarra parameterization introduced in Ref.~\cite{Lopez-Pavon:2015cga}.
However, very large mixings can only be realized without fine-tuning in the symmetry protected scenarios described in Sec.~\ref{SymmetryProtectedScenario}, in which radiative corrections are also suppressed 
by the small the parameters $\upmu$ and $\upepsilon$ defined in Eq.~(\ref{B-L_parameters}).
\item \emph{$0\nu \beta \beta$ decay} - 
If one assumes that there is no  special structure in the matrices $\lambda$ and $M_N$, then the non-observation of $0\nu \beta \beta$ decay by far imposes the strongest bounds on $U_e^2$ for $N_i$ that are heavier than the kaon \cite{Atre:2009rg,Blennow:2010th}. 
However, like all lepton number violating observables, these constraints can easily be avoided in the symmetry protected scenario described in Sec.~\ref{SymmetryProtectedScenario}. 
For $n=2$ and  $\upmu,\upepsilon\ll 1$,
the effective Majorana mass $m_{\beta\beta}$ that affects the rate of the $0\nu \beta \beta$ decay can be expressed as
\begin{eqnarray}\label{oderplus}
|m_{\beta\beta}|_{\rm NH}&\simeq& \left| [1-f_A(\bar{M})]m_{\beta\beta}^\nu+f_A^2(\bar{M})\frac{\bar{M}^2}{\Lambda^2}
\frac{\upmu}{\upepsilon}
|\Delta m_{\rm atm}|
e^{-2 i(\theta+\delta)}\right|,  \\ \nonumber
|m_{\beta\beta}|_{\rm IH}&\simeq& \bigg| [1-f_A(\bar{M})]m_{\beta\beta}^\nu\\
&&+ \ f_A^2(\bar{M})\frac{\bar{M}^2}{\Lambda^2}
\frac{\upmu}{\upepsilon}
|\Delta m_{\rm atm}| \cos^2\uptheta_{13}\\
&&
\quad \times
e^{-2 i  \theta}(\xi e^{ i  \alpha_2/2}\sin \uptheta_{12}+ i  e^{ i \alpha_1/2}\cos \uptheta_{12})^2
\bigg|,\nonumber
\end{eqnarray}
with $\xi=\pm1$.
Here  $m_{\beta\beta}^\nu=\sum_i (U_\nu)_{ei}^2m_i$ is the standard contribution from light neutrino exchange. 
The function $f_A(M)\simeq \frac{\Lambda^2}{\Lambda^2+\bar{M}^2}$ quantifies the virtuality of the exchanged $N_i$, where $\Lambda^2\sim (100 \ {\rm MeV})^2$ is the momentum exchange in the decay, whose exact value depends on the isotope.
It is well-known that the contribution from $N_i$ exchange can be sizeable \cite{Bezrukov:2005mx,Blennow:2010th,Ibarra:2010xw,Asaka:2011pb,Mitra:2011qr,LopezPavon:2012zg},
but it was long believed that the requirement $\mu\ll 1$ in the context of leptogenesis suppresses the last term in Eq.~(\ref{oderplus}), so that $|m_{\beta\beta}|\simeq | [1-f_A(\bar{M})]m_{\beta\beta}^\nu|$ is insensitive to the heavy neutrino parameters (except $\bar{M}$) \cite{Bezrukov:2005mx,Asaka:2011pb,Gorbunov:2014ypa}. 
Recently it has been pointed out that this argument only applies to the model with $n=2$, where $\upmu\ll1$ is a necessary requirement for leptogenesis.
For $n=3$, no mass degeneracy is required \cite{Drewes:2012ma}, and a large $m_{\beta\beta}$ can easily be realized within the leptogenesis parameter region \cite{Drewes:2016lqo}. 
Moreover, even in the model with $n=2$, 
there exists a corner in the parameter space in which the observed baryon asymmetry can be reproduced while the term $\propto \mu/\epsilon$ in $m_{\beta\beta}$ dominates \cite{Drewes:2016lqo,Hernandez:2016kel,Asaka:2016zib}. 
A part of this region that is not constrained by any other experiment has been ruled out by the constraints from GERDA \cite{Agostini:2013mzu} and KAMLAND-Zen \cite{KamLAND-Zen:2016pfg} experiments, making them the strongest constraint on $N_i$ properties in this regime \cite{Drewes:2016lqo}.
Possibly even more importantly, this allows to constrain the parameter $\theta$ even if $\upmu$ is so small that direct searches cannot kinematically distinguish the two heavy neutrinos, at least in the region where $\Lambda/\bar{M}>0.1$ \cite{Hernandez:2016kel,Drewes:2016jae}.
\end{itemize}
All other indirect constraints are currently not sensitive to the small mixing angles required for leptogenesis. 
For completeness, we shall nevertheless mention the most important observables that can constrain the low-scale seesaw model with $M_i$ below the TeV-scale, even though most of these are very far from being able to constrain the parameter region where low-scale leptogenesis is possible.
\begin{itemize}
\item \emph{Lepton universality} - The existence of heavy neutrinos violates the universality of the strength at which different SM neutrinos couple to the weak interaction. This e.g. leads to deviations of the relative size of the branching ratios at which mesons decay into muon and electron final states. 
If one tries to explain the observed data within the minimal seesaw model, the best fit value actually suggests $U_e^2\neq0$ \cite{Antusch:2014woa}.
However, the significance of this preference is rather low, and it would be difficult to make this result consistent with other observables. 
\item \emph{CKM unitarity} - The CKM matrix is unitary in the minimal seesaw model. However, its elements are experimentally determined from processes involving leptons, whose branching ratios are modified by the effect of the heavy neutrinos, cf. Eq.~(\ref{WeakWW}). 
If $N_i$ exists and one interprets the data in the framework of the SM, then one will incorrectly conclude that there is a violation of unitarity in the mixing of quarks. In reality this could be of course the effect of the non-unitarity in the light neutrino mixing matrix $U_\nu$.
The constraints from CKM data are currently sub-dominant, but almost saturated by the largest mixings $|U_{\alpha i}|^2$ allowed by direct searches across the entire mass range considered \cite{Drewes:2015iva}. 
\item \emph{Electroweak precision data} - The modification of the weak currents (\ref{WeakWW}) also affects the relations between the weak gauge boson masses, the Weinberg angle, the observed Fermi constant and the fine structure constant at the $Z$ pole. These relations are measured with high precision. 
In Ref.~\cite{Akhmedov:2013hec} it was reported that data provides tentative evidence for the existence of TeV-scale heavy neutrinos. In most of the mass range below the $W$ boson we consider here, the constraints from electroweak precision data are sub-dominant compared to direct searches. 
Also observations of the invisible Z decay width are not sensitive to heavy neutrinos that are much lighter than the Z boson because, as a result of the unitarity of the total (light and heavy) neutrino mixing matrix or ``GIM suppression", the additional decay channels into $N_i$ are exactly compensated by a reduction of the decay amplitude into light neutrinos (as long as $M_i$ can be kinematically neglected).
However, for masses near or above the W mass, where the constraints from DELPHI and L3 begin to fade (cf. Fig.~\ref{MvsU2}), electroweak precision data currently imposes the strongest upper bound on $U_\alpha^2$. 
%near mW
\item \emph{Flavor violating lepton decays} - The lepton flavor violating interactions with the $N_i$ mediate lepton flavor violating decays, such as $\mu\rightarrow e \gamma$, $\mu\rightarrow eee$, $\tau \rightarrow \mu \gamma$ and $\tau \rightarrow e \gamma$. 
These processes are subject of the GIM suppression as long as $M_i$ is much smaller than the weak gauge boson masses. For larger masses, where the GIM suppression does not apply and the direct search bounds are weaker, the upper bound on the $U_\alpha^2$ from non-observation of these lepton flavor violating decays are comparable to those from electroweak precision observables.
At the moment $\mu\rightarrow e\gamma$ is the process that has been measured most precisely. In the near future experiments measuring $\mu \rightarrow e$ conversion in nuclei will become a serious competitor \cite{Alonso:2012ji}, though it seems unlikely that they can reach the baryogenesis region in the minimal $n=2$ model \cite{Canetti:2013qna}.
\item \emph{Lepton dipole moments} - In Refs.~\cite{Abada:2015trh,Abada:2016awd} it has been pointed out that heavy neutrinos with masses above the W mass can make a sizable contribution to the electric dipole moments of charged leptons. 
\end{itemize}

\subsubsection{Cosmological constraints}
Leptogenesis from neutrino oscillations in principle can generate the observed baryon asymmetry for $M_i$ as small as $10$ MeV \cite{Canetti:2010aw,Canetti:2012vf,Canetti:2012kh}.
However, such small masses are strongly disfavored in the context of the seesaw mechanism because they lead to $N_i$ lifetimes that are long enough that their decay in the early Universe disturbs Big Bang Nucleosynthesis (BBN) \cite{Dolgov:2000jw,Dolgov:2000pj} unless their mixing angles are very large. 
More precisely, the requirement to decay before nucleosynthesis imposes a lower bound on $U_i^2=\sum_a |U_{\alpha i}|^2$ for given $M_i$, which becomes stronger than the lower bound from neutrino oscillation data for $M_i$ below a few hundred MeV. The exact point where the bound from nucleosynthesis becomes stronger depends on the flavor structure and neutrino mass hierarchy. Combining these consideration with the negative result of various past experimental searches \cite{Atre:2009rg,Ruchayskiy:2011aa,Drewes:2015iva,Drewes:2016jae}, one finds that the lower bound on $M_i$ is roughly 100 MeV \cite{Ruchayskiy:2012si,Asaka:2011pb,Hernandez:2014fha} in the minimal model with $n=2$. 
To reach this conclusion, one has to bear in mind that right handed neutrinos that decay after BBN leave an imprint in the effective number of relativistic degrees of freedom $N_{\rm eff}$, either because they are relativistic during BBN or because their decay disturbs the phase space distribution of SM particles at later times, which would have a similar effect on the properties of temperature fluctuations in the Cosmic Microwave Background as extra relativistic particles. 
A more careful analysis reveals that this conclusion can be avoided for some of the heavy neutrinos for $n>2$ \cite{Hernandez:2014fha}. However, the states with masses below 100 MeV are required to have very feeble mixing angles $U_{\alpha i}$ in order to avoid thermalization in the early Universe. While this opens up the interesting possibility that these particle compose the Dark Matter \cite{Dodelson:1993je,Shi:1998km}, which is e.g. realized in the \emph{Neutrino Minimal Standard Model} ($\nu$MSM) \cite{Asaka:2005an,Asaka:2005pn} (see Ref.~\cite{Boyarsky:2009ix} for a review), it means that the contribution of these states to the seesaw mechanism and to leptogenesis is negligible. We will therefore not further discuss this case and only consider masses $M_i>100$ MeV. For a recent review of the sterile neutrino DM scenario, we refer the interested reader to Ref.~\cite{Adhikari:2016bei}.
Leptogenesis does not provide an upper bound on the magnitude of the $M_i$, but heavy neutrinos with masses heavier than a few hundred GeV cannot be produced in sizable numbers at any existing experiment. 
%then along lines of September

\subsubsection{Global Constraints} Within a given model, it is possible to obtain ``global constraints" on the properties of the heavy neutrinos by combining the various observables mentioned above.  
Various authors have done this under different assumptions about the data sets that should be included, the number $n$ of heavy neutrinos, the mass $m_{\rm lightest}$ of the lightest neutrinos and the validity of cosmological constraints (which obviously depend on at least moderate assumptions about the thermal history of the universe) 
\cite{Atre:2009rg,
Blennow:2016jkn,
Kusenko:2009up,
Ibarra:2011xn,
Ruchayskiy:2011aa,
Asaka:2011pb,
Abazajian:2012ys,
Canetti:2012kh,Canetti:2012vf,
Asaka:2013jfa,
Abada:2013aba,
Drewes:2013gca,
Hernandez:2014fha,
Antusch:2014woa,
Asaka:2014kia,
Gorbunov:2014ypa,
Abada:2014vea,
Abada:2014kba,
Abada:2015oba,
Drewes:2015iva,
Deppisch:2015qwa,
Escrihuela:2015wra,
Fernandez-Martinez:2015hxa,
deGouvea:2015euy,
Lopez-Pavon:2015cga,
Fernandez-Martinez:2016lgt,
Rasmussen:2016njh,
Abada:2016awd,
Drewes:2016lqo,
Drewes:2016jae,
Das:2017nvm}.
Most of these efforts have been focused on the minimal model with $n=2$, where the number of free parameter is rather low and the combination of different bounds imposes considerably stronger constraints than simply superimposing them in the $M_N$-$U_\alpha^2$ plane. The gray area in Fig.~\ref{MvsU2} corresponds to the global constraints for $n=2$ as reported in Ref.~\cite{Drewes:2016jae}.
It is shown along with the minimal and maximal $U_\alpha^2$ compatible with successful leptogenesis.
In Ref.~\cite{Drewes:2015iva} a similar analysis has been performed for $n=3$. The leptogenesis parameter space has not been explored systematically for $n=3$ to date.

\subsection{Probing CP violation in meson decays}
\label{ch5:sec2:CPMesonDecays}
\begin{figure}[htb] %\unitlength=1mm
\begin{minipage}[b]{.48\linewidth}
\includegraphics[width=59mm]{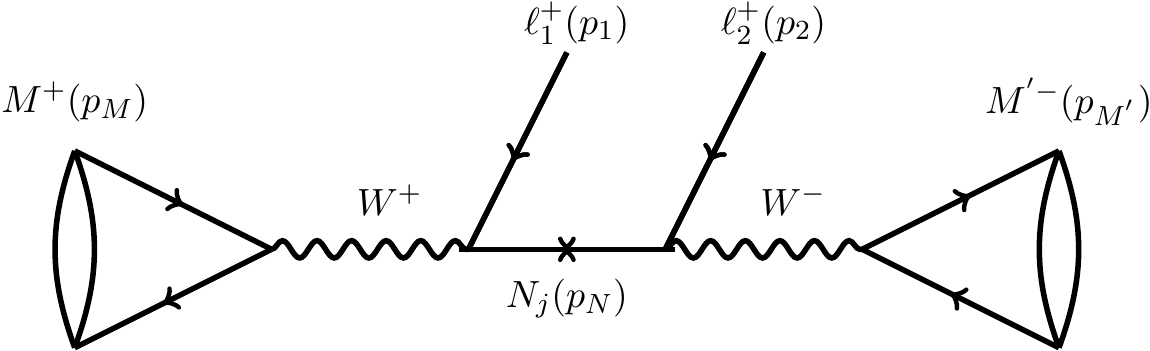}
\end{minipage}
\begin{minipage}[b]{.48\linewidth}
\includegraphics[width=59mm]{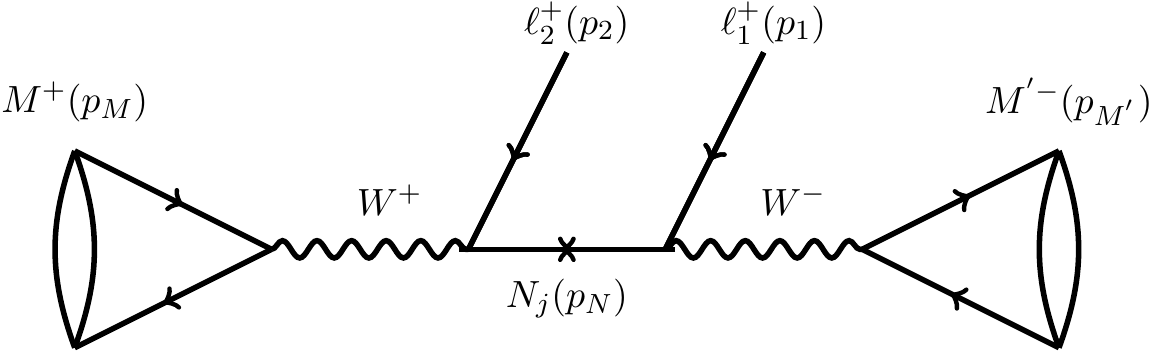}
\end{minipage}
\caption{Diagrams contributing to the LNV semileptonic meson decay
$M^+(p_M) \to \ell^+_1(p_1) \ell^+_2(p_2) M^{' -}(p_{M'})$ mediated
by a Majorana neutrino:  direct (\textbf{a}) and  crossed (\textbf{b}) channels.}
\label{fig2:FigMMp}
\end{figure}

We investigate the possibility of measuring CP violation in the rare
lepton number violating (LNV) semileptonic decays of
charged pseudoscalar mesons within the low-scale type I seesaw extension of the SM.
In particular, we consider    $M^{\pm} \to \ell_1^{\pm} \ell_2^{\pm} M^{' \mp}$,
where $M=K, D, D_s, B, B_c$ and $M^{'}=\pi, K, D, D_s$. The charged
leptons are $\ell_{1,2} = e, \mu$. The amplitude of this process receives two contributions, 
depicted by the direct (\textbf{a}) and crossed (\textbf{b}) channels in \fref{fig2:FigMMp}.

\begin{figure}[t]
\centering\includegraphics[width=90mm]{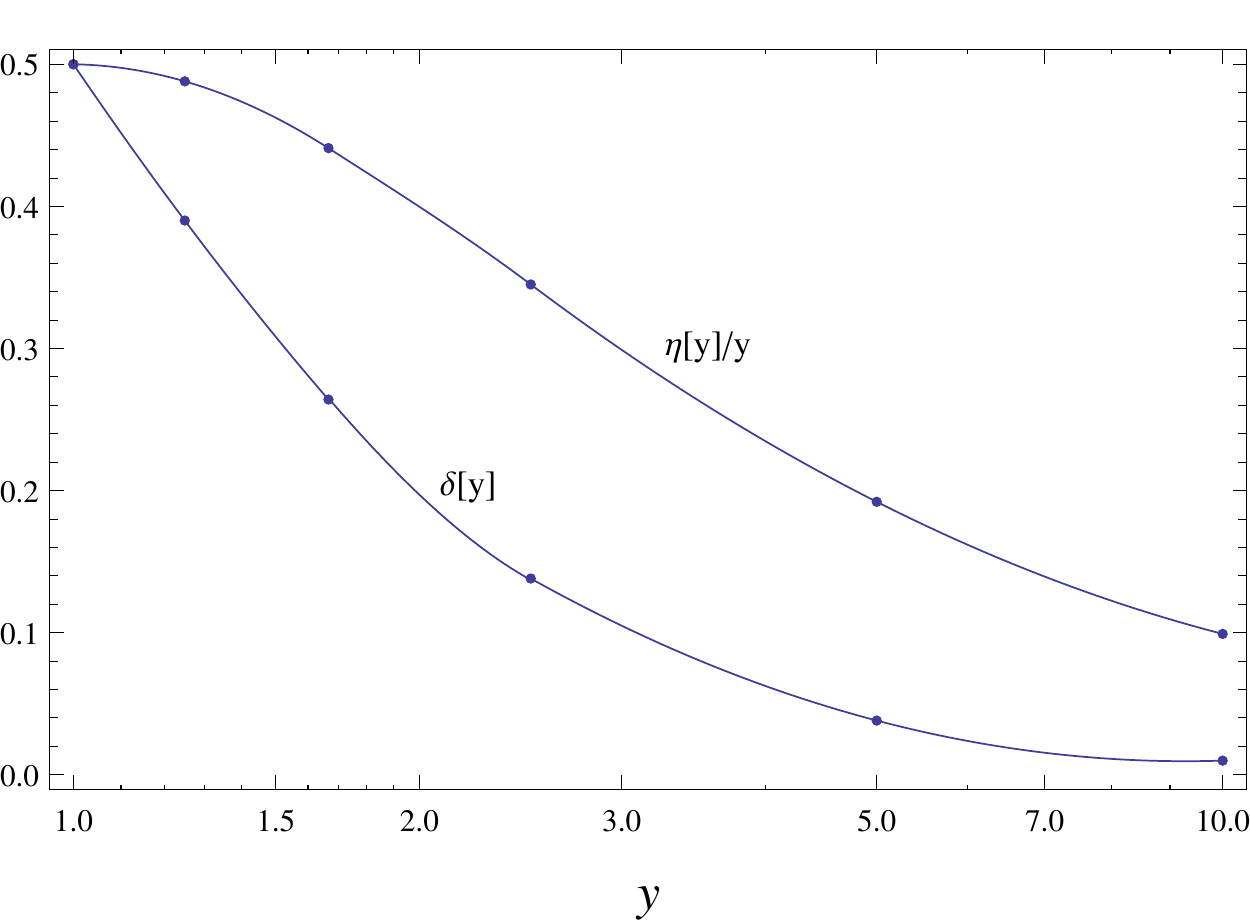}
\caption{The suppression factors $\eta(y)/y$ and $\delta(y)$, due to the
overlap of the propagators of $N_1$ and $N_2$,
as a function of $y\equiv 2|M_2-M_1|/(\Gamma(N_1)+\Gamma(N_2))$. Figure taken from \cite{Cvetic:2014nla}.}
\label{fig2:etadelfig}
\end{figure}

The  signal of CP violation in such decays is enhanced when the heavy Majorana neutrinos are on-shell and almost degenerate in mass.\footnote{Similar effects can be observed in decays as presented in \cite{Cvetic:2013eza,Cvetic:2015naa,Cvetic:2015ura,Zamora-Saa:2016qlk,Zamora-Saa:2016ito}.}
For concreteness, we consider below scenarios  with only two  sterile neutrinos
$N_1$ and $N_2$.
Then, the CP asymmetry in meson semileptonic decays is defined as follows \cite{Cvetic:2014nla,Cvetic:2015naa}
\begin{equation}
\begin{split}
\epsilon _{M} &
\equiv \;
\frac{ \Gamma(M^- \to \ell_1^- \ell_2^- M^{' +}) - \Gamma(M^+ \to \ell_1^+ \ell_2^+ M^{' -})}
{ \Gamma(M^- \to \ell_1^- \ell_2^- M^{' +}) + \Gamma(M^+ \to \ell_1^+ \ell_2^+ M^{' -})} \\
&= \; \frac{ \sin \tilde{\theta}_{21} }{
\left\{
\frac{1}{4} \left[
\kappa_{\ell_1} \kappa_{\ell_2} \left(1 + \frac{\Gamma(N_1)}{\Gamma(N_2)} \right)
+ \frac{1}{\kappa_{\ell_1} \kappa_{\ell_2}} \left(1 + \frac{\Gamma(N_2)}{\Gamma(N_1) } \right)
\right]  + \delta[y] \cos \tilde{\theta}_{21}
\right\}
}
\;  \frac{\eta[y]}{y} \ ,
\label{eq2:ACP}
\end{split}
\end{equation}
where $\Gamma(N_j)$ is the total decay width of the heavy Majorana neutrino $N_j$ ($j=1,2$). The relevant CP-violating phase $\tilde{\theta}_{21}$  in  \eref{eq2:ACP} is defined in terms of the phases of the active-sterile neutrino mixing, $U_{\ell j}\equiv |U_{\ell j}|e^{i \phi_{\ell j}}$, 
i.e. $\tilde{\theta}_{21}\equiv \phi_{\ell_1 2}+ \phi_{\ell_2 2}- \phi_{\ell_1 1}- \phi_{\ell_2 1}$. We also introduce the effective couplings  $\kappa_{\ell_i}\equiv |U_{\ell_i 2}|/ |U_{\ell_i 1}|$ and
the overlap functions $\delta[y]$ and $\eta[y]/y$, where  $y\equiv 2|M_2-M_1|/(\Gamma(N_1)+\Gamma(N_2))$.
The latter are computed numerically and shown in  \fref{fig2:etadelfig} as a function of $y$ in the range $1\leq y\leq 10$. It turns out that $\delta[y]$ and $\eta[y]$ are in practice independent of
the type of the pseudoscalar meson ($M^\pm$, $M^{\prime\mp}$) and of the final lepton flavor ($\ell_{1,2}=e,\mu$)  \cite{Cvetic:2014nla}.
We see that when the difference of the masses of the two sterile neutrinos becomes small enough,
comparable to the total decay widths of these neutrinos,
i.e. $y\approx 1$, a large CP asymmetry in meson decays, $|\epsilon_M| \lesssim 1$, is possible, provided 
CP-violating difference $\tilde{\theta}_{21}$ of the phases of
active-sterile neutrino mixings is not suppressed.  
In experiments one has to consider the acceptance (suppression) factor in the detection of these decays. It
results from the small length of
the detector in comparison to the relatively large
lifetime of the (on-shell) sterile neutrinos $N_j$.
Most of the on-shell neutrinos, produced via the
decay $M^{\pm} \to \ell_1^{\pm} N_j$,
could indeed survive a long enough time to decay (into $\ell_2^{\pm} M^{' \mp}$) outside the
detector.
This effect suppresses the number of detected decays
and has to be taken into account. For details of regarding the acceptance factor, we refer to \cite{Cvetic:2012hd,Bonivento:2013jag,Helo:2013esa,Cvetic:2013eza,Dib:2014iga}.
This effect may be absent only if $N_j$ is heavy ($M_j \gtrsim 3$ GeV) and thus can decay fast, i.e., in the rare decays of $B^{\pm}$ and $B_c^{\pm}$.

\section{General mechanisms for low scale leptogenesis from out-of-equilibrium decays}
%\section{Conditions for testable leptogenesis}
% I put it as a working title so far. 
\label{ch5:sec1:concepts}
After focusing on probing GeV-scale leptogenesis via oscillations, we want to review the testability of leptogenesis from out-of-equilibrium decays. As already described in \sref{ch5:sec0:intro}, in the latter type of leptogenesis an asymmetry is created by $L$-violating out-of-equilibrium decays of new heavy degrees of freedom $X$ that provide a new source of CP violation via loop effects. As long as these decays generate a net asymmetry above the electroweak scale, when electroweak sphalerons are in thermal equilibrium, baryogenesis via leptogenesis~\cite{Fukugita:1986hr} becomes possible. For more detailed reviews about leptogenesis from out-of-equilibrium decays, the reader is referred to Refs.~\cite{Buchmuller:2004nz,Trodden:2004mj,Strumia:2006qk,Chen:2007fv,Davidson:2008bu} and references therein.

Before addressing the possibility of probing specific models, we review in this section the main difficulties which arise in order to achieve thermal baryogenesis from heavy particle decays at low temperatures (around or below the TeV scale) and briefly explain the three known mechanisms to avoid the most crucial one. Let us note that the discussion also applies to baryogenesis from particle annihilations. 

Thermal baryogenesis from particle decays or annihilations at low energy scales faces two main problems: 
\begin{itemize}
\item[(a)] \emph{Small neutrino masses.} The CP violation relevant for baryogenesis is parameterized by the CP asymmetry $\epsilon$ per decay (or annihilation) of heavy particles. In some leptogenesis models $\epsilon$ is proportional to the masses of both, the light (mostly active) neutrinos and the heavy ones. Given the tiny value of light neutrino masses (compared to the electroweak scale), the leptogenesis scale set by the mass of the heavy neutrinos has to be very high in order to generate a large enough lepton asymmetry. Let us study this in more detail, considering the vanilla mechanisms of type I, type II, and type III seesaw, where the respective new degrees 
of freedom are singlet fermions, $SU(2)_L$ triplet scalars and $SU(2)_L$ triplet fermions, respectively. Here, the CP-violating decays $X \to \ell \phi$ (type I and III) or $X \to \ell \ell$ (type II), generate a CP asymmetry defined by
%%%
\begin{eqnarray}
\epsilon^{\rm I,III} &=& \frac{\Gamma(X \to \ell \phi) - \Gamma(X \to \ell^c \phi^c)}
{\Gamma(X \to \ell \phi) + \Gamma(X \to \ell^c \phi^c)}, \label{eq3:epsI/III}\\
\epsilon^{\rm II} &=& \frac{\Gamma(X \to \ell \ell) - \Gamma(X \to \ell^c \ell^c)}
{\Gamma(X \to \ell \ell) + \Gamma(X \to \ell^c \ell^c)},\label{eq3:epsII}
\end{eqnarray}
%%%
where $\ell, \phi$ indicate the physical states and  $\ell^c, \phi^c$ their charged conjugate. $\Gamma (a)$ denotes the decay width for process $a$.
Nonzero $\epsilon^{\rm I,II,III}$ can appear first at one loop. 
From dimensional analysis, the CP parameter originating from the decay of the lightest degree of freedom 
with mass $M_X$ can be estimated as
%%%
\begin{equation}
|\epsilon^{\rm I,II,III}| \sim \frac{1}{16\pi}\frac{m_3 M_X}{v^2}\,,
\label{eq3:general_CP_parameter}
\end{equation}
%%%
where $m_3$ denotes the largest of the light neutrino masses.

\parindent=5mm The generated lepton asymmetry can be parameterized as
%%%
\begin{equation}
Y_{\Delta L} = \eta \,\epsilon^{\rm I,II,III} \, Y_X^{\rm eq},
\label{eq3:lepton_asymmetry}
\end{equation}
%%%
where $Y_{a} \equiv n_a/s$ denotes the number density of $a$ normalized to
the entropy density $s$, $\eta \leq 1$ is the efficiency factor taking
into account possible washout, and $Y_X^{\rm eq} \sim 10^{-3}$ indicates
the thermal abundance.

\indent From electroweak sphalerons, the induced baryon asymmetry is approximated by
$Y_{\Delta B} \sim -Y_{\Delta L}/3$. The precise relation depends
on the degrees of freedom in the thermal bath during the freeze-out of the
electroweak sphalerons. Requiring $|Y_{\Delta B}| \gtrsim 10^{-10}$,  
in accordance with observations, we obtain the lower bound
on the mass of the heavy new degrees of freedom%%%
\begin{equation}
M_X \gtrsim {\cal O}(1) \times 10^{10} 
\,\left(\frac{0.1}{\eta}\right)
\,\left(\frac{0.1\,{\rm eV}}{m_\nu}\right)\,{\rm GeV}.
\label{eq3:leptogenesis_mass_bound}
\end{equation}
%%%
\indent This back-of-the-envelope estimation \cite{Barbieri:1999ma,Buchmuller:1998zf,Hambye:2001eu,Branco:2002kt} agrees with the exact bound firstly
derived by Davidson and Ibarra in the context of type I
leptogenesis, cp. \eqref{eq0:davidsonibarra}~\cite{Davidson:2002qv}.  The main difference between type I compared to type II and III leptogenesis is that the heavy degrees
of freedom in the former are singlet under $SU(2)_L$, while those for
the latter two are triplets and hence they will be maintained close to
thermal equilibrium by gauge interactions. Nevertheless, the bound
obtained in these models does not differ very much from the estimation in
\eref{eq3:leptogenesis_mass_bound}.  For a more detailed discussion the reader is referred to \sref{ch5:sec3:extended_seesaw-typeII} and the reviews~\cite{Hambye:2012fh,Fong:2013wr}. %, see section 6.4 and 6.5 in 
%a recent review~\cite{Fong:2013wr}). 
The bound on the mass of the new heavy degree of freedom, can be relaxed if lepton flavor
effects~\cite{Barbieri:1999ma,Abada:2006fw,Nardi:2006fx,Abada:2006ea}
are taken into account. In the context of type I leptogenesis, it was
shown in Ref.~\cite{Racker:2012vw} that the bound can be lowered to $10^6$ GeV.

\indent Although present in well motivated scenarios, this problem is not generic and does not arise in neutrino mass models like the inverse seesaw or the radiative ones. 

\item[(b)] \emph{Interplay between CP asymmetry and washout.} This problem is intrinsic to all models for thermal baryogenesis from particle decays (or annihilations). In order to generate a CP asymmetry, both a CP odd and a CP even phase are required. The former comes from complex couplings while the latter from the absorptive parts of one-loop contributions to the decay processes, which are always proportional to the amplitude of $B$ or $L$-violating scatterings~\cite{Nanopoulos:1979gx}, as depicted in \fref{fig5.1:cpasym}. Therefore the size of the CP asymmetry is tied to the strength of processes capable of erasing the asymmetry when they are faster than the expansion rate of the Universe $H$. Given that $H$ decreases with the temperature $T$ ($H \propto T^2/M_{\rm Pl}$), it is difficult to have both, small washout and large CP asymmetries (see \fref{fig5.1:washout}) when the baryogenesis scale is very low relative to $M_{\rm Pl}$.
\begin{figure}[t]
\begin{center}
\includegraphics[width=1.0 \textwidth]{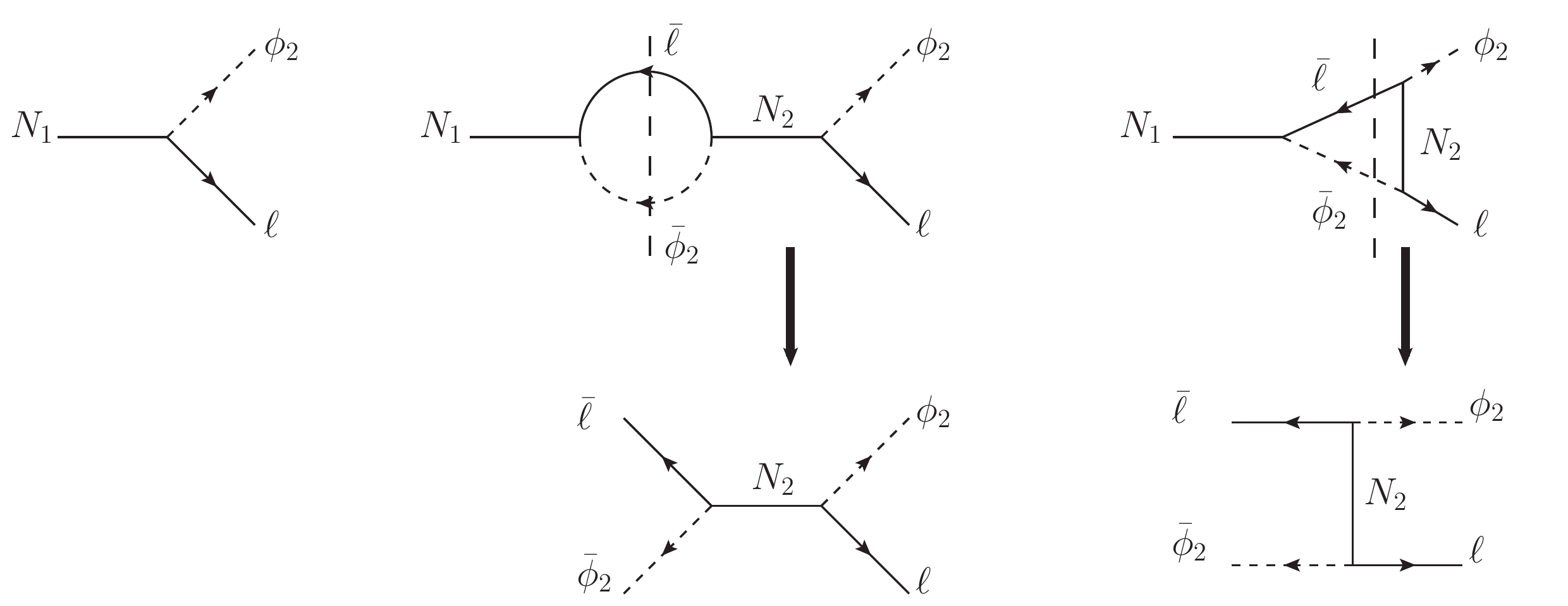}
\end{center}
\caption{At leading order the CP asymmetry in $N_1$ decays arises from the interference of tree-level and one-loop diagrams (upper part). The vertical cuts through the loops point out that the propagating particles can go on shell -a requisite for having CP even phases-. This entails the existence of $L$-violating processes, which might be very efficient at erasing the lepton asymmetry (lower part). Here the inert doublet model with sterile neutrinos has been used as an example.}
\label{fig5.1:cpasym}
\end{figure}
In this review, this issue will be discussed more specifically in \sref{ch5:sec3:baryogenesis-scale}, taking an extension to type I seesaw as an example. In \cite{Racker:2013lua}, the tight connection between CP asymmetry  and washout was discussed extensively within the Inert Doublet Model (IDM) with singlet neutrinos~\cite{Ma:2006km},\footnote{In this model, the SM is extended by a Higgs doublet, $\phi_2$, and some sterile neutrinos, $N_k$, with masses $M_k$. The new fields are odd under a conserved $Z_2$ discrete symmetry.} which circumvents problem (a) by having radiatively generated neutrino masses. The lower limit on the heavy neutrino mass was found to be $M_1 \gtrsim \mathcal{O}\bigl(10^2\bigr)$~TeV when none of the low scale mechanisms for baryogenesis explained next are incorporated. As discussed in~\cite{Racker:2013lua}, this bound is arguably quite model independent, e.g. it is similar within an order of magnitude in the inverse seesaw model~\cite{Racker:2012vw} and for cloistered baryogenesis discussed in \sref{ch5:sec3:baryogenesis-scale}. 
\begin{figure}[t]
\begin{center}
\includegraphics[width=0.5\textwidth,angle=270]{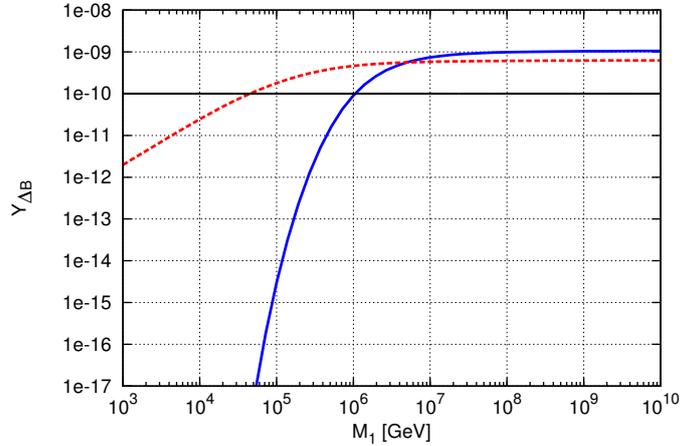}
\end{center}
%\vspace{-0.8cm}
\caption{The final baryon asymmetry in the inert doublet model normalized to the entropy density, $Y_{\Delta B}$, plotted as a function of $M_1$, including the washout scattering processes $\phi_2 \ell \leftrightarrow \bar \phi_2 \bar \ell$ and $\ell \ell \leftrightarrow \bar \phi_2 \bar \phi_2$ (solid blue curve) or neglecting them (dashed red curve). Here the mass of $\phi_2$, $M_{\phi_2}$, is taken to be negligible compared to $M_1$. The value of $M_2$ and the Yukawa couplings are chosen in order to keep $\epsilon_1$ and $\frac{\Gamma_1}{H(T=M_1)}$ constant for all $M_1$ (with $\Gamma_1$ being the $N_1$ decay width). It is apparent that for $M_1 \lesssim 10^7$~GeV, $Y_B$ falls exponentially as $M_1$ decreases. However if the washout is omitted, as sometimes done -wrongly- in the literature, the baryon asymmetry only decreases linearly and its value at the TeV scale is many orders of magnitude above the true one. This figure has been taken from Ref.~\cite{Racker:2014yfa}.}
\label{fig5.1:washout}
\end{figure}
\end{itemize}

In order to allow for leptogenesis at or below the TeV scale, the aforementioned two problems have to be addressed. In the following we would like to discuss the three known ways to avoid problem (b), which is the one common to all scenarios of thermal baryogenesis from particle decays or annihilations:
\begin{itemize}
\item {\it Almost degenerate particles}: In case of decaying particles almost degenerate in mass, the CP asymmetry can be enhanced up to $O(1)$ values~\cite{Flanz:1996fb,Covi:1996fm,Pilaftsis:1997jf}. As an example, let us consider two sterile neutrinos $N_{k}\, (k=1,2)$, with masses $M_{k}$, Yukawa couplings $\lambda_{\alpha k}$, and decay widths $\Gamma_k$. When the resonant condition $\Delta M \equiv M_2 - M_1 = \Gamma_2/2$ is fulfilled, the CP asymmetry in $N_1$ decays, $\epsilon_1$, can be as large as 1/2. But even if this condition does not hold exactly, the CP asymmetry is enhanced by a large ratio of $1/\delta \equiv M_1/\Delta M$, $\epsilon_1 \propto (\lambda^\dag \lambda)_{2 2}/\delta$, as long as  $\Gamma_{1,2} \ll \Delta M \ll M_1$.  This way, the key washout processes can be diminished by choosing small values of $(\lambda^\dag \lambda)_{2 2}$, while compensating the associated decrease in $\epsilon_1$ with a tiny value for $\delta$.\footnote{The required level of degeneracy $\delta$ when taking flavor effects into account has been studied in detail in Ref.~\cite{Racker:2013lua}. It dependens on the hierarchy among Yukawa couplings and on $M_1$.} This approach, known as ``resonant leptogenesis'' mechanism is thoroughly discussed in the accompanying chapter of this review \cite{leptogenesis:A03}. 

\item  {\it Late decay}: The washout processes inherent to the CP asymmetry in $N_1$ decays have a rate proportional to $(T/M_2)^a$ for $T \lesssim M_2$, where $a=2\, (4)$ for a Majorana (scalar or Dirac) mediator with mass $M_2$. Rising $M_2$ relative to $M_1$ is not helpful for lowering the baryogenesis scale, since both, washout and CP asymmetry, decrease by the same amount. Nevertheless, the previous relation shows that the washout effects can be diminished by letting $N_1$ decay very late. Two conditions must be fulfilled for this ``late-decay'' mechanism to be successful:
\begin{enumerate}
\item a small decay width $\Gamma_1 \ll H(T=M_1)$ such that the $N_1$'s begin to decay at $T \ll M_1$.
\item an interaction to populate the Universe with the $N_1$'s which is different from the one inducing the decay (as this one is too weak in order to fulfill (1)), e.g.~a new gauge interaction~\cite{Plumacher:1996kc,Racker:2008hp}. This interaction should be effective only at $T \gtrsim M_1$ such that the $N_1$'s disappear via the CP-violating processes. 
\end{enumerate}
With this setup, it is possible to achieve TeV-scale leptogenesis (as well as baryogenesis at much lower scales when the decays violate $B$, see~\cite{Racker:2013lua} for more details). 

\item {\it Massive decay (annihilation) products}: In general the washout rate depends not only on the amplitude of the corresponding process, but also on the number density of the particles involved. Therefore, if one of these particles has a mass such that it becomes non-relativistic in the baryogenesis era, the washout is Boltzmann suppressed. Under the non-trivial conditions explained next, this mechanism of washout suppression was demonstrated to enable baryogenesis at the TeV scale from DM annihilation in~\cite{Cui:2011ab} and from heavy particle decays later in~\cite{Racker:2013lua}. It also allows for baryogenesis well below the electroweak scale if $B$ is violated perturbatively.
%However, one has to be careful in applying this statement.\\

\parindent=5mm For clarity purposes we consider a specific model, the IDM with sterile neutrinos heavier than the exotic Higgs doublet $\phi_2$. The decays of the heavy neutrino $N_1$ generate an asymmetry in both, leptons and  $\phi_2$ (which plays the role of the massive decay product with mass $M_{\phi_2}$).  In turn, the asymmetries in each of these sectors ($\ell$ and $\phi_2$) induce washout of the lepton asymmetry (e.g.~more $\ell$'s than $\bar \ell$'s, but also more $\phi_2$'s than $\bar \phi_2$'s, lead to an increase of the rate of $\ell \phi_2 \rightarrow \bar \ell \bar \phi_2$ relative to $\bar \ell \bar \phi_2 \rightarrow \ell \phi_2$). Under kinetic equilibrium this observation leads to the following terms in the Boltzmann equation
\begin{eqnarray}
\label{eq5.1:wo}
\frac{\mathrm{d} Y_{\Delta_\ell}}{\mathrm{d} z} &=& \frac{-1}{sHz} \left\{ \frac{Y_{\Delta_\ell}}{Y_\ell^{\rm eq}} + \frac{Y_{\Delta \phi_2}}{Y_{\phi_2}^{\rm eq}} \right\} \gamma^{\rm eq}\left(\ell \phi_2 \leftrightarrow \bar \ell \bar \phi_2\right)+ \dots \\
&\equiv & \frac{-Y_{\Delta_\ell}}{sHz} \left[w_1(z) + w_2(z) \right]+ \dots \label{eq5.1:wob}
\end{eqnarray}
with $Y_{\Delta X} \equiv Y_X - Y_{\bar X} \equiv \tfrac{n_X - n_{\bar X}}{s}$, $z \equiv M_1/T$, and $\gamma (\dots)$ is the density rate of the corresponding process. For simplicity we omit flavor indices. The superscript ``eq'' denotes the value of a given quantity for equilibrium phase space distributions with zero chemical potentials. To understand the washout effects on $Y_{\Delta_\ell}$, we have introduced the functions $w_i(z)$ in \eref{eq5.1:wob}. While $w_1(z) \propto \gamma^{\rm eq}(\ell \phi_2 \leftrightarrow \bar \ell \bar \phi_2)/Y_{\ell}^{\rm eq}$ and is thus Boltzmann suppressed by $e^{-M_{\phi_2}/T}$, the case of $w_2(z)$ is more tricky as its effect depends on how $Y_{\Delta \phi_2}$ is related to $Y_{\Delta \ell}$. When $Y_{\Delta \phi_2}$ and $Y_{\Delta \ell}$ are linked by a conservation law implying $Y_{\Delta \phi_2} = {\rm number} \times Y_{\Delta \ell}$, then $w_2(z) \propto \gamma^{\rm eq}(\ell \phi_2 \leftrightarrow \bar \ell \bar \phi_2)/Y_{\phi_2}^{\rm eq}$, and consequently there is no Boltzmann suppression (the factor $e^{-M_{\phi_2}/T}$ cancels in the ratio). In this case the main problem for achieving baryogenesis at low energy scales persists. Instead, 
if the asymmetry in $\phi_2$ can be erased by some fast interaction different from the one that sources baryogenesis, with chemical equilibrium implying that $Y_{\Delta \phi_2}$ is Boltzmann suppressed relative to $Y_{\Delta \ell}$, then a way opens for thermal baryogenesis at the TeV scale.
 
\indent These possibilities can be illustrated by one of the terms in the scalar potential of the IDM, $\lambda_5 (\phi^\dag \phi_2)^2/2$. If $\lambda_5 = 0$ there is a conserved lepton number ($\phi_2$ is assigned $L=-1$), whereas for large values of $\lambda_5$, new fast $\phi_2$-depleting interactions exist. With $\lambda_5$ modeling a non-trivial $w_2(z)$, \fref{fig5.1:lam5} demonstrates the crucial dependence of the baryon asymmetry on the size of these interactions.

\indent Different ways of implementing consistently heavy decay or annihilation products in order to achieve baryogenesis at low scales have been developed in Refs.~\cite{Cui:2011ab,Bernal:2012gv,Bernal:2013bga,Frigerio:2014ifa}. A more radical approach to avoid the potential problem related to the asymmetry in the heavy sector is to make a Majorana fermion play the role of the heavy decay or annihilation product~\cite{Racker:2014uga}. In this way there is no asymmetry in the heavy field and the aforementioned problem is completely avoided. An implementation of this idea developed in~\cite{Racker:2014uga} is to have baryogenesis from DM annihilation, with the sterile neutrinos responsible for the masses of the SM neutrinos playing a key role in baryogenesis and the freeze-out of DM annihilations.     
\begin{figure}[t]
\begin{center}
\includegraphics[width=0.5\textwidth,angle=270]{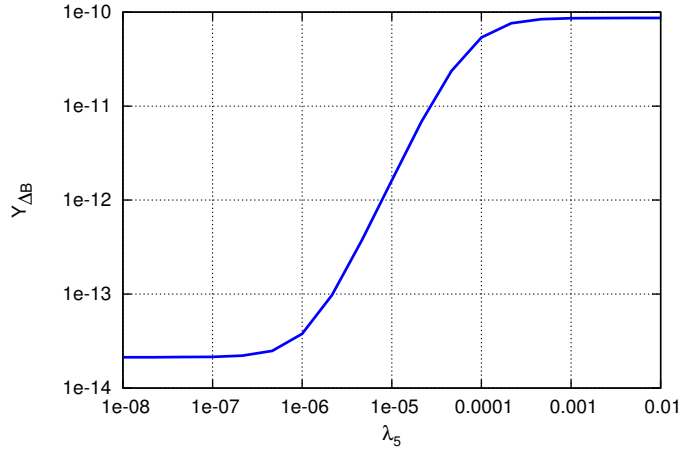}
\end{center}
%\vspace{-0.8cm}
\caption{The final baryon asymmetry normalized to the entropy density, $Y_{\Delta B}$, as a function of $\lambda_5$ for $M_1=2.8$~TeV and $M_{\phi_2}=0.6 M_1$ (the remaining parameters of the IDM have fixed values irrelevant for the discussion). It is apparent that the BAU can only be generated if $\lambda_5$ is large enough, in order for the processes $\phi_2 \bar \phi \leftrightarrow \bar \phi_2 \phi$ to be fast during the crucial time for baryogenesis. This figure has been taken from~\cite{Racker:2014yfa}.}
\label{fig5.1:lam5}
\end{figure}
\end{itemize}

To conclude, we have summarized the main difficulties which have to be overcome to achieve TeV-scale leptogenesis and discussed the three known mechanisms which allow to circumvent the main problem common to all models of thermal baryogenesis from particle decays (or annihilations). In the following sections, we will discuss their implementation in the context of specific models giving us the possibility to probe TeV-scale leptogenesis experimentally.

\section{Testability of leptogenesis in extended seesaw models}
\label{ch5:sec3:extended_seesaw}

As we have seen in the previous section, specific conditions have to be fulfilled in order to obtain TeV-scale leptogenesis. As this is difficult for the ``vanilla" seesaw models, as shown in \eref{eq3:leptogenesis_mass_bound}, we want to concentrate on extensions to these models which allow for testable scenarios. First, we focus on extensions to the standard seesaw models by new matter fields only, without considering any new gauge interactions.

\subsection{Extensions to the type I seesaw model}
\label{ch5:sec3:extensions-typeI}

In type I seesaw, at least two heavy Majorana singlet fermions $N_k$ are introduced,  
resulting in the Lagrangian given in \eref{eq0:lagrangiantypeI_seesaw}.
For definiteness, we consider three generations of RH neutrinos
$k=1,2,3$.
As illustrated in the previous section,
\eref{eq3:leptogenesis_mass_bound}, the mass of the lightest heavy neutrino
$N_1$ is bounded from below, essentially due to constraints implied by
the light neutrino mass scale.  In order to evade these
constraints, we can extend the type I seesaw by coupling $N_k$ to other
SM fermions. The type of new interactions is restricted by gauge
invariance such that only new scalars with the same quantum numbers
as SM fermions (generically denoted as
$\psi$) can be involved~\cite{Fong:2013gaa}:
%%%
\begin{equation}
  \label{eq3:newcoupling}
  -{\mathcal  L_{\tilde \psi}}=  \eta_{m k} \bar \psi_{L m}  N_k \> 
\tilde\psi +   
 \sum_{\psi'\,\psi''}   y_{m n} \bar{\psi}'_{L m} 
\psi''_{R n} \> \tilde\psi\,\  +\  {\rm h.c.}\,,
\end{equation}
%%%
where $\psi_L,\,\psi_L'$ denote the SM left-handed fermion
fields $\ell,\,e^c,\,Q,\,d^c,\,u^c$,
$\psi''_R = \ell^c,\,e,\,Q^c,\,d,\,u$ denote the corresponding SM RH fields, 
%$N^c=N^c_L$ denote the left-handed $SU(2)$ singlet heavy neutrinos, 
and $\tilde\psi$ denote scalars that must match the gauge quantum numbers
of $\psi_L$ in the first term (we borrow the usual supersymmetric
notation with $\tilde\psi = \tilde \ell\,,\,\tilde e,\,\tilde Q,\, \tilde d,\,\tilde u$).
The second term in \eref{eq3:newcoupling} represents
possible new couplings of a new scalar to the SM fermion bilinears.
\tref{tab3:1} lists the possible scalars, their couplings to
SM fermions allowed by gauge invariance and (as long as they can be
consistently given) the assignments that render the Lagrangian in \eref{eq3:newcoupling} $L$ and $B$ conserving.  The last two
columns indicate the amount of $L$ and $B$ violation of the
$ \bar \psi_L N\tilde\psi$ term with $L(N)=0$.  

The two cases of $\tilde u$ and $\tilde d$ yield $B$ violation, and
will lead to too fast nucleon decays. For instance, after
$SU(2)\times U(1)$ spontaneous symmetry breaking, a mixing between the
RH and light neutrinos of order $\sqrt{m_\nu/M_N}$ is induced and
gives rise to the dimension 6 operator
$\frac{1}{M_{\tilde u}^2}\,\sqrt{\frac{m_\nu}{M_N}} \left(\bar
  d^cd\right)\,\left(\bar \nu u\right)$.
Taking $m_{\nu}\sim 10^{-2}\,$eV and
$M_{\tilde u}\sim M_{N}\sim 1\,$TeV, this results in the following nucleon
lifetime:
\begin{eqnarray}
  \label{eq3:nucleonlifetime}
\tau_{N\to \pi \nu}  &\sim&  10^{32} 
\left(\frac{10^{-19}}
{y_{dd\tilde u}\,
\eta_{Nu\tilde u}}\right)^2\, {\rm years}\,.   
\end{eqnarray}
To satisfy the experimental limits~\cite{Patrignani:2016xqp} $\tau_{p\to
  \pi \nu} > 3.90\times 10^{32}\,$yr and $\tau_{n\to \pi \nu} >
11.0\times 10^{32}\,$yr, the couplings $y $ and $\eta$ need to be
extremely small,  
or additional symmetries like baryon number must be imposed to forbid them. 
For example, in \sref{ch5:sec3:cloistered-baryogenesis} we will focus on 
the case of $\tilde u$ by imposing baryon number conservation at the perturbative level, 
which forbids the interaction $\overline{d^c} d\, \tilde u $. This results 
in a novel mechanism known as cloistered baryogenesis~\cite{Sierra:2013kba}.

The cases $\tilde \ell$, $\tilde e$, and $\tilde Q$, instead, are
  viable without additional assumptions.  If these new scalars are
  around the TeV scale, they could be produced at colliders in principle.
  Notice that $\tilde \ell$ has the same electroweak quantum numbers
  as the SM Higgs doublet and as such, their presence can generally
  induce flavor-changing-neutral currents at tree-level~\cite{Georgi:1978ri}. However, experimental limits require that either
  $\tilde \ell$ is very heavy, or that its couplings are very
  small~\cite{Branco:2011iw}. A TeV scale $\tilde e$ can be pair produced at the LHC via electroweak processes, while
  $\tilde Q$ (leptoquark) can be produced with a large rate via QCD
  processes~\cite{Grifols:1981aq,Hewett:1987yg,Blumlein:1996qp,Kramer:1997hh,Kramer:2004df}.
  Their production modes are discussed in Ref.~\cite{Fong:2013gaa} and will
  not be discussed further in this review.

%%%
\begin{table}[t]
\tbl{
  Five types of scalars can be coupled to the RH neutrinos and to one 
  type of SM fermions ($\epsilon=i\tau_2$ is the $SU(2)$ 
  antisymmetric tensor). The third and fourth column list the assignments 
  that render these couplings $B$ and $L$ conserving.  
  $\tilde \ell$ is a (down-type) second Higgs, $\tilde e$ is a lepton, 
  $\tilde Q$ is a leptoquark,   
  $\tilde u$ is a baryon.      
  For $\tilde d$   no  $B$ and $L$ conserving 
  assignments are possible. The last two columns give the  amount of 
  $L$ and $B$ violation in the couplings to RH neutrinos with $L(N)=0$.
\label{tab3:1}
}
{\begin{tabular}{|c|c|c|c|c|c|}
  \hline 
  Scalar field & Couplings & $B$ & $L$  & $\Delta B$  &  $\Delta L$ 
  \\ 
  \hline 
  \phantom{$\Big|$}$\tilde \ell $ & $\bar \ell e\,(\epsilon \tilde \ell^*),\  \, 
  \bar Q d\, (\epsilon \tilde \ell^*),\ \, \bar Q u\, \tilde \ell  $  & $0$  & $0$ 
&$0$ & $-1$ \\
  \hline 
  \phantom{$\Big|$}$\tilde e  $  & $\bar \ell (\epsilon \ell^c)\,\tilde e  $  &0 &+2 
&$0$ & $+1$ \\
  \hline 
  \phantom{$\Big|$}$\tilde Q$ & $\bar  \ell d\,  (\epsilon \tilde Q^*) $&$+1/3$ &$-1$ 
&$0$ & $-1$ \\
  \hline 
  \phantom{$\Big|$}$\tilde u$ &$\overline{d^c} d\, \tilde u $  &$-2/3$ & 0 
&$ -1 $ & $ 0 $ \\
  \hline 
  \phantom{$\Big|$}$\tilde d $ &$\bar \ell (\epsilon Q^c)\,\tilde d,\ 
\overline{Q^c}(\epsilon Q)\,\tilde d,\ 
  \bar u e^c\,\tilde d,\ \overline{u^c}d\,\tilde d $  & $-$ & $-$ 
&$ - $ & $ - $ \\
  \hline 
\end{tabular}
}
\end{table} 
%%%

\begin{figure}[t]
\begin{center}
\includegraphics[width=1.0\textwidth]{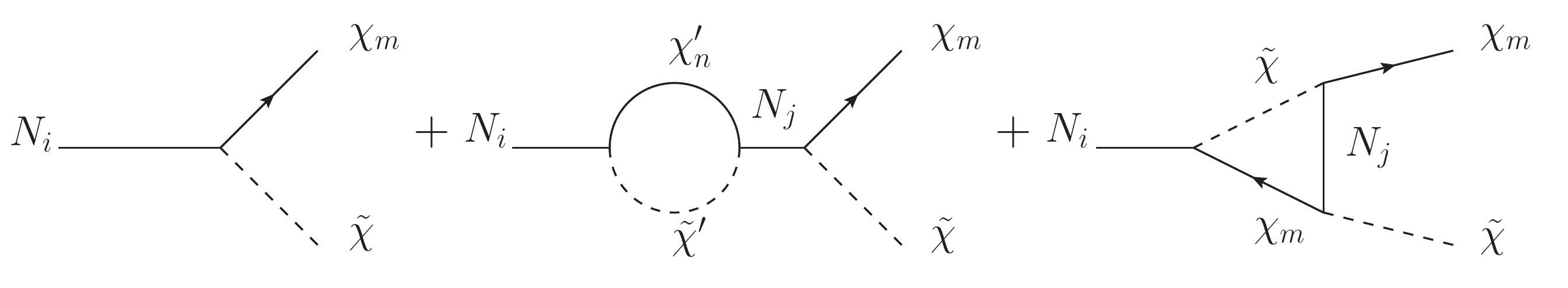}
\caption{
The one-loop self-energy and vertex diagrams with
$\chi^{(\prime)}_m= \ell_{\alpha},\,(\psi_{m}) $
and $\tilde{\chi}^{(\prime)}= H,\,(\tilde{\psi}) $
which generate CP asymmetries in 
$N_k \to \chi_m \tilde\chi$ decays. Figure
taken from Ref.~\cite{Fong:2013gaa}. 
\label{fig3:CP_violation_diagrams} }
\end{center}
\end{figure}

From the new interactions, we can also have new contributions to 
CP violation as shown in \fref{fig3:CP_violation_diagrams}.  
%%%
%%%
Assuming $M_j>M_{1}>M_{\tilde\psi}$ ($j=2,3$) and summing over final state
flavors, the self-energy and vertex contributions to the CP violation 
in $N_1\to \bar \ell \phi, \bar \psi\tilde\psi$ decays are:
%%%
\begin{eqnarray}
\label{eq3:self}
\epsilon_{1\chi}^{S} & = & \frac{\kappa_\chi}{16 \pi D_1}\sum_{j\neq 1}
\sum_{\chi'}\kappa_{\chi'}\,{\rm Im}
\left[\left(\xi_{\chi'}^{\dagger}\xi_{\chi'}\right)_{1j} 
\left(\xi_\chi^{\dagger}\xi_\chi\right)_{1j} \right]
f^{S}\left(\frac{M_{j}^{2}}{M_{1}^{2}}\right)\,, \\
\label{eq3:vertex}
 \epsilon_{1\chi}^{V} & = &  \frac{\kappa_\chi}{8 \pi D_1} \sum_{j\neq 1}
\,{\rm Im}\left[\xi_\chi^{\dagger}\xi_\chi
 \right]^2_{1j} f^{V}\left(\frac{M_{j}^{2}}{M_{1}^{2}}\right)\,, 
\end{eqnarray}
%%%
where $D_1=16\pi \Gamma_1/M_1$ with $\Gamma_1$ being the total $N_1$ decay width.
$\chi,\chi'=\{\ell,\psi\}$ denote the SM fermions in the final states and
in the loops, while $\xi_{\chi},\,\xi_{\chi'}=\{\lambda,\eta\}$ and
$\kappa_{\chi},\kappa_{\chi'}$ are the corresponding Yukawa couplings
and gauge multiplicities. The self-energy and vertex loop functions
are given by:
%%%
\begin{equation}
f^{S} = \frac{\sqrt{x}}{1-x},\qquad\qquad
f^{V} = \sqrt{x}\left[1-(1+x)\ln\frac{1+x}{x}\right].
\end{equation}
%%%

% {\bf 
In \sref{ch5:sec3:baryogenesis-scale}, we will show
that in the case of hierarchical RH neutrinos, the bound on $M_1$ (cf. \eref{eq3:leptogenesis_mass_bound}) can
be lowered down to $\sim 10^7$ GeV within extended seesaw models. This result does not apply only to
extensions to the type I seesaw, but is valid in general.  However,
due to the interplay between CP violation and washout scatterings,
leptogenesis with hierarchical RH neutrinos in the TeV range, as suggested in\cite{Fong:2013gaa},
cannot be reached. The reason, in case of hierarchical RH
neutrinos, is the following: In order keep the washout scatterings under control at the
TeV scale, the amount of CP violation will be too small for successful
leptogenesis. When considering, however, a scenario with quasi-degenerate RH neutrinos and resonantly
enhanced CP asymmetries, TeV leptogenesis can still be realized.

In \sref{ch5:sec3:cloistered-baryogenesis}, 
we will then discuss cloistered baryogenesis~\cite{Sierra:2013kba} that, as
was already mentioned above, is based on an extension to
the type I seesaw with $\tilde u$, where conservation of baryon number
at the perturbative level is imposed. 
In particular, we will prove that even in the presence of fast $B-L$
violating interactions, an asymmetry can still be obtained.

%%%%%%%%%%%%%%%%%%%%%%%%%%%%%%

\subsubsection{The scale for extended type I seesaw baryogenesis}
\label{ch5:sec3:baryogenesis-scale}

In this section we show that assuming a non degenerate RH neutrino
spectrum, we can derive a lower bound on the temperature that allows
for successful baryogenesis via leptogenesis, and specifically also
for cloistered baryogenesis.  In that case the bound follows from the
interconnection between the CP asymmetry and the requirement that the
cloistered sector will remain secluded from the active sector or, in
other words, from the requirement that the washout processes are kept under
control. This argument was first put forth in~\cite{Racker:2013lua}, in
the context of the inert Higgs doublet model (the case of
$\tilde \ell$ in \tref{tab3:1}), and in relation to the
derivation of a bound on the mass of the lightest RH neutrino.

In cloistered baryogenesis, one is interested in the case in which a
Majorana RH neutrino decays in a SM $u$-type quark and in the complex
conjugate of a new scalar $\tilde u$ of equal baryon charge.  However,
the argument does not apply only to the inert Higgs doublet model or
to cloistered baryogenesis, but it is quite general.  Although below
we will present it referring to a $B$ asymmetry, we are confident that
its wide range of validity will be easily understood.  Consider a
generic $U(1)_B$ invariant interaction between two self conjugate
particles $X_i = X_i^c\;(i=1,2)$ and two other complex fields $Y$ and
$Z$ carrying opposite $U(1)_B$ charges
%%%
\begin{equation}
-{\cal L} \supset \sum_{i=1,2} g_i X_i Y Z + {\rm h.c.},
\end{equation}
%%%
where $g_1$ and $g_s$ are two relatively complex parameters
${\rm Arg}(g_1^*g_2) \neq 0$. In general $X_i$ can be Majorana
fermions while $Y$ and $Z$ a pair of complex scalar and fermion or
$X_i$ can be real scalars while $Y$ and $Z$ a pair of fermions or
complex scalars.  In the first two cases $g_i$ are dimensionless
couplings while in the last case they have mass dimension one. Now let
us assume $M_{X_2} > M_{X_1} > M_Y + M_Z$ such that the decays
$X_i \to Y Z, \bar Y \bar Z$ are kinematically allowed.  These decays
are in general CP-violating, implying that $B$ asymmetries in $Y$ and
$Z$ that are equal in size and opposite in sign can be generated.  Let
us further assume that $X_1$ decays while electroweak sphalerons are
still active and that $Y$ has in-equilibrium chemical reactions with
electroweak sphalerons while $Z$ remains chemically decoupled from the
thermal bath, i.e. cloistered. The $B$ asymmetry in $Z$, denoted by
$\Delta B_Z$, remains constant, while the $B$ asymmetry carried by $Y$
gets quickly distributed among the SM particles through electroweak
sphalerons. Due to partial conversion to $L$ asymmetry, the $B$
asymmetry in the thermal bath is no longer balanced by the one stored
in $Z$ i.e. $\Delta B_{\rm SM} \neq - \Delta B_Z$.  Eventually after
electroweak sphalerons switch off, the $B$ conserving decays of $Z$
into the SM particles give rise to a nonzero total $B$ asymmetry
$\Delta B = \Delta B_{\rm SM} + \Delta B_Z \neq 0$.

The CP asymmetry between $Y$ baryons and $\bar Y$ anti-baryons produced 
in $X_1$ decays can be defined as the differences in their decay widths 
normalized over their sum:
%%%
\begin{equation}
  \label{eq3:CP-asymm-generic}
  \epsilon_{X_1}= 
  \frac{\Gamma\left(X_1\to Y Z\right)
    -\Gamma\left(X_1\to \bar Y\bar Z\right)}
  {\Gamma\left(X_1\to Y Z\right)
    +\Gamma\left(X_1\to \bar Y\bar Z\right)} \ .
\end{equation}
%%%
$\epsilon_{X_1}$ can be computed from the interference between
tree-level and one-loop vertex and wave-function diagrams.
Considering the hierarchical case $M_{X_1} \ll M_{X_2}$, for the cases
of a fermion $X_1$ decays into fermion-scalar pair and of a scalar $X_1$
decaying into fermion pairs or scalar pairs, we obtain respectively
the following leading terms:
\begin{eqnarray}
  \label{eq3:CP-asymm-generic-specific}
  \epsilon_{X_1}^{(fs)}   
   &\simeq& -\frac{\left|g_{2}\right|^{2}}{8\pi}\,
  \frac{M_{X_1}}{M_{X_2}}\sin\phi\ , \\
  \label{eq3:CP-asymm-generic-scalars}
  \epsilon_{X_1}^{(ff')} 
  &\simeq& -\frac{\left|g_{2}\right|^{2}}{8\pi}\,
  \frac{M_{X_1}^2}{M_{X_2}^2}\sin\phi\ , \\
  \epsilon_{X_1}^{(ss')} 
  &\simeq& -\frac{1}{8\pi}\,
 \frac{\left|g_{2}\right|^{2}}{M_{X_2}^2}\sin\phi\ ,
\end{eqnarray}
with $\phi={\rm Arg}\left[(g_1^*\,g_2)^2\right]$.  In the following, we
set $\sin\phi \sim 1$ to maximize the CP asymmetries.  We see that in
all the three cases they increase with $|g_2|$.  The value of $|g_2|$,
however, cannot be arbitrarily large because at some point
$YZ\leftrightarrow \bar Y\bar Z$ scatterings mediated by $X_2$ would
become sufficiently fast to enforce chemical equilibrium between $Y$
and $Z$, i.e.  $\mu_Y + \mu_Z = 0$, rendering cloistered baryogenesis
ineffective.  For the three cases above, the $2\leftrightarrow 2$
scattering rates are given by:
%%%
\begin{eqnarray}
  \label{eq3:two-to-two-rate-decay}
  \gamma^{(fs)}(YZ\leftrightarrow \bar Y\bar Z)
  &\simeq& \frac{1}{\pi^3}\frac{T^3}{M_{X_2}^2}
  \left|g_2\right|^4 \ , \\
  \gamma^{(ff')}(YZ\leftrightarrow \bar Y\bar Z)
  &\simeq& \frac{1}{\pi^3}\frac{T^5}{M_{X_2}^4}
  \left|g_2\right|^4  \ , \\
  \gamma^{(ss')}(YZ\leftrightarrow \bar Y\bar Z)
  &\simeq& \frac{1}{\pi^3}\frac{T}{M_{X_2}^4}
  \left|g_2\right|^4 \ .
\end{eqnarray}
%%%

For the temperature relevant to leptogenesis $T \sim M_{X_1}$,
all the expressions above can be rewritten as 
%%%
\begin{equation}
\gamma^{(a)}(YZ\leftrightarrow \bar Y\bar Z) 
\simeq \frac{64}{\pi}M_{X_1}
  \left(\epsilon^{(a)}_{X_1}\right)^2,
\end{equation}
%%%
with $a=(fs),(ff'),(ss')$. Hence we see that for all the cases the
equilibrating scattering rates are proportional to the square of the
respective (maximum) CP asymmetries.  Requiring that at
$T\sim M_{X_1}$ these scatterings are out of equilibrium, that is
$\gamma^{(a)}(YZ\leftrightarrow \bar Y\bar Z)\lesssim H( M_{X_1})$,
with $H( M_{X_1}) \sim 17 M_{X_1}^2/M_{\rm Pl}$ the Universe expansion
rate and $M_{\rm Pl}$ the Planck mass, $M_{\rm Pl}=1.2\times 10^{19}$ GeV, when

\begin{equation}
  \label{eq3:MX1-lower-limit}
  M_{X_1}\  \gtrsim \   10^{19}\times 
\left(\epsilon^{(a)}_{X_1} \right)^2 \; {\rm GeV}\,.
\end{equation}
Since a CP asymmetry smaller than $\epsilon^{(a)}_{X_1}\sim 10^{-6}$
could hardly explain the observed baryon asymmetry, \eref{eq3:MX1-lower-limit} implies
$ M_{X_1}\ \gtrsim \ 10^7\;$GeV, which constitutes a necessary
condition for successfully preserving an asymmetry.
%(See the next section.)
Let us note that in deriving the bound above, we have ignored the
masses of the final states.  It was argued in
Ref.~\cite{Racker:2013lua} that, by taking a final state mass of the
order of $M_{X_1}$, due to Boltzmann suppression of the washout
scatterings it is possible to lower the scale $M_{X_1}$ down to the
TeV region.  Arranging for such a condition, the leptogenesis
scenarios proposed in Ref.~\cite{Fong:2013gaa} becomes viable also at
the TeV scale. (See \sref{ch5:sec1:concepts} for details about the 
conditions required for lowering the scale of leptogenesis from
out-of-equilibrium decays.)

\subsubsection{Cloistered baryogenesis}
\label{ch5:sec3:cloistered-baryogenesis}

In the following, we will present a simple model which realizes
cloistered baryogenesis~\cite{Sierra:2013kba}.  
We consider the fourth model of \tref{tab3:1}, 
in which besides three RH singlet fermions
$N_k$ a new scalar $\tilde u$ with the SM quantum numbers of the
$SU(2)$ singlet RH  up-type quark $u$ is also introduced.
The relevant terms involving $\tilde u$ are
%%%
\begin{equation}
-{\cal L} \supset -\frac{1}{2} M_{\tilde u}^2 \tilde u^* \tilde u 
+ \left( \eta_{ak} \bar u_a^c N_{k} \tilde u^* + {\rm h.c.} \right),
\label{eq3:general-Lag-1}
\end{equation}
%%%
where we have imposed global $U(1)_B$ conservation to forbid the term
$y_{ab} \bar d_a^c d_b \tilde u$ in order to prevent nucleon decay, 
and we have assigned $B(\tilde u) = 1/3$. 

\begin{figure}[t]
  \centering
  \includegraphics[width=0.9\textwidth]{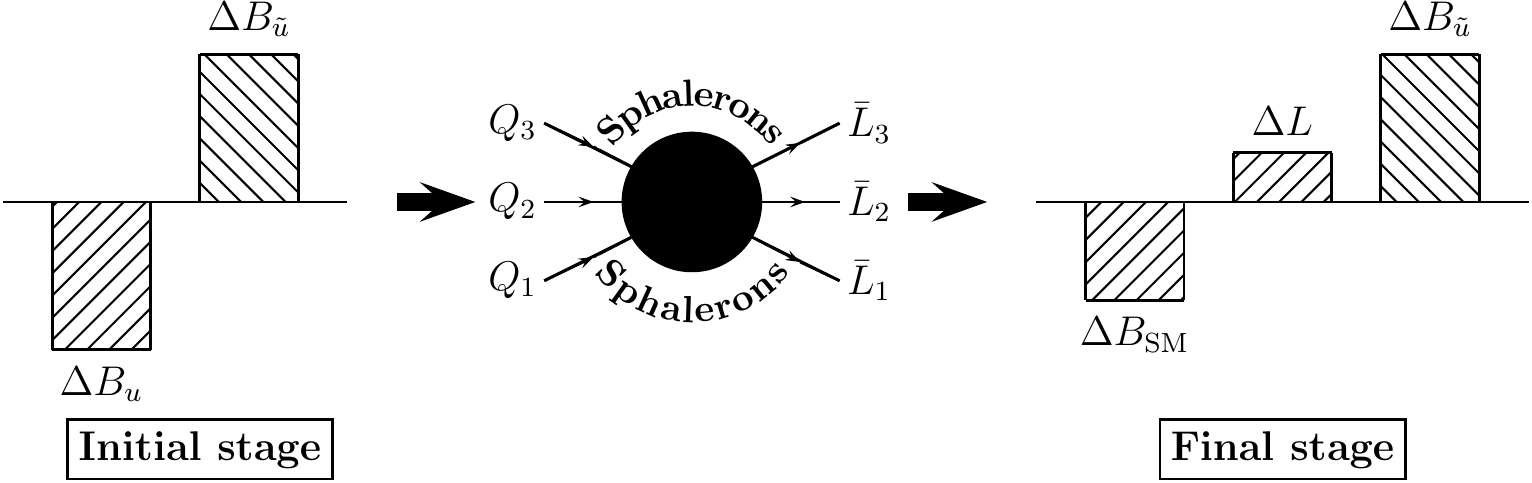}
  \caption{
   In the initial stage, the decays of $N_1$ generate an equal and opposite sign 
   $B$ asymmetries respectively in $u$ and $\tilde u$ 
  denoted by $\Delta B_{u}$ and $\Delta B_{\tilde u}$. 
  At electroweak sphaleron decoupling the $B$ asymmetry in SM particles  
     $\Delta B_{\rm SM}$ is no longer equal in magnitude 
    to the opposite sign asymmetry $\Delta B_{\tilde u}$ due to the
    electroweak sphaleron processes which distribute the initial $\Delta B_{u}$
    among the SM particles in which part of the asymmetry resides in the lepton sector. 
     As a result, we obtain a net nonzero baryon asymmetry
    $\Delta B_{\rm SM}+\Delta B_{\tilde u}\neq 0$. Figure
taken from Ref.~\cite{Sierra:2013kba}. 
  }
  \label{fig3:b-conserving-baryogenesis-sketch}
\end{figure}

Fixing $M_1 \sim 10^7$ GeV, baryogenesis can proceed at $T \sim 10^7$
GeV through out-of-equilibrium, $B$ conserving and CP-violating decays
$N_1 \to u_a \tilde u^*$ as long as $\tilde u$ remains cloistered
until after electroweak sphalerons freeze-out. Given that
astrophysical arguments rule out the possibility of cosmologically
stable heavy colored particles ~\cite{Nardi:1990ku,DiLuzio:2016sbl,DiLuzio:2017pfr},
$\tilde u$ must eventually decay. It is a feature automatically
embedded in our model that $\tilde u$ will decay only after
electroweak symmetry breaking. Decays can in fact proceed through 
the mixing between active and RH neutrinos ($\nu$--$N$), 
which opens up the decay channel
$\tilde u \to u_a \nu $.  After electroweak symmetry breaking and
switch off of electroweak sphaleron processes, the
$\Delta B_{\tilde u}$ released into the thermal bath from $\tilde u$
decays remains largely unbalanced by the baryon asymmetry already
present in the thermal bath, resulting in a net nonzero baryon
asymmetry. The mechanism for this scenario is sketched in \fref{fig3:b-conserving-baryogenesis-sketch}.  Notice that at
this temperatures the CP asymmetry for the standard $N_1$ leptogenesis
(assuming a hierarchical RH neutrino mass spectrum) is way too
small \cite{Davidson:2002qv}.  Here for simplicity we will assume that
the branching ratio for $N_1 \to \ell_\alpha \phi$ is much smaller than
that of $N_1 \to u_a \tilde u^*$, so that we can normalize the CP
asymmetries to the sum of $N_1 \to u_a \tilde u^*$.  In this case,
while the type I seesaw Lagrangian still accounts for neutrino masses
and mixing, it plays no role in baryogenesis.

%%%%%%%%%%%%%%
% Conditions
%%%%%%%%%%%%%%

\begin{figure}[t]
  \centering
  \includegraphics[width=0.9\textwidth]{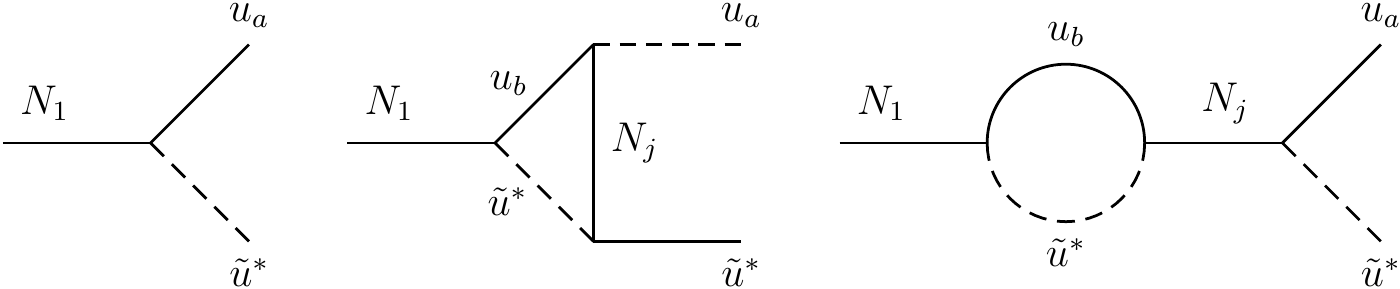}
  \caption{Tree-level and one-loop diagrams responsible for the CP asymmetry in
    the decays $N_1 \to u_a \tilde u^*$. Figure
taken from Ref.~\cite{Sierra:2013kba}. 
   }
  \label{fig3:cp-asymm}
\end{figure}

The CP asymmetry in $N_1$ decays results from the interference between 
the tree-level and the one-loop diagrams in \fref{fig3:cp-asymm}.
Assuming a hierarchical RH neutrino mass spectrum ($M_1 \ll M_2 \ll M_3$)
and summing over quark flavors, the CP asymmetry for $\sum_a (N_1 \to u_a \tilde u^*)$ is given as
%%%
\begin{equation}
  \label{eq3:CP-asymm-model-dependent}
  \epsilon^{\tilde u}_{N_1}\simeq -\frac{1}{4\pi}
  \frac{1}{(\eta^\dagger\eta)_{11}}
  \sum_{k\neq 1}\text{Im}
  \left[\left(\eta^\dagger\eta\right)^2_{1k}\right]
  \frac{M_1}{M_k}\ .
\end{equation}
%%%

In addition to sufficiently large CP asymmetry, we require
the following additional conditions:
\begin{itemize} 
\item[ ($i$)] The decays $\Gamma_{N_1}\equiv \sum_a\Gamma(N_1\to
  u_a\tilde u^*) $ should occur out of equilibrium (we recall that for simplicity  
  we have assumed $\sum_\alpha\Gamma(N_1\to \ell_\alpha \phi) \ll \Gamma_{N_1}$).
\item[($ii$)] The scalars $\tilde u$ should remain chemically
  decoupled from the thermal bath until after electroweak sphalerons switch off 
  at $T_{\rm fo}$, although strong interactions will keep them in kinetic equilibrium.
\item[($iii$)] The decays $\tilde u\to u_a\nu$ after electroweak
  symmetry breaking, which eventually fix the final magnitude and sign
  of the baryon asymmetry, should occur well before the Big Bang
  Nucleosynthesis (BBN) era. % at $T_{\rm BBN}$.
\end{itemize} 
For $M_1 \sim M_k/10 \sim 10^7$ GeV ($k=2,3$), 
condition (i) gives $|\eta_{a1}| \lesssim 10^{-5}$,
while condition (ii) gives $|\eta_{ak}||\eta_{bk}| \lesssim 10^{-4}$. 
On the other hand, condition (iii) requires at least one 
coupling $|\eta_{ak}| \gtrsim 10^{-4}$.
Hence, it is possible to have  $\tilde u\to u_a \nu$ 
decays at a sufficiently early stage from the contributions 
of $\nu$--$N_{2,3}$ mixing. For example, taking $\eta_{a(2,3)}\sim 0.03$ 
and $M_{2,3} \sim 10^8$ GeV, the lifetime of the colored scalar is 
$\tau_{\tilde u}\sim 10^{-4}\,$s, which ensures 
that all $\tilde u$ decay much before the onset of BBN.

Using the method advocated in Ref.~\cite{Fong:2015vna}
once we identified the effective $U(1)$ symmetries of the system, 
the final baryon asymmetry generated can be easily related to the asymmetry 
stored in $\tilde u$. The crucial point is that $U(1)_{\tilde u}$ 
remains an effective symmetry once the baryogenesis is completed,  
until after the freeze-out of electroweak sphalerons.
The detailed generation of an asymmetry in $\tilde u$ can be
studied quantitatively using  Boltzmann equations.

Since the asymmetry is generated and stored in $\tilde u$, $B-L$ only
plays a spectator role, with at most ${\cal O}(1)$
effects \cite{Buchmuller:2001sr,Nardi:2005hs}. Let us look at the two limiting cases
below. Assuming negligible $B-L$ violation, i.e. setting the total
$B-L$ charge to zero, after all $\tilde u$ decays, we
obtain the final baryon asymmetry\footnote{The factor of 3 with respect to~\cite{Sierra:2013kba}
  is because here $Y_{\Delta \tilde u}$ denotes abundance where we
  have summed over color.}
%%%
\begin{equation} 
Y_{\Delta B} = \frac{21}{79} Y_{\Delta \tilde u}.
\end{equation}
%%%
On the other hand, assuming fast $B-L$ violation, % such that it is not conserved, 
after all $\tilde u$ decays, we have
\begin{equation} 
Y_{\Delta B} = \frac{7}{33} Y_{\Delta \tilde u}.
\end{equation}
Comparing the two limiting cases above confirms that $B-L$ violation
acts as spectator.  In particular, the observation of fast
$L$-violation at colliders in Ref.~\cite{Deppisch:2013jxa} will not be
able to invalidate cloistered baryogenesis.

Let us conclude by pointing out that the discovery at the LHC or at
future colliders of new, relatively long-lived scalars, with similar
properties to the ones discussed in this review, will support the
possibility that baryogenesis arises from extended seesaw models of
type I. To avoid issues with BBN, the lifetime of these scalars should be
less than about $10^{-2}\,$s. 
For the case of $\tilde u$, choosing parameters consistent with cloistered baryogenesis, 
we have a typical lifetime of $\tau_{\tilde u} \sim 10^{-4}$ s, for which the decay length 
is still few $\times 10^4$ m. Clearly, these relatively heavy colored
scalars would hadronize into the so-called $R$-hadron $\tilde u \bar u$ 
and give rise to spectacular events right inside the detectors (see Ref.~\cite{Mackeprang:2009ad} 
and the references therein).  
Nevertheless, to probe directly that these scenarios are truly
responsible for the cosmological baryon asymmetry would require
producing also $N_k$. However, as was shown in ~\cite{Fong:2013gaa},
bringing down the RH neutrino mass scale to energies accessible at
foreseeable colliders remains quite challenging, and it would require,
for example, quasi-degenerate $N_k$ to resonantly enhance the CP asymmetries.

\subsection{Extensions to type II and III seesaw models}
\label{ch5:sec3:extended_seesaw-typeII}

As well-known and already emphasized above, to generate the neutrino masses at tree-level, an exchange 
of RH neutrinos between SM lepton and scalar doublets is not the only possibility.
There exist two alternatives, from the exchange of one or more scalar triplets  with hypercharge $Y=2$ (type II seesaw) and from the exchange of self-conjugate triplets of fermions (type III seesaw).
These scenarios offer the possibility to generate the lepton asymmetry from the decay of these heavy scalar triplets (type II leptogenesis \cite{Ma:1998dx,Hambye:2000ui,Hambye:2003ka,Hambye:2005tk,Hambye:2012fh,Felipe:2013kk,Sierra:2014tqa,Lavignac:2015gpa}) or of these fermion triplets (type III leptogenesis \cite{Hambye:2003rt,Fischler:2008xm,Strumia:2008cf,AristizabalSierra:2010mv,Hambye:2012fh}).
In order to discuss possibilities of testing/falsifying these scenarios it is necessary to briefly review how they work, discussing in particular the effects of gauge scatterings.
For a detailed review, see Ref.~\cite{Hambye:2012fh}.

The relevant Lagrangians for the type II and type III seesaw models are given in \tref{Tab:type-IIandIII}, together with their contribution to neutrino masses and the definition of the total lepton number CP asymmetries they lead to.
%
%\vspace{2cm}
\begin{table}[!t]
\tbl{Lagrangian, neutrino mass contribution and CP asymmetry definition for type II and type III seesaw model (in matrix notation, a summation over the lepton flavor and heavy state indices is implicit, except for the CP asymmetries where it is defined for each of the 3 triplet components separately), with $\ell_\alpha=(\nu_{L \alpha},e_{L \alpha})^T$, $\phi=(\phi^+,\phi^0)^T$ and $\tilde{\phi}=i\tau_2 \phi^*$.
For the scalar triplet $\Gamma_\Delta=\Gamma_{\bar{\Delta}}=\Gamma(\Delta_L\rightarrow\bar{\ell}\bar{\ell}) +\Gamma(\Delta_L\rightarrow {\phi} {\phi})$.\label{Tab:type-IIandIII}}
%\begin{center}
{\begin{tabular}{|c||c|c|} 
\hline
& \multicolumn{2}{c|}{Type of seesaw model} \rule[-5 pt]{0pt}{18 pt}\\
\cline{2-3}
  & Type II & Type III  \rule[-5 pt]{0pt}{18 pt}\\
\hline
\hline
 {{Seesaw states} } & { $\Delta_L=
\left(
\begin{array}{ cc}
   \delta^+/\sqrt{2}  &   \delta^{++} \\
     \delta^0 &  -\delta^+/\sqrt{2} 
\end{array}
\right)$ }& { $\Sigma=
\left(
\begin{array}{ cc}
   \Sigma^0/\sqrt{2}  &   \Sigma^+ \\
     \Sigma^- &  -\Sigma^0/\sqrt{2} 
\end{array}
\right)$ }  \rule[-18 pt]{0pt}{44 pt} \\ 
\hline
 { {Kin. term} }&  { $\text{Tr}\left[(D_\mu \Delta_L)^\dagger (D^\mu \Delta_L)\right]$ } & { $\text{Tr}\left[ \overline{\Sigma} i \slashed{D}  \Sigma \right] $ }\rule[-14 pt]{0pt}{34 pt}\\
 \hline
 { {Mass term}} & { $-M^2_\Delta \text{Tr}\left[\Delta_L^\dagger \Delta_L\right]$} & {$-\frac{1}{2} \text{Tr}\left[\overline{\Sigma}  M_\Sigma \Sigma^c 
                +\overline{\Sigma^c} M_\Sigma^* \Sigma\right] $ } \rule[-14pt]{0pt}{34pt}\\
                 \hline
 { {Interactions}}  & {$-\ell^T \lambda_\Delta C i \tau_2\Delta_L \ell+\mu \tilde{\phi}^Ti \tau_2 \Delta_L \tilde{\phi}$ } & { $- \tilde{\phi}^\dagger \overline{\Sigma} \sqrt{2} \lambda_\Sigma \ell 
-  \overline{\ell}\sqrt{2} {\lambda_\Sigma}^\dagger  \Sigma \tilde{\phi} $ } \rule[-14 pt]{0pt}{34 pt}\\
                 \hline
    { {$\nu$ masses}} & { $ M_\nu^\Delta=2 \lambda_\Delta v_{\Delta_L}=\lambda_\Delta\mu^*\frac{v^2}{M^2_\Delta}$ } & { $ M_\nu^\Sigma=-\frac{v^2}{2} \lambda_\Sigma^T\frac{1}{M_\Sigma} \lambda_\Sigma$ } \rule[-14 pt]{0pt}{34 pt}\\
                 \hline
 { CP asym.} & { $\epsilon_\Delta\equiv2\, \frac{\Gamma(\bar{\Delta}_L\rightarrow \ell \ell)-\Gamma(\Delta_L\rightarrow\overline{\ell}\overline{\ell})}{\Gamma_\Delta+\Gamma_{\bar{\Delta}}}$ }& { $\epsilon_\Sigma\equiv \frac{\Gamma(\Sigma \rightarrow \ell \phi)-\Gamma(\overline{\Sigma}\rightarrow\overline{\ell}\bar{\phi})}{\Gamma(\Sigma\rightarrow \ell \phi)+\Gamma(\overline{\Sigma} \rightarrow\overline{\ell}\bar{\phi})}$}
\rule[-14 pt]{0pt}{34 pt}\\
\hline
\end{tabular}
}
%\end{center}
\end{table}
Type II or type III seesaw states appear naturally in many gauge extensions of the SM. These extensions include the usual left-right (L-R) models and renormalizable SO(10) models, which on top of RH neutrinos involve a type II $SU(2)_L$ scalar triplet (related, together with its $SU(2)_R$ partner,  to the generation of the RH neutrino masses). Other examples are the $SU(5)$ models with an adjoint fermion representation
\cite{Bajc:2006ia,Dorsner:2006fx,Bajc:2007zf,Fischler:2008xm,Blanchet:2008cj,Kamenik:2009cb}, which contains both a type I and a type III seesaw state. Seesaw models with an extended gauge sector will be discussed in more detail in \sref{ch5:sec4:extendedgauge} and in \sref{ch5:sec5:soft-leptogenesis} in the context of Supersymmmetry.
\begin{figure}[t]
\centering
\includegraphics[width=1.0\textwidth]{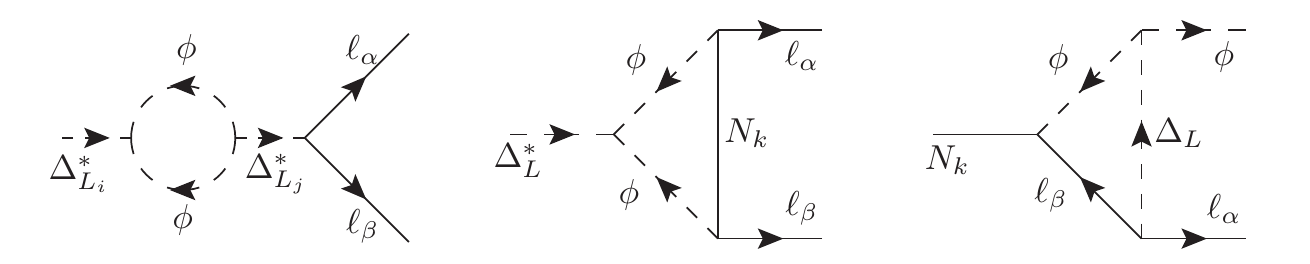}\\
\caption{Diagrams contributing to the asymmetry
created by the decay of a $\Delta_L$, if there are several scalar triplets, or a RH neutrino $N$. The third diagram also contribute to the asymmetry of a $N$ in presence of a heavier scalar triplet.}
\label{fig33typeII}
\end{figure}

A single seesaw state cannot lead to successful leptogenesis because such a state cannot generate alone the necessary CP asymmetry.
But, if at least a second seesaw state exists, as in all these GUT inspired models, this is possible.
For instance, as shown in \fref{fig33typeII}, from the decay of a scalar triplet one can generate a CP asymmetry from a diagram involving any heavier seesaw state, for example a heavier second scalar triplet \cite{Ma:1998dx,Hambye:2000ui} or a heavier RH neutrino \cite{ODonnell:1993obr,Hambye:2003ka}. Similarly, even if the lightest state is a RH neutrino, still, the presence of a heavier triplet can modify the type I leptogenesis scenario \cite{ODonnell:1993obr,Hambye:2003ka}, because it will contribute to the CP asymmetries of the RH neutrinos, see the last diagram of \fref{fig33typeII}. 
For the pure type III leptogenesis scenario the diagrams are the same as for the type I case, replacing each of the RH neutrinos in these diagrams by a $SU(2)_L$ triplet \cite{Hambye:2003rt}. 

Similarly to the type I case, the type II and III leptogenesis scenarios highly depend on the type of seesaw state mass spectrum.
By far the simplest case arises if the seesaw states have a hierarchical spectrum. It is easy to show that in this case, the total lepton number CP asymmetries depend on the heavy seesaw states (which are virtual in the one-loop diagram) only through their contributions to the neutrino mass matrix \cite{Antusch:2004xy}, whereas they depend on the decaying state through both its contribution to the neutrino mass matrix and its mass. For the type II and III cases, for each of the 3 triplet components, the CP asymmetry takes the form\cite{Hambye:2003ka,Hambye:2005tk,Hambye:2003rt}
%reduce to a product of the contributions of each seesaw state to the neutrino mass matrix
\begin{eqnarray}
\epsilon_{\Delta}&=&-\frac{1}{16\pi^2}\frac{{M}_\Delta^3}{\Gamma_\Delta v^4}\text{Im}[({M}^\Delta_\nu)_{\beta\alpha}({M}^H_\nu)^\dagger_{\alpha\beta}]\quad\label{epsIIhierarc}\\
\epsilon_{\Sigma}&=&-\frac{1}{32\pi^2}\frac{{M}_\Sigma^3}{\Gamma_\Sigma v^4}\text{Im}[({M}^\Sigma_\nu)_{\beta\alpha}({M}^H_\nu)^\dagger_{\alpha\beta}]\quad \label{epsIIIhierarc}
\end{eqnarray}
with ${M}^\Sigma_{\nu}$, ${M}^\Delta_{\nu}$ and ${M}^H_{\nu}$ indicating the neutrino mass matrix induced by the decaying $\Sigma$, the decaying $\Delta_L$ and by whatever heavier state in the loop diagram present, respectively. For type III leptogenesis the result above is the same as for type I, replacing $N\leftrightarrow \Sigma$.
Thus, similarly to the type I leptogenesis scenario, for a hierarchical spectrum of seesaw states, the produced lepton asymmetry  depends only on the mass and interactions of the  lightest state, and on the contribution of the heavy seesaw states to the neutrino mass matrix, no matter what these heavy states are. As these equations show, the smaller the mass of the decaying seesaw state is, the smaller is the CP asymmetry. This implies that successful leptogenesis requires this mass scale to be large enough.
As mentioned already in  \sref{ch5:sec1:concepts}, this lower bound is of order $10^{10}$~GeV.
For a quasi-degenerate spectrum of seesaw states, these simple picture and bounds do not hold. In this case, self-energy diagrams involving two states of the same type, such as in the first diagram of \fref{fig33typeII}, are resonantly enhanced. This allows successful leptogenesis for much smaller mass scales. Thus, as for type I leptogenesis, it is the quasi-degenerate case, which is relevant for possible tests, as already generally discussed in \sref{ch5:sec1:concepts}.

Triplets, unlike RH neutrinos, undergo gauge interactions, leading in particular to $X\overline{X}\leftrightarrow \text{SM SM}$ transitions.
These processes do not cause any washout effect, since they conserve lepton number, but they do thermalize the triplets. As a result they put the triplet number density $n_X$ close to the equilibrium distribution one, $n_X^{eq}$. As the lepton asymmetry production proceeds  proportionally to $(n_X-n_X^{eq})\cdot \epsilon_X$, this leads to a suppressed result.
At temperature $T\gg M_X$ the gauge scatterings  are not in equilibrium because the corresponding reaction rate over the Hubble rate, $\gamma_A/n_X^{eq}H$, scales as $M_{\rm Pl}/T$. At $T\ll M_X$ these processes are not in equilibrium either, because they involve two heavy seesaw states in the initial or final state, which means that the $\gamma_A$ reaction density is doubly Boltzmann suppressed,
\begin{equation}
	\gamma_A^{\rm II/III}(X\,\overline{X}\leftrightarrow \text{SM SM})\;\approx\; \frac{M_X\,T^3\,e^{-2\,M_X/T}}{32\,\pi^3}\,\left[ c_s^{\rm II/III}\,+\,\mathcal{O}\left(\frac{T}{M_k}\right)\right]\,,
\end{equation}
with \cite{Hambye:2005tk,Strumia:2008cf}
\begin{eqnarray*}
		c_s^{\rm III} \; = \; \frac{111\,g^4}{8\,\pi}\,,\quad\quad\quad\quad\quad c_s^{\rm II} \;=\; \frac{9\,g^4\,+\,12\,g^2\,g^{\prime^2}\,+\,3\,g^{\prime^4}}{2\,\pi}\,.
\end{eqnarray*}
 However, between these two regimes, at $T\sim M_X$, the gauge scatterings are in thermal equilibrium as soon as $M_X\lesssim 10^{14}$~GeV, because $\gamma_A/n_X^{eq}H|_{T\sim M_X} \simeq 10^{14}\,\hbox{GeV}/M_X$. In practice, as long as the gauge reaction rate  $\gamma_A/n_X^{eq}$ is both larger\cite{Hambye:2012fh} than the Hubble rate and larger than the triplet decay reaction rate $\gamma_D/n_X^{eq}$ (so that the triplets scatter before they decay), the asymmetry production is suppressed by a factor given by the ratio of both rates, $ \gamma_A/\gamma_D$. Afterwards, when one of these two conditions is not fulfilled anymore, say at $T\equiv T_A$, the gauge scatterings have a negligible effect.  As the $\gamma_A/\gamma_D$ ratio is Boltzmann suppressed at $T<M_X$ (due to the double Boltzmann suppression of $\gamma_A$), this decoupling of the gauge scatterings always happens for $M_X/T_A$ not larger than a few. This is why at $T>T_A$ no substantial asymmetry can be created, but afterwards it can, because the number of triplets remaining at this time and decaying afterwards, which is given by $\sim n_X^{eq}\propto e^{-M_X/T_A}$, can be substantial.
This explains\cite{Hambye:2000ui,Hambye:2005tk} why type II or type III leptogenesis can still be successful for $M_X$ much smaller than $10^{14}$~GeV.
However, the smaller $M_X$ is, the more the gauge scatterings are in thermal equilibrium at $T\sim M_X$. The larger $M_X/T_A$ is, the smaller is the asymmetry produced. 
In practice, see the details in Ref.~\cite{Hambye:2012fh}, the final asymmetry depends on whether at $T\simeq T_A$ the decay/inverse decay rate is still in thermal equilibrium. If they are not in thermal equilibrium, the final asymmetry produced is not affected by any washout effect from inverse decays and one lies in the so-called ``gauge regime" \cite{Hambye:2012fh,AristizabalSierra:2010mv}. In the opposite case, one lies in the ``Yukawa regime", where, at $T\simeq T_A$, one enters in a strong washout regime, in case the final asymmetry does not depend on the fact that there were gauge scatterings occurring at $T>T_A$ \cite{Hambye:2012fh,AristizabalSierra:2010mv}.

The type III leptogenesis scenario differs from the type I scenario mainly by the fact that it involves this gauge scattering effect (besides the fact that it involves a number of decaying and virtual states in the one-loop diagrams which is three times larger than for the type I case).
As a result of this effect, the lower bound on the lightest triplet for a hierarchical mass spectrum of seesaw states is a factor of about two orders of magnitude\cite{Hambye:2003rt} larger than for the type I case, $M_\Sigma\gtrsim 3\cdot 10^{10}$~GeV. 
Below this mass scale, the number of triplets remaining at the temperature $T_A$, whose subsequent decays produce the asymmetry, is already too Boltzmann suppressed to produce a large enough lepton asymmetry (given the CP asymmetry in this case, \eref{epsIIhierarc}).
For a quasi-degenerate spectrum of triplet fermions, the resonance of the self-energy diagram allows for much larger CP asymmetries. Still, the lower $M_\Sigma$ is, the larger is the gauge reaction rate with respect to the Hubble rate at $T\sim M_\Sigma$, the larger is $M_\Sigma/T_A$, the more the remaining number of triplets at $T_A$ is Boltzmann suppressed, the less an asymmetry is produced. For instance, for $M_\Sigma=1$~TeV, one gets $M_\Sigma/T_A\sim 25$. At this temperature the number of triplets remaining is so suppressed that even with CP asymmetries  of order unity successful leptogenesis cannot proceed.
One obtains the lower bound from taking a CP asymmetry of order one \cite{Strumia:2008cf}
\begin{equation}
M_\Sigma>1.6\,\,\hbox{TeV}\,.
\end{equation}
Note that flavor effects have no impact on this bound. These effects for type III leptogenesis have been incorporated in Ref.~\cite{AristizabalSierra:2010mv}. For the regime where the inverse decays do have an impact on the asymmetry (Yukawa regime), they can be relevant, by reducing the washout effect from these inverse decays, similarly to the flavor effects in type I leptogenesis. In contrast, for the regime where inverse decays  can be neglected (gauge regime), the suppression comes only from the gauge scatterings. These processes and the resulting suppression are flavor blind: gauge scatterings put $n_X$ close to $n_X^{eq}$ no matter Boltzmann equations are considered for individual lepton flavors or only for the total lepton number. The 1.6~TeV bound above is clearly obtained deeply in the gauge regime.

The gauge scatterings have the same effects in type II and type III leptogenesis. This is why the bound we get with a quasi-degenerate spectrum, i.e.~with CP asymmetries of order unity from the self-energy diagram involving at least two different scalar triplets, is basically the same\cite{Strumia:2008cf}
\begin{equation}
M_\Delta>1.6 \,\,\hbox{TeV}\,.
\end{equation}
However, the dynamics operating for type II leptogenesis for the creation of the asymmetry is quite different from the one  in the fermion seesaw cases. One important difference is that a scalar triplet, unlike the type I and type III states, is not self-conjugated. This implies that one must distinguish the number of triplets and antitriplets. Both number densities do not necessarily evolve in the same way. As a result, with respect to the other cases, the type II leptogenesis scenario involves one more Boltzmann equation, determining the triplet-antitriplet  asymmetry number density. 
Moreover, unlike the fermion seesaw states, which involve only one type of couplings (the Yukawa couplings), the type II seesaw is based on two completely different interactions: the coupling of the scalar triplet to two Higgs doublets and to two lepton doublets, whose coexistence breaks lepton number. This leads naturally to situations where one decay channel could proceed much faster than the other, resulting in an approximate conservation of lepton number, even if the triplets decay at a rate much larger than the Hubble rate.
All in all, this implies that the interplay of Boltzmann equations can be very different from the fermion cases, resulting in a different dynamics. For instance, this allows a lepton asymmetry to be effectively stored in a large triplet-antitriplet asymmetry which is created first, and which will lead to a lepton asymmetry afterwards, when these states will decay. 
Details can be found in Refs.\cite{Hambye:2005tk,Hambye:2012fh}, including the lower bound on the scalar triplet mass in the hierarchical case (which around $10^{10}$~GeV depends on the contribution of the heavier seesaw states to the neutrino mass matrix). Similarly, these effects and the fact that a single scalar triplet can induce all three neutrino masses (unlike a fermion seesaw state, which can induce only one) lead to different flavor dynamics when the flavor effects are incorporated\cite{Felipe:2013kk,Sierra:2014tqa,Lavignac:2015gpa}, see also the related discussion in \cite{leptogenesis:A01}.

All the discussion above, in particular on the role of the gauge scatterings, lead us to a situation which is representative of the difficulties one has to face 
when one tries to find ways to test leptogenesis.
The fact that the type II and type III seesaw states undergo gauge interactions is welcome from the point of view of discovering the seesaw states at the LHC. These states, unlike the type I seesaw fermions, can be pair produced via SM gauge interactions\cite{Ma:2002pf,Bajc:2006ia,Bajc:2007zf,Franceschini:2008pz,Perez:2008ha,Melfo:2011nx}, and they are intensively searched for at the LHC experiments. 
The seesaw states can also lead in addition to specific signatures, in particular displaced vertices, since at LHC energies the seesaw mechanism naturally predicts small values of the seesaw couplings, leading to slow decays, i.e.~characteristic displaced vertices \cite{Franceschini:2008pz}.
However, as discussed above, the very same interactions that allow seesaw states production also imply that these states will be pretty close to thermal equilibrium when they decay, leading to very large suppression of the produced asymmetry at these energies.
In particular, even in the best situation where the CP asymmetry is of order unity, one cannot go below the $\sim1.6$~TeV lower bound above, due to this effect.
This scale turns out to be already a bit too high to be reachable at LHC.
But at the very least this offers clear possibilities of falsifying these leptogenesis scenarios or even any leptogenesis scenario producing the asymmetry at higher energies. For instance a triplet which would be discovered at the LHC would not be able to produce the asymmetry through its decay but could washout a previously produced asymmetry, if it has large enough interactions (to all lepton flavors). More generally if these triplets are present at LHC energies they could come with other associated states, such as in the L-R model which, on top of the scalar triplets, involves a  $Z_R$ or a $W_R$. The discovery of such associated states, especially a $W_R$ (because its thermalization effect is extremely efficient at these energies, see \sref{ch5:sec4:extendedgauge} below) could rule out type I leptogenesis, in case at the seesaw level one would be left with the type II or type III options.

Extensions of the low scale  type II/III leptogenesis scenarios usually involve the addition of new scalars and fermions, which either implement the resonant enhancement of the asymmetry or provide an
additional/dominant contribution to the neutrino mass generation (see, e.g., \cite{Hambye:2001eu,Gu:2007dr,Gu:2010ye,Hati:2015hvq,Borah:2014bda,Kim:2016xyi,AristizabalSierra:2011ab,Gu:2009hu,JosseMichaux:2011ba,JosseMichaux:2012wj,Kashiwase:2015pra}). In some scenarios (see Ref.\cite{Gu:2009hu} for an example of TeV scale type II and Refs.\cite{JosseMichaux:2011ba,JosseMichaux:2012wj} for TeV scale type I/III (inverse) seesaw mechanisms), 
an initial asymmetry is dynamically produced from \textit{lepton number conserving} decays of new electroweak singlet fields
 and in a second step transferred to the SM leptons through the same Yukawa interactions which contribute to
neutrino mass generation via the seesaw mechanism. All in all, 
this separation between the scale involved in the generation of the lepton/baryon asymmetry and the one responsible for neutrino masses, typically, increases the phenomenology and testability of this class of models.
In particular, the new TeV scale fields carrying gauge interactions provide specific collider signatures, testable in the current run of LHC. Note that for type I leptogenesis low scale alternatives\cite{Hambye:2001eu,Hambye:2009pw,Racker:2013lua,Boubekeur:2004ez} to the resonant mechanism are to consider a radiative generation of the neutrino masses, with either a hierarchy of couplings for the various seesaw states or with seesaw states decaying to three-body decays rather than to two-body decays. To generate in those ways the baryon asymmetry from the decay of type II or III states does not help much because, still in these cases,  too large gauge scatterings heavily suppress the result at low scale.

 \section{Testability of leptogenesis with extended gauge sectors}
\label{ch5:sec4:extendedgauge}
So far we just covered leptogenesis within seesaw models that are based on an extended matter sector with particles obeying the SM gauge groups. In the following, we open up the discussion for models with an extended gauge sector and allow for the observation of an additional gauge boson. As an example, we discuss the $Z^\prime$ and the L-R symmetric model.

\subsection{TeV leptogenesis in $Z^\prime$ models}
\label{ch5:sec4:extendedgauge-Zprime}
RH neutrinos realizing the seesaw mechanism may be connected to the presence of an extra gauge symmetry  
$U(1)^\prime$ and an extra gauge boson $Z'$ at the TeV scale, which can be searched for at colliders.  
However, such $U(1)'$ seesaw models  have a difficulty to provide successful baryogenesis
as the resulting lepton asymmetry is highly suppressed by efficient gauge annihilations through $Z^\prime$ mediation.
Such a strong suppression may be compensated by a resonant enhancement, which generically involves 
an unpleasant fine-tuning. Nonetheless, it would be interesting to see if leptogenesis can work successfully 
in $Z^\prime$ models, or to set a limit on the $Z'$ mass (or the $U(1)'$ gauge coupling, $g'$) for successful leptogenesis.
Assuming the standard gauge coupling strength $\alpha'\equiv g'^2/4\pi=1/60$ and maximal CP asymmetry,
the $Z'$ mass has to be larger than $2-3$ TeV \cite{Chun:2005tg}. 
It will be an interesting task to find such a heavy $Z^\prime$ and the associated 
large CP violation required for TeV leptogenesis at colliders.

\medskip

\begin{figure}[t]
\begin{center}
\includegraphics[width=0.75\textwidth]{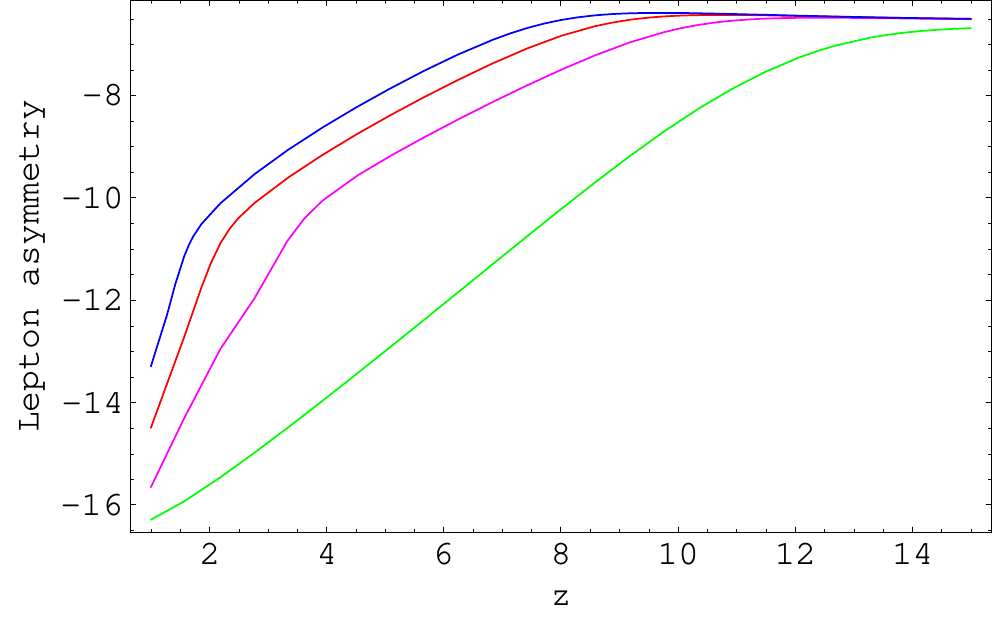}
\end{center}
\caption{\label{fig4:figLA} The lepton asymmetry in unit of $\epsilon$
is calculated as a function of  $z= M_X/T$ with
$M_X=0.3$ TeV and $K=500$. The curves, $\log(Y_{\Delta L}/\epsilon)$, are for
$M_{Z^\prime}=1,2,3$ and $4$ TeV from below. Figure taken from Ref.~\cite{Chun:2005tg}.  }
\end{figure}

For a definite discussion, we assume a $Z^\prime$ model with $U(1)_\chi$  \cite{Hewett:1988xc, Langacker:1991zr, Leike:1998wr}.
 In terms of the $SU(5)$
notation, the SM fermions and the singlet field $N$ carry the
$U(1)_\chi$ charges as follows:
$$ {\bf 10}\, ({-1 \over\sqrt{40}}),\quad {\bf \overline{5}}\,
({3\over\sqrt{40}}), \quad {\bf 1}\, ({5\over\sqrt{40}}).$$
The Lagrangian of the RH neutrino sector reads as follows:
\begin{equation}
 - {\cal L} = \lambda \ell \phi_u N + {1\over2} \lambda S  N N + h.c.\,,
\end{equation}
where $\ell, \phi_u$ and $N$ denote the  lepton, up-type Higgs and RH
neutrino field (in chiral notation), respectively. The supersymmetric version has a similar superpotential, 
promoting all fields to the corresponding superfields. The spontaneous breaking of $U(1)'$ by 
the vacuum expectation value, $\langle S \rangle$, generates the $Z'$ mass $M_{Z'} \sim g' \langle S \rangle$
and the RH neutrino mass $M_k= \lambda_k \langle S \rangle$. 
In the non-supersymmetric seesaw models, resonant leptogenesis occurs, when two RH neutrinos $N_{1,2}$ are 
close in mass, $\Delta M  = M_2 -M_1 \ll M_{1,2}$, and satisfy $\Delta M \approx \Gamma_{N_{1,2}}$.  
In the supersymmetric version, leptogenesis can work even with one RH neutrino via ``soft leptogenesis" 
and the resonance condition $B \approx \Gamma_{\tilde N}$ for the RH sneutrino $\tilde N$ (fore more details see \sref{ch5:sec5:soft-leptogenesis}, in particular \sref{ch5:sec5:typeII-soft-leptogenesis}).
The decay of a heavy seesaw particle $X$ (which can be $N$ or $\tilde N$)  generating a CP asymmetry $\epsilon$, will produce a lepton asymmetry in unit of the entropy density $s$,  $Y_{\Delta \ell}=(n_{\ell} - n_{\bar \ell})/s$,  following the 
Boltzmann equation:
\begin{eqnarray} \label{eq4:boltz}
 {d Y_{\Delta X} \over d z} &=& - z K \left[ \gamma_D (Y_{\Delta X}-Y_{\Delta X}^{eq}) +
 \gamma_A {(Y_{\Delta X}^2-Y_{\Delta X}^{eq\,2})\over Y_{\Delta X}^{eq}}  \right] \nonumber\\
 {d Y_{\Delta \ell} \over d z} &=& 2 z K \gamma_D\left[ \epsilon (Y_{\Delta X}-Y_{\Delta X}^{eq})
 - {Y_{\Delta X}^{eq} \over 2 Y_{\Delta \ell}^{eq} } Y_{\Delta \ell} \right]\,,
\end{eqnarray}
where $K\equiv \Gamma_X/H(M)$ and $H(T)=1.66 \sqrt{g_*}
T^2/M_{\text{Pl}}$.  Note that the above Boltzmann equation can be applicable for both, the non-supersymmetric and the supersymmetric (realized by soft leptogenesis) version taking $X$  as either almost mass degenerate RH neutrinos $N_{1,2}$ or sneutrinos $\tilde N_{1,2}$, respectively. 
In both cases, we take the SM value
$g_*=106.75$. 
 In \eref{eq4:boltz}, $\gamma_D = K_1(z)/K_2(z)$ is
the usual contribution from RH (s)neutrino decays
involving the neutrino Yukawa coupling $\lambda$, and $\gamma_A$
accounts for the annihilation of $X$ to the
light SM fermions mediated by the heavy $Z^\prime$: $X \bar X \to Z^\prime \to f\bar{f}$.
The gauge annihilation contribution is given by
\begin{equation}
\gamma_A = {5\over \pi} {\alpha^{\prime 2} M_X\over K H(M_X)}
\int^\infty_1\!\! d t\, {K_1(2zt)\over K_2(z)} {t^3 (t^2-1)^{3/2}
\over (t^2 - {1\over 4}r^2)^2 + {1\over16} u^2}\,,
\end{equation}
with $r\equiv M_{Z^\prime}/M_X$ and $u\equiv r
\Gamma_{Z^\prime}/M_X$.  Estimating
$\Gamma_X=h^2 M_X/4\pi$, we get $ K \sim {m_\nu/
5\!\times\!10^{-4} \mbox{eV} } \sim 100 $ for $m_\nu = 0.05$ eV,
independently of $M_X$. Note that the annihilation gets very strong with a huge factor, $M_{X}/H(M_X\sim 1 \mbox{TeV})$. 
As one expects from such large $K$ and an even larger gauge annihilation
effect, the RH sneutrinos follow closely the thermal
equilibrium distribution until very low temperature, and the
lepton asymmetry freezes out at large $z$, that is, $T <100$ GeV.
Such a behavior is shown in \fref{fig4:figLA}, in which  the evolution of the
lepton asymmetry, $\log(Y_{\Delta L}/\epsilon)$, is plotted.  The lepton asymmetry increases for a larger $M_{Z^\prime}$, as annihilation processes are suppressed for heavy $M_{Z^\prime}$. For lager $z$, the asymmetry reaches an
asymptotic value for which the annihilation effect eventually
drops out due to Boltzmann suppression.

Let us note that the lepton asymmetry is still increasing during
the electroweak phase transition at $T\sim 100$ GeV, which is well
before the asymptotic value is reached. Therefore, it is important
to include the sphaleron effect after phase transition to
obtain the right value for the  baryon asymmetry.  To do this, let
us define $\tilde{Y}_{\Delta \ell}= Y_{\Delta L} - Y_{\Delta B}$ where the $\tilde{Y}_{\Delta \ell}$ is the
solution of the above Boltzmann equations \eref{eq4:boltz} as
shown in \fref{fig4:figLA}. The baryon
asymmetry can be expressed as
\begin{equation}
 Y_{\Delta B}(z) = -\int_0^z\! dy\, A_1(y) \tilde{Y}_{\Delta \ell}(y) \exp[- \int^z_y\!
 dx\,
 A_2(x)]\,,
\end{equation}
with
\begin{eqnarray}
A_1(z)  &=& z {\Gamma_{\rm EWsp} \over H_1} \left( {28\over 51}
 + {108\over 561} {v^2(T)\over T^2}  \right) \nonumber\\
A_2(z) &=& z {\Gamma_{\rm EWsp} \over H_1} \left( {79\over 51} +
{333\over 561} {v^2(T)\over T^2}\right) .\nonumber
\end{eqnarray}
$\Gamma_{\rm EWsp}$ denotes the sphaleron interaction rate and $v(T)=v_0
(1-T^2/T_c^2)^{1/2}$ with $v_0=246$ GeV.  Above the electroweak
phase transition $T>T_c$, $v(T)=0$ and $\Gamma_{\rm EWsp}$ can be taken
to be infinitely large so that the standard result, $Y_{\Delta B} =
-{28\over 51} Y_{\Delta L} = -{28\over 79} \tilde{Y}_{\Delta \ell}$, is recovered.
Below $T_c$, several calculations for $\Gamma_{\rm EWsp}$ have been made
within the validity range of $ M_W(T) \ll T \ll M_W(T)/\alpha_w$
\cite{Khlebnikov:1988sr, Kuzmin:1985mm, Arnold:1987mh, Carson:1990jm}, which indicates that the sphaleron interaction is
still very active just below $T_c$ and its freeze-out happens
somewhat later \cite{Pilaftsis:2005rv}.  One finds that $\Gamma_{\rm EWsp}$ is an
extremely steep function of $T$ and thus it is a fairly good
approximation to calculate the final baryon asymmetry as follows:
\begin{equation} \label{eq4:basymmetry}
Y_{\Delta B} \approx -{A_1\over A_2} \tilde{Y}_{\Delta \ell}\Big|_{T_{\rm EWsp}} \approx
-{1\over3} \tilde{Y}_{\Delta \ell} \Big|_{T_{\rm EWsp}}\,,
\end{equation}
with $T_{\rm EWsp}$ being the sphaleron freeze-out temperature.  There is
an uncertainty in determining $T_{\rm EWsp}$, which arises from the
limited knowledge in calculating the sphaleron rate given by
\begin{equation}
\gamma_{\Delta (B+L)}= 4\pi \kappa \omega_- {\cal N}_{\rm tr} 
{\cal N}_{\rm rot}  \alpha_3^{-6} \left( \alpha_W T\over 4\pi\right)^3
e^{-E_{\rm EWsp}/T}.
\end{equation}
The most uncertain quantity, $\kappa$, is known to  lie in the
range $\kappa=(10^{-4}-1)$ \cite{Khlebnikov:1988sr, Kuzmin:1985mm, Arnold:1987mh, Carson:1990jm}.  Adopting the parameters given
in Ref.~\cite{Pilaftsis:2005rv}, we find $T_{\rm EWsp}=80-90$ GeV, which is in the
region of   $T>M_W(T)$ to which the above sphaleron calculation (\eref{eq4:basymmetry}) may be well extended. Current lattice calculations point us, however, to higher values for the electroweak sphaleron freeze-out temperature $T_{\rm EWSp} = 131.7 \pm 2.3$ GeV~\cite{DOnofrio:2014rug}.

\begin{figure}[t]
\begin{center}
\includegraphics[width=0.75\textwidth]{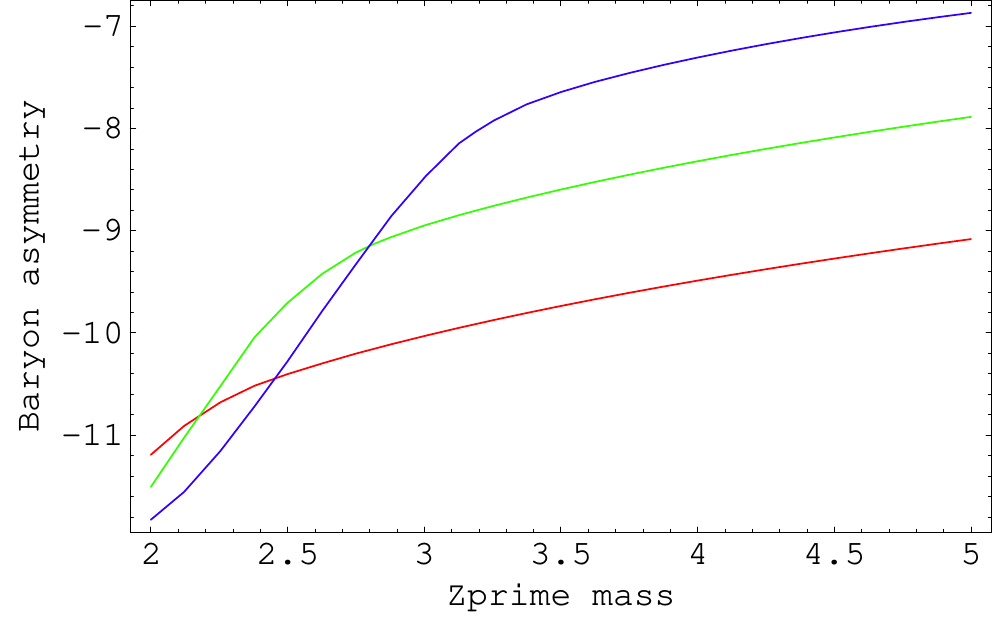}
\end{center}
\caption{\label{fig4:figBA} The baryon asymmetry in unit of $\epsilon$
is shown with respect to $M_{Z^\prime}/\mbox{TeV}$ for $K=100$.
The curves, $\log(Y_{\Delta B}/\epsilon)$, are for $M_X=0.3, 0.6$ and $0.9$ TeV from
below on the right-hand side.  Figure taken from Ref.~\cite{Chun:2005tg}.}
\end{figure}

Following the above prescription, the baryon asymmetry,
$\log(Y_{\Delta B}/\epsilon)$, can be calculated as a function of
$M_{Z^\prime}$ for various values of $M_X$ and is shown in \fref{fig4:figBA}, taking
$T_{\rm EWsp}=90$ GeV. As one can see, $Y_B$ depends strongly on the
masses, $M_{Z^\prime}$ and $M_X$. It is, however, almost unaffected by a variation of $K$ in the range: $K=10-10^3$.
Requiring $Y_{\Delta B}/\epsilon>10^{-10}$, one obtains a lower bound on $M_{Z^\prime}$:
$M_{Z^\prime} > (2.3-3)$ TeV for $M_X= (0.3-0.9)$ TeV.
Taking $T_{\rm EWsp}=80$ GeV, the bound becomes weaker: $M_{Z^\prime} >
(2.1-2.6)$ TeV.

\medskip

Current LHC experiments are already probing the  multi-TeV region of $Z'$ models.
For the $U(1)_\chi$ model, the latest bound on the $Z'$ mass is about 3.7 TeV \cite{ATLAS:2016cyf}, which requires $M_X\lesssim 1$ TeV according to \fref{fig4:figBA}.  A crucial test of TeV leptogenesis in $Z'$ models 
is by observing CP asymmetry in the RH neutrino (sneutrino) decay, $N\to \ell^\pm W^\mp (\tilde N \to \ell^\pm  \tilde \chi^\mp)$ 
following the production $pp \to Z' \to NN (\tilde N \tilde N^\dagger)$ (here and in the following, $\ell^\pm$ indicates specifically the charged leptons $e^\pm, \mu^\pm, \tau^\pm$).
RH neutrinos, being Majorana particles, decay to both-sign leptons $N \to \ell^\pm W^\mp$ leading to clean signals of same-sign di-leptons at the LHC. 
The CP asymmetry $\epsilon$ relevant for the leptogenesis can be measured by 

\begin{equation}
\epsilon = {1\over2} { {\cal N}(\ell^- \ell^-) - {\cal N}(\ell^+ \ell^+) \over 
{\cal N}(\ell^- \ell^-)  + {\cal N}(\ell^+ \ell^+) }.
\end{equation}
For the sneutrinos $\tilde N$ and $\tilde N^\dagger$, having the definite decay mode 
$\tilde N \to \ell^+ \tilde \chi^-$ and $\tilde N^\dagger \to \ell^- \tilde \chi^-$, a similar relation arises through $\tilde N$--$\tilde N^\dagger$ oscillation: $pp \to Z' \to \tilde N \tilde N^\dagger \Rightarrow \tilde N \tilde N~ (\tilde N^\dagger \tilde N^\dagger) \to \ell^+ \ell^+ ~(\ell^- \ell^-)$. 
Furthermore, the sneutrino oscillation parameter, $x\equiv \Delta M_{\tilde N}/\Gamma_{\tilde N} = B/\Gamma_{\tilde N}$, could be measured by 
\begin{equation}
{ {\cal N}(\ell^\pm \ell^\pm)  \over  {\cal N}(\ell^- \ell^+)  } = {r \over 1+r^2}\,,
\end{equation}
where $r\equiv x^2/(2+x^2)$.  Similar, but more spectacular phenomena can occur in the resonant 
type II seesaw, which will be discussed in \sref{ch5:sec5:typeII-soft-leptogenesis}.

To summarize, $U(1)^\prime$ seesaw models can lead to successful  leptogenesis for a $Z^\prime$ mass above $\sim$ 3 TeV provided a resonant condition.  Such a realization could be tested  by observing large CP violation in the same-sign di-lepton final states as well as by sneutrino-antisneutrino oscillation phenomena.

\subsection{Leptogenesis in left-right symmetric models}
\label{ch5:sec4:extendedgauge-LR}
In the context of the SM electroweak gauge group $SU(2)_L\times U(1)_Y$, the two main features of the seesaw mechanism~\cite{Minkowski:1977sc,Mohapatra:1979ia,Yanagida:1979as,GellMann:1980vs,Glashow:1979nm}, namely, the existence of RH neutrinos and endowing them with a Majorana mass which breaks the accidental global $B-L$ symmetry of the SM, do not follow from any underlying principle. Extending the EW gauge group to the L-R symmetric gauge group  $SU(2)_L\times SU(2)_R\times U(1)_{B-L}$~\cite{Pati:1974yy, Mohapatra:1975jm, Senjanovic:1975rk} provides a simple ultraviolet-complete seesaw model. Here, the RH neutrinos are a necessary part of the model guaranteed by the gauge symmetry and anomaly cancellations, and do not have to be added ab initio to implement the seesaw mechanism~\cite{Minkowski:1977sc, Mohapatra:1979ia, Yanagida:1979as, GellMann:1980vs, Glashow:1979nm}. An important point is that the RH neutrinos acquire a Majorana mass as soon as the $SU(2)_R$ symmetry is broken at a scale $v_R$. This is quite analogous to the way the charged fermions get masses in the SM by the Higgs mechanism when the $SU(2)_L$ gauge symmetry is broken at the EW scale $v$. The Higgs field that gives mass to the RH neutrinos becomes the analogue of the 125 GeV Higgs boson discovered at the LHC. 

Just as in the minimal seesaw with RH neutrinos, the observed baryon asymmetry of the Universe can be explained in the L-R seesaw by the mechanism of leptogenesis. For this to be both efficient and testable in laboratory experiments, we need two main ingredients: (i) low RH scale $v_R$ not far above the TeV scale, so that the RH neutrino effects are experimentally accessible; (ii) quasi-degeneracy between at least two RH neutrinos, with the mass difference comparable to their decay widths in order to have a resonant enhancement of the CP asymmetry~\cite{Pilaftsis:1997dr, Pilaftsis:1997jf, Pilaftsis:2003gt}, thus avoiding the Davidson-Ibarra lower bound of $\sim 10^9$ GeV on the RH neutrino mass~\cite{Davidson:2002qv, Buchmuller:2002jk} for successful leptogenesis (see \eref{eq0:davidsonibarra}). Unlike in the minimal seesaw case, we do not necessarily require the Yukawa couplings to be large in L-R seesaw for observable RH neutrino effects, thanks to the purely RH gauge interactions. For instance, both the production and decay of the RH neutrinos at colliders can proceed via the RH charged-current process $pp\to W_R\to N\ell^\pm \to \ell^\pm \ell^\pm jj$~\cite{Keung:1983uu, Nemevsek:2011hz, Das:2012ii, Chen:2013fna, Deppisch:2015qwa, Dev:2015kca, Ruiz:2017nip}, regardless of the size of the Yukawa couplings. Similarly, the low-energy rare processes, such as lepton flavor violation~\cite{Riazuddin:1981hz, Cirigliano:2004mv, Barry:2013xxa, deGouvea:2013zba} and $0 \nu \beta \beta$ decay~\cite{Mohapatra:1980yp, Hirsch:1996qw, Tello:2010am, Rodejohann:2011mu, Dev:2013vxa} receive additional contributions from the RH gauge sector which could enhance these signals for low-scale L-R seesaw.  

However, the same RH gauge interactions that lead to new experimental signals in the L-R seesaw could have potentially dangerous effects on successful leptogenesis~\cite{Carlier:1999ac, Frere:2008ct, Dev:2014iva, Dev:2015vra, Dhuria:2015cfa}. This is because a single RH neutrino can interact through $W_R$ exchange with SM fermions which are all in thermal equilibrium, thereby inducing very efficient scatterings, such as $Ne_R\leftrightarrow d_R\bar{u}_R$, $Nu_R\leftrightarrow e_R\bar{d_R}$, $Nd_R\leftrightarrow e_Ru_R$, as well as decays $N\to W_R^*\ell \to \ell jj$, which are CP-conserving and therefore contribute to the dilution of the lepton asymmetry generated from the CP-violating decay of the RH neutrinos $N\to \ell \phi, \ell^c\phi^c$. Since in generic versions of the TeV-scale seesaw model, sub-eV neutrino masses would require the Dirac Yukawa couplings to be $\lambda \lesssim 10^{-5}$,
the additional dilution effects would lead to an efficiency factor $\kappa \sim \frac{\lambda^2 M^4_{W_R}}{g_R^4 M^4_N}$ ($g_R$ being the $SU(2)_R$ gauge coupling, which is assumed to be equal to the $SU(2)_L$ gauge  coupling in the minimal L-R seesaw), which is too small for a low-scale leptogenesis to work. 
Through a detailed analysis of the relevant Boltzmann equations,  it was concluded in~\cite{Frere:2008ct} that even for maximal CP asymmetry $\epsilon\sim {\cal O}(1)$, the observed 
value of the baryon asymmetry can be explained by leptogenesis in the minimal L-R model, only if $M_{W_R}\geq 18 $ TeV. Turning this argument around, if a positive signal for $W_R$ is observed at the LHC, this will falsify leptogenesis in L-R seesaw framework. It is worth noting here that the leptogenesis constraint on $W_R$ is significantly more stringent than that on $Z'$ (see Section~\ref{ch5:sec4:extendedgauge-Zprime}) mainly because the $Z'$-induced RH neutrino scattering processes are doubly Boltzmann suppressed, and therefore, their interaction rates drop more quickly than the decay/inverse decay rate and the $W_R$ induced scattering, which are simply Boltzmann suppressed.  

Since this is such an important issue for the survival of the L-R seesaw leptogenesis, it is worth examining the robustness of the lower bound on $M_{W_R}$ to see if there exists {\em any} allowed parameter space in TeV-scale L-R seesaw models with successful leptogenesis for smaller values of $M_{W_R}$. The basic idea is that the only source term for leptogenesis comes from the CP-violating RH neutrino decays $N\to \ell \phi, \ell^c\phi^c$, which depends on the Dirac Yukawa coupling. So, for a class of the L-R seesaw models with larger Yukawa couplings~\cite{Dev:2013oxa}, it is in principle possible to overcome the large dilution effect induced by a lighter $W_R$, thereby weakening the lower bound on $M_{W_R}$~\cite{Dev:2014iva, Dev:2015vra}. Of course, a TeV-scale type I seesaw with large Yukawa couplings comes with the price of fine-tuning in the seesaw matrix in order to be consistent with the light neutrino oscillation data \cite{Olive:2016xmw}. However, such special textures of the Dirac and Majorana neutrino mass matrices can be motivated from some flavor symmetry in the leptonic sector, which ensures their stability against large radiative corrections~\cite{Kersten:2007vk, Dev:2013oxa}. For such L-R seesaw models with large Dirac Yukawa couplings and taking into account flavor effects, it was found that successful leptogenesis requires $M_{W_R}\geq 9.9$ TeV for the maximal CP asymmetry~\cite{Dev:2015vra}. A similar estimate of this absolute lower bound was also given in Ref.~\cite{Frere:2008ct}. This absolute lower bound is not too far from the generic bound of 18 TeV, which is consistent with the expectation that flavor effects cannot significantly alter the results because gauge scatterings are flavor-blind.  For comparison, the $\sqrt s=14$ TeV LHC with the high-luminosity option can probe $W_R$ masses only up to 5-6 TeV~\cite{Ferrari:2000sp}. Therefore, a positive signal for $W_R$ at the LHC  will indeed falsify leptogenesis as a mechanism for understanding the origin of matter in L-R seesaw framework.  

\subsubsection{Boltzmann Equations}
The basic steps in our analysis are as follows: first we calculate the flavored CP asymmetry
\begin{align}
\epsilon_{k \alpha} \ = \ \frac{1}{\Gamma_{N_k}} [\Gamma (N_k \to \ell_\alpha \phi) - \Gamma (N_k \to \ell_\alpha^c \phi^c)] \; ,
\label{eq4:epsilon}
\end{align}
where the partial decay widths are given in terms of the resummed Yukawa couplings in the RH neutrino mass eigenbasis~\cite{Pilaftsis:2003gt, Dev:2014laa},
\begin{align}
  \Gamma(N_{k}\to \ell_\alpha \phi) \
  = \  \frac{M_{k}}{16\pi}{\lambda}_{\alpha k}{\lambda}^*_{\alpha k}\; , \qquad   \Gamma(N_{k}\to \ell^c_\alpha \phi^c ) \
  = \  \frac{M_{k}}{16\pi}{\lambda^c}_{\alpha k}{\lambda}^{c*}_{\alpha k}\; .
  \label{eq4:gamma}
\end{align}

and the total decay width is given by 
\begin{align}
\Gamma_{N_k} \ = \ \sum_{\alpha}\left[\Gamma(N_k \to \ell_\alpha \phi) + \Gamma(N_k\to \ell_\alpha^c\phi^c) \right] + 2\: \Gamma(N_k\to \ell_R q_R\bar{q}'_R) \; .
\label{eq4:tot_width}
\end{align}
Note that the last term in \eref{eq4:tot_width} is due to the three-body decay mediated by $W_R$: 
\begin{eqnarray}
\Gamma(N_k\to \ell_R q_R\bar{q}'_R)  &=& \ \Gamma(N_k\to \bar{\ell}_R \bar{q}_R q'_R) \nonumber \\
&=& \ \frac{3g_R^4}{2^9\pi^3 M_{k}^3} \int_0^{M_{k}^2} ds \frac{M_{k}^6-3M_{k}^2 s^2+2 s^3}{(s-M_{W_R}^2)^2+M_{W_R}^2\Gamma_{W_R}^2} \; ,
\label{eq4:3body}
\end{eqnarray}
where $\Gamma_{W_R}\simeq (g_R^2/4\pi)M_{W_R}$ is the total decay width of $W_R$, assuming that all three heavy neutrinos are lighter than $W_R$. The CP asymmetry~\eref{eq4:epsilon} in our case is determined by the RH neutrino Majorana mass matrix, as well as the magnitudes of the Yukawa couplings, and these parameters are constrained by the light neutrino and charged lepton mass and mixing parameters in the L-R model~\cite{Dev:2013oxa}. 

The next step is to calculate the thermodynamic evolution of the normalized heavy neutrino and lepton doublet number densities $\eta^N_k$ and $\eta_{\alpha}^{\Delta L}$ respectively, where $\eta^X\equiv (n^X - n^{\bar{X}})/n_\gamma$ and $n_\gamma=2M_{1}^3\zeta(3)/(\pi^2 z^3)$ is the photon number density, $\zeta(x)$ being the Riemann zeta function. For simplicity, we use the flavor-diagonal Boltzmann equations:
\begin{align}
\frac{d\eta^N_k}{dz} \ & = \ -\left(\frac{\eta^N_k}{\eta^N_{\rm eq}}-1\right)(D_k+S_k) \;, \label{eq4:be1} \\
\frac{d\eta^{\Delta L}_\alpha}{dz} \ & = \ \sum_k \epsilon_{k \alpha}\left(\frac{\eta^N_k}{\eta^N_{\rm eq}}-1\right)\widetilde{D}_k -\frac{2}{3} \eta^{\Delta L}_\alpha W_\alpha \;, \label{eq4:be2}
\end{align}
where %$k,l = e,\mu,\tau$ are the lepton indices, $k=1,2$ the heavy neutrino indices, 
$z=M_{1}/T$ is a dimensionless variable ($T$ being the temperature of the Universe) and
%\begin{eqnarray}
$\eta^N_{\rm eq}  \equiv  n^N_{\rm eq}/n_\gamma  =  z^2 K_2(z)/2\zeta(3) $
%\label{etaeq}
%\end{eqnarray} 
is the heavy neutrino equilibrium number density, $K_n(x)$ being the $n$-th order modified Bessel function of the second kind.  The various decay ($D_k,~\widetilde{D}_k$), scattering ($S_k$), and washout ($W_\alpha$) rates appearing in \eref{eq4:be1} and \eref{eq4:be2} are given by 
\begin{align}
		\widetilde{D}_k & \ = \ \frac{z}{n_\gamma H_N} \sum_\beta \widetilde{\gamma}^D_{k \beta}, \label{eq4:Dtilde} \\
	D_k & \ = \ \frac{z}{n_\gamma H_N} \sum_\beta \gamma^D_{k \beta}, \\
	S_k & \ = \ \frac{z}{n_\gamma H_N} \sum_\beta (\gamma^{S_L}_{k \beta} + \gamma^{S_R}_{k \beta}) ,\\
W_\alpha  & \ = \ \frac{z}{n_\gamma H_N} \left[ \sum_k \left( B_{k \beta} \sum_\beta \gamma^D_{k \beta} + \widetilde{\gamma}^{S_L}_{k \alpha} + \widetilde{\gamma}^{S_R}_{k \alpha} \right) + \sum_\beta \left( \gamma^{(\Delta L = 2)}_{\alpha \beta} + \gamma^{(\Delta L = 0)}_{\alpha \beta} \right) \right] \nonumber \\
	& \ \equiv \  \frac{1}{2 \zeta (3)} z^3 K_1 (z) K_{\alpha}^\text{eff} (z) \; , \label{eq4:wash}
\end{align}
where $H_N\equiv H(z=1)\simeq 17 M_{1}^2/M_{\rm Pl}$ is the Hubble parameter at $z=1$, assuming only SM degrees of freedom in the thermal bath, and $B_{k \alpha}$ is the branching fraction of the RH neutrino decays relevant for the generation of CP asymmetry: 
\begin{align}
		B_{k \alpha} \ = \ \frac{1}{\Gamma_{N_k}} [\Gamma (N_k \to \ell_\alpha \phi) + \Gamma (N_k \to \ell_\alpha^c \phi^c)] \; . 
\label{eq4:BR}
\end{align}
The various $\gamma$'s appearing in \eref{eq4:Dtilde} to \eref{eq4:wash} represent the reaction rates that involve the RH neutrino decays and inverse decays as well as other $2\leftrightarrow 2$ scattering processes in the model (see Ref.~\cite{Dev:2014iva} for details). Only the two-body decays of the RH neutrinos involving complex Yukawa couplings are responsible for building up the CP asymmetry, whereas all other processes lead to dilution/washout effects. 

\begin{figure}[t]
	\centering
		\includegraphics[scale=0.49]{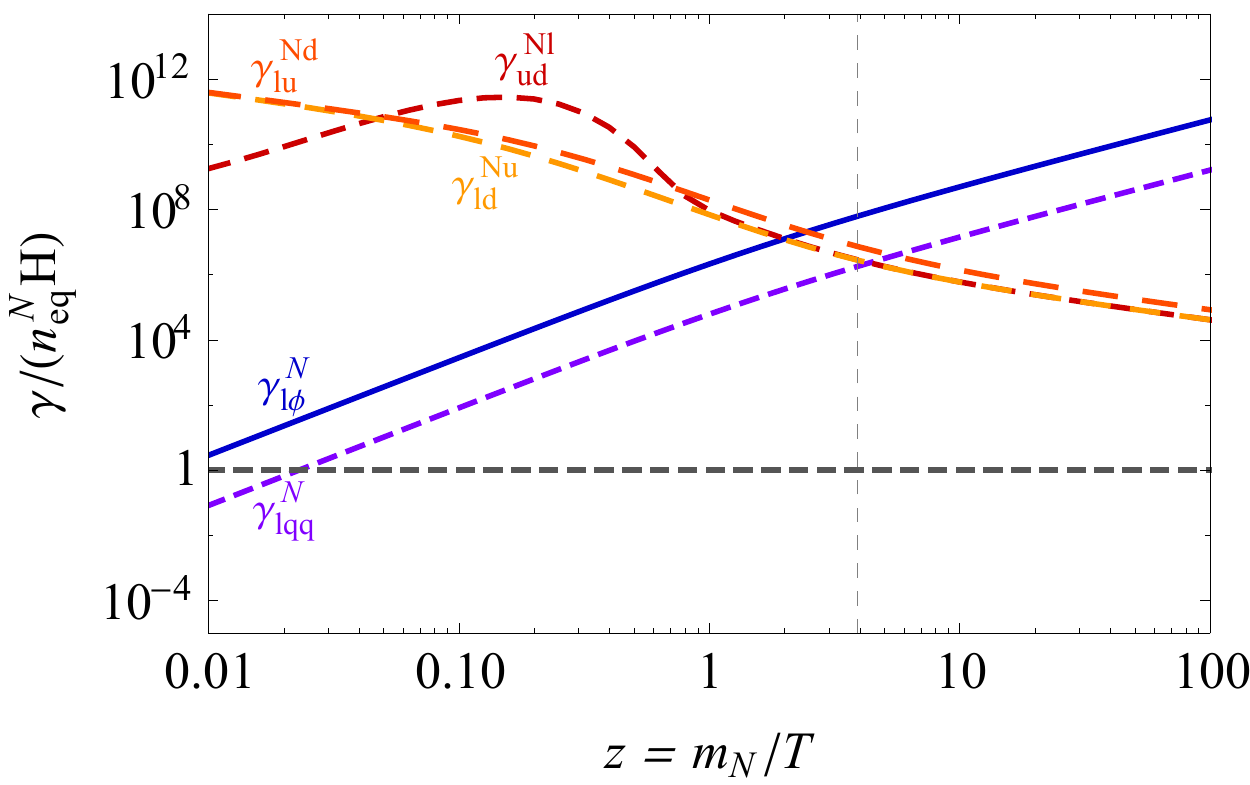} 
		\includegraphics[scale=0.49]{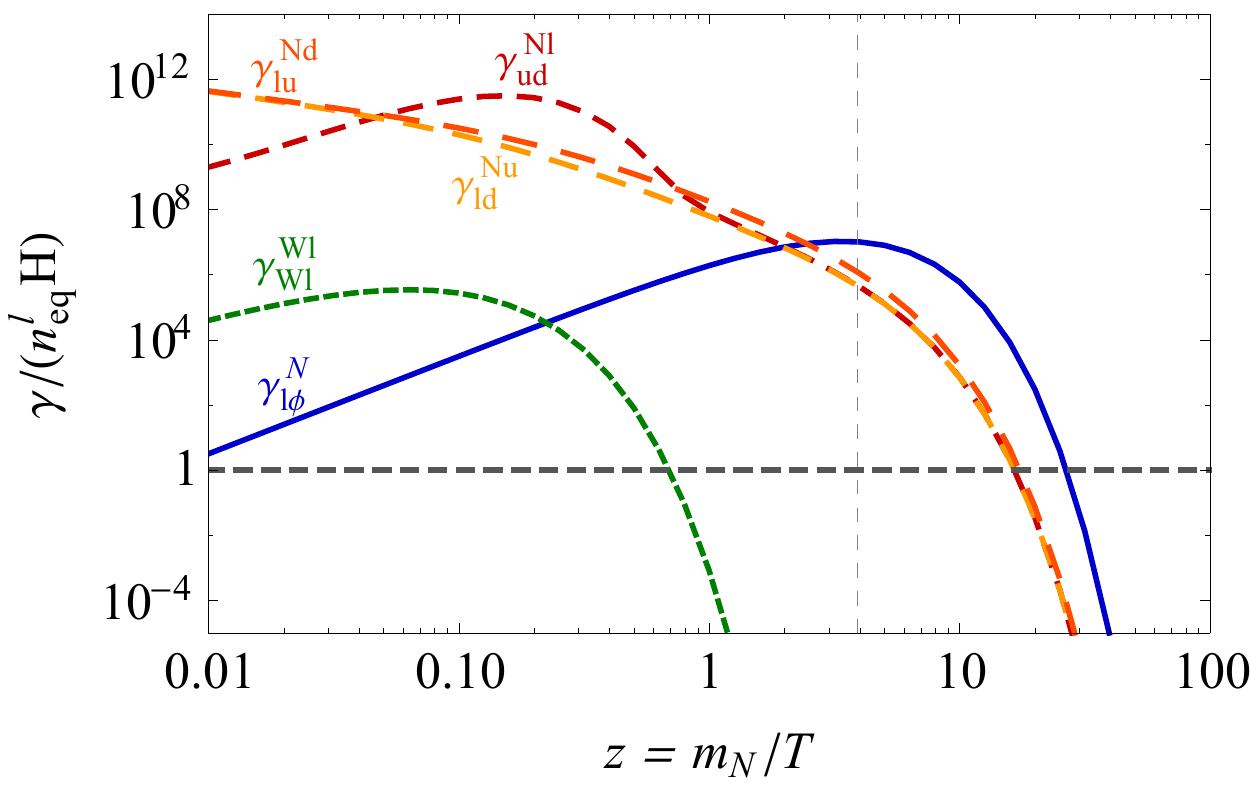} 
		\caption{Thermodynamic evolution of various reaction rates in the L-R seesaw leptogenesis. $m_N$ indicates in this plot the mass scale of the heavy neutrino $M_N$. Figure taken from Ref.~\cite{Dev:2015vra}.}
	\label{fig4:gnH}
\end{figure}

We present in \fref{fig4:gnH} the evolution of the various collision rates, where the left panel shows the evolution of the normalized heavy neutrino number density \eref{eq4:be1}, while the right panel shows it for the lepton doublet number density \eref{eq4:be2}. The vertical dashed line indicates the critical value $z=z_c$ beyond which the sphaleron processes become ineffective. The horizontal dashed line is shown for easy comparison with the Hubble rate. %Here we have used $m_{W_R}=13.1$ TeV which is the lowest value of $m_{W_R}$ we found in this model with an {\em explicit} neutrino fit that satisfies the leptogenesis constraints. 
%Also we have  varied the average heavy neutrino mass $M_N$ and the overall scale of the Dirac mass matrix $M_D$ to keep the neutrino fit intact. 

\subsubsection{Lower bound of $M_{W_R}$}
Now we discuss our procedure to derive a lower bound of $M_{W_R}$ which is compatible with successful leptogenesis in the L-R seesaw model with all three RH neutrino masses being quasi-degenerate. In the strong washout regime, the flavored lepton asymmetry is approximately given by~\cite{Buchmuller:2004nz}
\begin{align}
	\eta^{\Delta L}_{k \alpha}(z) \ \simeq \  \frac{3}{2 z K_\alpha^\text{eff} (z)} \epsilon_{k \alpha} \frac{\widetilde{D}_k (z)}{D_k (z) + S_k (z)} \; ,
	\label{eq4:etaL}
\end{align}
where $K_\alpha^{\rm eff}$ is the effective washout parameter as defined in \eref{eq4:wash} and $\widetilde{D}/(D+S)$ is the effective dilution factor. The total lepton asymmetry is given by \eref{eq4:etaL}, which can be rewritten using \eref{eq4:wash} as
\begin{align}
	\eta^{\Delta L} (z) &= \sum_{k \alpha} \eta^{\Delta L}_{k \alpha} = \frac{3}{4 \zeta (3)} z^2 K_1 (z) \sum_{k \alpha} \frac{1}{W_\alpha (z)}\epsilon_{k \alpha}  \frac{\widetilde{D}_k (z)}{D_k (z) + S_k (z)}   \; .
\label{eq4:etaL1}
\end{align}
This is to be compared with $\eta^{\Delta L}_{\rm obs} = -(2.47\pm 0.03) \times 10^{-8}$, which reproduces the observed value of the baryon asymmetry $\eta^{\Delta B}$ as in \eref{eq0:etaBobs}, after taking into account the sphaleron conversion rate and the entropy dilution factors. 

For the purpose of deriving a lower bound on $M_{W_R}$, we assume that the dominant dilution effect comes from the $W_R$-mediated scattering processes, i.e., $S_k \approx \sum_\beta S_{R k} = 3 S_{R k}$ where $S_{R k} \equiv z\widetilde{\gamma}^{S_R}_{k \beta}  / (n_\gamma H_N)$. This follows from the fact that  gauge interactions are flavor-blind,  and hence, $\widetilde{\gamma}^{S_R}_{k \beta}$ has no dependence on lepton flavors. Similarly, when all the RH neutrino masses are quasi-degenerate, we can write $S_R \equiv \sum_k S_{R k} \approx 3 S_{R k}$. We also assume that $D = \sum_k D_k \approx 3 D_k$ for simplicity. Note that $S_R$ is not summed over lepton flavors while $D$ is. We further define $\epsilon^Y_{k \alpha}$ and $B^Y_{k \alpha}$ as the CP asymmetry and branching ratio, respectively, without the three-body decay width included in the denominator of \eref{eq4:epsilon} and \eref{eq4:BR}. For simplicity, we further assume that the branching ratios of the decay process $N_k \to \ell_\alpha \phi$ are the same for all the lepton flavors. With all the above-mentioned reasonable assumptions, we can approximate the total lepton asymmetry in \eref{eq4:etaL1} at the critical temperature  as
\begin{align}
	|\eta^{\Delta L} (z_c)| &\approx \frac{3 z_c^2 K_1 (z_c)}{4 \zeta (3)} \sum_\alpha \frac{1}{\sum_k [B_{k \alpha} D_k (z_c) + S_{R k} (z_c)]} \left| \sum_k \epsilon_{k \alpha} \frac{\widetilde{D}_k (z_c)}{D_k (z_c) + 3 S_{R k} (z_c)} \right| \nonumber \\
	&\approx \frac{9}{4 \zeta (3)} \frac{z_c^2 K_1 (z_c)}{S_R (z_c)} \frac{r_s r_d^2}{(3 + r_s r_d) (3 + r_s)} \epsilon^Y_\text{tot}\,,
	\label{eq4:etaL2}
\end{align}
where the ratios $r_s \equiv D (z_c) / S_R (z_c)$ and $r_d \equiv \widetilde{D} (z_c) / D (z_c)$ parameterize the relative dilution effect for the two-body decay, $W_R$-mediated decay and $W_R$-mediated scattering processes. The values of $r_s$ and $r_d$ depend on $M_N$, $M_{W_R}$, and the Yukawa coupling $\lambda$. Assuming a specific value of $\lambda$ and calculating the two-body decay width simply as $\Gamma (N_k \to \ell_\alpha \phi) = \Gamma (N_k \to \ell_\alpha^c \phi^c) = \lambda^2 M_k / 16 \pi$, we can evaluate these parameters as functions of $M_k$ and $M_{W_R}$.

\begin{figure}[t]
	\centering
		\includegraphics[scale=0.6]{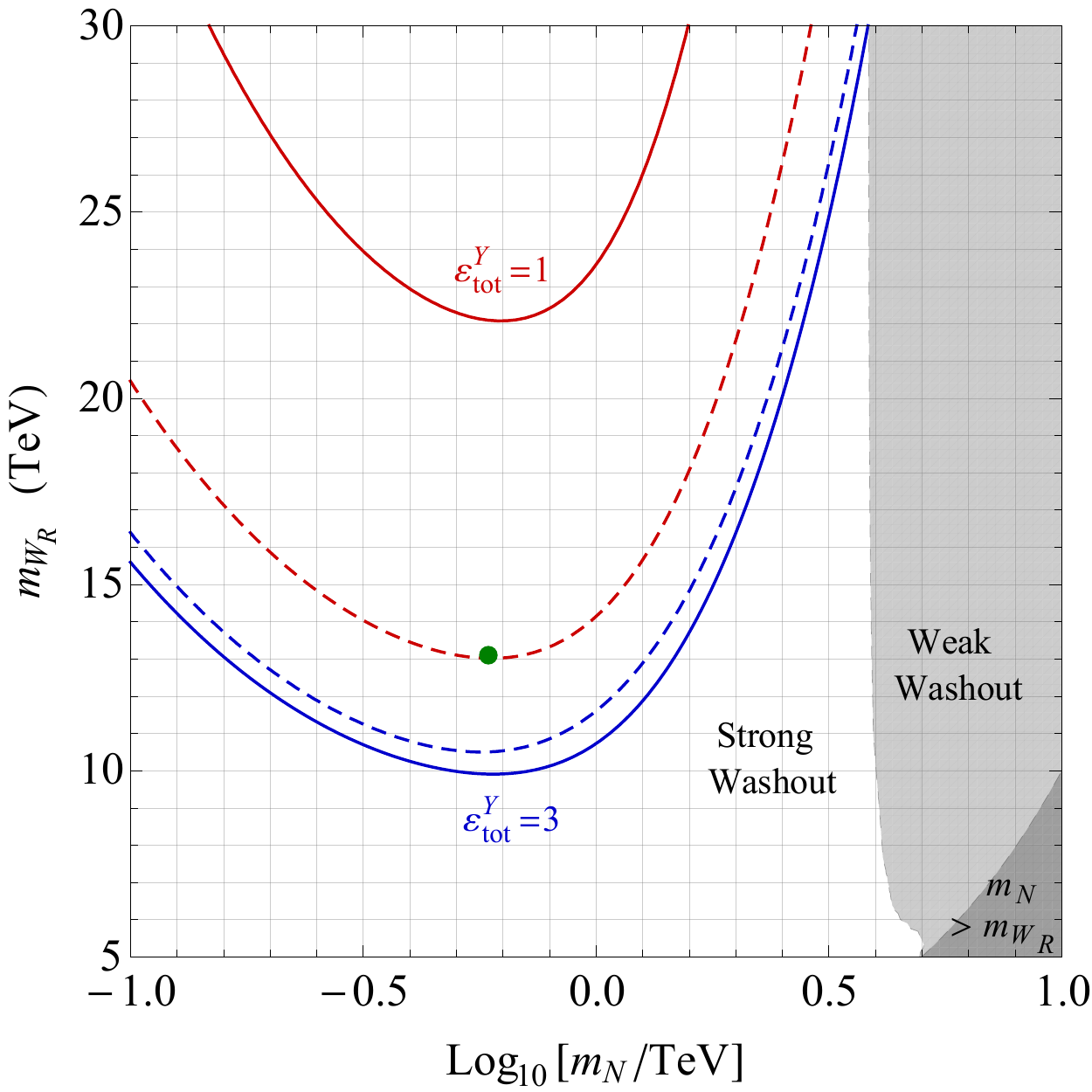} 
	\caption{Contour plots of $|\eta^{\Delta L} (z_c)| = 2.47 \times 10^{-8}$ for $\lambda = 10^{-3.8}$ (dashed lines) and $\lambda = 10^{-3.5}$ (solid lines) with $\epsilon^Y_{\rm tot}=1$ (red lines) and  $\epsilon^Y_{\rm tot}=3$ (blue lines). The non-trivial dependence on the Yukawa coupling is due to the interplay of decay and dilution/washout terms, which is further illustrated in \fref{fig4:3D}. $m_N$ indicates in this plot the mass scale of the heavy neutrino $M_N$. The green dot corresponds to the example fit value presented in Ref.~\cite{Dev:2015vra} from which this plot is taken.}
	\label{fig4:etaLCon}
\end{figure}

\Fref{fig4:etaLCon} shows the contour plots of $|\eta^{\Delta L} (z_c)| = 2.47 \times 10^{-8}$ for two different Yukawa couplings $\lambda = 10^{-3.8}$ and $\lambda = 10^{-3.5}$. The red curves correspond to $\epsilon^Y_\text{tot} = 1$, which is the total CP asymmetry in the example fit given in Ref.~\cite{Dev:2015vra}. Any region outside the red curve is incompatible with successful leptogenesis under the assumptions we introduced to obtain it, i.e. $M_{1} \approx M_{2} \approx M_{3}$, $B^Y_{k 1} \approx B^Y_{k 2} \approx B^Y_{k 3} \approx 1/3$, $\epsilon^Y_\text{tot} = 1$ for the two specific values of $\lambda$. For $\lambda = 10^{-3.8}$,  we find that the lowest value of $M_{W_R}$ allowed for  $\epsilon^Y_\text{tot} = 1$ is 13 TeV at around $M_k$ = 580 TeV. 
If we use the same expression and take the maximal CP asymmetry allowed in principle, i.e. $\epsilon^Y_\text{tot} \equiv \sum_{k \alpha} \epsilon^Y_{k \alpha} = 3$, then we obtain the blue curves \fref{fig4:etaLCon}. With this maximal CP asymmetry, we have found that the Yukawa coupling $\lambda = 10^{-3.5}$ gives the lower bound of $M_{W_R}$ = 9.9 TeV at $M_k$ = 630 GeV.

\begin{figure}[t]
	\centering
		\includegraphics[scale=0.5]{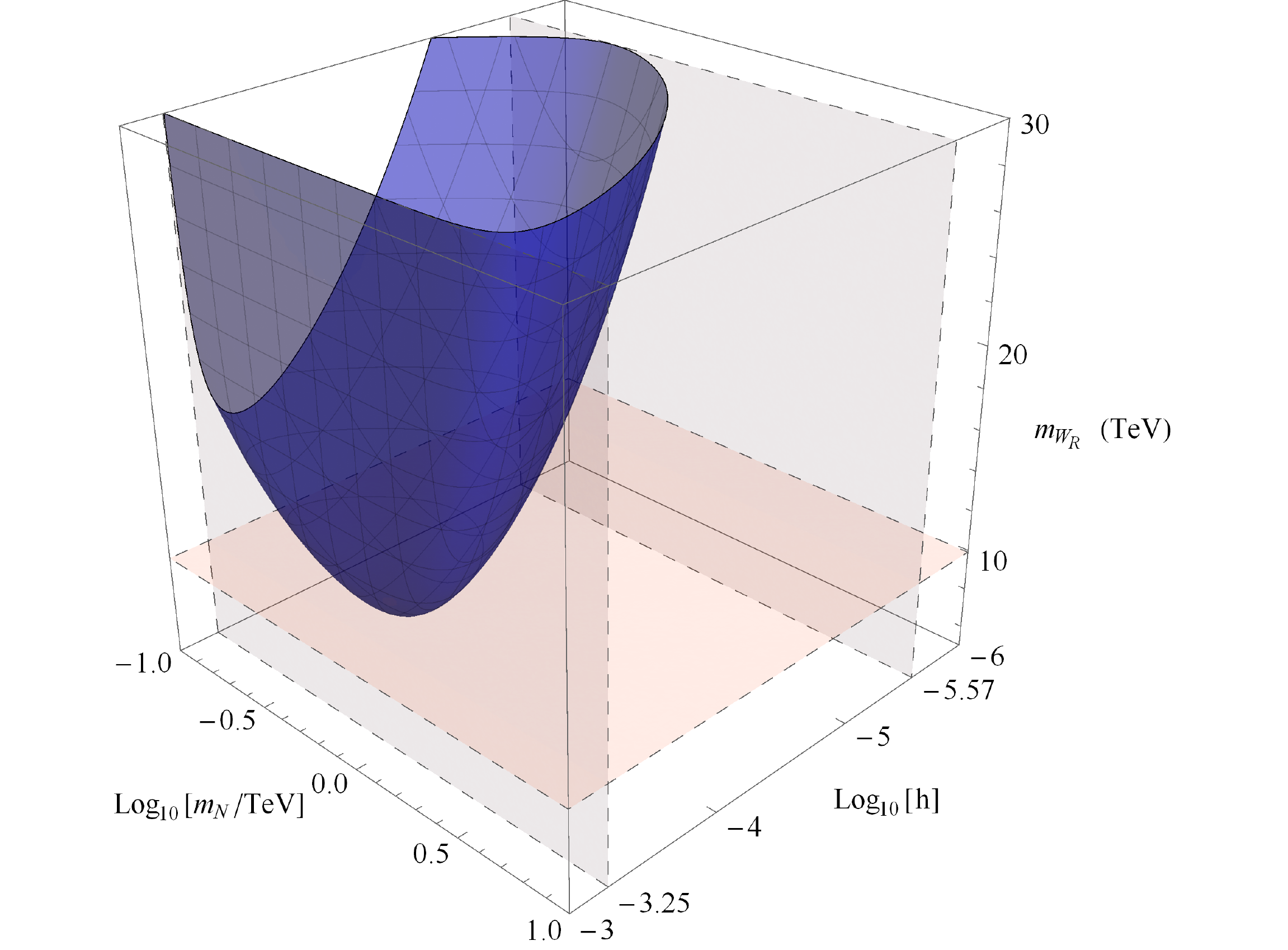} 
	\caption{The allowed region of parameter space (blue shaded region) yielding successful leptogenesis in the L-R seesaw model of Ref.~\cite{Dev:2015vra} from which this plot is taken. The vertical gray surfaces show the bound on the Yukawa couplings ($h=\lambda$), while the horizontal pink surface shows the lower bound on $M_{W_R}$. $m_N$ indicates in this plot the mass scale of the heavy neutrino $M_N$.}
	\label{fig4:3D}
\end{figure}

The Yukawa couplings cannot be increased arbitrarily without spoiling the lepton asymmetry, since not only the source term due to the two-body decay of the RH neutrinos, but also the dilution/washout effects due to inverse decay $\ell \phi\to N$ and $\Delta L=2$ scattering $\ell \phi\leftrightarrow \ell^c\phi^c$ increase with the Yukawa couplings.  Similarly, for very small values of the Yukawa couplings, the branching fraction of the two-body decay mode becomes comparable or smaller than the three-body decay mode due to $W_R$ interactions, and therefore, the dilution/washout effect again increases. Thus, successful leptogenesis works only in a range of the Yukawa coupling parameter space. This is shown in the three-dimensional plot given in \fref{fig4:3D}, where we see that leptogenesis constraints in the given L-R model require the Yukawa coupling to be $10^{-5.6}\leq \lambda \leq 10^{-3.2}$. The robustness of the lower bound on $M_{W_R}$ obtained in \fref{fig4:etaLCon} can also be verified from \fref{fig4:3D}.  

In summary, observing a heavy $W_R$ gauge boson at the LHC would directly exclude leptogenesis in the L-R symmetric model, as for $M_{W_R}<10~\mathrm{TeV}$ the dilution effect due to $W_R$-induced scattering processes is too strong to yield the observed baryon asymmetry. Hereby, the precise limit on the $W_R$ mass depends highly on the assumption of Dirac Yukawa couplings as was discussed above, but the statement on the falsifiability of leptogenesis by the LHC holds true in all cases studied here.

\section{Testability of supersymmetric leptogenesis}
\label{ch5:sec5:soft-leptogenesis}
% \section{Testable soft leptogenesis}

When supersymmetry (SUSY) is imposed on leptogenesis, 
new (undesirable, neutral and desirable) features arise: 

\begin{enumerate}

\item \emph{Undesirable.}
The main undesirable feature is the gravitino problem~\cite{Pagels:1981ke,Weinberg:1982zq}:\footnote{To include gravity, SUSY should be local.}
The thermal production of gravitinos increases with the reheating temperature $T_{\rm R}$. While the limit on its abundance places an upper bound on $T_{\rm R}$ (in order not to overclose the Universe), thermal production of RH neutrinos for leptogenesis 
requires a minimum $T_{\rm R}$. The severity of this tension is 
model-dependent: besides $T_{\rm R}$ and the gravitino mass, 
it also depends on the mass spectrum of supersymmetric particles.
For stable gravitinos (a form of dark matter) $T_{\rm R} \lesssim 10^9$ GeV~(see a recent study~\cite{Heisig:2013sva} and the references therein) 
while for unstable gravitinos $T_{\rm R} \lesssim 10^5 - 10^{10}$ GeV with stringent bounds coming mainly from BBN constraints~(see Ref.~\cite{Kawasaki:2008qe} and the references therein).
For hierarchical RH neutrinos, however, a dedicated study~\cite{Antusch:2006gy} 
in supersymmetric type I leptogenesis gives $T_{\rm R} \gtrsim 10^9$ GeV.
(See \sref{ch5:sec3:extended_seesaw} for possibilities of 
lowering the scale of leptogenesis by extending seesaw models.)

\item \emph{Neutral.}
Doubling degrees of freedom and new effective global symmetries at 
high temperature are neutral features \cite{Ibanez:1992aj}. In type I leptogenesis, the impact of doubling 
degrees of freedom~\cite{Campbell:1992hd,Covi:1996wh,Plumacher:1997ru}
and new global symmetries~\cite{Fong:2010qh} 
give rise to ${\cal O}(1)$ modifications with respect to non-SUSY scenarios.

\item \emph{Desirable.}
A desirable feature of SUSY leptogenesis is the introduction of new sources of CP violation, arising from the soft SUSY breaking terms (or just soft terms)~\cite{Boubekeur:2002jn,
Grossman:2003jv,DAmbrosio:2003nfv,Boubekeur:2004ez}. 
Via dimensional analysis, one can naively estimate the
leading CP parameter from the soft term of scale $m_{\rm soft}$ to be
%%%
\begin{equation}
|\epsilon| \sim \frac{1}{16\pi}\frac{m_{\rm soft}}{M_1}.
\end{equation}
%%%
Requiring large enough CP violation (typically 
$|\epsilon| \gtrsim 10^{-6}$, see \sref{ch5:sec3:extended_seesaw}), one can derive the following condition for the lightest heavy neutrino mass 
%%%
\begin{equation}
M_1 \lesssim 10^7\left(\frac{m_{\rm soft}}{1\,{\rm TeV}}\right)\,{\rm GeV}.
\end{equation}
%%%
This relaxes or completely avoids the gravitino problem.
In a particular realization known as 
soft leptogenesis~\cite{Grossman:2003jv,DAmbrosio:2003nfv}, it was shown that utilizing the new effective global symmetries, 
an enhancement as large as ${\cal O}(100)$ can be obtained~\cite{Fong:2010bv}, 
turning the previously neutral feature to a desirable one.

\end{enumerate}

\subsection{Type I soft leptogenesis}
\label{ch5:sec5:typeI-soft-leptogenesis}

Here we will work within the framework of type I seesaw soft leptogenesis 
based on~\cite{Grossman:2003jv,DAmbrosio:2003nfv} (see a recent review~\cite{Fong:2011yx}). 
Since the CP violation comes from soft terms, leptogenesis proceeds through decays of 
RH sneutrinos $\widetilde N_k$ and its antiparticles $\widetilde N_k^*$. 
In particular, soft leptogenesis can proceed even within one generation of RH sneutrinos 
since the otherwise degenerate masses of $\widetilde N_k$ and $\widetilde N_k^*$ 
within a single generation will be split due to the presence of soft bilinear $B_k$ terms. 
In the following we will assume that the RH neutrinos are hierarchical 
such that only the lightest RH sneutrino $\widetilde N_1$ is relevant for soft leptogenesis.
Henceforth, we will drop the family index of RH neutrinos, e.g. $N \equiv N_1$ and $\lambda_{\alpha} \equiv \lambda_{\alpha 1}$, etc..

In the original proposal~\cite{Grossman:2003jv,DAmbrosio:2003nfv} 
a soft trilinear term $A_{\alpha}$ which is proportional to the neutrino Yukawa coupling
$A_{\alpha} = A \lambda_{\alpha}$ was assumed. As a result, the leading CP violation 
vanishes in the limit of zero temperature, while only with thermal effects, 
the cancellation between asymmetries from the decays of RH sneutrinos 
to leptons and sleptons is avoided, resulting in a finite, but suppressed 
total lepton asymmetry. Furthermore, a small soft bilinear term $B \ll m_{\rm soft}$ 
is required to resonantly enhance the CP violation to achieve successful leptogenesis.

In the following, however, we will focus on a generic trilinear soft $A_{\alpha}$ 
term~\cite{Adhikari:2015ysa} in soft leptogenesis~\cite{Fong:2010zu}, which results in interesting consequences:
\begin{enumerate}
\item[(a)] The leading CP violation can be nonzero even when thermal corrections are neglected, 
implying a \emph{nonthermal} and enhanced CP violation.

\item[(b)] The mixing CP violation away from the resonance is of the order of 
$\lambda_\alpha A_\alpha/B$ and large enough for leptogenesis.
As a result, a natural $B \sim m_{\rm soft}$ can be maintained.

\item[(c)] The generic $A_{\alpha}$ gives contributions to charged lepton flavor violating 
processes which are close to the sensitivities of present and future experiments.
\end{enumerate}
In order to discuss the aforementioned consequences, we assume the temperature range $T \lesssim 10^8 $ GeV for $m_{\rm soft} \sim 1$ TeV, 
where the effective global symmetries discussed in Ref.~\cite{Ibanez:1992aj} are 
absent. The enhancement effects of effective global symmetries 
in soft leptogenesis are discussed in Ref.~\cite{Fong:2010bv} and will not be considered further.

The superpotential for type I seesaw is given by 
%%%
\begin{eqnarray}
W_N & = & \frac{1}{2}M_N\hat{N^{c}}\hat{N^{c}}
+\lambda_{\alpha}\hat{\ell}_{\alpha}\hat{N^{c}}\hat{\phi}_{u},
\end{eqnarray}
%%%
where $\hat{N^c}$, $\hat{\ell}_{\alpha}$ and $\hat{\phi}_{u}$ denotes 
the chiral superfields of the RH neutrinos, the lepton doublet and the up-type Higgs doublet, respectively, 
and $\alpha$ indicates the RH neutrino lepton flavor index.
We have left the contraction between $SU(2)_L$ doublets
$\hat{\ell}_{\alpha} \hat{\phi}_{u} = \epsilon_{ab} \hat{\ell}_{\alpha}^a \hat{\phi}_{u}^b$ implicit.
The corresponding soft terms are
%%%
\begin{eqnarray}
-{\cal L}_{{\rm soft}} & = & \widetilde{M}_N^{2}\widetilde{N}^{*}\tilde{N}
+\left(\frac{1}{2}BM_N\widetilde{N}\widetilde{N}
+A_\alpha \widetilde{N}\widetilde{\ell}_{\alpha} \phi_{u}+{\rm h.c.}\right),
\label{eq5:soft}
\end{eqnarray}
%%%
where in the above and in the following, 
$\tilde x$ denotes the superpartner of the corresponding SM field $x$.

Through field redefinitions, it can be shown that we have three physical phases
%%%
\begin{eqnarray}
\Phi_\alpha = \arg\left(A_\alpha \lambda_\alpha^* B^*\right),
\label{eq5:phases}
\end{eqnarray}
%%%
and without loss of generality, the phases can be assigned to $A_\alpha$ 
and all other parameters will be taken to be real and positive.\footnote{Here we do 
not assume the proportionality of $A_\alpha$ to the neutrino Yukawa couplings 
($A_\alpha = A \lambda_\alpha$) as has been done in Refs.~\cite{Grossman:2003jv,DAmbrosio:2003nfv,Fong:2008mu}
which results in only one physical phase $\Phi = \arg(AB^*)$.}
Due to the bilinear $B$ term, $\widetilde{N}$ and $\widetilde{N}^{*}$ 
mix to form mass eigenstates
%%%
\begin{equation}
\widetilde{N}_{+} = \frac{1}{\sqrt{2}}\left(\widetilde{N}+\widetilde{N}^{*}\right),
\;\;\;\;\;
\widetilde{N}_{-} = -\frac{i}{\sqrt{2}}\left(\widetilde{N}-\widetilde{N}^{*}\right),
\label{eq5:mass_eigenstates}
\end{equation}
%%%
with the corresponding masses given by $\widetilde{M}_{\pm}^{2} =  M_N^{2}+\widetilde{M}_N^{2}\pm BM_N$.
We assume $B < M_N + \widetilde {M}_N^2/M_N$ to avoid a tachyonic mass which in turn implies 
that $\widetilde{N}_{-}$ will develop a vacuum expectation value.
Rewriting the interactions in terms of mass eigenstates $\widetilde{N}_{\pm}$,
we have
%%%
\begin{eqnarray}
-{\cal L} & \supset & 
\frac{1}{\sqrt{2}}\left\{ \widetilde{N}_{+}\left[ \lambda_{\alpha} \overline{\widetilde{\phi}_{u}^{c}}\ell_{\alpha}
+\left(A_\alpha + M_N \lambda_\alpha\right)\widetilde{\ell}_{\alpha}\phi_{u}\right]\right.\nonumber \\
 &  & \left.+i \widetilde{N}_{-}\left[ \lambda_{\alpha} \overline{\widetilde{\phi}_{u}^{c}}\ell_{\alpha}
+\left(A_\alpha - M_N \lambda_\alpha\right)\widetilde{\ell}_{\alpha}\phi_{u}\right]+{\rm h.c.}\right\} .
\label{eq5:lag}
\end{eqnarray}
%%%

%%%%%%%%%%%

The total decay width for $\widetilde{N}_\pm$ is given by
%%%
\begin{equation}
\Gamma_{\pm} \simeq \frac{M_N}{4\pi}
\sum_\alpha \left[\lambda_\alpha^2 + \frac{|A_\alpha|^2}{2M_N^2} 
\pm \frac{\lambda_\alpha {\rm Re}(A_\alpha)}{M_N}\right],
\label{eq5:decay_width}
\end{equation}
%%%
where we have expanded up to 
${\cal O}(\lambda_\alpha^2,m_{\rm soft}^2/M_N^2,\lambda_\alpha m_{\rm soft}/M_N)$ 
and ignored the final state phase space factors. 
We will impose the restriction that $|A_\alpha|, B < M_N$ and $\lambda_\alpha < 1$ 
to ensure that we are always in the perturbative regime. 

Implying the out-of-equilibrium condition for leptogenesis ($\Gamma_{\pm} \lesssim H(T = M_N)$)
%where the Hubble expansion rate is given by 
%$H = 1.66\sqrt{g_\star}\, T^2/M_N_{\rm Pl}$ with Planck mass 
%$M_N_{\rm Pl} = 1.22 \times 10^{19}$ GeV and $g_\star$ the relativistic degrees of freedom.
with the MSSM relativistic degrees of freedom $g_\star = 228.75$, one gets the following relation
%%%
\begin{equation}
\sqrt{ \sum_\alpha\left[\lambda_\alpha^2 + \frac{|A_\alpha|^2}{2M_N^2} 
\pm \frac{\lambda_\alpha {\rm Re}(A_\alpha)}{M_N}\right] }
\lesssim 1.6 \times 10^{-5} \left(\frac{M_N}{10^7\,{\rm GeV}}\right)^{1/2}.
\label{eq5:out-of-equilibrium}
\end{equation}
%%%
If $M_N \sim $ TeV, we require
$|A_\alpha| \lesssim 10^{-4}$ GeV and the mass splitting 
between $\widetilde N_+$ and $\widetilde N_-$ to be of the order of
their decay widths to resonantly enhance the CP violation. 
To avoid such tuning, we consider $|A_\alpha| \sim $ TeV in the following such that \eref{eq5:out-of-equilibrium} implies $M_N \gtrsim 4\times 10^7 {\rm GeV}$.

%%%%%%%%%%%

\begin{figure}[t]
\begin{center}
\includegraphics[width=0.98\textwidth]{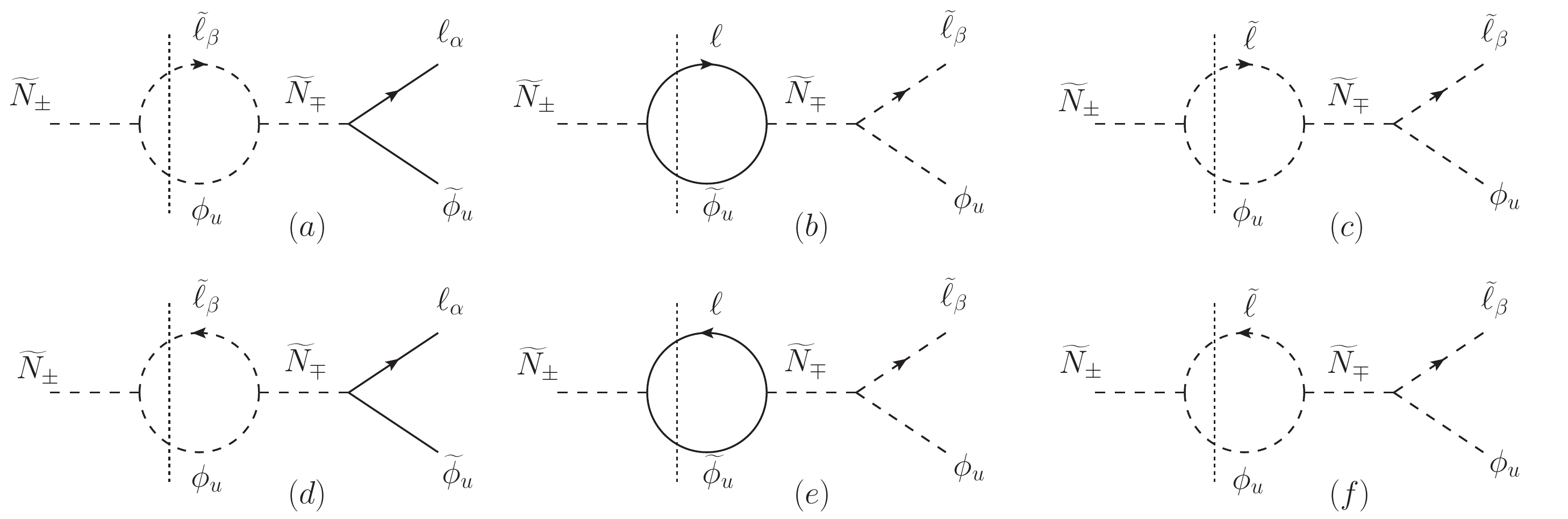}
\caption{One-loop self-energy diagrams for the decays 
$\widetilde N_\pm \to \ell_\alpha \widetilde \phi_u$ [$(a)$,$(d)$]
and $\widetilde N_\pm \to \widetilde\ell_\alpha \phi_u$ [$(b)$,$(c)$,$(e)$,$(f)$].
The arrow indicates the flow of lepton number. 
The dotted vertical lines indicate the possibility of on-shell propagators.
The diagrams with fermionic loop and fermionic 
final states do not contribute to the CP violation since they do not 
involve the soft couplings $A_\alpha$. Figure
taken from Ref.~\cite{Adhikari:2015ysa}. 
\label{fig5:selfenergy_diagrams}}
\end{center}
\end{figure}

Further, we define the CP asymmetry arising from the decays $\widetilde N_\pm \to a_\alpha$, with
$a_\alpha = \{\widetilde\ell_\alpha \phi_u, \ell_\alpha \widetilde \phi_u\}$, as
%%%
\begin{equation}
\epsilon_{\pm\alpha} \equiv 
\frac{\gamma(\widetilde N_\pm \to a_\alpha) 
- \gamma(\widetilde N_\pm \to \overline {a_\alpha})}
{\sum_{a_\beta;\beta}\left[\gamma(\widetilde N_\pm \to a_\beta) 
+ \gamma(\widetilde N_\pm \to \overline {a_\beta})\right]},
\label{eq5:CP_asym}
\end{equation}
%%%
where $\overline {a_\alpha}$ indicates the CP conjugate of $a_\alpha$
and $\gamma(i \to j)$ is the thermal averaged reaction density of the process $i \to j$. We will only consider CP violation arising from interference between tree-level and one-loop self-energy diagrams as shown in \fref{fig5:selfenergy_diagrams}.Vertex contributions 
were shown to be negligible~\cite{Adhikari:2015ysa} and will not be considered further. As pointed out in Refs.~\cite{Grossman:2003jv,DAmbrosio:2003nfv}, 
thermal effects are crucial to prevent the cancellation of leading CP asymmetries 
from the decays of $\widetilde N_\pm$ to leptons and sleptons. However, terms higher order in $m_\mathrm{soft}/M_N$ can dominate when considering generic
$A_\alpha$~\cite{Adhikari:2015ysa}. 

In the following, we will focus on two interesting limits of \eref{eq5:CP_asym}, the full expressions can be found in Ref.~\cite{Adhikari:2015ysa}.

\begin{itemize}

\item[(i)]
In the limit $\lambda_\alpha \gg A_\alpha/M_N$, \eref{eq5:CP_asym} leads to
%%%
\begin{eqnarray}
\epsilon_{\pm \alpha}
& \simeq & \frac{1}{4\pi}
P_\alpha \sum_\beta \lambda_\beta \frac{{\rm Im} (A_\beta)}{M_N}
\frac{4 B M_N}{4 B^2 + \Gamma_\lambda^2} \frac{c_F(T) - c_B(T)}{c_F(T) + c_B(T)} r_B(T),
\label{eq5:ep_S_tot_i}
\end{eqnarray}
%%%
where we define $P_\alpha \equiv \lambda_\alpha^2/\lambda^2$ with $\sum_\alpha P_\alpha = 1$ and $\lambda \equiv \sum_{\alpha} \lambda_{\alpha}$. We define $\Gamma_\lambda \equiv \frac{\lambda^2 M_N}{4 \pi}$, terms higher order in $m_{\rm soft}/M_N$ are neglected. 
The parameters $r_{B}(T)$ and $c_{B,F}(T)$ go to one for $T \to 0$ such that the CP asymmetry in \eref{eq5:ep_S_tot_i}, being proportional to $c_F(T) - c_B(T)$, goes as well to zero for $T \to 0$. The explicit expressions 
of $r_{B}(T)$ and $c_{B,F}(T)$ are given in the Appendix A of Ref.~\cite{Adhikari:2015ysa}.
\item[(ii)]
In the limit $\lambda_\alpha \ll A_\alpha/M_N$, 
we have
%%%
\begin{eqnarray}
\epsilon_{\pm \alpha} 
& \simeq & \frac{1}{4\pi}\frac{|A_\alpha|^2}{\sum_\delta |A_\delta|^2}  
\sum_\beta \lambda_\beta \frac{{\rm Im} (A_\beta)}{M_N} 
\frac{4 B M_N}{4 B^2 + \Gamma_A^2} r_B(T),
\label{eq5:ep_S_tot_ii}
\end{eqnarray}
%%%
with $\Gamma_A \equiv \sum_\alpha \frac{|A_\alpha|^2}{8\pi M_N}$.
The CP asymmetries \eref{eq5:ep_S_tot_ii} do not vanish 
for $T = 0$ and represents \emph{nonthermal} CP violation. By 
thermal (nonthermal) CP violation, we refer to the case 
where CP violation does (not) vanish as $T \to 0$.
Although thermal effects are always present, the fact that CP violation 
is nonvanishing for $T = 0$ implies less suppression compared 
to case (i).

\end{itemize}

\begin{figure}[t]
\begin{center}
\frame{\includegraphics[width=0.49\textwidth]{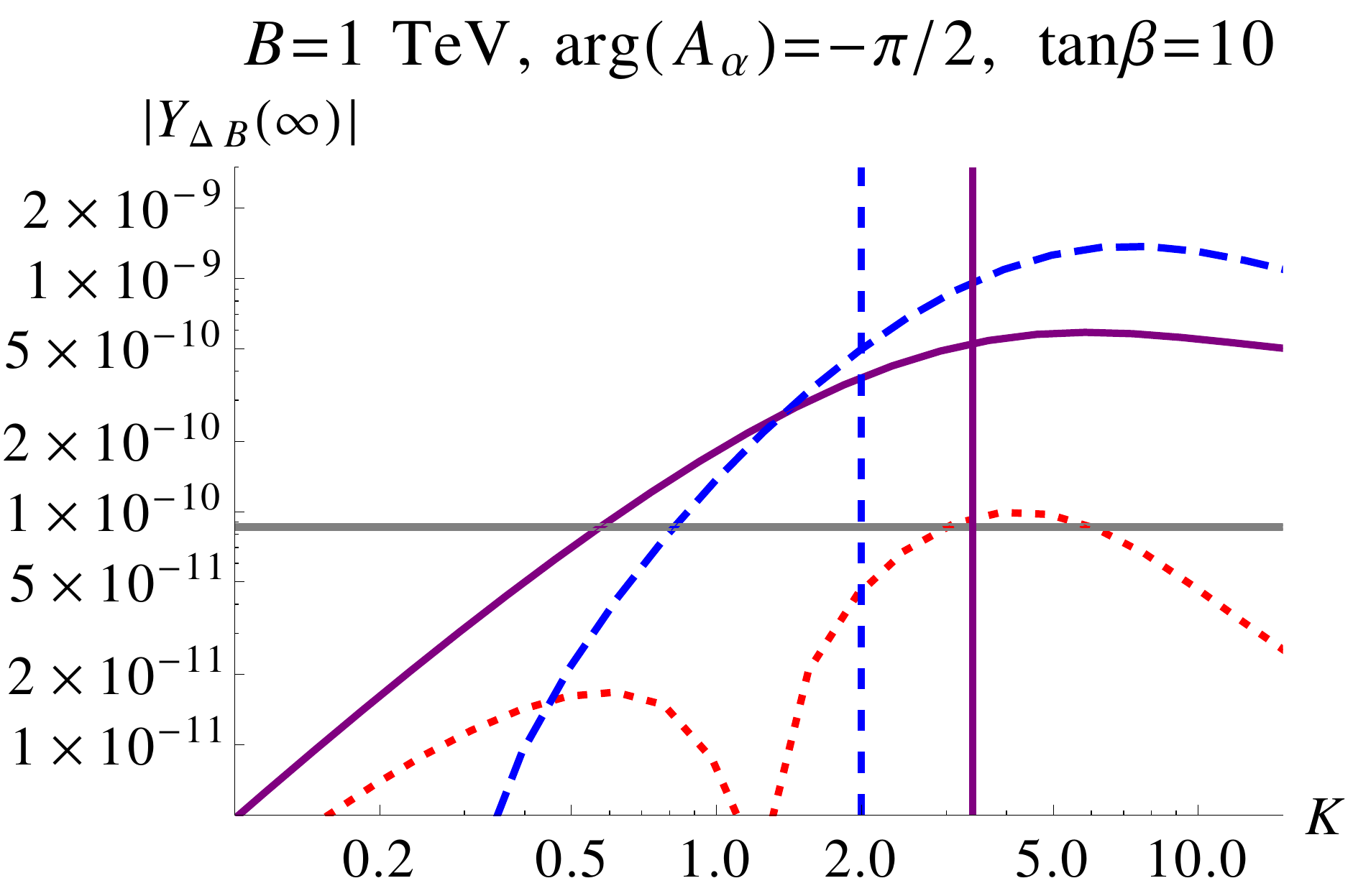}}
\frame{\includegraphics[width=0.49\textwidth]{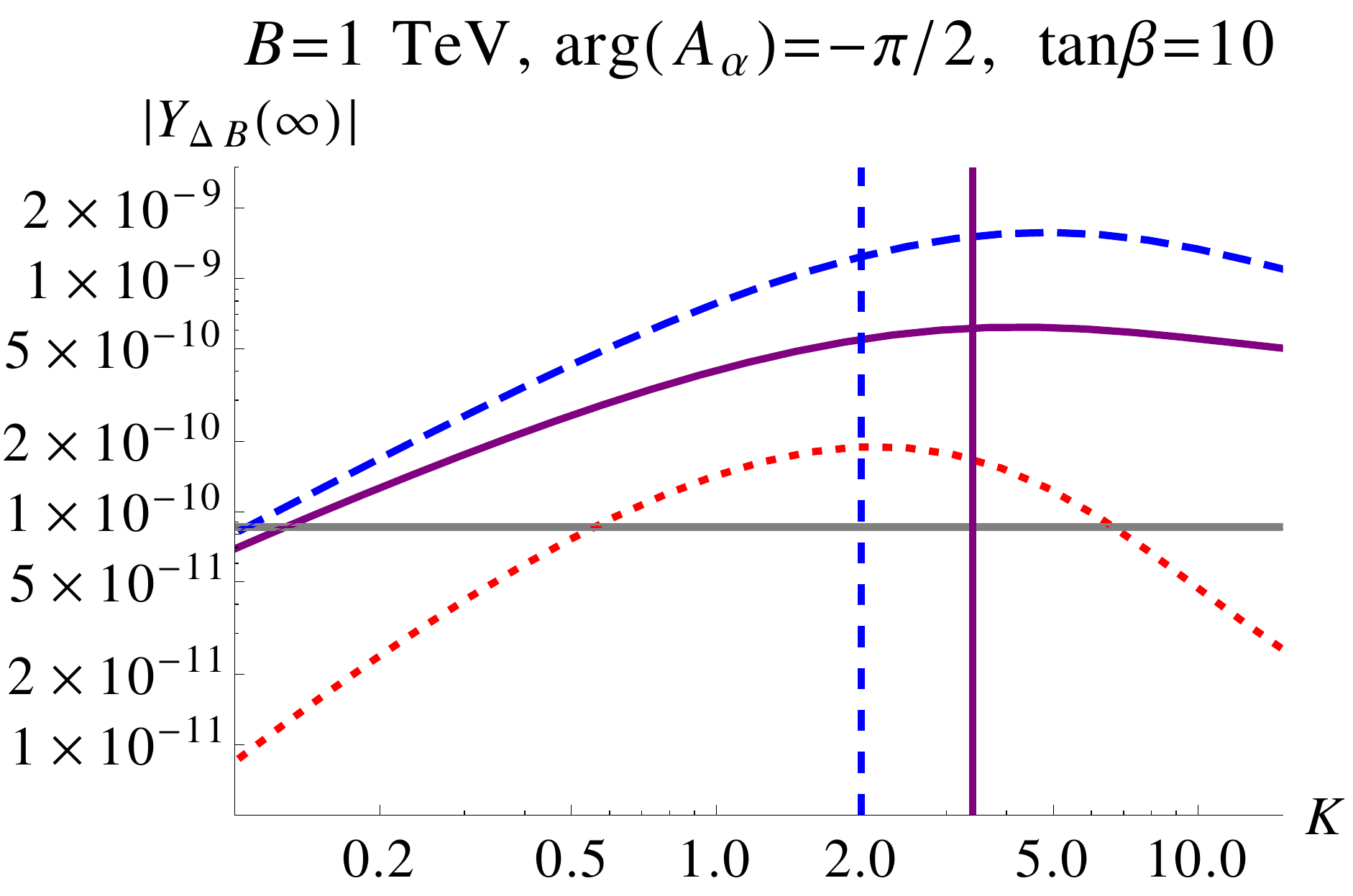}}
\caption{
The absolute value of the final baryon asymmetry $|Y_{\Delta B}(\infty)|$
as a function of the washout parameter $K \equiv \Gamma_\pm/H(T=M_N)$ for
$M_N = 5 \times 10^7$ GeV for the three scenarios described in the
text: NTD (blue dashed), TD (red dotted), and MIX (purple solid).
The left plot corresponds to the case of zero initial number densities 
of $N$ and $\widetilde N_\pm$, while the right plot corresponds 
to the case of thermal initial number densities of $N$ and $\widetilde N_\pm$.
The regions to the left of the blue dashed and purple solid 
vertical lines correspond to $K$ values when $A_\alpha < 5$ TeV 
for the NTD and MIX scenarios, respectively, while for the TD scenario
we always have $A_\alpha < 5$ TeV in the range of the plot.
The gray band represents the recent combined Planck and WMAP CMB
 measurements of cosmic baryon asymmetry~\cite{Ade:2013zuv,Bennett:2012zja} at 2$\sigma$.
 The dip in the TD scenario in the left plot refers to a change in the sign of the baryon asymmetry. 
 Figure taken from Ref.~\cite{Adhikari:2015ysa}. 
\label{fig5:K_YB}}
\end{center}
\end{figure}

Now, we numerically solve the Boltzmann equations using the full expression for the 
asymmetry parameter in \eref{eq5:CP_asym}, fixing 
$M_N = 5 \times 10^7$ GeV, $\tan \beta = 10$, $\arg(A_\alpha) = -\pi/2$, and $B = 1$ TeV.  
We consider the following three scenarios: 

\begin{enumerate}[label={\textbf{(\alph*)}},leftmargin=1.25cm,align=left]

\item[(NTD)] \emph{Nonthermal-dominated}. 
In this scenario, we choose {\boldmath$A$}$/M_N=(10^{-4},10^{-2},1) w$ 
and {\boldmath$\lambda$}$=(10^{-5},10^{-3},10^{-1}) w$.

\item[(TD)] \emph{Thermal-dominated}.
In this scenario, we choose {\boldmath$A$}$/M_N=(10^{-5},10^{-3},10^{-1}) w$ 
and {\boldmath$\lambda$}$=(10^{-4},10^{-2},1) w$.

\item[(MIX)] \emph{Mixed}. 
In this scenario, we choose {\boldmath$A$}$/M_N=(10^{-4},10^{-2},1) w$ 
and {\boldmath$\lambda$}$=(10^{-4},10^{-2},1) w$.

\end{enumerate}
The couplings are written as 3-vectors {\boldmath$A$} and {\boldmath$\lambda$}. A scan through the parameter space from 
$K \equiv \left.\Gamma_\pm \right|_{T=M_N} = 0.1$ to $K = 15$ is realized by varying the dimensionless variable $w$ between $10^{-6}$ and $10^{-4}$. Since scattering processes only give an $\mathcal{O}(1)$ correction in the intermediate to strong washout regime (which we are considering), we take into account for simplicity only decays and inverse decays $\widetilde N_\pm$.
The numerical result, the absolute final baryon asymmetry
$|Y_{\Delta B}(\infty)|$ as a function of $K$ is shown in \fref{fig5:K_YB}.

We conclude that it is possible to have successful leptogenesis
for TeV-scale $A_\alpha$ and $B \gg \Gamma_\pm$ far away from the resonant regime. 
We see that nonthermal CP violation can significantly enhance the efficiency of soft leptogenesis.
Notice that we have assumed some hierarchies in {\boldmath$A$} and {\boldmath$\lambda$} 
to illustrate lepton flavor effects. Though not crucial for leptogenesis, 
the hierarchy in {\boldmath$A$} is needed to satisfy phenomenological bounds, which
we will discuss next.

%%%%%%%%%%%%%%%%%%%%%

In the following, we will discuss the phenomenology of having generic $A_\alpha$.
Since we focus on the case $M_N\gtrsim 10^7$ GeV such that $|A|, B$ can remain at around TeV-scale, 
the production of RH sneutrinos is beyond the energy range of current 
colliders.\footnote{Even if $M_\pm \sim$ TeV, the bound on the Yukawa couplings from the requirement 
of out-of-equilibrium decays of $\widetilde N_\pm$ in \eref{eq5:out-of-equilibrium} 
makes $\widetilde N_\pm$ challenging to be produced at colliders~\cite{Bambhaniya:2014hla}.}
On the other hand, the soft terms relevant for soft leptogenesis $A_\alpha$, $B$ and $\widetilde M_N$ 
can contribute to Electric Dipole Moments (EDM) of leptons and 
to Charged Lepton Flavor Violating (CLFV) interactions. 
Under the assumption of universality, soft trilinear couplings $A_\alpha = A \lambda_\alpha$, 
the analysis of Ref.~\cite{Kashti:2004vj} showed that the contributions to lepton EDM and CLFV 
processes are much below the experimental bounds. Here we will repeat the analysis by considering 
generic $A_\alpha$. This, as we will show below, can enhance the contributions to lepton EDM 
and CLFV processes.\footnote{Reference~\cite{Dedes:2007ef} considered in details 
the phenomenological consequences of the soft terms considering three generations of RH neutrino chiral 
superfields. Here, we will focus only on parameters related to $N_1$ that are relevant for 
soft leptogenesis, i.e. $B$, $\widetilde M_N$, $A_\alpha$, $\lambda_\alpha$ and $M_N$.}

\begin{itemize}

 \item \emph{Electric dipole moment of the electron.}
The contributions of $A_\alpha$ and $B$ to the EDM of the electron can be estimated by \cite{Kashti:2004vj}
%%%
\begin{equation}
|d_e|\approx \frac{e\,m_e \tan\beta}{16 \pi m^2_{\tilde\nu}}
\left|\frac{m_\chi \lambda_\alpha}{M_N^2}\right|
\left(
|A_\alpha| + {B \lambda_\alpha} \right)\,,
\label{eq5:EDM}
\end{equation}
%%%
where $m_e$ is the electron mass, $m_{\tilde\nu}^2$ the squared mass 
of the light sneutrino, and $m_\chi$ the mass of chargino (${\cal O}(1)$ contributions of the phases and mixing angles in the chargino sectors 
are assumed). For generic $A_\alpha$, the first term in the bracket 
in \eref{eq5:EDM} dominates. 
Assuming $m_{\tilde\nu} = m_\chi = m_{\rm soft}$ and taking into account 
\eref{eq5:out-of-equilibrium}, we have
%%%
\begin{equation}
|d_e| \lesssim 5 \times 10^{-38} \left(\frac{\tan\beta}{10}\right)
\left(\frac{10^7\,{\rm GeV}}{M_N}\right)^{3/2} 
\left(\frac{1\,{\rm TeV}}{m_{\rm soft}}\right) e\,{\rm cm}.
\end{equation}
%%%
The bound above is about nine orders of magnitude below the current experimental bound 
$|d_e|_{\rm exp} < 8.7 \times 10^{-29}e\,\rm{cm}$ \cite{Baron:2013eja}. 
We can estimate the contributions to $\mu$ and $\tau$ EDMs
by replacing $m_e$ in \eref{eq5:EDM} by $m_\mu$ 
and $m_\tau$, respectively. However, the current experimental constraints
on them are even weaker:
$|d_\mu|_{\rm exp} < 1.9 \times 10^{-19}e\,\rm{cm}$ \cite{Bennett:2008dy} 
and $|d_\tau|_{\rm exp} < 5.1 \times 10^{-17}e\,\rm{cm}$ \cite{Inami:2002ah}.

%%%%%%%%%%%%%%%%%

\item \emph{Charged lepton flavor violating interactions.}
The branching ratio for CLFV due to the soft mass matrix of the doublet sleptons $m_{\tilde \ell}^2$ 
is given by \cite{Kashti:2004vj,Hirsch:2012ti}
\begin{equation}
\mbox{BR}(\ell_\alpha \rightarrow \ell_\beta \gamma) \approx \frac{\alpha_{\rm EM}^3}{G_F^2}
\frac{ \left| (m_{\tilde \ell}^2)_{\alpha\beta} \right|^2  }{m_{\rm soft}^8} \tan^2\beta,
\label{eq5:CLFV}
\end{equation}
where $\alpha_{\rm EM}$ is the electromagnetic fine structure constant.
Generically, the off-diagonal elements of $m_{\tilde \ell}^2$
will induce too large CLFV rates. As usual, we assume mSUGRA
boundary conditions at the Grand Unified Theory (GUT) scale, where 
the off-diagonal elements of $m_{\tilde \ell}^2$ vanish. 
Considering only off-diagonal elements that 
are generated due to renormalization effects, $m_{\tilde \ell}^2$ evolves 
from the GUT scale $M_{\rm GUT}$ to the RH neutrino mass scale $M_N$~\cite{Hisano:1995cp} as
%%%
\begin{equation}
(m_{\tilde \ell}^2)_{\alpha\beta} 
\approx -\frac{1}{8\pi^2} A_\alpha^* A_\beta 
\ln \left(\frac{M_{\rm GUT}}{M_N}\right)
\label{eq5:od_slepton_masses}
\end{equation}
%%%
for $\alpha \neq \beta$.

The strongest constraint on the rare decay $\mu \rightarrow e \gamma$ 
comes from nonobservation at the MEG experiment \cite{Adam:2011ch,Adam:2013mnn}:
%%%
\begin{equation}
\mbox{BR}(\mu \rightarrow e \gamma)_{\rm exp} < 5.7 \times 10^{-13}.
\label{eq5:mu_to_egamma}
\end{equation}
%%%
Substituting \eref{eq5:od_slepton_masses} in \eref{eq5:CLFV} and applying the constraint \eref{eq5:mu_to_egamma}, we obtain
%%%
\begin{equation}
|A_\mu^* A_e| \lesssim 5 \times 10^3\,{\rm GeV}^2
\left(\frac{m_{\rm soft}}{1\,{\rm TeV}}\right)^4
\left(\frac{10}{\tan\beta}\right),
\label{eq5:CLFV_mu}
\end{equation}
%%%
where we set $M_{\rm GUT} = 10^{16}$ GeV and $M_N = 10^7$ GeV.
Similarly, using the experimental bounds on CLFV processes in $\tau$ decays,
$\mbox{BR}(\tau \rightarrow e \gamma)_{\rm exp} < 3.3 \times 10^{-8}$ and
$\mbox{BR}(\tau \rightarrow \mu \gamma)_{\rm exp} < 4.4 \times 10^{-8}$~\cite{Aubert:2009ag},
we obtain
%%%
\begin{equation}
|A_\tau^* A_e| \approx |A_\tau^* A_\mu| \lesssim 1 \times 10^6\,{\rm GeV}^2 
\left(\frac{m_{\rm soft}}{1\,{\rm TeV}}\right)^4
\left(\frac{10}{\tan\beta}\right).
\label{eq5:CLFV_tau}
\end{equation}
%%%
For $m_{\rm soft} \sim $ TeV, the bound in \eref{eq5:CLFV_tau} can be satisfied
with all $|A_\alpha|$ at the TeV-scale while the stronger bound in \eref{eq5:CLFV_mu}
requires the product of $|A_\mu^* A_e|$ to be \emph{smaller} than the TeV$^2$-scale. 
This requires some hierarchy in $A_\alpha$, consistent with what 
was assumed in \fref{fig5:K_YB}. 

Finally, nonzero $(m_{\tilde \ell}^2)_{\alpha\beta}$ 
can also give rise to other CLFV processes like 
$\mu \rightarrow 3 e$ and $\mu - e $ conversion. 
If such processes are dominated by the dipole type operator 
for relatively large $\tan \beta$, $\mbox{BR}(\mu \rightarrow 3 e)$ 
and the rate of  $\mu - e $ conversion rate $R_{\mu e}$ are proportional to 
$\mbox{BR}(\mu \rightarrow e \gamma)$ and are approximately given by \cite{Ellis:2002fe}
%%%
\begin{eqnarray}
\mbox{BR}(\mu \rightarrow 3 e) \sim 6.6 \times 10^{-3} \mbox{BR}(\mu \rightarrow e \gamma),
\end{eqnarray}
%%%
and, for the $^{27}_{13}$Al nucleus, by \cite{Kitano:2002mt}
%%%
\begin{eqnarray}
R_{\mu e} \sim 2.5 \times 10^{-3} \mbox{BR}(\mu \rightarrow e \gamma).
\end{eqnarray}
%%%
Currently these constraints are less stringent than those coming from $\mu \rightarrow e \gamma $. 
In future experiments, however, the sensitivity for such processes 
could reach the presently allowed parameter space. For instance,
the future Mu3e experiment \cite{Blondel:2013ia} could reach 
$\mbox{BR}(\mu \rightarrow 3 e) \sim 10^{-15} - 10^{-16}$.
As for $\mu -e$ conversion process, the sensitivities of Mu2e \cite{Abrams:2012er} and COMET  \cite{Kuno:2012pt} 
experiments could reach $R_{\mu e} \sim 10^{-17} $ for 
the $^{27}_{13}$Al nucleus while the PRISM/PRIME \cite{Kuno:2012pt} project 
may have even greater sensitivity by two orders of magnitude.

\end{itemize}

%%%%%%%%%%%%%%%%%

In conclusion, in the framework of supersymmetric type I seesaw,  
soft leptogenesis is an attractive mechanism to explain the cosmological matter-antimatter asymmetry 
since it works at a lower temperature regime $T \lesssim 10^9$ GeV, 
where the tension caused by overproduction of gravitinos can be relaxed 
or even evaded. By considering generic soft trilinear $A_\alpha$ couplings, 
we reach both, interesting theoretical and phenomenological consequences. 
Theoretically, nonthermal CP violation can dominate and successful leptogenesis 
can be achieved with all relevant soft parameters taking \emph{natural} values at the around the TeV-scale.
Phenomenologically, such large $A_\alpha$ couplings allow significant contributions to 
CLFV processes that are close to sensitivities of present and future experiments. 
While the direct probe of soft leptogenesis with generic $A_\alpha$ terms 
will require the production of $\tilde N$, which is not possible
with current or foreseeable particle colliders, the detection of CLFV processes 
consistent with this framework will provide a boost of such a realization.

%%%%%%%%%%%%%%%%%%%%%%%
%%%%%%%%% TYPE II %%%%%%%%%%
%%%%%%%%%%%%%%%%%%%%%%%

\subsection{Type II soft leptogenesis}
\label{ch5:sec5:typeII-soft-leptogenesis}

The type II seesaw, in which  a Higgs triplet boson is responsible for the observed neutrino
masses and mixing, enjoys also the basic features of soft leptogenesis \cite{DAmbrosio:2004rko,Chun:2005ms}. 
Contrary to the type I leptogenesis, the type II leptogenesis suffers from rather efficient gauge annihilation processes, which strongly suppresses the resulting lepton asymmetry (see the detailed discussion in \sref{ch5:sec3:extended_seesaw-typeII}).  
The gauge annihilation effect becomes weaker
for higher Higgs triplet mass, which, however, suppresses the lepton asymmetry in soft leptogenesis 
as it is inversely proportional to the Higgs triplet mass. 
For TeV-scale seesaw, the huge suppression of lepton asymmetry by gauge annihilations 
can be overcome by the effect of soft SUSY breaking terms such that successful TeV-scale leptogenesis
can be obtained \cite{Chun:2005ms}.

In the supersymmetric type II seesaw a vector-like pair of
$\Delta=(\Delta^{++},\Delta^+,\Delta^0)$ and $\Delta^c=(\Delta^{c--},
\Delta^{c-},\Delta^{c0})$ with hypercharge $Y=1$ and $-1$ is introduced to construct
the following renormalizable superpotential:
\begin{equation} \label{eq5:WtypeII}
W= \lambda_{\Delta} \ell \ell \Delta + \lambda_1 H_1 H_1 \Delta  + \lambda_2 H_2 H_2
\Delta^c  + M_\Delta \Delta \Delta^c\,.
\end{equation}
The term $\lambda_{\Delta} \ell\ell \Delta$ contains the neutrino mass operator $h \nu \nu
\Delta^0$. The relevant soft SUSY breaking terms are
\begin{eqnarray} \label{eq5:LsoftII}
{-\cal L}_{soft} &=& \left( \lambda_{\Delta} A \ell\ell \Delta + \lambda_1 A H_1
H_1 \Delta \right.
\nonumber\\
&+&\! \left. \lambda_2 A  H_2 H_2 \Delta^c + BM_\Delta \Delta \Delta^c
+ \mathrm{h.c.} \right) \nonumber\\
& +& m_0^2  |\Delta|^2 + m^2_{0} |\Delta^c|^2 .
\end{eqnarray}
Note, we have used the same capital letters to denote
superfields and their scalar components, and considered
the universal soft masses, $A$ and $m_0$.
In the limit of  $M_\Delta \gg m_0 ,A$, the Higgs triplet vacuum
expectation value (vev) $\langle \Delta^0 \rangle= \lambda_2 \langle
H_2^0 \rangle^2/M_\Delta$ gives the neutrino mass
\begin{equation}
(M_\nu)_{\alpha\beta} =2 (\lambda_\Delta)_{\alpha\beta} \langle \Delta^0 \rangle
= 2 (\lambda_\Delta)_{\alpha\beta} \lambda_2 {v_2^2 \over M_\Delta} \,.
\end{equation}
The mass eigenstates of the scalar triplets are given by
$
\Delta_{\pm} = {1\over\sqrt{2}} (\Delta \pm \bar{\Delta}^{c}) 
$
with the mass-squared values $M^2_{\pm}=M_\Delta^2+ m_0^2 \pm BM_\Delta$, and the
mass-squared difference $\Delta M^2= 2 B M_\Delta$.  
From  \eref{eq5:WtypeII} and \eref{eq5:LsoftII}, one easily finds 
the scalar triplet couplings of the mass eigenstates 
for the leptonic
final states ${\ell}{\ell}, \bar{\tilde{\ell}}\bar{\tilde{\ell}}$, and
the Higgs final states ${H}_i {H}_i,{\tilde{H}}_i {\tilde{H}}_i $ with $i=1,2$.

%\begin{eqnarray}
%{-\cal L} &=& {1\over \sqrt{2}} \Delta_{\pm} \big[ h\,\tilde{L}\tilde{L} +
%h(A \pm M)\, LL \nonumber\\
%&& \lambda_1\, \tilde{H}_1 \tilde{H}_1 + \lambda_1 (A\pm M)\, H_1 H_1 \\
%&&  \pm \lambda_2^*\, \bar{\tilde{H}}_2 \bar{\tilde{H}}_2 \pm
%\lambda_2^*(A^*\pm M)\,\bar{H}_2 \bar{H}_2 \big] + h.c.
%\nonumber
%\end{eqnarray}

%
\begin{figure}
\begin{center}
\includegraphics[width=0.85\textwidth]{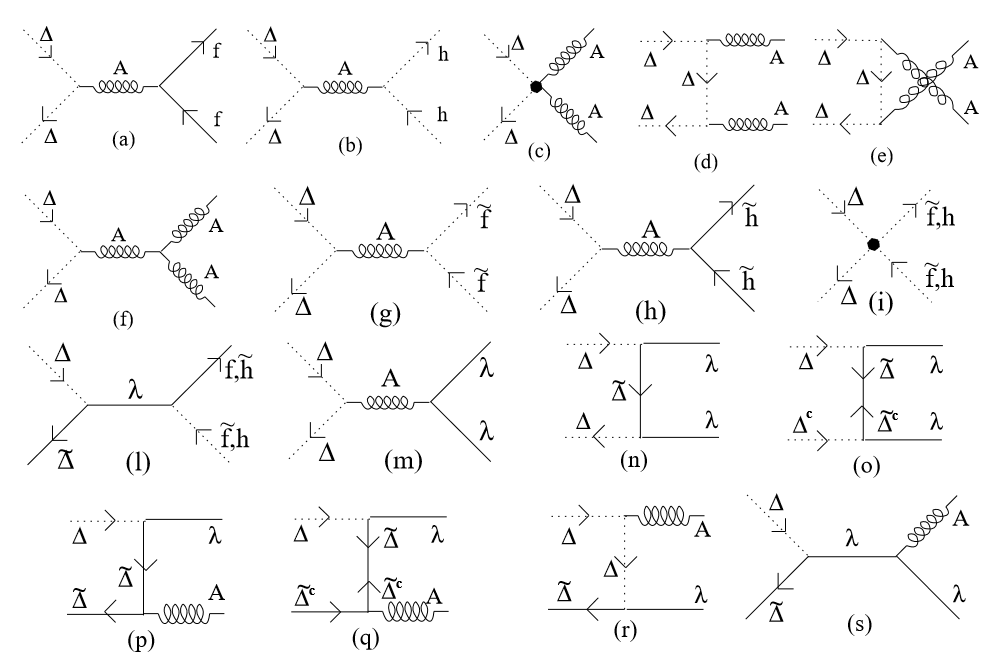}
\end{center}
\caption{\label{fig5:fig2_last} Diagrams contributing to the
gauge--annihilation amplitude of triplet particles.
$\tilde{\Delta}$, $\tilde{\Delta_c}$ represent the fermionic
partners of $\Delta$ and $\Delta_c$, respectively, while $A$
indicates a gauge boson, $\lambda$ a gaugino, $h$ a Higgs
particle, $\tilde{h}$ a higgsino, $f$ a fermion and $\tilde{f}$ a
sfermion. Figure taken from Ref.~\cite{Chun:2005ms}. }
\end{figure}

Lepton asymmetry generation in the supersymmetric triplet seesaw model can be described by 
a general system of a charged particle $X$ ($\bar{X}$) decaying to a
final state $j$ ($\bar{j}$) and generating tiny CP asymmetric
number densities, $n_X-n_{\bar{X}}$ and $n_j-n_{\bar{j}}$.   The
relevant Boltzmann equations in the approximation of
Maxwell--Boltzmann distributions are
\begin{eqnarray} \label{eq5:boltzmann}
 {d Y_{X} \over d z} &=& - z K \left[ \gamma_D (Y_{X}-Y_{X}^{eq}) +
 \gamma_A {(Y_{X}^2-Y_{X}^{eq\,2})\over Y_{X}^{eq}}  \right] \nonumber\\
 {d Y_{\Delta x} \over d z} &=& - z K \gamma_D \left[ Y_{\Delta x}-
 \sum_k 2 B_k {Y_{X}^{eq}\over Y_{\Delta k}^{eq}} Y_{\Delta k} \right] \\
 {d Y_{\Delta j} \over d z} &=& 2 z K \gamma_D\left[ \epsilon_j  (Y_{X}-Y_{X}^{eq})
 + B_j ( Y_{\Delta x} - 2 {Y_{X}^{eq} \over Y_{\Delta j}^{eq} } Y_{\Delta j} ) \right]\,, \nonumber
\end{eqnarray}
where $Y$'s are the number densities in unit of the entropy
density $s$ as defined by   $Y_{X}\equiv n_X/s \approx
n_{\bar{X}}/s$, $Y_{\Delta x} \equiv (n_X-n_{\bar{X}})/s$ and $Y_{\Delta j}\equiv
(n_j-n_{\bar{j}})/s$.  Here, the CP asymmetry $\epsilon_j$ in the
decay $X\to j$ is defined by
\begin{equation}
\epsilon_j \equiv {\Gamma(X\to j) -\Gamma(\bar{X}\to \bar{j} )
\over \Gamma_X }.
\end{equation}
In \eref{eq5:boltzmann}, $K\equiv \Gamma_X/H(T=M_X)$, and $B_j$ is the branching ratio of the decay $X\to j$.
%The
%relativistic degrees of freedom in thermal equilibrium is $g_*=228.75$ 
%in the Supersymmetric Standard Model.
%
The evolution of the $X$ abundance is determined by the decay and
inverse decay processes, as well as by the annihilation effect
described by the diagrams of \fref{fig5:fig2_last}, and are
accounted for by the functions $\gamma_D$ and $\gamma_A$,
respectively.  Note that unitarity requires $2 Y_{\Delta x} + \sum_j Y_{\Delta j}\equiv 0$, and thus
the equation for $Y_{\Delta x}$ can be dropped by the replacement:
$
Y_{\Delta x}=-{1\over2} \sum_j Y_{\Delta j}
$
in the last line of \eref{eq5:boltzmann}.

In our model, the heavy particle $X$ can be one of the six
charged particles; $X=\Delta^{++}_{\pm}, \Delta^{+}_{\pm}$ or
$\Delta^0_{\pm}$. Each of them follows the first Boltzmann
equation in \eref{eq5:boltzmann} where $\gamma_D$ and $\gamma_A$
are given by
\begin{eqnarray}
\gamma_D &=& {K_1(z) \over K_2(z)}\,, \label{eq5:gamma_d}\\
\gamma_A &=& {\alpha_2^2 M_X \over  \pi K H_1}
 \int^\infty_1\!\! dt\, \frac{K_1(2zt)}{K_2(z)}\, t^2 \beta(t)\, \sigma(t)\,,
\label{eq5:sigmat_int}
\end{eqnarray}
with
\begin{eqnarray}
&&\sigma(t)=(14+11 t_w^4)(3+\beta^2)+(4+ 4 t_w^2+t_w^4)\left [
16+4(-3-\beta^2 + \frac{\beta^4+3}{2\beta}\ln
\frac{1+\beta}{1-\beta})\right ]\nonumber \\ &&+4 \left
[-3+\left(4-\beta^2+\frac{(\beta^2-1)(2-\beta^2)}{\beta}\ln\frac{1+\beta}{1-\beta}\right)
\right ]. \label{eq5:sigmat}
\end{eqnarray}
Here $t_w\equiv\tan(\theta_W)$ with $\theta_W$ being the Weinberg angle,
and $\beta(t)\equiv \sqrt{1-t^{-2}}$.  For the region of relevance, $z\equiv M_X/T >10$, the  Boltzmann
approximation for the  decay and inverse decay amplitudes 
is in good agreement with the full numerical calculation, 
where Bose--Einstein and Fermi--Dirac distributions, as well as
thermal masses, are included.

Given $\gamma_D$ and $\gamma_A$ in the Boltzmann approximation, 
one can figure out the behavior of $Y_{X}$ in an analytic way.
First of all, the inverse decay freezes out at $z_f\approx 9$ satisfying the relation $K
z_f^{5/2} e^{-z_f}=1$ with $K=32$.  On the other hand,
the thermal annihilation and decay rates can be compared as
$${<\Gamma_A> \over <\Gamma_D>} (z_f)\simeq 2 {\alpha^2\over
\alpha_X} z_f^{-3/2}e^{-z_f} \approx {2\times 10^8\mbox{ GeV}\over
M_X },$$ where $\alpha_X = K H/M_X$.    Thus,  the annihilation
effect is less effective for higher $M_X$ and becomes irrelevant for $M_X \gtrsim 10^8$ GeV. 
But, the CP asymmetry in soft leptogenesis is inversely proportional to $M_X$
(${\epsilon}_\ell \propto A/M_X$ as shown explicitly later in \eref{eq5:epspm}). Thus, there is a tension between
these two effects, and lower values of $M_X$ turn out to be favored.
As the annihilation processes freeze out at $z\approx 20$, $Y_{X}$ follows more closely its
equilibrium density $Y_{X}^{eq}$ 
with a deviation of order $10^{-3}$.
Since decoupling occurs indeed at high
$z$, one can safely approximate
\begin{equation} \label{eq5:YXapp}
Y_{X}-Y_{X}^{eq} = {-Y_{X}^{eq
\prime}\over zK(\gamma_D+2\gamma_A)}\,.
\end{equation}

Concerning the evolution of  $Y_{\Delta j}$ with $j=\ell\ell$ and $\tilde{\ell}\tilde{\ell}$, let us 
recall that  the combined lepton asymmetry vanishes  in the supersymmetric limit, $Y_{\Delta \ell}\equiv
Y_{\Delta \ell\ell}+Y_{\Delta \tilde{\ell} \tilde \ell}=0$,
as the CP asymmetries
in the bosonic and fermionic final states takes the opposite sign,
$\epsilon_{\ell\ell}=-\epsilon_{\tilde{\ell} \tilde \ell}$.
A nonvanishing lepton asymmetry arises
after taking into account the SUSY breaking effect by temperature and by the soft terms. 
The temperature  effect can be well accounted for by a slight
modification of the last Boltzmann equation of \eref{eq5:boltzmann}, resulting from the extension of the usual
Maxwell--Boltzmann approximation to the second order. The total lepton asymmetry density $Y_{\Delta \ell}$ turns out to follow the
approximate Boltzmann equation (in the limit of $|\lambda_{\Delta}|=|\lambda_2|\gg |\lambda_1|$):
\begin{equation} \label{eq5:boltzII}
{d Y_{\Delta \ell} \over d z} = 2
g_\Delta z K \gamma_D \left[ \tilde{\epsilon}_\ell \delta(z)
 (Y_{X}-Y_{X}^{eq})
 - {Y_{X}^{eq} \over Y_{\Delta \ell}^{eq} } Y_{\Delta \ell} \right]\,,
\end{equation}
where $g_\Delta=6$ counts the total number of triplet components
generating the lepton asymmetry.
The function  $\delta(z)\equiv
\delta_{BF}(z)+\delta_{soft}$ accounts for the SUSY breaking effect with 
\begin{equation}
\delta_{BF}(z)\equiv 2\sqrt{2} {K_1(\sqrt{2}z) \over K_1(z)}\,,\quad
\delta_{soft} = {m_0^2 + |A|^2 \over M_X^2}
\label{eq5:delta_bf}
\end{equation}
describing the difference between the Bose--Einstein and
Fermi--Dirac distributions and  the soft SUSY breaking effect, respectively.
The CP asymmetry parameter $\tilde \epsilon_\ell$ is given by
\begin{equation} \label{eq5:epspm}
\tilde \epsilon_\ell \equiv { 4 B \Gamma_\pm \over 4B^2 + \Gamma_\pm^2} {4 |\lambda_{\Delta}|^2 |\lambda_2|^2
\over (|\lambda_{\Delta}|^2 + |\lambda_1|^2 + |\lambda_2|^2)^2} {\mbox{Im}(A)
\over M_X}.
\end{equation}
In \eref{eq5:boltzII},  the number
$K=\Gamma_\pm/ H$ takes the minimal value of  $K=32$ for
$|\lambda_{\Delta}|=|\lambda_2| \gg |\lambda_1|$ following from the relation:
\begin{equation}
K=32 \,{|\lambda_{\Delta}|^2+|\lambda_2|^2 \over 2 |\lambda_{\Delta}| |\lambda_2|}
\left({|m_\nu| \over 0.05\mbox{ eV}}\right) \,.
\label{eq5:K}
\end{equation}
Notice that  the maximal value of
$\tilde{\epsilon}_\ell = {\mbox{Im}(A)\over M_X}$ is obtained at the resonance point
of $B=\Gamma_{\pm}$.

\begin{figure}[t]
\begin{center}
  \includegraphics[width=0.65\textwidth]{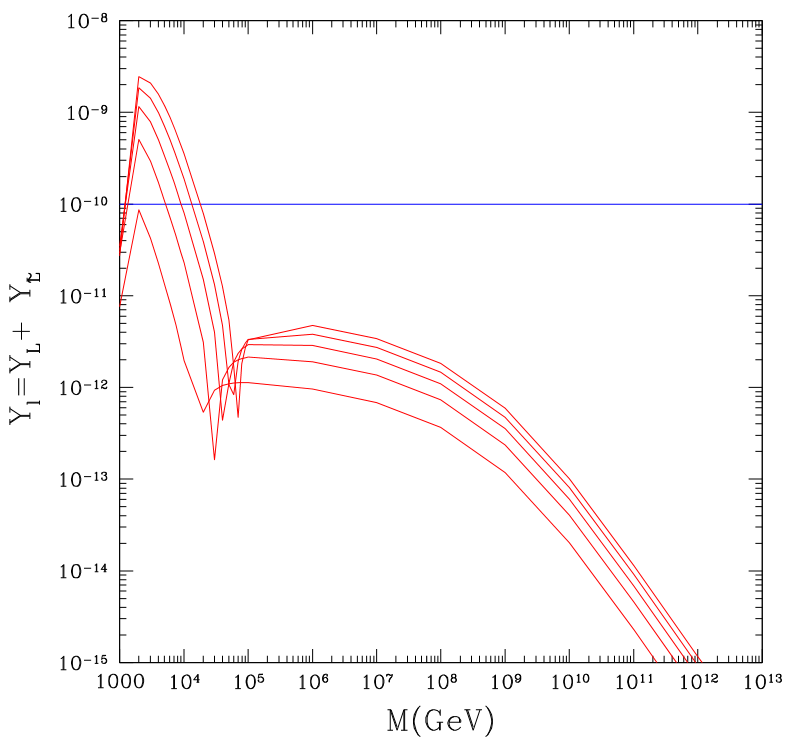}
\end{center}
\caption{\label{fig5:fig6_last} Final lepton asymmetry produced by
triplet decay as a function of $M_X$.  Different curves refer to
$\mbox{Im}(A)=1,2,3,4,5$ TeV from bottom to top. Figure taken from Ref.~\cite{Chun:2005ms}.}
\end{figure}

Combining the two simplified Boltzmann equations, \eref{eq5:YXapp} and \eref{eq5:boltzII}, 
it is straightforward  to get solutions for the lepton asymmetry $Y_{\Delta \ell}$.
\Fref{fig5:fig6_last} shows
the final lepton asymmetry as a function of the
triplet mass $M_X$, varying $\mbox{Im}(A)$ from 1 to 5 TeV.  One finds that 
the required baryon asymmetry can be reached whenever $A$ and $M_X$ are in
the multi-TeV region for which the soft SUSY breaking term $\delta_{soft}$ 
plays an important role. 

\smallskip

A doubly charged boson in type II seesaw can lead to clean signals of same-sign di-lepton resonances at colliders: $\Delta^{\pm\pm} \to \ell^\pm_\alpha \ell^\pm_\beta$.  
The LHC13 data from the integrated luminosity of  13.9 fb$^{-1}$  puts bounds on the doubly charged boson mass of 570 GeV assuming   Br($\Delta^{\pm\pm} \to e^\pm e^\pm$)=1 \cite{ATLAS:2016pbt}.
Observation of a same-sign di-lepton resonance would be a hint for type II seesaw. A concrete confirmation could come from measuring branching ratios and checking if they are consistent with the observed neutrino mass matrix as the type II seesaw relation requires $B_{\alpha\beta}\equiv$ Br($\Delta^{\pm\pm} \to \ell^\pm_\alpha \ell^\pm_\beta$) $\propto |(M_\nu)_{\alpha\beta}|^2$ \cite{Chun:2003ej,Dinh:2012bp}. Assuming for example a normal hierarchy with no CP phase, the current neutrino oscillation data \cite{Olive:2016xmw} determines the ratio of the branching ratios as follows:
\begin{equation}
B_{ee}: B_{e\mu}: B_{e\tau}: B_{\mu\mu}: B_{\mu\tau}: B_{\tau\tau} =
0.62: 5.11: 0.51: 26.8: 35.6: 31.4\,,
\end{equation} 
which still allows for a triplet mass much lower than 570 GeV.

\begin{figure}[t]
\begin{center}
  \includegraphics[width=0.75\textwidth]{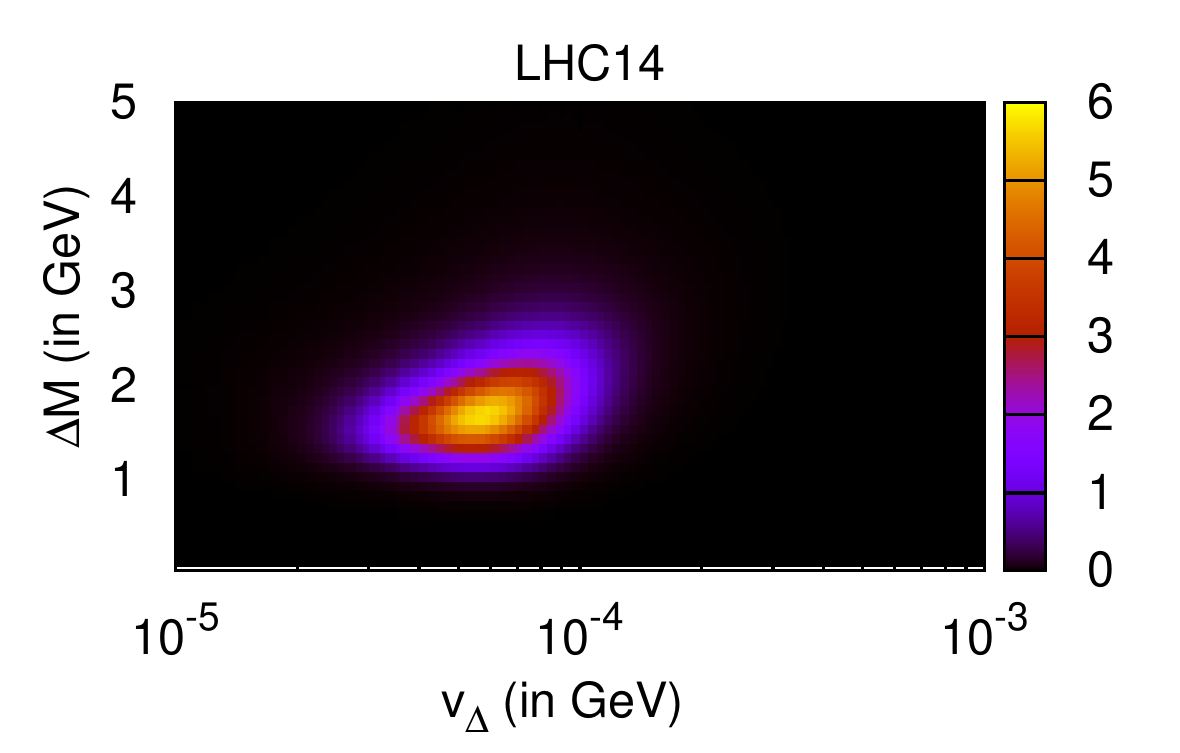}
\end{center}
\caption{\label{fig5:xfac-lhc14} Cross sections (in fb) of same-sign tetra-lepton signals  
for $M_{\Delta^{\pm\pm}}=400$ GeV. Figure taken from Ref.~\cite{Chun:2012zu}.}
\end{figure}

A more striking feature of type II seesaw is triplet-antitriplet oscillation, which can lead to a signal of same-sign tetra-leptons \cite{Chun:2012zu}. Observation of such signals can tell us about the triplet vacuum expectation value correlated with a tiny neutral triplet mass splitting driven by the neutrino mass \cite{Chun:2012zu}.
The triplet vev $v_\Delta \equiv \langle \Delta^0 \rangle/\sqrt{2}$ breaks the lepton number as well as the Higgs triplet boson number and thereby generates a neutrino masses as well as a correlated mass splitting $\delta M_{HA}$ between two neutral triplet boson states $H^0$ and $A^0$ of, e.g., $\Delta^0=(H^0 + i A^0)/\sqrt{2}$.  Such a small mass gap induces oscillation between $\Delta^0$ and $\bar{\Delta}^0$ and thus, an initial production of a triplet-antitriplet pair oscillates into a triplet (or antitriplet) pair leads to same-sign tetra-lepton production in the following chain reaction:
$$ pp \to \Delta^0 \bar{\Delta}^0 \to  \Delta^0 \Delta^0  (\bar{\Delta}^0 \bar{\Delta}^0)
\to \Delta^\pm \Delta^\pm 2 W^{\mp,*} \to \Delta^{\pm\pm} \Delta^{\pm\pm} 4 W^{\mp,*}
\to 4\ell^\pm  4 W^{\mp,*}\,. $$
As usual, the oscillation probability is proportional to $x_{HA}^2$ with the oscillation parameter $x_{HA} \equiv \delta M_{HA}/\Gamma_{\Delta^0}$, where $\delta M_{HA} \equiv M_{H^0}-M_{A^0}$ and
$\Gamma_{\Delta^0} \equiv \Gamma(\Delta^0\to \Delta^\pm W^{\mp,*} )$ such that
\begin{equation}
\delta M_{HA}  \approx  2 {v_\Delta^2\over v_{EW}^2} M_{H^0}
~~\mbox{and}~~
\Gamma_{\Delta^0} \approx {G_F^2 \Delta M^5 \over \pi^3} 
\end{equation}
with $\Delta M \equiv M_{\Delta^0}-M_{\Delta^\pm}$.  \Fref{fig5:xfac-lhc14} shows cross sections for same-sign tetra-lepton signals in the plane of $(v_\Delta,\Delta M)$ taking $M_{\Delta^{\pm\pm}}=400$ GeV. 
One finds that the cross-section is maximized for $v_\Delta \approx 7\times10^{-5}$ GeV and $\Delta M\approx 1.5$ GeV for which $x_{HA}\sim 1$.

\section{Model independent falsification of high-scale leptogenesis}
\label{ch5:sec6:highscale}
While the previous sections were mainly focused on probing a certain model of leptogenesis, we review now the possibility to falsify leptogenesis in a model independent approach. Especially, when the generation mechanism of the lepton asymmetry is at high scale, and thus difficult to access experimentally, such a method can be a powerful tool to narrow down possible underlying leptogenesis mechanisms.

Soon after scenarios of high energy baryogenesis, such as leptogenesis were devised, it was realized that baryon and lepton number violating effective operators would washout the generated asymmetry \cite{Nelson:1990ir,Campbell:1990fa} (for further details we refer to \sref{ch5:sec1:concepts}). In Refs.~\cite{Deppisch:2013jxa,Deppisch:2015yqa,Harz:2015fwa}\,, it was demonstrated that an observation of $\Delta L = 2$ washout processes via non-standard contributions at experiments would directly imply a sizable washout rate. Due to the significant washout, leptogenesis models that generate a lepton asymmetry above the scale of observation could be excluded, leading to an insufficient small baryon asymmetry at the electroweak scale.

\begin{figure}[t]
\centering
\includegraphics[clip,width=0.243\linewidth]{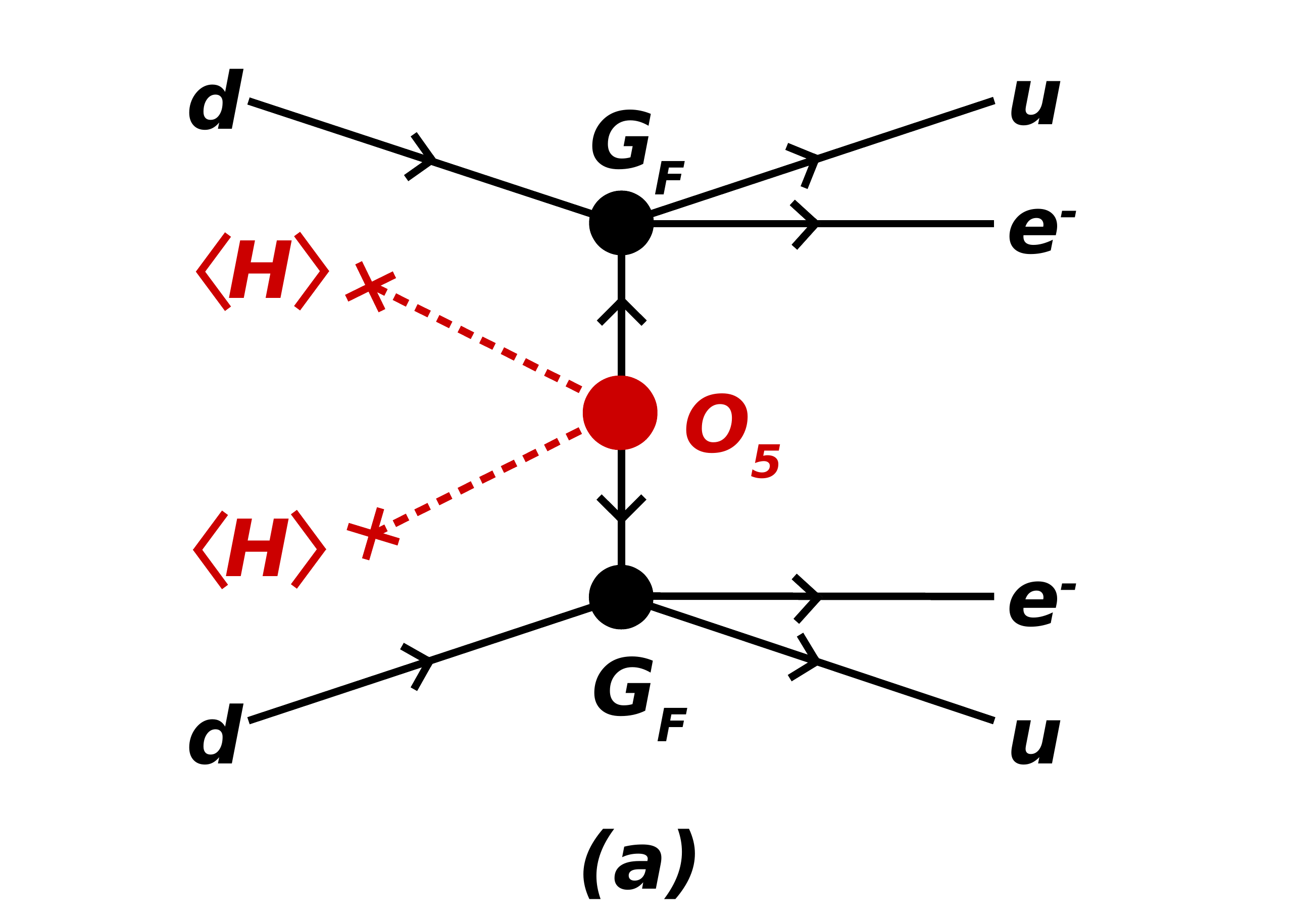}
\includegraphics[clip,width=0.243\linewidth]{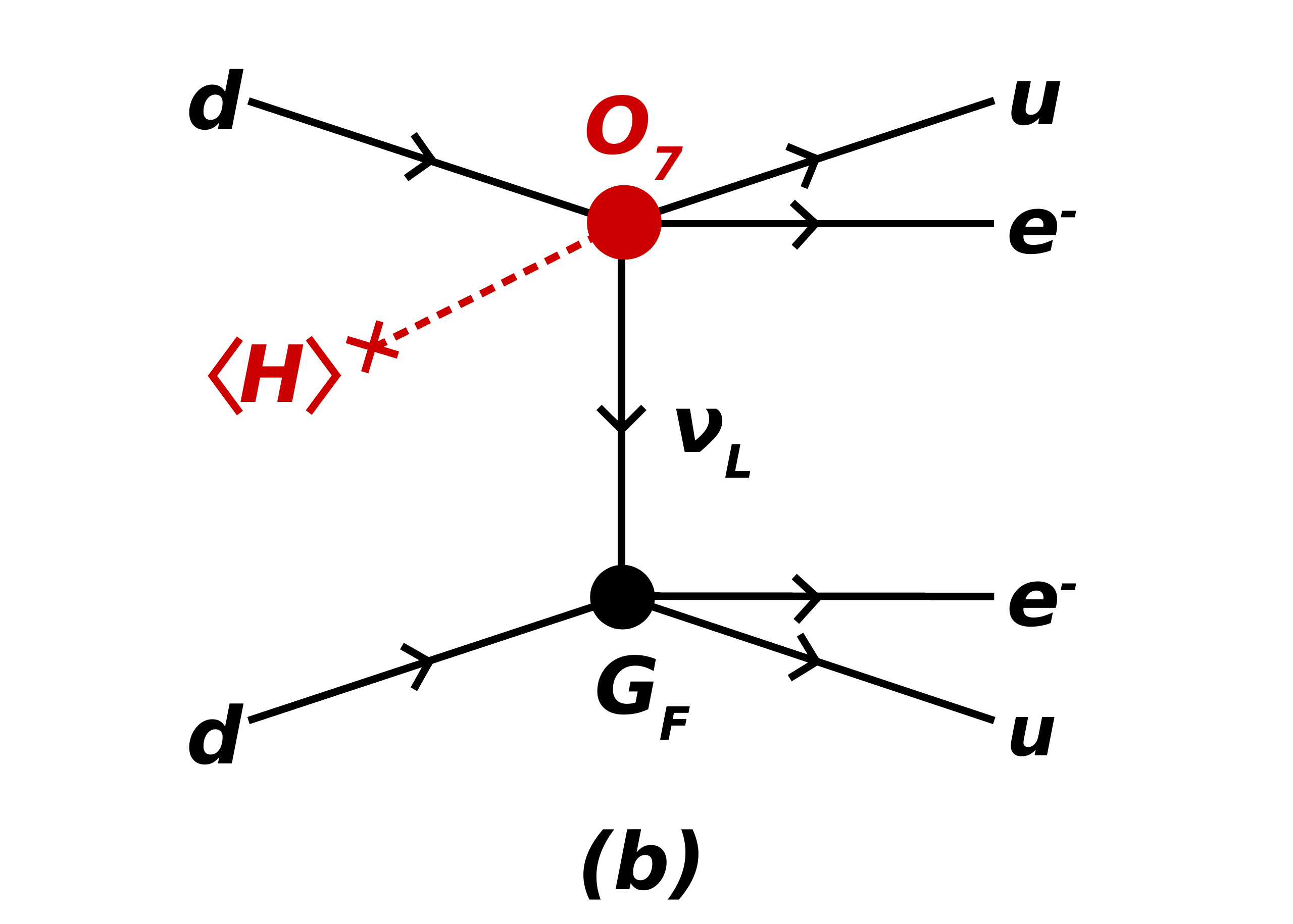}
\includegraphics[clip,width=0.243\linewidth]{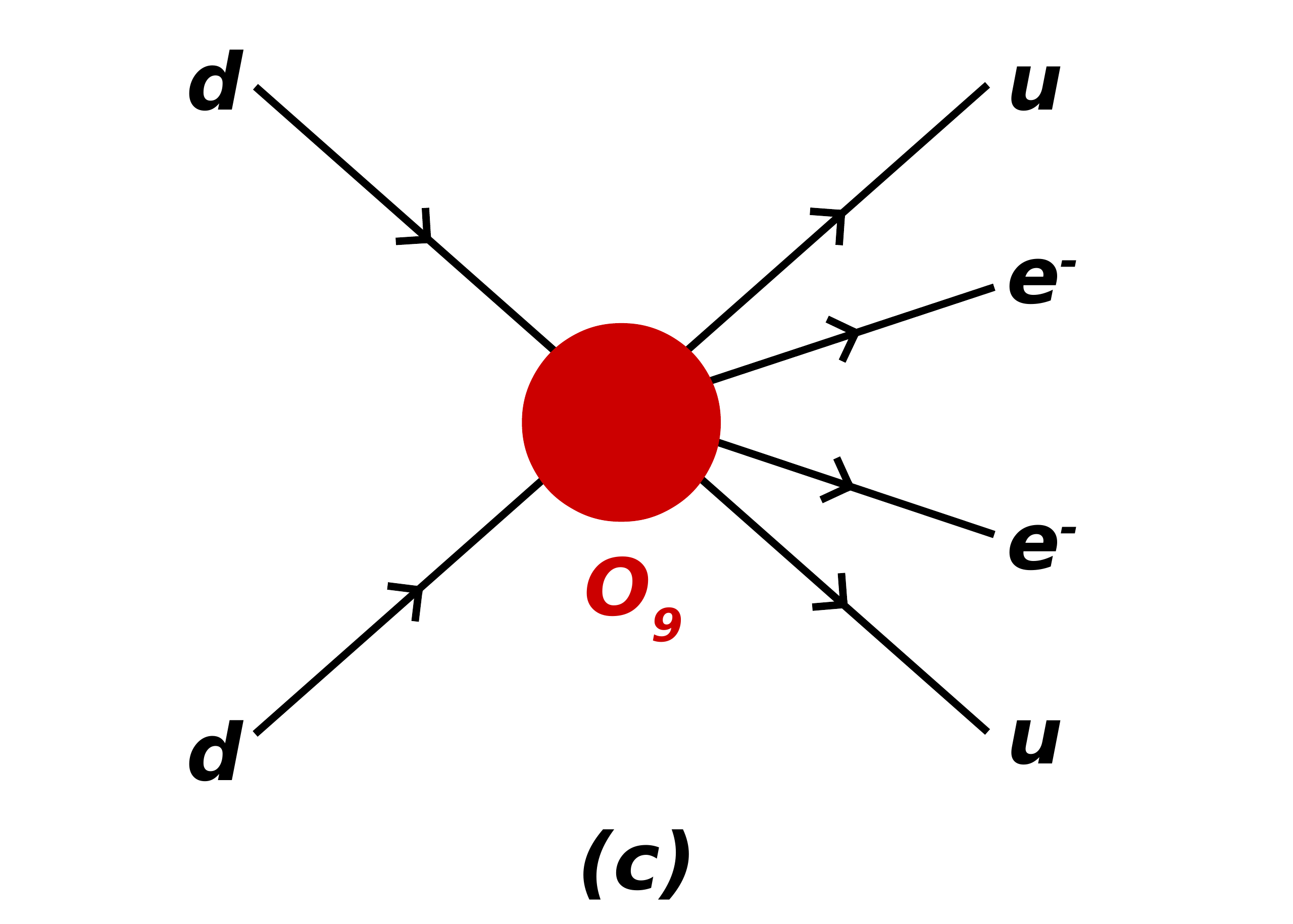}
\includegraphics[clip,width=0.243\linewidth]{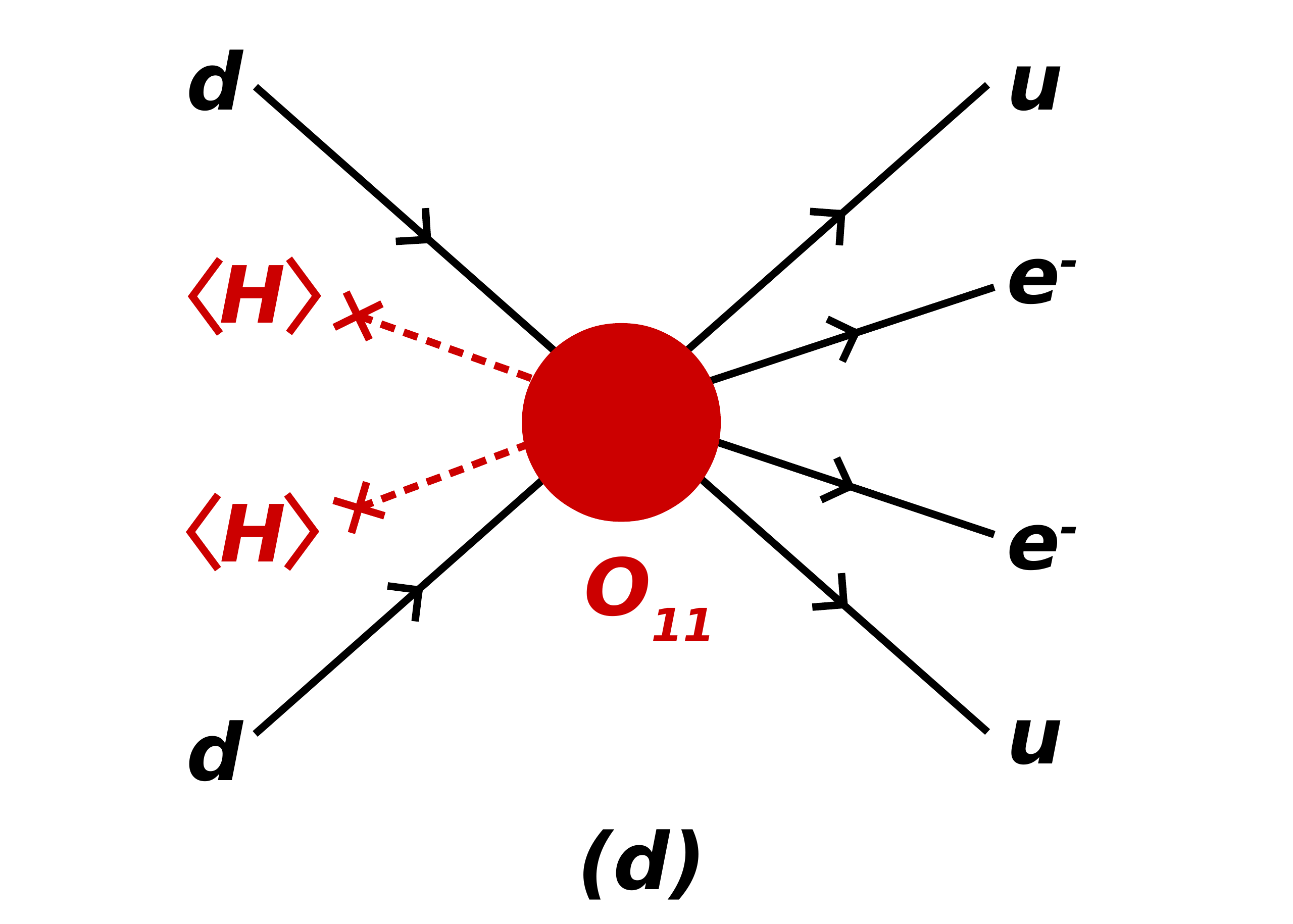}
\caption{Contributions to $0\nu\beta\beta$ decay: Standard Weinberg operator (a), long-range contribution (b), and short-range contribution (c,d). Figure taken from Ref.~\cite{Deppisch:2015yqa}.}
\label{fig6:0vbbcontrib}  
\end{figure}

The low energy effects can be described by $\Delta L = 2$ operators of odd mass dimension, assuming no other light particles beyond the SM at or below the electroweak scale. A full list of all possible 129 $\Delta L = 2$ operators up to 11-dim is given in Ref.~\cite{deGouvea:2007qla}\,, extending the work of Ref.~\cite{Babu:2001ex}\,. These operators can be probed in different experimental set ups, e.g. at the LHC \cite{Deppisch:2013jxa}, neutrinoless double beta decay \cite{Deppisch:2015yqa}, meson decays \cite{Liu:2016oph,Wilson:2013uv,Goudzovski:2011rw,Seyfert:2012ug}, or other low energy experiments \cite{Berryman:2016slh,Renga:2012ba} (for an overview see e.g. Ref.~\cite{Jungmann:2000qp}). Especially sensitive to such $\Delta L = 2$ operators are $0\nu \beta \beta$ decay experiments.
As depicted in \fref{fig6:0vbbcontrib}, $0\nu \beta \beta$ decay could be realized not only by the standard Weinberg operator (\fref{fig6:0vbbcontrib} (a)), but also via a long- (\fref{fig6:0vbbcontrib} (b)) or short-range contribution (\fref{fig6:0vbbcontrib} (c,d)). We quote one operator for each possible contribution as example:\footnote{For easier comparison and readability the commonly used convention of Weyl-spinor-doublets are used with $\ell=(\nu_L, e_L)^T$, $Q=(u_L, d_L)^T$ and $\phi=(\phi^+, \phi^0)^T$ being the left-handed $SU(2)_L$ doublets and ${e}^c$,${u}^c$,${d}^c$ being the charge conjugates of the $SU(2)_L$ right-handed charged fermion operators. The bar notation indicates Weyl-spinor conjugation in contrast to the Dirac-adjoint.}
\begin{align}
	\label{eq6:operators}
	\mathcal{O}_5    &= (\ell^i \ell^j) \phi^k \phi^l \epsilon_{ik} \epsilon_{jl},          % 1
	&& &&\mathcal{O}_7    = (\ell^i d^c) (\bar{e^c} \bar{u^c}) \phi^j \epsilon_{ij},     \\  % 8
	\mathcal{O}_9    &= (\ell^i \ell^j) (\bar{Q}_i \bar{u^c}) (\bar{Q}_j \bar{u^c}), \nonumber % 12a
	&& &&\mathcal{O}_{11} = (\ell^i \ell^j) (Q_k d^c) (Q_l d^c) \phi_m \bar{\phi_i} 
	\epsilon_{jk}\epsilon_{lm}. % 24a
\end{align}
Assuming one operator dominant at a time, the effective coupling of this operator can be related to the half-life of $0\nu \beta \beta$ decay
\begin{align}
T_{1/2}^{-1} = \omega_i^2 G_i |\mathcal{M}_i|^2,
\end{align}
where $G_i$ is the nuclear phase space factor and $|\mathcal{M}_i|$ the nuclear matrix element. Therefore, an observation of $0\nu \beta \beta$ decay would allow one to pin down the effective coupling $\omega_i$ of this operator\cite{Pas:1999fc,Deppisch:2012nb} and thus as well to estimate the corresponding operator scale $\Lambda_D$
\begin{gather}
\label{eq6:epsilons}
	                m_e \omega_5          = \frac{g^2 v^2}{2 \Lambda_5},   \,\,
	\frac{G_F \omega_7}{\sqrt{2}}         = \frac{g^3 v}{2 \sqrt{2} \Lambda_7^3}, \,\,
	\frac{G_F^2 \omega_{\{9,11\}}}{2 m_p} = \{\frac{g^4}{\Lambda_9^5},
	                                            \frac{g^6 v^2}{2\Lambda_{11}^7} \},
\end{gather}
with $m_e$ being the mass of the electron, $m_p$ the proton mass, and $G_F$ the Fermi constant. The coupling $g$ indicates the expected scaling of a UV complete model, set to 1 for simplicity. The very same operator that triggers $0\nu \beta \beta$ decay would contribute as well to the washout processes in the early Universe. Applying the condition of efficient washout ($\Gamma_W / H > 1$), a lower limit on the temperature can be derived above which scale the washout is highly effective:
\begin{align}
\label{eq6:temp_limit}
  \Lambda_D \left( \frac{\Lambda_D}{c_D' \Lambda_\text{Pl}} \right)^{\frac{1}{2D-9}} 
	\equiv \lambda_D \lesssim T \lesssim \Lambda_D.
\end{align}
The operator dependent constant $c_D' = \pi^2 c_D/(3.3 \sqrt{g_*}) \approx 0.3 c_D$ with $c_{\{5,7,9,11\}} = \{ 8/\pi^5, 27/(2\pi^7), 3.2\times 10^4/\pi^9, 3.9\times 10^5/\pi^{13}\}$ arises from the calculation of the reaction density integrated over the phase space including all possible permutations of particles in the initial and final state. The upper limit indicates up to which scale the effective operator approach is valid for estimating the washout rate reliably, given by the scale of the corresponding operator.

A more precise estimate can be performed by solving the Boltzmann equation for a temperature $\hat{\lambda}_D$ at (above) which a lepton asymmetry of order one could still have existed and would have been washed out to (less than) the observed baryon asymmetry $\eta_B^{\mathrm{obs}}$ at the electroweak scale $v$:
\begin{align}
	\hat\lambda_D \approx 
	\left[(2D-9) \ln\left(\frac{10^{-2}}{\eta_B^\text{obs}}\right) \lambda_D^{2D-9} + (v/\sqrt{2})^{2D-9}\right]^{\frac{1}{2D-9}}\!\!\!.
\end{align}
Assuming an observation of $0\nu \beta \beta$ decay at the current (\fref{fig6:ranges}, left bars) and future sensitivities (\fref{fig6:ranges}, right bars), the corresponding temperature interval for a highly efficient washout can be obtained. This leads to the main result, directly extractable from \fref{fig6:ranges}:
\begin{figure}[t]
\centering
\includegraphics[clip,width=0.65\linewidth]{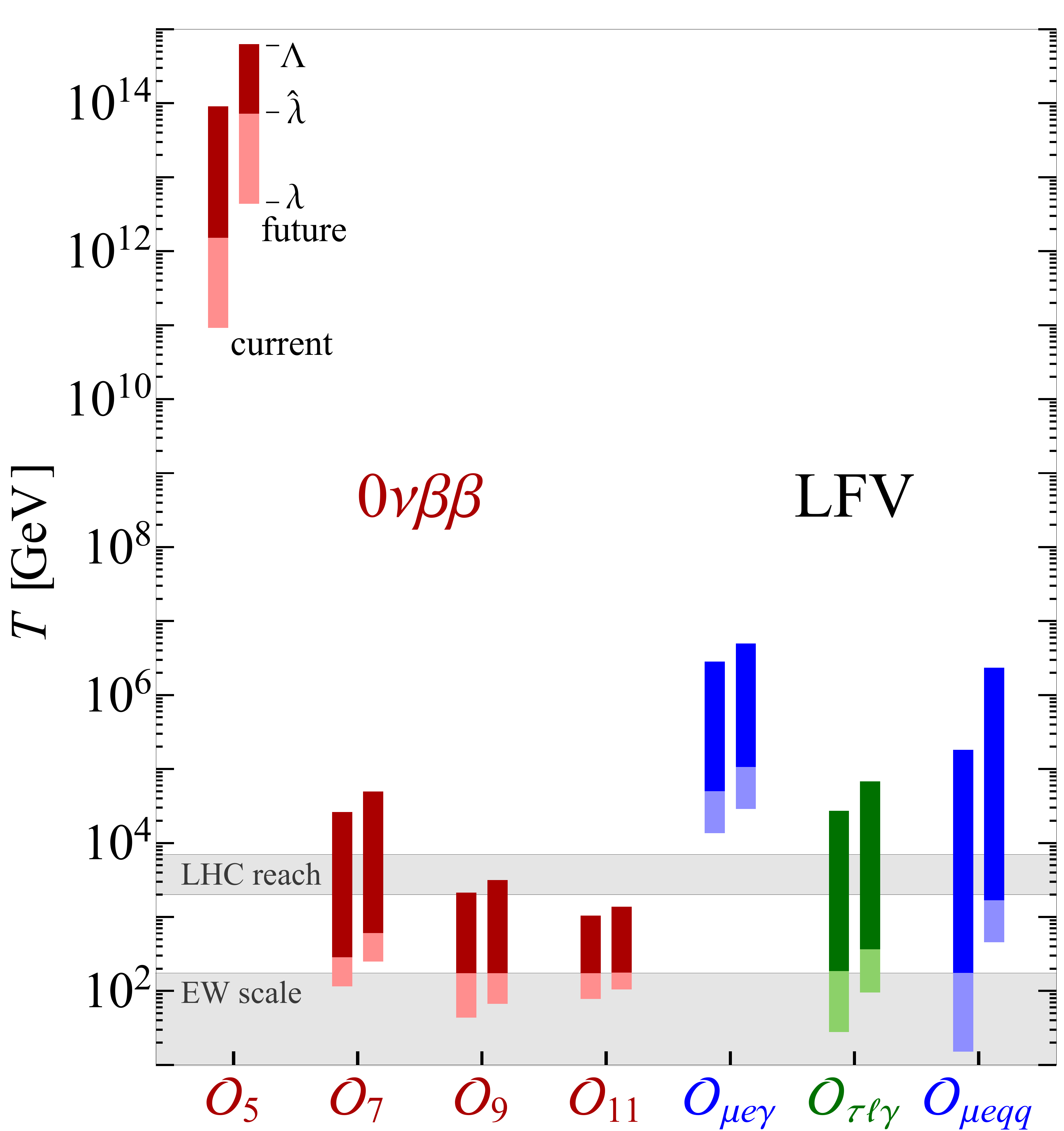}
\caption{Temperature intervals in which the given LNV and LFV operators are in equilibrium, defined by washout scales $\lambda, \hat\lambda$ and the operator scale $\Lambda$. All scales are evaluated assuming an observation at current (left bar) and future (right bar) sensitivities. Figure taken from Ref.~\cite{Deppisch:2015yqa}.}
\label{fig6:ranges}  
\end{figure}
If $0\nu \beta \beta$ was discovered and was not triggered by the standard Weinberg operator, this will imply that any baryogenesis mechanism will be excluded above the scale $\hat{\lambda}_D$, implying anywhere but close to the electroweak scale.

This, however, makes it necessary to being able to distinguish between the Weinberg operator and other higher dimensional operators. Although being a challenge, this is not an impossible task: 
\begin{itemize}
 \item $0\nu \beta \beta$ decay triggered by a 9-dim or 11-dim operator hints towards a visible signature at the LHC. In contrast, the scale of a 7-dim or 5-dim operator is already pushed towards too high values in order to be observable at the LHC \cite{Deppisch:2013jxa}. Possible signatures include an invariant mass peak corresponding to the mass of the new physics particle created on-shell or a possible asymmetry in the rate of $e^+ e^+$ and $e^- e^-$ final states~\cite{Bonnet:2012kh,Helo:2013dla,Helo:2013ika}.
 \item Specifically the 7-dim operator can be probed with the future SuperNEMO experiment, as it will be able to measure the angular and energy distribution of the two electrons involved in $0\nu \beta \beta$. As $0\nu \beta \beta$ decay realized by a long-range contribution features a signature of a left- and a right-handed electron in contrast to two left-handed ones for the standard mass mechanism, SuperNEMO will be able to distinguish them~\cite{Arnold:2010tu}.
 \item A comparison with cosmology can as well reveal a new physics contribution. A discrepancy between the sum of the neutrino masses from cosmology and $|m_{\beta \beta}|$ from $0\nu \beta \beta$ decay experiments would imply that $0\nu \beta \beta$ is not realised by the Weinberg operator.
 \item A discrepancy when comparing the experimental ratio of the $0\nu \beta \beta$ half-life of different isotopes with the theoretical prediction could as well hint towards new physics contributions. While systematic uncertainties and effective couplings cancel by considering the ratio $T_{1/2}({}^{76}\mathrm{Ge}) / T_{1/2}({}^{A}X)= (|\mathcal{M}({}^{76}\mathrm{Ge})|^2 G({}^{76}\mathrm{Ge})) / (|\mathcal{M}({}^{A}X)|^2 G ({}^{A}X))$, the ratio of the half-life, however, still depends on the model dependent matrix element and phase space factor and is thus sensitive to new physics~\cite{Deppisch:2006hb}.
 \item As demonstrated in Ref.~\cite{Hirsch:1994es}, the comparison between $0\nu\beta^- \beta^-$ decay and $0\nu\beta^{+}/\mathrm{EC}$ or $0\nu\beta^-\beta^-$ and $0\nu\beta^+\beta^+$ decay can point us as well towards new physics. Similarly, the comparison between $0\nu\beta\beta$ decay to the ground state and an excited state is interesting with that respect~\cite{Simkovic:2001qf}.
\end{itemize}
In order to guarantee that no lepton asymmetry is hidden in another flavor sector, LNV should be observed as well in the $\mu \mu$ and $\tau \tau$ sector or in lepton flavor violating (LFV) $\Delta L =0$ rare decays. In order to estimate above which scale an equilibration of an asymmetry between flavors is highly efficient, a similar analysis as outlined above was performed in Ref.~\cite{Deppisch:2015yqa}. To this end, the corresponding operators $\mathcal{O}_{ll\gamma} = \mathcal{C}_{ll\gamma} \bar \ell_l \sigma^{\mu\nu} \bar{l^c} H F_{\mu\nu}$ and $\mathcal{O}_{ll q q} = \mathcal{C}_{ll q q} (\bar{l} \, \Pi_1 l) (\bar{q} \, \Pi_2 q)$ (the $\Pi_i$ represent possible Lorentz structures)~\cite{Raidal:2008jk}, with $l=e,\mu,\tau$ were considered. Assuming an observation of the corresponding rare decays at their current and future sensitivity, the operator scale can be evaluated $\mathcal{C}_{ll\gamma} = \frac{e g^3}{16\pi^2 \Lambda^2_{ll\gamma}}, \quad \mathcal{C}_{ll q q}    = \frac{g^2}{\Lambda^2_{ll q q}}$ and the corresponding equilibration interval derived. As depicted in \fref{fig6:ranges}, an observation of $\tau^\pm \rightarrow \ell^\pm \gamma$ or $\mu -e$ conversion will guarantee the equilibration of an asymmetry between different flavors. The constraint on $\mu \rightarrow e \gamma$, however, is already too sensitive.

Although this approach is conservative and model independent, certain limitations exist, e.g. new conserved quantum numbers or hidden sectors can act as protection mechanism~\cite{Weinberg:1980bf,Dimopoulos:1988jw,Antaramian:1993nt}, see as well \sref{ch5:sec3:cloistered-baryogenesis} for an example.

\begin{figure}[t]
\centering
\vskip-6mm
\includegraphics[clip,width=0.47\linewidth]{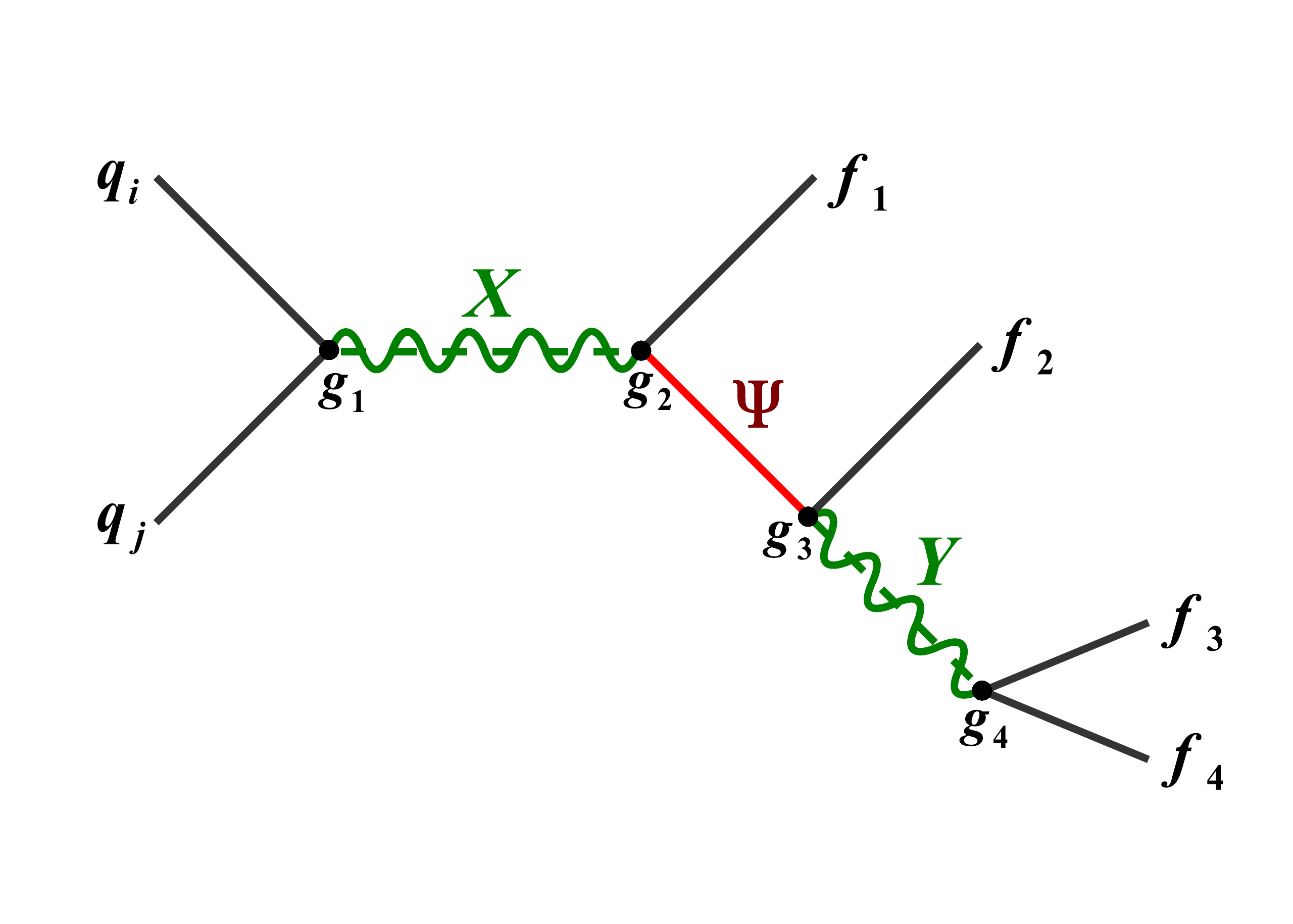}
\includegraphics[clip,width=0.47\linewidth]{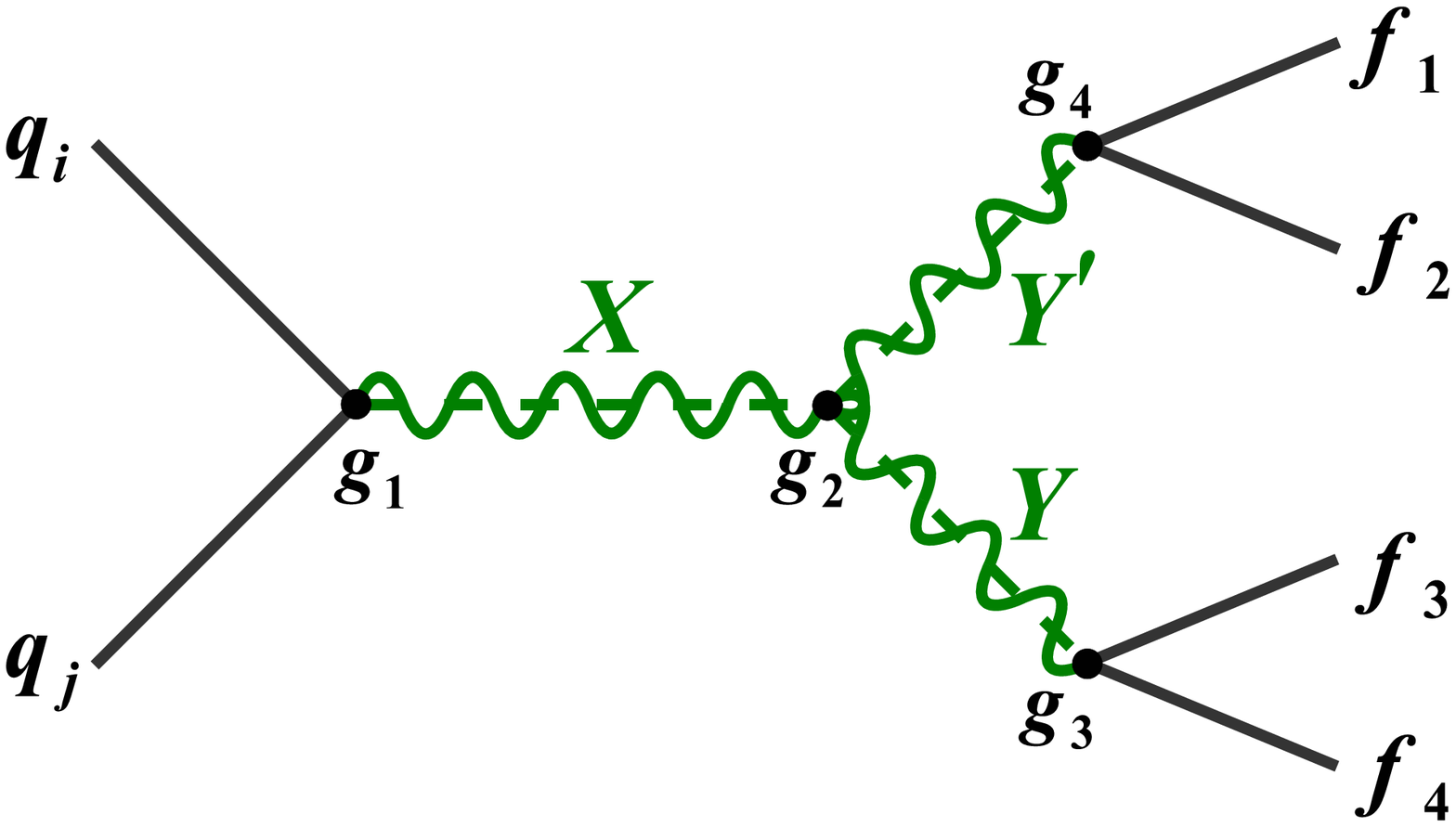}
\vskip-5mm
\caption{Possible diagrams contributing to the resonant same-sign di-lepton
signal $pp \to \ell^\pm
  \ell^\pm q q$. The intermediate particles $X$ and $Y^{(')}$ indicate
different vector or scalar bosons, $\Psi$ denotes a fermion. Generally, any two of the four fermions $f_i$ can be leptons. Figures taken from~\cite{Deppisch:2013jxa}.}
\label{fig6:decompositions}  
\end{figure}

\begin{figure}[t]
\centering
\includegraphics[clip,width=0.49\linewidth]{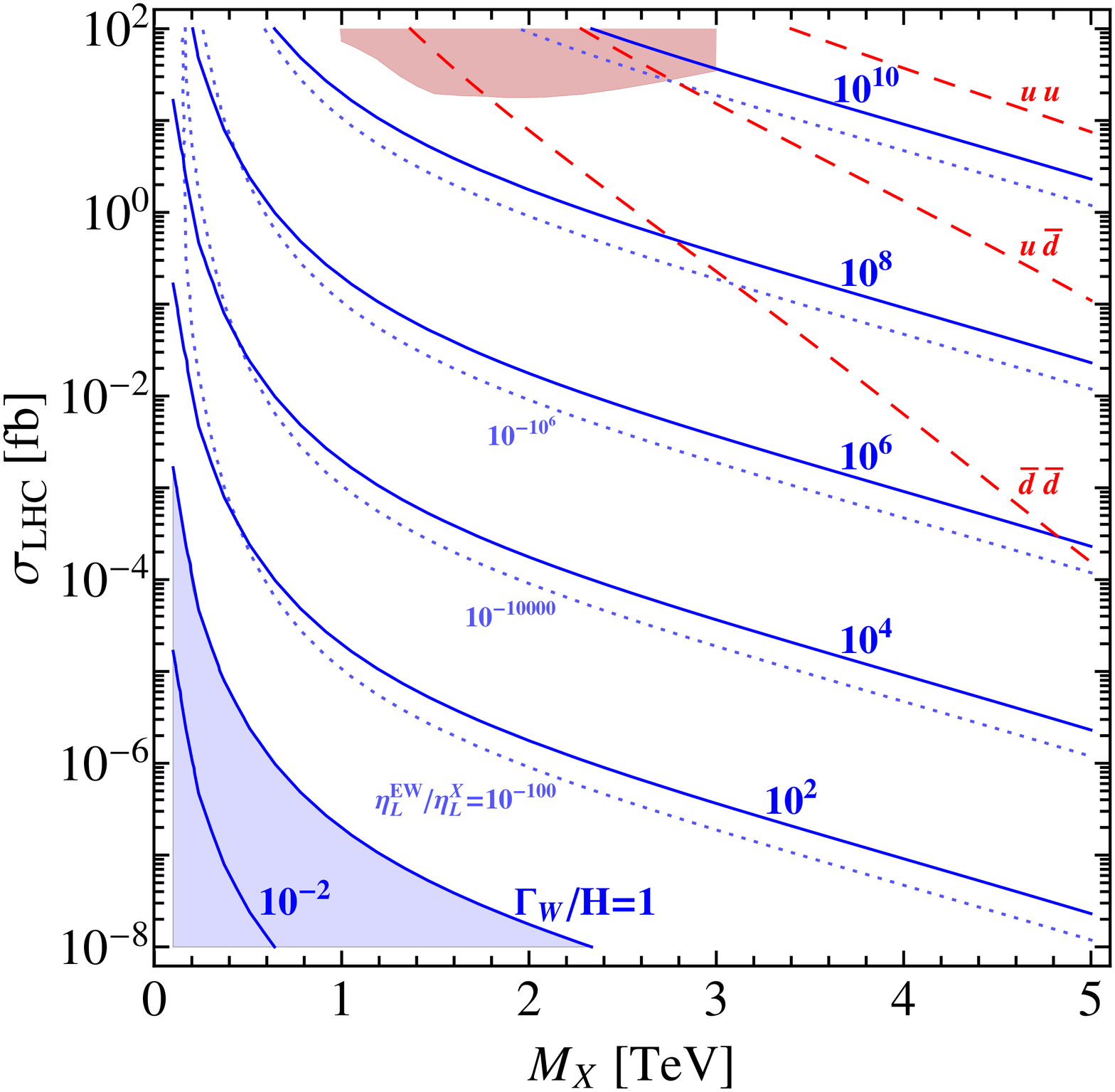}
\includegraphics[clip,width=0.49\linewidth]{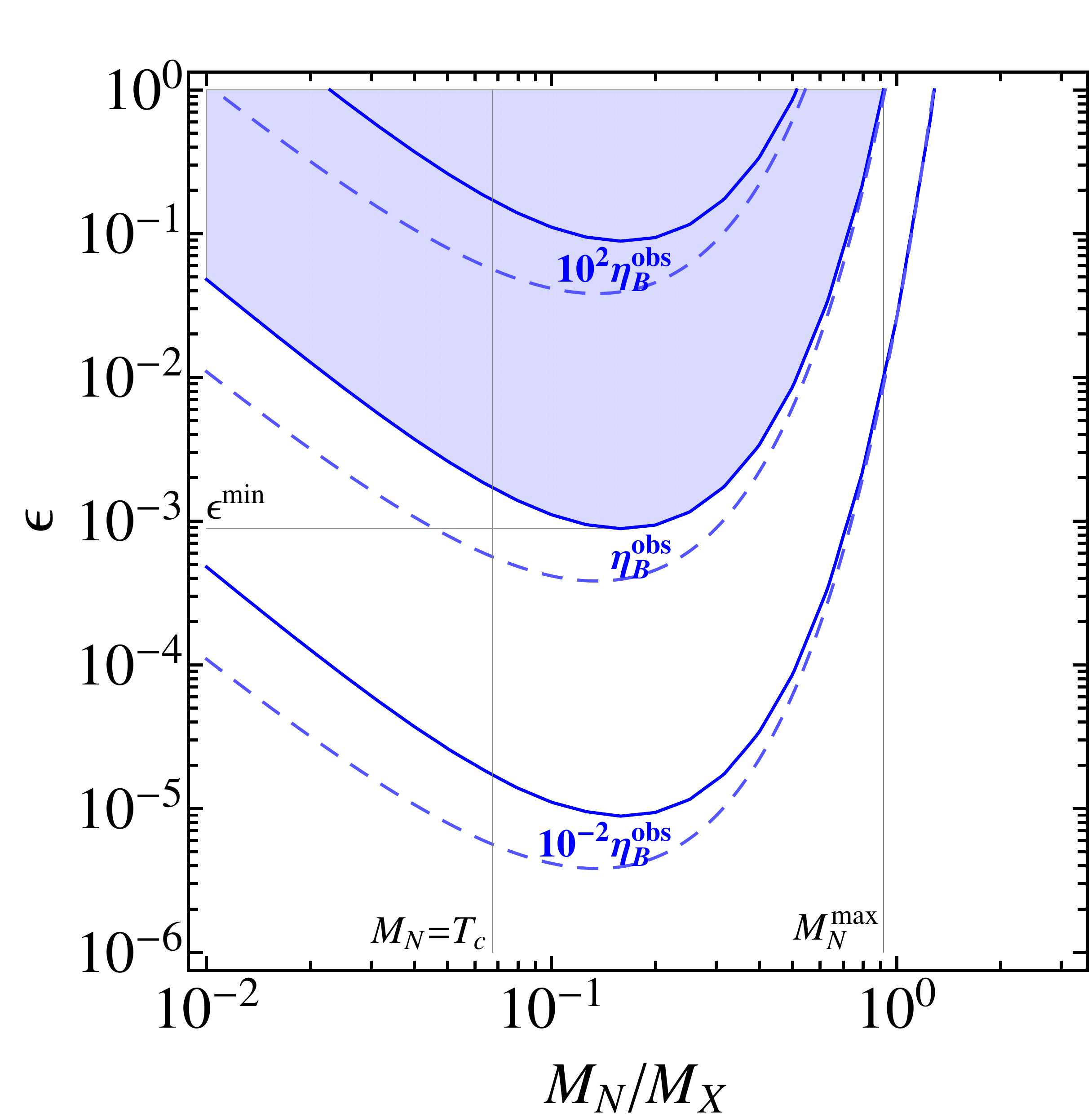}
\caption{Left: Washout rate $\Gamma_W/H$ at $T = M_X$ as a function of $M_X$
  and $\sigma_\text{LHC}$ (solid blue contours). The dotted light blue 
	contours show the surviving lepton asymmetry at the EW
	scale relative to its value at $M_X$ (
	$\eta_L^\text{EW}/\eta_L^X$).
	The red dashed curves indicate typical cross sections of the process 
	$pp \to \ell^\pm \ell^\pm q q$. The red shaded region is excluded
according to LHC searches for resonant same-sign di-leptons \cite{CMS:2012uaa}. Right: Baryon asymmetry $\eta_B$ as a function of $M_N/M_X$ and
  $\epsilon$ for $M_X = 2$~TeV and $\sigma_\text{LHC} = 0.1$~fb (solid
  contours). The intermediate contour corresponds to the observed
  value $\eta_B^\text{obs}$. Figures taken from~\cite{Deppisch:2013jxa}.
}
\label{fig6:GammaW_mX_sigma}
\end{figure}

Besides $0\nu \beta \beta$ decay, an observation of $\Delta L =2$ processes at the LHC would similarly allow one to falsify model independently high-scale leptogenesis models \cite{Deppisch:2013jxa}, e.g. via the resonant process $pp \rightarrow \ell^{\pm} \ell^{\pm} q q$ (cf. \fref{fig6:decompositions}).
Based on the observed LHC cross section $\sigma_{\mathrm{LHC}}$ and the resonant mass $M_X$, the washout can be estimated \cite{Deppisch:2013jxa}
\begin{align}
\label{eq6:washout_factor_related}
  \frac{\Gamma_W}{H} &= 
  \frac{0.028}{\sqrt{g_*}}
  \frac{M_\text{P}M_X^3}{T^4}
  \frac{K_1\left( M_X/T \right)}
  {f_{q_1 q_2}\left( M_X / \sqrt{s} \right)}
  \times(s\sigma_\text{LHC}),
\end{align}
where the parton distribution function  $f_{q_1q_2}$ has to be singled out for calculating the reaction density of the washout process in the early Universe. As shown in \fref{fig6:GammaW_mX_sigma} (left), the observation of an LNV signal at the LHC would imply a significant washout such that high-scale leptogenesis models would be excluded. Similarly, this was discussed in \sref{ch5:sec4:extendedgauge-LR} within the context of the L-R symmetric model, where a discovery of a heavy $W_R$ with a mass below $\approx 10~\mathrm{TeV}$ would exclude viable leptogenesis.
If the CP asymmetry is generated below the scale of the resonant mass $M_X$, a falsification is not necessarily possible, however, a lower limit on the CP asymmetry can be set in order to create the observed baryon asymmetry $\eta_B^{\mathrm{obs}}$ with the washout present. In order to demonstrate this, a scale $M_N$ at which the CP asymmetry of order $\omega$ is generated was considered \cite{Deppisch:2013jxa}
\begin{align}
  \label{eq6:etaBestimation}
	\log_{10} \left|\frac{\eta_B}{\eta_B^\text{obs}}\right| &\lesssim 
	2.4\, \frac{M_X}{\text{TeV}} \left( 1 - \frac{4}{3} \frac{M_N}{M_X}
\right)\nonumber \\
&+ \log_{10} \left[ |\epsilon|\, \left(
\frac{\sigma_\text{LHC}}{\text{fb}} \vphantom{\frac{4}{3}
\frac{M_N}{M_X}}\right)^{-1} \left( \frac{4}{3} \frac{M_N}{M_X} \right)^2
\right].
\end{align}
The results are shown in \fref{fig6:GammaW_mX_sigma} (right). For $M_N<M_X$, there exists a lower limit on the CP asymmetry $\epsilon >
\epsilon^\text{min} \approx 10^{-3}$, which could strongly constrain resonant leptogenesis models. In concordance with the discussion before, for $M_X<M_N$, high-scale leptogenesis models would be excluded.
Similarly to the discussion of the limitations with respect to $0 \nu \beta \beta$ decay, it should be guaranteed that no asymmetry is stored in another flavor. To this end, different flavor combinations should be observed to unambiguously exclude high-scale leptogenesis models, meaning $pp \to \ell^\pm \ell^\pm q q$ for either $\ell \ell = e e$, $\mu\mu$ and $\tau\tau$, or for $e\mu$ and $e(\mu)\tau$. Again, one has to be aware of possible protection mechanisms like a hidden lepton number, which is converted into a baryon number below the scale $M_X$.

However, generally, this method is a powerful approach in order to falsify high-scale leptogenesis and baryogenesis models with current experiments and should encourage experiments to refine their analyses and set-ups with respect to $\Delta L = 2$ searches.

\section{Summary and conclusions}
In this chapter, we have reviewed the possibility to probe leptogenesis models at experiments. In \tref{summaryprobe} we show an overview of experimental signatures probing different leptogenesis models, in \tref{summaryfalsify} signatures are summarized which allow for falsifying certain models.

We commenced this chapter with GeV-scale leptogenesis, discussing experimental prospects of the type I seesaw model that allows for a baryon asymmetry via the ARS-mechanism. In \sref{ch5:sec2:ARS-seesaw}, we reviewed the experimental prospects of the minimal seesaw model with $n=2$ RH neutrinos in the oscillatory regime. We showed that the input from neutrino oscillation experiments, neutrinoless double beta decay searches, and future searches for heavy leptons at SHiP or high intensity $e^+e^-$ colliders might be sufficient in order to fully constrain the model by the observed baryon asymmetry of our Universe and is thus possible to be probed. In contrast to the naive seesaw regime, we reviewed in \sref{ch5:sec2:ARS-seesaw} the symmetry protected scenario in the overdamped regime. It can accommodate successful leptogenesis for much larger mixing angles and leads hence to larger branching ratios. As for relatively large $U^2$, the requirement $U_\alpha^2/U^2<<1$ has to be met for at least one flavor in order to generate successfully the correct baryon asymmetry, a comparison of all three $U_\alpha^2$ can be used as a test for leptogenesis. As leptogenesis via the ARS-mechanism requires light RH neutrinos, they can be searched for at experiments. The current constraints on RH neutrino was reviewed in \sref{ch5:sec2:nhlcons} including direct and indirect searches, cosmological and ``global'' constraints. Another hint for RH neutrinos are lepton-number violating meson decays. The possibility to measure CP violation in these rare decays was discussed in \sref{ch5:sec2:CPMesonDecays}.

After focusing on the testability of leptogenesis from oscillations, we discussed the possibility to probe leptogenesis from out-of-equilibrium decays. In \sref{ch5:sec1:concepts}, we summarized the problems that have to be addressed in order to obtain TeV-scale leptogenesis. We outlined three mechanisms in order overcome the destructive interplay between CP asymmetry generation and washout processes and to achieve models testable at current experiments: almost degenerate particles, late decays or massive decay products.

\begin{table}
\tbl{Overview of experimental signatures of different leptogenesis/baryogenesis models.
Besides ARS leptogenesis where the mass scale of new particles is around GeV,
other leptogenesis/baryogenesis models from decays require new particles to be above
the electroweak sphalerons freeze-out temperature $T_{\rm EWSp}$ (lattice calculation gives
$T_{\rm EWSp} = 131.7 \pm 2.3$ GeV~\cite{DOnofrio:2014rug}) such that a nonzero baryon asymmetry
can be induced. In general, low-scale realizations of the second type of models face two issues due
to constraints from (a) the light neutrino mass scale (b) washout from scatterings as described
in \sref{ch5:sec1:concepts}. The scale can be lowered down to $T_{\rm EWSp}$ by evoking
resonant enhancement from almost degenerate decaying particles (see \sref{ch5:sec1:concepts} for details).\label{summaryprobe}}
{\begin{tabular}{|c|c|l|c|}
  \hline
  model & lowest scale & experimental signatures & section\\
  \hline
\multirow{7}{*}{ARS leptogenesis} & \multirow{7}{*}{GeV} & RH neutrino production in fixed  & \multirow{7}{*}{\sref{ch5:sec2:ARS}}\\
 {}& {} &  target experiments (beam dump or  & {}\\
 {} & {} &  peak searches) or colliders, & {}\\
 {}& {} &  RH neutrino decays & {}\\
 {} & {} &   (displaced vertices, LNV, LFV), &{} \\
  {}  &{}  &  $\nu$ oscillations, $0\nu\beta\beta$ decay,& {}\\
 {} & {} &   meson decays& {}\\
  cloistered baryogenesis &  $ T_{\rm EWsp}$ & long-lived scalars at hadron colliders & \sref{ch5:sec3:cloistered-baryogenesis}\\
leptogenesis in $Z'$ models & $ T_{\rm EWsp}$ & same-sign di-lepton final states & \sref{ch5:sec4:extendedgauge-Zprime}\\
soft leptogenesis with generic $A$ term & $ T_{\rm EWsp}$  & charged LFV & \sref{ch5:sec5:typeI-soft-leptogenesis}\\
\multirow{2}{*}{soft type II leptogenesis} & \multirow{2}{*}{$T_{\rm EWsp}$}  & same-sign di-lepton resonances & \multirow{2}{*}{\sref{ch5:sec5:typeII-soft-leptogenesis}}\\
 {} & {}  &  same-sign tetra-leptons & {}\\
  \hline
\end{tabular}
}
\end{table}
%%%

%%%
\begin{table}[]
\tbl{Experimental signatures that could falsify leptogenesis/baryogenesis models which rely on $B-L$ violation. \label{summaryfalsify}}
{\begin{tabular}{|c|l|c|}
  \hline
  Models & experimental signatures  & section\\
  \hline
  low-scale type II/III seesaw model  & $M_\Delta,~M_\Sigma < 1.6$ TeV & \sref{ch5:sec3:extended_seesaw-typeII}\\
 L-R symmetric model  & $M_{W_R} < 10$ TeV & \sref{ch5:sec4:extendedgauge-LR}\\
 \multirow{2}{*}{model independent}  & LNV at LHC, $0\nu\beta\beta$ decay via &  \multirow{2}{*}{\sref{ch5:sec6:highscale}}\\
 {}   &  non-standard contributions, charged LFV & {} \\
  \hline
\end{tabular}
}
\end{table}

We then reviewed different models which allow for viable TeV-scale leptogenesis. We started with extensions of the standard, vanilla seesaw models involving new matter fields. After discussing the different possibilities to extend the type I seesaw model with scalar particles charged under SM quantum numbers, and after identifying the corresponding complications to achieve low-scale leptogenesis, we highlighted cloistered baryogenesis in \sref{ch5:sec3:extensions-typeI}.  
In this mechanism, a global $U(1)_B$ is imposed. The decays of heavy states $N$ generate an equal
in size and opposite in sign baryon asymmetry in the two daughter particle species $u$ and $\tilde u$ ($\Delta B_{u} = -  \Delta B_{\tilde u}$).
While the $u$-sector couples to electroweak sphalerons, the $\tilde u$ sector remains chemically
decoupled (cloistered), and since baryon asymmetry in $u$ gets partially converted to lepton asymmetry,
it does no longer balance the asymmetry stored in $\tilde u$. Eventually $\tilde u$ particles decay into
SM particles but only after electroweak sphalerons freeze out, so that a non-vanishing total
baryon asymmetry $\Delta B = \Delta B_u + \Delta B_{\tilde u} \neq 0$ results. This scenario requires
a new long-lived particle carrying color and baryon number, which can be looked for at hadron colliders.

In \sref{ch5:sec3:extended_seesaw-typeII}, we then discussed limitations and possible extensions to type II and type III seesaw models. Triplets, unlike heavy neutrinos, undergo gauge interactions. These do not lead to any washout, however, they get thermalized which leads to a suppression of the lepton asymmetry: If the gauge interaction rate is larger than the Hubble rate and the decay rate, the asymmetry production is suppressed by the ratio of these two rates. This leads to a lower bound on the triplet mass $M_{\Sigma} > 3 \times 10^{10}\mathrm{GeV}$. Assuming a quasi-degenerate spectrum and a CP asymmetry of order one, the bound can be lowered to $M_{\Sigma} > 1.6~\mathrm{TeV}$ for type III, holding as well for $M_{\Delta}$ in type II. This opens up the possibility to falsify these scenarios by discovering lighter states. Unlike type I seesaw states, the new heavy states in type II and type III seesaw can be pair produced at the LHC and are intensively searched for. Also characteristic displaced vertices could be probed. By extending these models by new scalars and fermions, either the resonant enhancement takes place or an additional contribution to the neutrino mass generations mechanism is added. This leads to a separation between the scales of leptogenesis and neutrino mass generation, and opens up a new versatile phenomenology and testability.

We then focused on leptogenesis generated by seesaw models involving an extended gauge sector. In \sref{ch5:sec4:extendedgauge-Zprime}, we discussed (resonant) $U(1)^\prime$ models. Here, LHC bounds already strongly constrain the mass of the $Z^\prime$ ($M_{Z^\prime} > 3.7~\mathrm{TeV}$) and thus the corresponding heavy neutrino mass. Observation of large CP violation in same-sign di-lepton final states at the LHC would be a hint for this scenario. Another popular scenario is the L-R symmetric model, which was reviewed in \sref{ch5:sec4:extendedgauge-LR}. With the mass of the $W_R$ being constrained to be above 10~TeV in order to have successful leptogenesis, it is difficult to test this scenario. Thus, we discussed to which extent it is possible to falsify this model if a $W_R$ with a mass below 10~TeV were to be discovered at the LHC.

We further reviewed supersymmetric leptogenesis models. While the direct probe of type I soft leptogenesis with generic trilinear couplings by the observation of heavy sneutrinos is not forseeable in the future, possible significant contributions to CLFV processes are close to sensitivities of present and future experiments. This was discussed in \sref{ch5:sec5:typeI-soft-leptogenesis}. As shown in \sref{ch5:sec5:typeII-soft-leptogenesis}, also in type II soft leptogenesis the heavy neutrinos are too heavy to be probed directly. However, signatures of doubly charged bosons decaying into same-sign leptons at the LHC are an indicator for this scenario, as well as a same-sign tetra-lepton signal from triplet-antitriplet oscillations.

Being unable to probe high-scale models directly at current experiments, we reviewed in \sref{ch5:sec6:highscale} the possibility to falsify high-scale leptogenesis by the observation of $\Delta L =2$ washout processes. We demonstrated that the observation of $pp \to \ell^\pm
  \ell^\pm q q$ without missing energy at the LHC would imply the falsification of certain high-scale models. Similarly, the observation of $0\nu\beta\beta$ by a non-standard operator would point us towards a low-scale leptogenesis model.

Generally, we could show that common experimental and theoretical effort are in place. It is important to continue this joint effort in order to pin down the underlying mechanism responsible for the generation of the observed cosmic baryon asymmetry.

\section*{Acknowledgments}
This work has been initiated at the Munich Institute for Astro- and Particle Physics (MIAPP) of the DFG cluster of excellence ``Origin and Structure of the Universe''.

\bibliographystyle{ws-rv-van-mod2}

\bibliography{chapter5.bib}

\end{document}